# Relative entropies and their use in quantum information theory

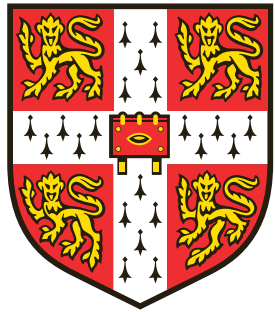

**Felix Leditzky**

Girton College, University of Cambridge

November 2016

This dissertation is submitted for the degree of Doctor of Philosophy.

*Für meine Familie, and for Leah*

# Contents

# Summary

This dissertation investigates relative entropies, also called generalized divergences, and how they can be used to characterize information-theoretic tasks in quantum information theory. The main goal is to further refine characterizations of the optimal rates for quantum source coding, state redistribution, and measurement compression with quantum side information via second order asymptotic expansions and strong converse theorems. The dissertation consists of a mathematical and an information-theoretic part.

In the mathematical part, we discuss two recently introduced relative entropies, the $\alpha$-sandwiched Rényi divergence and the information spectrum relative entropy. For the sandwiched Rényi divergence, we first investigate the limit $\alpha \to 0$ to determine whether this recovers the well-known 0-Rényi relative divergence. Furthermore, we extend the Rényi entropic calculus by proving various bounds for entropic quantities derived from the sandwiched Rényi divergence, including a useful bound in terms of the fidelity between two quantum states. We then focus on a particular property called the data processing inequality, which states that the sandwiched Rényi divergence cannot increase under quantum operations. We derive a necessary and sufficient condition for equality in the data processing inequality, and discuss applications of this condition to entropic bounds. For the information spectrum relative entropy, we demonstrate how to obtain its known second order asymptotic expansion when evaluated on pairs of independently and identically distributed quantum states.

In the information-theoretic part of the dissertation, we first focus on the task of visible quantum source coding. We introduce general mixed quantum sources, which provide an important toy model for an information-theoretic task employing a resource with memory, and derive the second order asymptotics of visible quantum source coding using such a mixed source. This result is obtained from the second order asymptotic expansion of the information spectrum relative entropy. As a special case, we also show how to obtain the second order asymptotics of visible quantum source coding using a single memoryless quantum source. The last part of the dissertation is dedicated to strong converse theorems, which establish an optimal rate of an information-theoretic task as a sharp threshold beyond which all codes fail with certainty. We focus on the tasks of state redistribution and measurement compression with quantum side information, proving strong converse theorems in each case. The key ingredients in proving these theorems are the aforementioned fidelity bounds on entropic quantities derived from the sandwiched Rényi divergence.

# Declaration

This dissertation is the result of my own work and includes nothing which is the outcome of work done in collaboration except as declared in the Preface and specified in the text.

It is not substantially the same as any that I have submitted, or, is being concurrently submitted for a degree or diploma or other qualification at the University of Cambridge or any other University or similar institution except as declared in the Preface and specified in the text. I further state that no substantial part of my dissertation has already been submitted, or, is being concurrently submitted for any such degree, diploma or other qualification at the University of Cambridge or any other University of similar institution except as declared in the Preface and specified in the text.

# Acknowledgements

First and foremost, I would like to express my deepest gratitude to my supervisor, Nilanjana Datta, for guiding me through my PhD. When I arrived in Cambridge three years ago, I had little knowledge of quantum information theory, and therefore I had to learn most things — in particular the information-theoretic part of it — from scratch. Nilanjana made me fit for research in no time, and it is through her valuable insights that I learned to appreciate our wonderful field. I am especially grateful to her for always keeping her door open for me, and more generally for teaching me what dedication to one's work truly means. I would also like to thank the Department of Pure Mathematics and Mathematical Statistics, in particular my academic home, the Statistical Laboratory, as well as the Centre for Quantum Information and Foundations in the Department of Applied Mathematics and Theoretical Physics, and Girton College, for providing funding as well as travel grants that enabled me to attend various conferences and workshops.

The research that led to this thesis is the product of fruitful collaborations with Salman Beigi, Nilanjana Datta, Cambyse Rouzé, and Mark M. Wilde. It was a pleasure to work and co-author papers with them, and I hope that there will be many more to follow. I also greatly enjoyed discussions with the following people, who, in one way or another, all contributed to my understanding of quantum information theory: Christian Arenz, Koenraad Audenaert, Johannes Bausch, Mario Berta, Fernando Brandão, Steve Brierley, Francesco Buscemi, Toby Cubitt, Ben Derrett, Frédéric Dupuis, Paul Erker, Christoph Harrach, Aram Harrow, Masahito Hayashi, Christoph Hirche, Min-Hsiu Hsieh, Felix Huber, Richard Jozsa, Will Matthews, Graeme Mitchison, Milán Mosonyi, Māris Ozols, Christopher Perry, Renato Renner, Graeme Smith, Sergii Strelchuk, Michał Studziński, David Sutter, Vincent Tan, Marco Tomamichel, Dave Touchette, and Andreas Winter. Special thanks go to Māris Ozols and Andreas Winter for agreeing to examine this thesis. Furthermore, I feel obliged to thank my master thesis advisor Harald Grosse at the University of Vienna, who planted in me the idea of pursuing a PhD in quantum information theory.

Apart from the quantum information theory community, I could share my time in Cambridge with many great people. My heartfelt thanks go to my dear friends and fellow Girtonians Sam Abujudeh, Francesco Brizzi, Constantine Capsaskis, Luca Matrá, Jose Reis, and Stefan Ritter, for sharing homes in Wolfson Court and 7A Parker St, and making forays into the fascinating world of cooking with vegetables; to the Girtonian Wolves, including all of the previously mentioned along with my friends Hamza El-Gomati, Shane Heffernan, and Abdul Zeybek, for playing, watching, discussing, and sharing our love for football (and teaching the local 5-a-side teams a lesson or two); to the B-block crew Rachel Knighton, Leah Oppenheimer, Manon Turban, and Sarah Whittam, for being the best neighbors one could wish for; to Jason Dudek, Shane Heffernan, Sarah Whittam, and many others, for those music sessions; and in general to the wonderful Girton MCR community, which made it so easy to feel at home in Cambridge.

Academia is a rather peculiar world, and sometimes it seems necessary to touch base with the real world: thanks to my old friends András Bárány, Johannes Felling, Victor Navas, and Bernhard Romstorfer for doing exactly that. Having shared most important milestones of our lives, including an almost identical academic record, I am especially indebted to András, not least for giving me the courage to apply in Cambridge, and helping me a great deal in getting through this experience. Moreover, he never grew tired of counseling me in formatting and LaTeX issues, and gave valuable comments on this thesis, which has also greatly benefited from proofreading by Rachel Knighton and Leah Oppenheimer. At this point, I cannot help but extend a warm thank you to the countless smart people at `tex.stackexchange.com` for having solutions for almost anything LaTeX-related.

I am very lucky to have a family that has always encouraged and enabled me to go my own way, however far that might be from Vienna. For this unconditional moral and financial support, I am forever grateful to my parents Brigitte and Walter, and my brothers Georg and Lorenz.

Last, but certainly not least, a huge thank you to Leah Oppenheimer, with whom I was fortunate to share the better part of my time in Cambridge. Thanks for thinking that we write papers in Greek, and worrying about the lemurs. On a more serious note, I can only say that her love and support, her praise in good times, and her consolation in not-so-good times were absolutely vital for me, and a never-ending source of encouragement to push through these three years.

# Notation and acronyms

| Acronyms | |
|---|---|
| CDF | Cumulative distribution function |
| CPTP | Completely positive and trace-preserving |
| DPI | Data processing inequality |
| MES | Maximally entangled state |
| POVM | Positive operator-valued measurement |
| QSI | Quantum side information |
| RRE | Relative Rényi entropy |
| SRD | Sandwiched Rényi divergence |
| c-q | Classical-quantum |
| i.i.d. | Independent and identically distributed |

| Linear algebra | |
|---|---|
| $\mathcal{H}, \mathcal{K}$ | Generic finite-dimensional Hilbert spaces |
| $\mathcal{B}(\mathcal{H})$ | Algebra of linear operators on $\mathcal{H}$ |
| $\mathrm{Herm}(\mathcal{H})$ | Set of Hermitian operators on $\mathcal{H}$ |
| $\mathcal{P}(\mathcal{H})$ | Set of positive semidefinite operators on $\mathcal{H}$ |
| $\mathcal{D}(\mathcal{H})$ | Set of quantum states (density operators) on $\mathcal{H}$ |
| $z^*$ | Complex conjugate of a complex number $z \in \mathbb{C}$ |
| $X^\dagger$ | Adjoint (or Hermitian conjugate) of an operator $X \in \mathcal{B}(\mathcal{H})$ |
| $\mathbb{1}_\mathcal{H}$ | Identity operator on $\mathcal{H}$ |
| $\Lambda^\dagger \colon \mathcal{B}(\mathcal{K}) \to \mathcal{B}(\mathcal{H})$ | Adjoint map of a linear map $\Lambda \colon \mathcal{B}(\mathcal{H}) \to \mathcal{B}(\mathcal{K})$ |
| $\mathrm{id}_\mathcal{H}$ | Identity map on $\mathcal{B}(\mathcal{H})$ |
| $\lambda_i(H)$ | $i$-th largest eigenvalue of $H \in \mathrm{Herm}(\mathcal{H})$ |
| $\lambda(H)$ | Vector of eigenvalues of $H \in \mathrm{Herm}(\mathcal{H})$ in decreasing order |
| $\nu(H)$ | Number of distinct eigenvalues of $H \in \mathrm{Herm}(\mathcal{H})$ |
| $\mathrm{rk}\,X$ | Rank of $X \in \mathcal{B}(\mathcal{H})$ |
| $\mathrm{supp}\,X$ | Support of $X \in \mathcal{B}(\mathcal{H})$ |
| $\Pi_X$ | Projection onto $\mathrm{supp}\,X$ |
| $X_+$ | Positive part of $X \in \mathrm{Herm}(\mathcal{H})$ |
| $\{X \geq Y\}$ | $\Pi_{(X-Y)_+}$ |
| $[X, Y]$ | Commutator of operators $X$ and $Y$ |

## Quantum information theory

| | |
|---|---|
| $\mathcal{H}_A$ | Hilbert space associated to quantum system $A$ |
| $|\psi\rangle_A$ | Pure state on $A$ |
| $\psi_A \equiv |\psi\rangle\langle\psi|_A$ | Density operator associated to $|\psi\rangle_A$ |
| $\rho_A$ | Mixed quantum state (density operator) on $A$ |
| $|\psi^\rho\rangle$ | Purification of a mixed state $\rho$ |
| $\Phi^k_{AA'}$ | Maximally entangled state of Schmidt rank $k$ between $A$ and $A'$ |
| $\pi_A$ | Completely mixed state on $A$ |
| $\Lambda\colon A \to B$, $\Lambda^{A\to B}$ | Quantum operation from $A$ to $B$, i.e., $\Lambda\colon \mathcal{B}(\mathcal{H}_A) \to \mathcal{B}(\mathcal{H}_B)$ |
| $U_\Lambda$ | Stinespring isometry of quantum operation $\Lambda$ |
| $\mathrm{tr}_A$ | Partial trace over $A$ |

## Norms, distances, and entropies

| | |
|---|---|
| $\|\cdot\|_1$ | Trace norm |
| $\|\cdot\|_p$ | Schatten $p$-norm |
| $F(\cdot, \cdot)$ | Fidelity between quantum states |
| $T(\cdot, \cdot)$ | Trace distance between quantum states |
| $D(\cdot\|\cdot)$ | Quantum relative entropy |
| $V(\cdot\|\cdot)$ | Quantum information variance |
| $S(\cdot)$ | von Neumann entropy |
| $\sigma(\cdot)$ | Information variance of a quantum source |
| $S(A)_\rho$ | von Neumann entropy of $\rho_A$ |
| $S(A|B)_\rho$ | Conditional entropy of a bipartite state $\rho_{AB}$ |
| $I(A;B)_\rho$ | Mutual information of a bipartite state $\rho_{AB}$ |
| $I(A;B|C)_\rho$ | Conditional mutual information of a tripartite state $\rho_{ABC}$ |

| | |
|---|---|
| $D_s^\varepsilon(\cdot\|\cdot)$ | Information spectrum relative entropy |
| $D_\alpha(\cdot\|\cdot)$ | $\alpha$-relative Rényi entropy |
| $\widetilde{D}_\alpha(\cdot\|\cdot)$ | $\alpha$-sandwiched Rényi divergence |
| $S_\alpha(A)_\rho$ | $\alpha$-Rényi entropy of $\rho_A$ |
| $\widetilde{S}_\alpha(A\|B)_\rho$ | $\alpha$-Rényi conditional entropy of $\rho_{AB}$ |
| $\tilde{I}_\alpha(A;B)_\rho$ | $\alpha$-Rényi mutual information of $\rho_{AB}$ |

## Probability theory

| | |
|---|---|
| $P, Q, \ldots$ | Probability distributions |
| $P^n$ | Product distribution (i.i.d.) of the probability distribution $P$ |
| $\Pr\{X \geq a\}$ | Probability of the random variable $X$ taking values $\geq a$ |
| $F_X$ | Cumulative distribution function of the random variable $X$ |
| $\mathbb{E}(X)$ | Expectation of the random variable $X$ |
| $\Phi(z)$ | CDF of a standard normal random variable |
| $D(\cdot\|\cdot)$ | Kullback-Leibler divergence |
| $V(\cdot\|\cdot)$ | Information variance |
| $P_{\rho,\sigma}, Q_{\rho,\sigma}$ | Nussbaum-Szkoła distributions of the quantum states $\rho$ and $\sigma$ |

## Miscellaneous

| | |
|---|---|
| $\mathbb{N}$ | Natural numbers |
| $\mathbb{R}$ | Real numbers |
| $\mathbb{C}$ | Complex numbers |
| $[m]$ | Set of natural numbers from 1 to $m \in \mathbb{N}$ |
| $a^n = (a_1, \ldots, a_n)$ | $n$-tuple with entries $a_j \in \mathcal{A}$ for $j = 1, \ldots, n$ |
| $\mathcal{A}^n$ | Set of $n$-tuples $a^n = (a_1, \ldots, a_n)$ where $a_j \in \mathcal{A}$ for $j = 1, \ldots, n$ |

# List of Figures and Tables

# 1. Introduction

## 1.1. Shannon entropy and von Neumann entropy

The storage and transmission of information through communication have proved essential for the development of modern civilization and technology. Strangely enough, a sound mathematical theory of communication was not developed until 1948, when Claude Shannon published his landmark paper [Sha48], in which he initiated what is now called information theory. At the heart of Shannon's formulation of a theory of communication is the idealization of any communication system into five basic parts: information source, transmitter, noise source, receiver, and destination. For the purpose of motivating the topics of this thesis, we focus on the first part of such a communication system.

An information source can be thought of as any physical device emitting *sequences of signals*, where the individual signals come from a fixed alphabet $\mathcal{A}$ which we assume to be finite for simplicity. Examples of such alphabets are the Latin alphabet $\{a, \ldots, z\}$, the Morse code alphabet $\{., -\}$, or the binary alphabet $\{0, 1\}$. However, the actual labels of these letters are irrelevant, and all we care about are the probabilities with which the individual letters are emitted. In the following, we consider an arbitrary enumeration of the elements of the alphabet $\mathcal{A}$, say, from 1 to $d = |\mathcal{A}|$, where $|\mathcal{A}|$ denotes the cardinality of the alphabet $\mathcal{A}$, and attach probabilities $p_i$ to each of the signals, where $i = 1, \ldots, d$. That is, the source emits the signal $i$ with probability $p_i$.

One of Shannon's most important insights in [Sha48] was to find a measure of how much information such an information source produces. To this end, he defined the *surprisal* of a signal $i$ as the logarithm of the reciprocal of the associated probability $p_i$. If a signal $i$ has probability $p_i = 1$ (i.e., we are certain that the source will always produce this same signal $i$), then the surprisal of $i$ is 0. The smaller the probability of a signal, the higher its surprisal. Shannon then defined his information measure as the average surprisal of the information source, eventually settling on the name *entropy* for this measure (cf. the quote at the beginning of this section):

$$H(P) := \sum_{i=1}^{d} p_i \log \frac{1}{p_i}, \tag{1.1}$$

where $P = (p_1, \ldots, p_d)$ denotes the underlying probability distribution of the source, and $0 \log 0 \equiv 0$ by convention.

The quantity $H(P)$ defined in (1.1) is now called *Shannon entropy*. The following two properties follow immediately from its definition: $H(P)$ is always non-negative, and 0 if and only if the probability distribution $P$ is concentrated on a single event, i.e., $P = (0, \ldots, 0, 1, 0, \ldots, 0)$. Furthermore, the maximal value of $H(P)$ is $\log d$, and this is the case if and only if all signals are equally likely, i.e., $P = (1/d, \ldots, 1/d)$ is the uniform distribution. These two properties already indicate that Shannon entropy is a reasonable measure of the information produced by a source: if a source always emits the same signal, we do not learn any new information upon receiving this signal. On the other hand, if all signals are equally likely, we gain the maximal amount of information upon learning which signal was actually emitted. In principle, there are many candidates for reasonable information measures. However, the Shannon entropy has a precise operational meaning in the information-processing task of *source coding*, which turns the Shannon entropy into a fundamental information measure. In the following, we explain the task of source coding and the role of Shannon entropy in it.

Let us assume that we receive a sequence of signals from an information source $(\mathcal{A}, P)$ with alphabet $\mathcal{A}$ and corresponding probability distribution $P$, where the successive signals emitted from the source are *independent and identically distributed* (i.i.d.). That is, each successive signal is emitted according to $P$, and is independent of previous signals. Furthermore, we assume that the source emits sequences of signals of arbitrary length. These two assumptions characterize the *asymptotic, memoryless setting*, which is a central concept in information theory. Informally, the goal in source coding is to compress sequences of signals of length $n$ to bit strings of length $N(n, \varepsilon)$ in such a way that the original sequence can be restored up to a chosen error $\varepsilon \in (0, 1)$. That is, each sequence $(a_1, \ldots, a_n)$ with $a_i \in \mathcal{A}$ is mapped to a string $(x_1, \ldots, x_N)$ with $N \equiv N(n, \varepsilon)$ and $x_i \in \{0, 1\}$. Shannon's *noiseless coding theorem* [Sha48] states that, for any error $\varepsilon \in (0, 1)$, the average number of bits needed for the compressed sequences approaches the Shannon entropy of the source in the limit $n \to \infty$:

$$\lim_{n \to \infty} \frac{N(n, \varepsilon)}{n} = H(P) \quad \text{for all } \varepsilon \in (0, 1). \tag{1.2}$$

The noiseless source coding theorem (1.2) makes precise how $H(P)$ measures the information produced by a source: any compression protocol attempting to compress a sequence of length $n$ to a string with fewer bits than $nH(P)$ will eventually incur a non-zero error (the *converse part*). On the other hand, there are reliable compression schemes with rate $H(P) + \delta$ for any $\delta > 0$ (the *achievability part*). Both parts together constitute the coding theorem, which proves that the Shannon entropy is the optimal rate of source compression.

The preceding discussion of source compression was based on the fact that the information source is *classical*, which essentially means that the receiver can perfectly distinguish the signals emitted by the source. If we instead consider an information source emitting *quantum* signals, we lose this ability to perfectly distinguish the signals, and the corresponding information-processing task of quantum source coding becomes more involved.

Postponing a detailed discussion of the intricacies of quantum source coding to Chapter 5, we focus here on its optimal rate, which is given by the quantum counterpart of the Shannon entropy: the von Neumann entropy, defined for a quantum state $\rho$ as $S(\rho) = -\operatorname{tr}(\rho \log \rho)$. This quantity has its origin in quantum statistical mechanics, and was named in honor of John von Neumann for his groundbreaking work in this field. Similar to the Shannon entropy, the von Neumann entropy is always non-negative, and zero if and only if the quantum state is *pure*, meaning that the quantum system is unambiguously described by this pure state. On the other hand, the von Neumann entropy is maximal for a *completely mixed state*, a quantum analogue of a uniform probability distribution.[1] The von Neumann entropy thus serves as a measure of the quantum information of a quantum system. Moreover, as mentioned above, it is equal to the optimal rate in quantum source coding, a result known as Schumacher's quantum coding theorem [Sch95], which we discuss in more detail in Section 5.1.2.

Quantum source coding is a fundamental task in quantum information theory, which, informally speaking, is a theory of communication involving physical systems described by quantum mechanics. Quantum mechanical systems exhibit a range of characteristic properties that fundamentally distinguish them from classical systems. The most striking of these is the phenomenon of *entanglement*. The physicist Erwin Schrödinger[2] described entanglement as the phenomenon of having complete knowledge of the total state of a compound system, without knowing the state of any one of its parts. Entanglement is a purely quantum concept and can act as a resource to facilitate quantum information-processing tasks such as the famous teleportation protocol [BBC+93], or state redistribution, where the goal is to redistribute part of a quantum system from one party to another. In Chapter 6, we discuss this protocol in detail.

---

[1] See Chapter 2 for a detailed explanation of the terms pure and mixed quantum state.

## 1.2. Refining optimal rates

In both classical and quantum information theory, coding theorems such as Shannon's source coding theorem (1.2) and Schumacher's quantum source coding theorem characterize the optimal rates of information-processing tasks in terms of entropic quantities. Proving coding theorems is one of the main goals of information theory. However, in certain contexts one might be interested in a more refined characterization of information-processing tasks beyond knowing the optimal rate.

One such context concerns the relaxation of the rather strict assumptions of the asymptotic, memoryless setting in which coding theorems are usually proved. In realistic scenarios we might not have access to an arbitrary number of uses of a resource such as an information source. We are then interested in approximations of the characteristic operational quantity of a task for a finite number of uses of the underlying resource. This is also called the *finite blocklength* regime and naturally leads to second order asymptotic expansions, one of the two main information-theoretic themes of this thesis.

Moreover, in Chapter 5 we also leave the memoryless scenario by analyzing the second order asymptotics of a simple toy model for an information-processing task employing a resource with memory: mixed quantum source coding. It is important to analyze such tasks, since the (idealizing) assumption of memoryless resources might not be justified in realistic scenarios due to memory effects or imperfections in the physical devices used in the task.

The second central information-theoretic aspect of this thesis is strong converses. Recall from the discussion of Shannon's noiseless coding theorem above that every coding theorem includes a converse part. Essentially, a converse tells us that coding at a rate beyond the optimal rate necessarily results in a non-vanishing error. However, no further implication about the nature or magnitude of this error can be made from the coding theorem alone. A *strong converse* refines the converse part of a coding theorem inasmuch as every such code with a rate beyond the optimal one fails *with certainty* in the asymptotic limit. This is established by proving strong converse theorems for information-processing tasks, which is the content of Chapter 6. Note that we are often able to determine the speed at which this convergence occurs by proving explicit lower bounds on the error incurred in the protocol.

---

[2]A pioneer of quantum mechanics, Schrödinger introduced the term 'entanglement' as a translation of the German term 'Verschränkung,' which he also coined.

## 1.3. Relative entropies

The main tools in deriving second order asymptotic expansions and strong converse theorems are *relative entropies*, or generalized divergences. These quantities are defined on pairs of positive operators, and usually required to be non-negative on pairs of states. Used in both classical and quantum information theory, relative entropies serve at least two purposes: first, they provide a notion of distance on the set of probability distributions or quantum states, albeit not in the strict mathematical sense (that is, they do not necessarily constitute a metric). Secondly, they act as parent quantities for entropic quantities such as the Shannon entropy or the von Neumann entropy. For example, the latter can be obtained from the *quantum relative entropy* defined in Definition 2.2.5, one of the most important examples of a relative entropy in the quantum setting.

A crucial property of relative entropies is the *data processing inequality*: if two probability distributions or quantum states are subjected to a transformation (or dynamical evolution), the relative entropy evaluated on this pair cannot increase. In the quantum setting, such a transformation is given by a *quantum operation* $\Lambda$, whose precise definition we give in Chapter 2. Denoting a relative entropy by $D(\cdot \| \cdot)$, the data processing inequality implies that for any two quantum states $\rho$ and $\sigma$, we have

$$D(\rho \| \sigma) \geq D(\Lambda(\rho) \| \Lambda(\sigma)).$$

In view of the interpretation of a relative entropy as a distance, the data processing inequality expresses the fact that two distributions or states cannot be distinguished any better after a transformation. This can be put in information-theoretically precise terms in the context of hypothesis testing, which we discuss in Chapter 4.

Optimal rates for quantum information-theoretic tasks, e.g., the von Neumann entropy in the case of Schumacher's quantum source coding theorem, are typically given in terms of entropic quantities derived from the quantum relative entropy. In order to achieve our goal of refining these optimal rates, we therefore need to consider relative entropies that generalize the quantum relative entropy in a specific way. This is the mathematical part of this thesis.

We first consider a one-parameter family of relative entropies called the *sandwiched Rényi divergence*, which reduces to the quantum relative entropy in a particular limit of said parameter. We then analyze the properties of entropic quantities derived from the sandwiched Rényi divergence. Characterizing information-processing tasks in terms of these *Rényi entropic quantities* allows us to obtain strong converse theorems for these tasks.

We also discuss another generalized divergence, the information spectrum relative entropy. Strictly speaking, this quantity should not be called a relative entropy, as it can be negative for certain pairs of quantum states, and does not satisfy the data processing inequality. However, the average of this quantity, when evaluated on pairs of i.i.d. quantum states, tends to the quantum relative entropy in the asymptotic limit. Hence, for pairs of sufficiently many copies of i.i.d. quantum states, the information spectrum relative entropy becomes non-negative and satisfies the data processing inequality, inheriting both properties from the quantum relative entropy. A more refined analysis of the convergence to the quantum relative entropy leads to finite blocklength approximations, viz. second order asymptotic expansions, which can in turn be used to obtain second order asymptotic expansions of optimal rates of information-processing tasks.

## 1.4. Main contributions and outline of this thesis

This thesis is based on the following publications and preprints:

[DL14]     Nilanjana Datta and Felix Leditzky. "A limit of the quantum Rényi divergence". *Journal of Physics A: Mathematical and Theoretical* 47.4 (2014), p. 045304. DOI: 10.1088/1751-8113/47/4/045304. arXiv: 1308.5961 [quant-ph].

[DL15]     Nilanjana Datta and Felix Leditzky. "Second-Order Asymptotics for Source Coding, Dense Coding, and Pure-State Entanglement Conversions". *IEEE Transactions on Information Theory* 61.1 (2015), pp. 582–608. DOI: 10.1109/TIT.2014.2366994. arXiv: 1403.2543 [quant-ph].

In the course of the thesis, the following original results are derived:

1.  In Chapter 3, we prove in Theorem 3.2.7 that the limit $\alpha \to 0$ of the $\alpha$-sandwiched Rényi divergence is equal to the 0-relative Rényi entropy $D_0(\rho\|\sigma)$ only if $\operatorname{supp}\rho = \operatorname{supp}\sigma$. This result was obtained in collaboration with Nilanjana Datta, and appeared in [DL14].

2.  For entropic quantities derived from the $\alpha$-sandwiched Rényi divergence, we prove various novel properties, including dimension bounds (Proposition 3.3.5), a bound relating the fidelity of two quantum states to the difference of Rényi entropic quantities (Theorem 3.3.6), and bounds concerning classical-quantum states (Proposition 3.3.7). These results were obtained in collaboration with Mark M. Wilde and Nilanjana Datta, and appeared in [LWD16].

3.  In Theorem 3.4.1 we prove a necessary and sufficient algebraic condition for equality in the data processing inequality for the $\alpha$-sandwiched Rényi divergence. We give applications of this result by determining conditions for equality in: (a) a Rényi version of the Araki-Lieb inequalities (Theorem 3.4.9); (b) a (newly derived) lower bound on the Rényi entanglement of formation (Theorem 3.4.13); (c) an upper bound on the entanglement fidelity in terms of the usual fidelity (Proposition 3.4.15). These results were obtained in collaboration with Cambyse Rouzé and Nilanjana Datta, and appeared in [LRD16].

4.  In Chapter 5, Theorem 5.2.8 we derive the second order asymptotics of visible quantum source coding using a general mixed source. To prove the achievability part, we introduce universal quantum source codes achieving second order rates for memoryless quantum sources (Proposition 5.2.3). Furthermore, in Corollary 5.2.11 we recover the second order asymptotics of visible quantum source coding using a memoryless source, which were derived in [DL15]. These results were obtained in collaboration with Nilanjana Datta, and appeared in [LD16].

5. In Chapter 6, Theorem 6.1.2 we derive a strong converse theorem for quantum state redistribution. We also extend this theorem to a feedback version of state redistribution (Theorem 6.1.4). Finally, in Theorem 6.2.2 we prove a strong converse theorem for the classical communication cost in measurement compression with quantum side information. These results were obtained in collaboration with Mark M. Wilde and Nilanjana Datta, and appeared in [LWD16].

The remainder of this thesis is structured as follows: Chapter 2 sets the notation used throughout the thesis, and aims to summarize the mathematical framework of quantum information theory in finite-dimensional Hilbert spaces. This summary is far from complete, and only intended to enable both experts and readers unfamiliar with quantum information theory to follow the arguments of the thesis. In addition, we collect a few useful results from matrix analysis.

The main part of the thesis is divided into a mathematical part and an information-theoretic part. The mathematical part begins in Chapter 3, where we introduce the $\alpha$-sandwiched Rényi divergence, and discuss some of its well-known properties. In the following sections, we first investigate special values and limits of the parameter $\alpha$, deriving Result 1 in the course. We then discuss entropic quantities derived from the sandwiched Rényi divergence, and prove Result 2. Finally, we analyze the proof of the data processing inequality by Frank and Lieb [FL13] in detail to obtain Result 3.

In Chapter 4, we discuss the information spectrum relative entropy introduced by Tomamichel and Hayashi [TH13], and review the derivation of its second order asymptotic expansion given in the same paper. Furthermore, in view of the discussion in Chapter 5 we give a simplified, direct proof of the second order expansion in the case when the second operator in the relative entropy is equal to the identity. We stress at this point that Chapter 4 does not contain any original material. However, in Appendix A we briefly discuss variants of the information spectrum relative entropies that were introduced in [DL15]. In contrast to the information spectrum relative entropy defined in [TH13], these quantities have the particular advantage of satisfying the data processing inequality.

Chapter 5 marks the beginning of the information-theoretic part of this thesis. We first discuss quantum sources and the tasks of visible and blind quantum source coding. We then derive Result 4, the second order asymptotics of visible quantum source coding using a mixed source. Furthermore, we compare the second order asymptotics of visible quantum source coding using a memoryless source to the second order asymptotic bounds on the minimal compression length in the blind encoding setting, which we derived in [DL15].

Finally, in Chapter 6 we focus on strong converse theorems and prove Result 5. We also show how state redistribution reduces to well-known information-processing tasks when considering special cases of the protocol. As a corollary, we recover known strong converse theorems for these tasks.

We conclude in Chapter 7 with a summary of the results obtained in this thesis, as well as a collection of open problems pointing to possible directions of future research.

# 2. Preliminaries

In this chapter we set the notation and recall mathematical results that are used throughout the thesis. We start with a review of basic concepts in linear algebra, which are then used to give a short introduction to the mathematical framework of quantum information theory in finite-dimensional Hilbert spaces. Finally, we collect some useful results from matrix analysis.

## 2.1. Notation and basics of linear algebra

We denote the set of natural numbers, real numbers, and complex numbers by $\mathbb{N}$, $\mathbb{R}$, and $\mathbb{C}$, respectively. For an arbitrary set $\mathcal{A}$ and $n \in \mathbb{N}$, we use the notation $\mathcal{A}^n := \{(a_1, \ldots, a_n) \colon a_i \in \mathcal{A} \text{ for } i = 1, \ldots, n\}$ for the set of $n$-tuples with entries in $\mathcal{A}$. For $d \in \mathbb{N}$ we define $[d] := \{1, \ldots, d\}$. The symmetric group of order $n \in \mathbb{N}$ is denoted by $\mathcal{S}_n$, and the unitary group of degree $n$ is denoted by $\mathcal{U}_n$. In this thesis all exponentials and logarithms are taken to base 2.

A Hilbert space $\mathcal{H}$ is a complex vector space equipped with an inner product $(\cdot, \cdot)_{\mathcal{H}} \colon \mathcal{H} \times \mathcal{H} \to \mathbb{C}$, such that it is complete with respect to the metric $d(x, y) := \sqrt{(x - y, x - y)}$ induced by $(\cdot, \cdot)_{\mathcal{H}}$. Note that every finite-dimensional normed vector space is automatically complete with respect to the induced metric. Unless specified otherwise, throughout this thesis $\mathcal{H}$ denotes a finite-dimensional Hilbert space.

We denote the algebra of linear operators acting on $\mathcal{H}$ by $\mathcal{B}(\mathcal{H})$. For $A \in \mathcal{B}(\mathcal{H})$, the adjoint operator $A^\dagger$ is the unique operator satisfying $(Ax, y)_{\mathcal{H}} = (x, A^\dagger y)_{\mathcal{H}}$ for all $x, y \in \mathcal{H}$. An operator $A \in \mathcal{B}(\mathcal{H})$ is called Hermitian if $A = A^\dagger$, and positive semidefinite (or positive in short) if $(x, Ax) \geq 0$ for all $x \in \mathcal{H}$. We denote the sets of Hermitian and positive semidefinite operators by $\mathrm{Herm}(\mathcal{H})$ and $\mathcal{P}(\mathcal{H})$, respectively, and note that $\mathcal{P}(\mathcal{H}) \subseteq \mathrm{Herm}(\mathcal{H})$. An operator $A$ is called normal if $[A, A^\dagger] = 0$, where $[X, Y] := XY - YX$ denotes the commutator. By the spectral theorem, every normal operator $A$ can be written as

$$A = \sum_i \lambda_i P_i,$$

where for every $i$ the operator $P_i$ projects onto the eigenspace corresponding to the eigenvalue $\lambda_i \in \mathbb{C}$, that is, for an eigenvector $v_i \in P_i\mathcal{H}$ we have $Av_i = \lambda_i v_i$. The eigenvalues $\{\lambda_i\}_i$ of a Hermitian operator are real, and if furthermore $A \in \mathcal{P}(\mathcal{H})$, then $\lambda_i \geq 0$ for all $i$. For $A \in \mathrm{Herm}(\mathcal{H})$ we define $\mathrm{supp}\,A$ as the span of the eigenvectors of $A$ corresponding to non-zero eigenvalues, and we write $\Pi_A$ for the projection onto $\mathrm{supp}\,A$. For $A, B \in \mathrm{Herm}(\mathcal{H})$ we write $A \not\perp B$ if $\mathrm{supp}\,A \cap \mathrm{supp}\,B$ contains at least one non-zero vector. We denote the identity operator on $\mathcal{H}$ by $\mathbb{1}_{\mathcal{H}}$, i.e., $\mathbb{1}_{\mathcal{H}}x = x$ for all $x \in \mathcal{H}$. For a tensor product space $\mathcal{H}^{\otimes n}$, we use the notation $\mathbb{1}_n \equiv \mathbb{1}_{\mathcal{H}^{\otimes n}}$.

The operator algebra $\mathcal{B}(\mathcal{H})$ is itself a Hilbert space when equipped with the Hilbert-Schmidt inner product, defined by $(A, B)_{\mathcal{B}(\mathcal{H})} := \mathrm{tr}(A^\dagger B)$ for $A, B \in \mathcal{B}(\mathcal{H})$. Let $\mathcal{K}$ be another Hilbert space and consider a linear map $\Lambda \colon \mathcal{B}(\mathcal{H}) \to \mathcal{B}(\mathcal{K})$. The adjoint map $\Lambda^\dagger$ is the adjoint in the above sense of the superoperator $\Lambda$ acting on the Hilbert space $\mathcal{B}(\mathcal{H})$. Explicitly, $\Lambda^\dagger$ is defined as the unique linear map satisfying $(\Lambda(X), Y)_{\mathcal{B}(\mathcal{K})} = (X, \Lambda^\dagger(Y))_{\mathcal{B}(\mathcal{H})}$ for all $X \in \mathcal{B}(\mathcal{H})$ and $Y \in \mathcal{B}(\mathcal{K})$. We denote the identity map on $\mathcal{B}(\mathcal{H})$ by $\mathrm{id}_{\mathcal{H}}$, i.e., $\mathrm{id}_{\mathcal{H}}(X) = X$ for all $X \in \mathcal{B}(\mathcal{H})$.

This thesis makes heavy use of Dirac's bra-ket notation: a *ket* $|\psi\rangle$ denotes an element (or vector) in $\mathcal{H}$, whereas a *bra* $\langle\phi|$ denotes an element (or linear functional) in $\mathcal{H}^*$, the dual space of $\mathcal{H}$. For vectors $|\psi\rangle, |\phi\rangle \in \mathcal{H}$, the scalar product $(|\psi\rangle, |\phi\rangle)_{\mathcal{H}}$ is conveniently written as $\langle\psi|\phi\rangle$, and the outer product of $|\psi\rangle$ and $|\phi\rangle$ is written as $|\psi\rangle\langle\phi|$. For $A \in \mathrm{Herm}(\mathcal{H})$, we can unambiguously write $\langle\psi|A|\phi\rangle$, where $A$ either acts to the left on $\langle\psi|$ or to the right on $|\phi\rangle$.

## 2.2. Mathematical framework of quantum information theory

The following section comprises a short introduction to the mathematical framework of quantum information theory in finite dimensions. On the one hand, it provides a 'vocabulary' of common terminology for readers with a quantum information theory background. On the other hand, it should equip those unfamiliar with the subject with enough ammunition to be able to follow the main points of this thesis.[1] Note that pedagogical clarity is sacrificed for brevity, and the following treatise only hints at the physical motivation behind the mathematical framework of quantum information theory. For a much more thorough and pedagogical introduction to quantum information theory, we refer to the excellent textbooks by Nielsen and Chuang

---

[1]The reader is however assumed to have a basic understanding of advanced mathematics.

[NC00], Watrous [Wat16], and Wilde [Wil16], and the lecture notes by Wolf [Wol12].

## 2.2.1. Systems and states

A quantum system $Q$ is associated with a Hilbert space $\mathcal{H}_Q$. In this thesis we are only concerned with finite-dimensional Hilbert spaces, that is, $\mathcal{H}_Q$ is isomorphic to $\mathbb{C}^{|Q|}$, where $|Q| = \dim \mathcal{H}_Q$ denotes the dimension of the underlying Hilbert space $\mathcal{H}_Q$. We call a system trivial if $|Q| = 1$ such that $\mathcal{H}_Q \cong \mathbb{C}$. Examples of physical systems modeled by finite-dimensional Hilbert spaces include:

- Polarization of a photon: Here, $\mathcal{H}_Q = \mathbb{C}^2$, and a basis is for example given by $\{|L\rangle, |R\rangle\}$, where $L$ and $R$ correspond to left- and right-handed circular polarization, respectively.

- Spin of an electron: Here, we again have $\mathcal{H}_Q = \mathbb{C}^2$, and a basis is given by $\{|0\rangle, |1\rangle\}$, where $|0\rangle$ and $|1\rangle$ are the eigenstates of the Pauli $Z$-operator $\sigma_z = \left(\begin{smallmatrix} 1 & 0 \\ 0 & -1 \end{smallmatrix}\right)$ corresponding to measuring the spin of the electron in the $z$-axis.

- Excitations of an atom: Here, we consider $\mathcal{H}_Q = \mathbb{C}^d$ for some $d \in \mathbb{N}$. The elements of the basis $\{|0\rangle, |1\rangle, \dots, |d-1\rangle\}$ correspond to the excitation levels of the atom, with $|0\rangle$ denoting the ground state.

In the following, $\mathcal{H}$ denotes a generic Hilbert space of dimension $d$ corresponding to a physical system with $d$ degrees of freedom. A *pure state* of this quantum system is described by a normalized vector $|\psi\rangle \in \mathcal{H}$ with $\langle\psi|\psi\rangle = 1$. More generally, a *mixed state* $\rho$ is a positive semidefinite, normalized operator acting on $\mathcal{H}$, that is, $\rho \in \mathcal{P}(\mathcal{H})$ with $\operatorname{tr} \rho = 1$. We denote the convex set of quantum states on $\mathcal{H}$ by $\mathcal{D}(\mathcal{H}) := \{\rho \in \mathcal{P}(\mathcal{H}) \colon \operatorname{tr} \rho = 1\}$, and identify a pure state $|\psi\rangle \in \mathcal{H}$ with the projection $|\psi\rangle\langle\psi| \in \mathcal{D}(\mathcal{H})$ onto the linear space spanned by $|\psi\rangle \in \mathcal{H}$. We also use the notation $\psi \equiv |\psi\rangle\langle\psi|$. Conversely, every mixed state $\rho$ of rank 1 corresponds to a pure state $|\psi\rangle \in \mathcal{H}$ that spans the one-dimensional eigenspace of $\rho$, and we have $\rho = |\psi\rangle\langle\psi|$. Physically, a mixed state $\rho$ reflects an uncertainty about the preparation of the system: Suppose a system is prepared in such a way that it is found in the state $|\psi_i\rangle$ with probability $p_i$. The state of the system is then given by $\rho = \sum_i p_i |\psi_i\rangle\langle\psi_i|$, which is not pure if at least two of the vectors $|\psi_i\rangle$ are linearly independent.

Given physical systems $A_1, \dots, A_k$ modeled by Hilbert spaces $\mathcal{H}_{A_1}, \dots, \mathcal{H}_{A_k}$, respectively, the composite system $A_1 \dots A_k$ is described by the tensor product $\mathcal{H}_{A_1 \dots A_k} \equiv \mathcal{H}_{A_1} \otimes \dots \otimes \mathcal{H}_{A_k}$.

Throughout this thesis, we imply such a tensor product structure when denoting Hilbert spaces as $\mathcal{H}_{AB...}$, with each upper-case subscript corresponding to one subsystem. For a bipartite system $AB$ in a state $\rho_{AB}$, the state of the system $A$ is given by the *marginal* $\rho_A := \mathrm{tr}_B \, \rho_{AB}$. Here, $\mathrm{tr}_B$ denotes the *partial trace* over the $B$ system, defined as the adjoint of the linear map $X_A \mapsto X_A \otimes \mathbb{1}_B$ for $X_A \in \mathcal{B}(\mathcal{H}_A)$ with respect to the Hilbert-Schmidt inner product on $\mathcal{B}(\mathcal{H}_{AB})$. Equivalently, for an arbitrary operator $X_{AB} \in \mathcal{B}(\mathcal{H}_{AB})$ the partial trace $X_A = \mathrm{tr}_B \, X_{AB}$ is the unique operator satisfying $\mathrm{tr}(X_{AB}(Y_A \otimes \mathbb{1}_B)) = \mathrm{tr}(X_A Y_A)$ for all $Y_A \in \mathcal{B}(\mathcal{H}_A)$. In terms of a basis $\{|i\rangle_B\}_{i=1}^{|B|}$ of $\mathcal{H}_B$, the partial trace can be expressed as

$$\mathrm{tr}_B \, X_{AB} = \sum_{i=1}^{|B|} (\mathbb{1}_A \otimes \langle i|_B) X_{AB} (\mathbb{1}_A \otimes |i\rangle_B).$$

Physically, the partial trace corresponds to discarding one of the constituent systems. We follow the convention that a dropped upper-case subscript (corresponding to a physical system) always implies a partial trace over that system: for a bipartite state $\rho_{AB}$ the states $\rho_A$ and $\rho_B$ are understood as the marginals on the systems $A$ and $B$ given by $\rho_A = \mathrm{tr}_B \, \rho_{AB}$ and $\rho_B = \mathrm{tr}_A \, \rho_{AB}$, respectively.

Every mixed state $\rho \in \mathcal{D}(\mathcal{H})$ can be regarded as the marginal of a pure state on a larger Hilbert space. To see this, consider the spectral decomposition $\rho = \sum_i \lambda_i |i\rangle\langle i|_{\mathcal{H}}$, and form the pure state

$$|\psi^\rho\rangle := \sum_i \sqrt{\lambda_i} |i\rangle_{\mathcal{H}} \otimes |i\rangle_{\mathcal{H}'} \in \mathcal{H} \otimes \mathcal{H}',$$

where $\mathcal{H}'$ is an auxiliary Hilbert space of dimension $\dim \mathcal{H}' = \mathrm{rk}\, \rho$, spanned by the (orthonormal) eigenvectors $\{|i\rangle\}_i$ of $\rho$. It follows that $\rho = \mathrm{tr}_{\mathcal{H}'} \, \psi^\rho$, and hence $\psi^\rho$ is called a *purification* of $\rho$. Any two purifications of $\rho$ can be obtained from each other by an isometry acting on the purifying system alone. The purification technique can also be applied to quantum operations (see Section 2.2.2), and has proved immensely useful in quantum information theory.[2]

A mixed state $\rho_{AB} \in \mathcal{D}(\mathcal{H}_{AB})$ is called *separable*, if it can be written as a convex combination of product states, that is,

$$\rho_{AB} = \sum_i p_i \omega_A^i \otimes \omega_B^i \tag{2.1}$$

for a probability distribution $\{p_i\}_i$ and states $\omega_A^i \in \mathcal{D}(\mathcal{H}_A)$ and $\omega_B^i \in \mathcal{D}(\mathcal{H}_B)$. A state is called

---

[2]The phrase "Going to the Church of the Larger Hilbert Space," coined by John Smolin, captures this fact.

*entangled*, if it cannot be written as in (2.1). A pure separable state is always of tensor product form, i.e., $|\psi\rangle_{AB} \in \mathcal{H}_{AB}$ is separable if and only if there are vectors $|\phi\rangle_A \in \mathcal{H}_A$, $|\chi\rangle_B \in \mathcal{H}_B$ such that $|\psi\rangle_{AB} = |\phi\rangle_A \otimes |\chi\rangle_B$. We often omit tensor products between vectors and simply write $|\psi\rangle_{AB} = |\phi\rangle_A |\chi\rangle_B$.

A useful tool in analyzing pure bipartite quantum states is the *Schmidt decomposition*, which we state below as a theorem. Its proof is straightforward and based on the singular value decomposition of a matrix (see for example [NC00] or [Wil16]).

**Theorem 2.2.1** (Schmidt decomposition).
*Let $|\psi\rangle_{AB} \in \mathcal{H}_A \otimes \mathcal{H}_B$ be a pure bipartite quantum state. There is an integer $d \leq \min\{|A|, |B|\}$, orthonormal sets of vectors $\{|i\rangle_A\}_{i=1}^d$ and $\{|i\rangle_B\}_{i=1}^d$, and positive real numbers $\{\lambda_i\}_{i=1}^d$ satisfying $\sum_{i=1}^d \lambda_i^2 = 1$, such that*

$$|\psi\rangle_{AB} = \sum_{i=1}^d \lambda_i |i\rangle_A \otimes |i\rangle_B.$$

*The integer $d$ is called the* Schmidt rank *of $|\psi\rangle_{AB}$, and $\{\lambda_i\}_{i=1}^d$ are the* Schmidt coefficients *of $|\psi\rangle_{AB}$. For the marginals $\psi_A = \mathrm{tr}_B \psi_{AB}$ and $\psi_B = \mathrm{tr}_A \psi_{AB}$, it holds that*

$$\psi_A = \sum_{i=1}^d \lambda_i^2 |i\rangle\langle i|_A \qquad\qquad \psi_B = \sum_{i=1}^d \lambda_i^2 |i\rangle\langle i|_B,$$

*that is, the sets of positive eigenvalues of both $\psi_A$ and $\psi_B$ are equal to $\{\lambda_i^2\}_{i=1}^d$.*

By Theorem 2.2.1, a pure bipartite state is separable if and only if its Schmidt rank is equal to 1. Given two Hilbert spaces $\mathcal{H}_A$ and $\mathcal{H}_{A'}$, a *maximally entangled state* (MES) $|\Phi^k\rangle_{AA'}$ of Schmidt rank $k \leq \min\{|A|, |A'|\}$ is a vector

$$|\Phi^k\rangle_{AA'} = \frac{1}{\sqrt{k}} \sum_{i=1}^k |i\rangle_A |i\rangle_{A'},$$

where $\{|i\rangle_A\}_{i=1}^k \subseteq \mathcal{H}_A$ and $\{|i\rangle_{A'}\}_{i=1}^k \subseteq \mathcal{H}_{A'}$ are orthonormal vectors. Hence, the Schmidt coefficients $\lambda_i$ from Theorem 2.2.1 are all equal to $1/\sqrt{k}$. If $|A| = |A'| = k$, then $|\Phi\rangle_{AA'} \equiv |\Phi^k\rangle_{AA'}$ is simply called an MES, and satisfies $\mathrm{tr}_{A'} \Phi_{AA'} = \pi_A$ and $\mathrm{tr}_A \Phi_{AA'} = \pi_{A'}$, where $\pi_A = \frac{1}{|A|}\mathbb{1}_A$ denotes the *completely mixed state* on $\mathcal{H}_A$.

We call a state $\rho$ *classical*, if it is diagonal with respect to a *fixed* basis $\{|x\rangle\}_{x \in \mathcal{X}}$, where $\mathcal{X}$ is an alphabet, i.e., a finite set of classical symbols. Conversely, any classical probability distribution

$P$ on $\mathcal{X}$ can be embedded into a classical state $\rho_X = \sum_{x \in \mathcal{X}} P(x)|x\rangle\langle x|_X$, where we labeled the classical system by $X$. If two states $\rho$ and $\sigma$ commute, the fixed classical basis is typically given by the common eigenbasis of $\rho$ and $\sigma$, and the information-theoretic task involving $\rho$ and $\sigma$ then reduces to a classical problem. We also consider *classical-quantum* (c-q) states $\rho_{XB}$ of the form

$$\rho_{XB} = \sum_{x \in \mathcal{X}} P(x)|x\rangle\langle x|_X \otimes \rho_B^x,$$

where $P$ is a probability distribution on $\mathcal{X}$, and $\rho_B^x \in \mathcal{D}(\mathcal{H}_B)$ for $x \in \mathcal{X}$.

### 2.2.2. Quantum operations and measurements

Given Hilbert spaces $\mathcal{H}$ and $\mathcal{K}$, a linear map $\Lambda \colon \mathcal{B}(\mathcal{H}) \to \mathcal{B}(\mathcal{K})$ is called *positive* if $\Lambda(\mathcal{P}(\mathcal{H})) \subseteq \mathcal{P}(\mathcal{K})$, *n-positive* if the map $\Lambda \otimes \mathrm{id}_{\mathbb{C}^n} \colon \mathcal{B}(\mathcal{H}) \otimes \mathcal{B}(\mathbb{C}^n) \to \mathcal{B}(\mathcal{K}) \otimes \mathcal{B}(\mathbb{C}^n)$ is positive, and *completely positive* if $\Lambda$ is $n$-positive for all $n \in \mathbb{N}$. The map $\Lambda$ is *trace-preserving* if $\mathrm{tr}(\Lambda(X)) = \mathrm{tr}\,X$ for all $X \in \mathcal{B}(\mathcal{H})$, and *unital* if $\Lambda(\mathbb{1}_{\mathcal{H}}) = \mathbb{1}_K$. The adjoint of a trace-preserving map is unital, and vice versa.

The dynamical evolution of an open quantum system with Hilbert space $\mathcal{H}$ is given by a *quantum operation* (or *quantum channel*), which is defined to be a linear, completely positive, trace-preserving (CPTP) map $\Lambda \colon \mathcal{B}(\mathcal{H}) \to \mathcal{B}(\mathcal{K})$. For a quantum operation $\Lambda \colon \mathcal{B}(\mathcal{H}_A) \to \mathcal{B}(\mathcal{H}_B)$ evolving the quantum system $A$ into $B$, we use the shorthand notation $\Lambda \colon A \to B$ or $\Lambda^{A \to B}$. It is natural to require a quantum operation to be trace-preserving and (at least) positive, as this maps quantum states (i.e., positive semidefinite, normalized operators) to quantum states. The stronger notion of complete positivity of a map $\Lambda \colon \mathcal{B}(\mathcal{H}) \to \mathcal{B}(\mathcal{K})$ ensures that this remains true if the quantum system is regarded as a part of a larger system. That is, the Hilbert space of the total system is $\mathcal{H} \otimes \mathcal{H}'$, where $\mathcal{H}'$ is a Hilbert space of arbitrarily large dimension, and the total evolution is given by the map $\Lambda \otimes \mathrm{id}_{\mathcal{H}'}$.

An important example of a CPTP map is the partial trace defined in Section 2.2.1. An example of a positive, but not completely positive map is transposition with respect to a fixed basis, which is not even 2-positive. The following alternative characterization of CPTP maps is due to Stinespring, and gives rise to a more intuitive, physical representation of quantum operations:

**Theorem 2.2.2** (Stinespring Representation Theorem [Sti55])**.**
*Let $\Lambda \colon \mathcal{B}(\mathcal{H}) \to \mathcal{B}(\mathcal{K})$ be a quantum operation. There exists a Hilbert space $\mathcal{H}'$, a pure state*

$|\tau\rangle \in \mathcal{H}' \otimes \mathcal{K}$, and a unitary $U$ on $\mathcal{H} \otimes \mathcal{H}' \otimes \mathcal{K}$ such that for all $\rho \in \mathcal{B}(\mathcal{H})$

$$\Lambda(\rho) = \mathrm{tr}_{12}\left(U(\rho \otimes \tau)U^{\dagger}\right),$$

where $\mathrm{tr}_{12}$ denotes the partial trace over $\mathcal{H} \otimes \mathcal{H}'$. Equivalently, we can write this action as

$$\Lambda(\rho) = \mathrm{tr}_{12}\left(V\rho V^{\dagger}\right) \tag{2.2}$$

with the Stinespring isometry (or Stinespring dilation) $V\colon \mathcal{H} \mapsto \mathcal{H} \otimes \mathcal{H}' \otimes \mathcal{K}$ defined by $V := U(\mathbb{1}_{\mathcal{H}} \otimes |\tau\rangle)$, and satisfying $V^{\dagger}V = \mathbb{1}_{\mathcal{H}}$.

The Stinespring Representation Theorem 2.2.2 implies that we can view a quantum operation as a unitary evolution of a quantum state embedded in a larger Hilbert space comprised of the quantum system and an environment, where the latter can be assumed to be initially in a pure state. The resulting action of the quantum operation on the original system is obtained by tracing out the state of the environment. Furthermore, Theorem 2.2.2 tells us that the partial trace is the prototypical quantum operation; every quantum operation can be written as a composition of tensoring with a fixed state, unitary evolution, and taking a partial trace. We make use of this fact in Chapter 3, where we investigate the behavior of relative entropies under quantum operations. Finally, we note that the isometric formulation (2.2) of Theorem 2.2.2 is intimately connected to purification of mixed states as described in Section 2.2.1, and both are often used in conjunction (see for example the proof of Lemma 6.1.1 in Chapter 6).

A generalized quantum measurement is defined in terms of a *positive operator-valued measure* (POVM) $\{E_i\}_i$, where for all $i$ the $E_i$ are positive operators forming a resolution of the identity, $\sum_i E_i = \mathbb{1}$. Given a state $\rho \in \mathcal{D}(\mathcal{H})$ of a quantum system with associated Hilbert space $\mathcal{H}$, the measurement with respect to the POVM $\{E_i\}_i$ yields the outcome $i$ with probability $\mathrm{tr}(E_i\rho)$. A special case of a POVM is a *projective measurement*, where the operators $E_i$ are furthermore orthogonal projections, i.e., $E_i E_j = \delta_{ij} E_i$ for all $i$ and $j$. Most common formulations of quantum mechanics assume the following *measurement postulate* to hold: for a quantum system in a (mixed) state $\rho \in \mathcal{D}(\mathcal{H})$, a projective measurement $\{P_i\}_i$ yields the outcome $i$ with probability $\mathrm{tr}(P_i\rho)$, and the post-measurement state of the system is

$$\frac{P_i \rho P_i}{\mathrm{tr}(P_i \rho)}.$$

This specification of the post-measurement state is the main (physical) difference between a projective measurement and a POVM, since the latter only determines the probabilities of the possible measurement outcomes.

A POVM is called *pure* if the $E_i$ are of rank one for all $i$, i.e., if there are (not necessarily orthogonal) vectors $|\psi_i\rangle \in \mathcal{H}$ such that $E_i = |\psi_i\rangle\langle\psi_i|$ for all $i$. For a pure projective measurement $\{|\phi_i\rangle\langle\phi_i|\}_i$, the vectors $\{|\phi_i\rangle\}_i$ form an orthonormal basis for $\mathcal{H}$. Conversely, every orthonormal basis of $\mathcal{H}$ gives rise to a projective measurement, and every *complete* set of (possibly non-orthogonal and subnormalized) vectors $\{|\psi_i\rangle\}_i$ gives rise to a pure POVM upon setting $E_i :=$ $|\psi_i\rangle\langle\psi_i|$. Completeness of the vectors $|\psi_i\rangle$ (i.e., $\sum_i |\psi_i\rangle\langle\psi_i| = \mathbb{1}$) ensures that the POVM $\{E_i\}_i$ forms a resolution of the identity. Every pure POVM can be implemented as a projective measurement on a larger Hilbert space. This is known as Naimark's Dilation Theorem and can be obtained as a corollary of Stinespring's Representation Theorem 2.2.2.

Finally, we note that every POVM $\{E_i\}_i$ gives rise to a certain quantum operation $\mathcal{M}_E$ called a *measurement channel*, defined by $\mathcal{M}_E(\rho) := \sum_i \mathrm{tr}(E_i\rho)|i\rangle\langle i|$. This is sometimes also called a quantum-classical channel, as it maps a quantum system in the state $\rho$ to a classical system corresponding to the possible outcomes of the POVM. Measurement channels give rise to c-q states in a natural way: suppose we have a bipartite quantum system $AB$ in the state $\rho_{AB}$ shared between Alice ($A$) and Bob ($B$), and Alice performs a measurement given by a POVM $\{E_x\}_{x\in\mathcal{X}}$ on her share of the system. The total state of the system after Alice's measurement is

$$\sigma_{XB} := (\mathcal{M}_E \otimes \mathrm{id}_B)(\rho_{AB}) = \sum_{x\in\mathcal{X}} p_x |x\rangle\langle x|_X \otimes \rho_B^x,$$

where $p_x = \mathrm{tr}((E_x \otimes \mathbb{1}_B)\rho_{AB})$ and $\rho_B^x = \mathrm{tr}_A((E_x \otimes \mathbb{1}_B)\rho_{AB})$ for $x \in \mathcal{X}$, and the classical system corresponding to Alice's measurement outcome is labeled by $X$. We consider c-q states obtained from a POVM in the above way in measurement compression with quantum side information, which we discuss in Section 6.2.

### 2.2.3. Distance measures and entropic quantities

In the following, we define the most important distance measures between quantum states and entropic quantities that are used throughout this thesis. To start with, we recall the *trace norm* of an operator $X \in \mathcal{B}(\mathcal{H})$, defined as

$$\|X\|_1 := \mathrm{tr}\sqrt{X^\dagger X} = \sum_i s_i(X),$$

where $\{s_i(X)\}_i$ are the singular values of $X$, i.e., the square roots of the eigenvalues of the positive semidefinite operator $X^\dagger X$. The trace norm gives rise to two important distance measures on $\mathcal{D}(\mathcal{H})$, the fidelity and the trace distance:

**Definition 2.2.3** (Fidelity and trace distance).
Let $\rho$ and $\sigma$ be quantum states.

(i) The *fidelity* $F(\rho, \sigma)$ is defined as

$$F(\rho, \sigma) := \left\| \rho^{1/2} \sigma^{1/2} \right\|_1.$$

(ii) The *trace distance* $T(\rho, \sigma)$ is defined as

$$T(\rho, \sigma) := \frac{1}{2} \| \rho - \sigma \|_1.$$

The trace distance $T(\cdot, \cdot)$ is a metric in the mathematical sense, i.e., $T(\rho, \sigma) \geq 0$ with equality if and only if $\rho = \sigma$, and $T(\cdot, \cdot)$ is symmetric and satisfies the triangle inequality. The fidelity is not a metric by itself, but can be used to define metrics such as the angular distance $A(\rho, \sigma) :=$ arccos $F(\rho, \sigma)$ [NC00], the Bures metric $B(\rho, \sigma) := \sqrt{2 - 2F(\rho, \sigma)}$ [Bur69], or the purified distance $P(\rho, \sigma) := \left(1 - F_{\text{gen}}(\rho, \sigma)^2\right)^{1/2}$, where $F_{\text{gen}}(\cdot, \cdot)$ denotes the generalized fidelity defined on subnormalized states (that is, $\rho \geq 0$ with $\text{tr}\, \rho \leq 1$) [TCR10].

The following result is useful:

**Theorem 2.2.4** (Uhlmann's theorem [Uhl76]).
*Let $\rho$ and $\sigma$ be quantum states. Then,*

$$F(\rho, \sigma) = \max_{\psi^\rho, \phi^\sigma} |\langle \psi^\rho | \phi^\sigma \rangle|,$$

*where the maximization is over all purifications $\psi^\rho$ and $\phi^\sigma$ of $\rho$ and $\sigma$, respectively.*

We now define two central quantities of this thesis:

**Definition 2.2.5** (Quantum relative entropy and quantum information variance).
Let $\rho \in \mathcal{D}(\mathcal{H})$ and $\sigma \in \mathcal{P}(\mathcal{H})$ be such that $\text{supp}\, \rho \subseteq \text{supp}\, \sigma$.

(i) The *quantum relative entropy* [Ume62] is defined as

$$D(\rho \| \sigma) := \text{tr}\left(\rho(\log \rho - \log \sigma)\right).$$

(ii) The *quantum information variance* [TH13] is defined as

$$V(\rho\|\sigma) := \operatorname{tr}\left(\rho(\log\rho - \log\sigma)^2\right) - D(\rho\|\sigma)^2.$$

Further, we define $\sigma(\rho\|\tau) := \sqrt{V(\rho\|\tau)}$ and

$$\sigma(\rho) := \sigma(\rho\|\mathbb{1}) = \sqrt{V(\rho\|\mathbb{1})}. \tag{2.3}$$

For states $\rho$ and $\sigma$, the quantum relative entropy satisfies $D(\rho\|\sigma) \geq 0$ with equality if and only if $\rho = \sigma$ (this is also known as Klein's inequality). Hence, $D(\cdot\|\cdot)$ defines a *premetric* on the set of quantum states. Note however that the quantum relative entropy is not symmetric, and does not satisfy a triangle inequality.

The quantum relative entropy acts as a parent quantity for the following fundamental information quantities:

**Definition 2.2.6** (von Neumann entropies)**.**
Let $\rho_{ABC} \in \mathcal{D}(\mathcal{H}_{ABC})$, and denote its marginals by $\rho_{AB}$ and $\rho_A$. Then we define:

(i) the *von Neumann entropy*

$$S(\rho_A) := -D(\rho_A\|\mathbb{1}_A) = -\operatorname{tr}\left(\rho_A \log \rho_A\right),$$

for which we also use the notation $S(A)_\rho \equiv S(\rho_A)$;

(ii) the *quantum conditional entropy*

$$S(A|B)_\rho := -D(\rho_{AB}\|\mathbb{1}_A \otimes \rho_B) = S(AB)_\rho - S(B)_\rho;$$

(iii) the *quantum mutual information*

$$I(A;B)_\rho := D(\rho_{AB}\|\rho_A \otimes \rho_B) = S(A)_\rho + S(B)_\rho - S(AB)_\rho;$$

(iv) the *quantum conditional mutual information*

$$I(A;B|C)_\rho := S(A|C)_\rho + S(B|C) - S(AB|C)_\rho.$$

### 2.2.4. Useful tools in quantum information theory

In the following we briefly discuss two useful tools in quantum information theory, namely, pinching and spectral projections.

**Pinching**

As indicated in Section 2.2.1, for two commuting operators a quantum information-theoretic problem can sometimes be reduced to a classical one by considering a common eigenbasis in which both operators are simultaneously diagonalized. In this sense, the non-commutativity of two operators captures the genuine quantum nature of the problem at hand. The pinching technique enforces commutativity (and hence a derived classical structure) between two states by slightly perturbing one of the operators in terms of the other one in a way such that the properties of both operators are approximately preserved.

**Definition 2.2.7** (Pinching).
Let $\sigma \in \mathcal{P}(\mathcal{H})$ with spectral decomposition given by $\sigma = \sum_{i=1}^{\nu(\sigma)} \lambda_i P_i$, where $\nu(\sigma)$ denotes the number of distinct eigenvalues of $\sigma$, and $P_i$ is the projection onto the eigenspace corresponding to the eigenvalue $\lambda_i$ for $i = 1, \ldots, \nu(\sigma)$. The *pinching map* $\mathcal{E}_\sigma$ is defined as

$$\mathcal{E}_\sigma(\rho) := \sum_{i=1}^{\nu(\sigma)} P_i \rho P_i.$$

The pinching map $\mathcal{E}_\sigma$ satisfies the following properties:

**Proposition 2.2.8.** *Let $\rho, \sigma \in \mathcal{P}(\mathcal{H})$.*

(i) *$\mathcal{E}_\sigma$ is a CPTP map.*

(ii) *We have $[\sigma, \mathcal{E}_\sigma(\rho)] = 0$.*

(iii) [OH04] *If $\operatorname{tr} \rho = 1$, then $\rho \leq \nu(\sigma)\mathcal{E}_\sigma(\rho)$.*

(iv) *Set $d = \dim \mathcal{H}$ and assume that $\sigma = \sum_{j=1}^{d} \mu_j \pi_j$, where for $j = 1, \ldots, d$ the $\mu_j$ are the eigenvalues of $\sigma$ listed in decreasing order and repeated according to their multiplicities, and $\pi_j$ are rank-1 projections. Then, it holds for $\rho$ with $\operatorname{tr} \rho = 1$ that*

$$\rho \leq d \sum_{j=1}^{d} \pi_j \rho \pi_j.$$

*Proof.* (i) Since the set $\{P_i\}_{i=1}^{\nu(\sigma)}$ of projections satisfies the relation

$$\sum_{i=1}^{\nu(\sigma)} P_i = \mathbb{1}, \qquad (2.4)$$

the Choi-Kraus Representation Theorem (see for example [Wat16; Wil16; Wol12]) implies that the pinching map $\mathcal{E}_\sigma$ is a CPTP map.

To show (iii), it suffices to prove the claim for pure states $\rho = |\phi\rangle\langle\phi|$. For arbitrary $|\psi\rangle \in \mathcal{H}$, we have

$$
\begin{aligned}
\langle\psi|\nu(\sigma)\mathcal{E}_\sigma(|\phi\rangle\langle\phi|)|\psi\rangle &= \nu(\sigma)\sum_{i=1}^{\nu(\sigma)}\langle\psi|P_i|\phi\rangle\langle\phi|P_i|\psi\rangle \\
&\geq \left|\sum_{i=1}^{\nu(\sigma)}\langle\psi|P_i|\phi\rangle\right|^2 \\
&= |\langle\psi|\phi\rangle|^2 \\
&= \langle\psi|\phi\rangle\langle\phi|\psi\rangle, \qquad (2.5)
\end{aligned}
$$

where the inequality follows from the Cauchy-Schwarz inequality applied to the vectors $(\langle\psi|P_1|\phi\rangle, \ldots, \langle\psi|P_{\nu(\sigma)}|\phi\rangle)$ and $(1, \ldots, 1)$, and the second equality follows from (2.4). Since (2.5) holds for all $|\psi\rangle \in \mathcal{H}$, we have $\nu(\sigma)\mathcal{E}_\sigma(|\phi\rangle\langle\phi|) \geq |\phi\rangle\langle\phi|$, and hence the claim follows.

(ii) follows immediately from Definition 2.2.7 of $\mathcal{E}_\sigma$, and (iv) follows from (iii) using Gram-Schmidt orthogonalization. □

Due to Proposition 2.2.8(ii), the pinched operator $\mathcal{E}_\sigma(\rho)$ and $\sigma$ are both diagonalized by the eigenbasis of $\sigma$, and hence they can be described by classical probability distributions. Another means of mapping two (non-commuting) quantum states to classical probability distributions is the Nussbaum-Szkoła distributions, which we discuss in Section 4.2.1.

**Spectral projections**

Let $H \in \operatorname{Herm}(\mathcal{H})$ be a Hermitian operator with spectral decomposition $H = \sum_i h_i|i\rangle\langle i|$. We define a projection $\{H \geq 0\}$ by

$$\{H \geq 0\} := \sum_{i:\ h_i \geq 0} |i\rangle\langle i|. \qquad (2.6)$$

That is, $\{H \geq 0\}$ projects onto the subspace of $\mathcal{H}$ spanned by eigenvectors of $H$ corresponding to non-negative eigenvalues. The projections $\{H > 0\}$, $\{H \leq 0\}$, and $\{H < 0\}$ are defined analogously by replacing the condition $h_i \geq 0$ in (2.6) with $h_i > 0$, $h_i \leq 0$, and $h_i < 0$, respectively. We use the notation $H_+ \coloneqq \{H \geq 0\}H\{H \geq 0\}$. For two Hermitian operators $A$ and $B$, we write $\{A \geq B\} \equiv \{A - B \geq 0\}$, and similarly for $>$, $\leq$, and $<$. The spectral projection satisfies the following properties:

**Lemma 2.2.9.** *Let $H \in \mathrm{Herm}(\mathcal{H})$.*

(i) *For any unitary $U$, we have $\left\{ UHU^\dagger \geq 0 \right\} = U\{H \geq 0\}U^\dagger$, and the same holds for $\{H > 0\}$, $\{H \leq 0\}$, and $\{H < 0\}$.*

(ii) *Let $T \geq 0$, then $\{H \otimes T \geq 0\} = \{H \geq 0\} \otimes \mathbb{1}$, and the same holds for $\{H \leq 0\}$.*

*Proof.* (i) follows immediately from the definition (2.6).

To prove (ii), let $H = \sum_i h_i |e_i\rangle\langle e_i|$ and $T = \sum_i t_i |f_i\rangle\langle f_i|$ be spectral decompositions of $H$ and $T$, respectively. Then $H \otimes T = \sum_{i,j} h_i t_j |e_i f_j\rangle\langle e_i f_j|$ is a spectral decomposition of $H \otimes T$, where $|e_i f_j\rangle \coloneqq |e_i\rangle \otimes |f_j\rangle$. Since $T \geq 0$, it holds that $t_j \geq 0$ for all $j$, and hence for all $i$ and $j$ we have $h_i t_j \geq 0$ if and only if $h_i \geq 0$. Therefore,

$$\{H \otimes T \geq 0\} = \sum_{(i,j):\, h_i t_j \geq 0} |e_i f_j\rangle\langle e_i f_j| = \sum_{i:\, h_i \geq 0} |e_i\rangle\langle e_i| \sum_j |f_j\rangle\langle f_j| = \{H \geq 0\} \otimes \mathbb{1}.$$

The same argument shows that $\{H \otimes T \leq 0\} = \{H \leq 0\} \otimes \mathbb{1}$. $\qquad\square$

We also use the following results:

**Lemma 2.2.10** ([ON00]). *For $A, B \in \mathcal{P}(\mathcal{H})$ and $0 \leq P \leq \mathbb{1}$, we have*

$$\mathrm{tr}(A - B)_+ = \mathrm{tr}(\{A \geq B\}(A - B)) \geq \mathrm{tr}(P(A - B)).$$

*Proof.* Let the spectral decomposition of $A - B$ be given by $A - B = \sum_i \lambda_i |i\rangle\langle i|$. Then, observe that

$$\mathrm{tr}(P(A - B)) = \sum_i \lambda_i \langle i|P|i\rangle \leq \sum_{i:\, \lambda_i \geq 0} \lambda_i \langle i|P|i\rangle \leq \sum_{i:\, \lambda_i \geq 0} \lambda_i = \mathrm{tr}(A - B)_+,$$

where we used $P \geq 0$ in the first inequality, and $P \leq \mathbb{1}$ in the second one. $\qquad\square$

**Lemma 2.2.11** ([BD06a]). *For $A, B \in \mathcal{P}(\mathcal{H})$ and a quantum operation $\Lambda \colon \mathcal{B}(\mathcal{H}) \to \mathcal{B}(\mathcal{K})$,*

$$\operatorname{tr}(A - B)_+ \geq \operatorname{tr}(\Lambda(A) - \Lambda(B))_+.$$

*Proof.* Since $\Lambda$ is completely positive and trace-preserving, the adjoint map $\Lambda^\dagger$ is completely positive and unital, that is, $\Lambda^\dagger(\mathbb{1}_\mathcal{K}) = \mathbb{1}_\mathcal{H}$. Hence, for the projection $X \coloneqq \{\Lambda(A) \geq \Lambda(B)\} \leq \mathbb{1}_\mathcal{K}$, we obtain

$$\Lambda^\dagger(X) \leq \Lambda^\dagger(\mathbb{1}_\mathcal{K}) = \mathbb{1}_\mathcal{H}.$$

By Lemma 2.2.10, we now have

$$\begin{aligned}
\operatorname{tr}(A - B)_+ &\geq \operatorname{tr}\left[(A - B)\Lambda^\dagger(X)\right] \\
&= \operatorname{tr}\left[(\Lambda(A) - \Lambda(B))X\right] \\
&= \operatorname{tr}(\Lambda(A) - \Lambda(B))_+,
\end{aligned}$$

where the second line follows from the definition of the adjoint map. $\qquad\square$

Finally, we record the following simple observation: Let $A, B, C \in \mathcal{P}(\mathcal{H})$ be pairwise commuting operators with $B \leq C$. Then we have $\{A \leq B\} \leq \{A \leq C\}$, which can easily be seen to be true by considering a common eigenbasis of $A$, $B$, and $C$ and checking the corresponding relation in the scalar case. We will use this result in the following special form:

**Lemma 2.2.12.** *For $a, b \in \mathbb{R}$ with $a \leq b$ and any $X \geq 0$, we have*

$$\left\{X \leq 2^{-b}\mathbb{1}\right\} \leq \left\{X \leq 2^{-a}\mathbb{1}\right\}.$$

## 2.3. Selected results from matrix analysis

We conclude this chapter with a collection of useful results from matrix analysis. An excellent introduction to this topic is the textbook by Bhatia [Bha97].

### 2.3.1. Schatten norms

For an arbitrary operator $X \in \mathcal{B}(\mathcal{H})$, we define $|X| \coloneqq \sqrt{X^\dagger X}$.

**Definition 2.3.1** (Schatten *p*-(quasi)norm)**.**
For $M \in \mathcal{B}(\mathcal{H})$ and $p > 0$, let

$$\|M\|_p := \left(\operatorname{tr} |M|^p\right)^{1/p}.$$

For $1 \leq p \leq \infty$, the functional $\|\cdot\|_p$ defines a norm, the *Schatten p-norm*. The Schatten 1-norm is the trace norm defined in Section 2.2.3.

In Chapter 3 we often make use of the identity

$$\|X\|_{2\alpha}^2 = \|X^\dagger X\|_\alpha, \tag{2.7}$$

which follows directly from Definition 2.3.1.

**Theorem 2.3.2** (Properties of Schatten norms)**.**
*Let $M, N \in \mathcal{B}(\mathcal{H})$.*

(i) *Hölder's inequality: For $1 \leq p \leq \infty$ let $q$ be the Hölder conjugate of $p$ defined by $\frac{1}{p} + \frac{1}{q} = 1$. Then,*

$$\|MN\|_1 \leq \|M\|_p \|N\|_q.$$

(ii) *McCarthy's inequalities: [McC67] If $p \in (0, 1)$, then*

$$\|M + N\|_p^p \leq \|M\|_p^p + \|N\|_p^p. \tag{2.8}$$

*[Sim05, Thm. 1.22] If $1 \leq p \leq \infty$ and $M, N \geq 0$, then*

$$\|M\|_p^p + \|N\|_p^p \leq \|M + N\|_p^p. \tag{2.9}$$

For $1 \leq p \leq \infty$, the functional $\|\cdot\|_p$ defines a norm, and hence satisfies the triangle inequality, which for the Schatten *p*-norms is also known as the *Minkowski inequality*:

$$\|M + N\|_p \leq \|M\|_p + \|N\|_p. \tag{2.10}$$

However, for $p \in (0, 1)$ the Minkowski inequality (2.10) fails to hold, and we have the weaker inequality (2.8) instead. Therefore, $\|\cdot\|_p$ only defines a quasinorm for this range.

### 2.3.2. Operator inequalities and trace inequalities

We define a partial order on Hermitian operators in the following way: For $A, B \in \mathrm{Herm}(\mathcal{H})$, we write $A \geq B$ if $A - B \in \mathcal{P}(\mathcal{H})$. For a function $f : \mathbb{R} \to \mathbb{R}$ and $A \in \mathrm{Herm}(\mathcal{H})$ with spectral decomposition $A = \sum_i a_i |i\rangle\langle i|$, we define $f(A) = \sum_i f(a_i)|i\rangle\langle i|$.

A function $g$ is called:

- *operator monotone*, if $A \geq B$ implies $g(A) \geq g(B)$ for all $A, B \in \mathrm{Herm}(\mathbb{C}^n)$ and $n \in \mathbb{N}$;

- *operator convex*, if

$$g(\lambda A + (1 - \lambda)B) \leq \lambda g(A) + (1 - \lambda)g(B)$$

  holds for all $A, B \in \mathrm{Herm}(\mathbb{C}^n)$, $\lambda \in [0, 1]$, and $n \in \mathbb{N}$;

- *operator concave*, if $-g$ is operator convex.

**Theorem 2.3.3** (Löwner's Theorem [Löw34])**.**

(i) *For $p \in [-1, 0]$ and $x \geq 0$, the function $x \mapsto -x^p$ is operator monotone and operator concave.*

(ii) *For $p \in [0, 1]$ and $x \geq 0$, the function $x \mapsto x^p$ is operator monotone and operator concave.*

(iii) *For $p \in [1, 2]$ and $x \geq 0$, the function $x \mapsto x^p$ is operator convex.*

The following standard result is for example proved in [Car10, Thm. 2.10].

**Theorem 2.3.4.** *Let $A$ be a Hermitian matrix whose eigenvalues lie in a set $\mathcal{D} \subseteq \mathbb{R}$, and let $g : \mathcal{D} \to \mathbb{R}$ be a continuous function.*

(i) *If $g$ is monotonically increasing, then so is $A \mapsto \mathrm{tr}\, g(A)$.*

(ii) *If $g$ is (strictly) convex, then so is $A \mapsto \mathrm{tr}\, g(A)$.*

We also use the well-known Araki-Lieb-Thirring inequalities.

**Theorem 2.3.5** (Araki-Lieb-Thirring inequalities [LT76; Ara90])**.**
*For $A, B \in \mathcal{P}(\mathcal{H})$ and $q, r \geq 0$,*

(i) $\mathrm{tr}\left(B^{1/2}AB^{1/2}\right)^{rq} \leq \mathrm{tr}\left(B^{r/2}A^r B^{r/2}\right)^q$ *if $r \geq 1$;*

(ii) $\mathrm{tr}\left(B^{1/2}AB^{1/2}\right)^{rq} \geq \mathrm{tr}\left(B^{r/2}A^r B^{r/2}\right)^q$ *if $r \in [0, 1)$.*

### 2.3.3. Eigenvalues of Hermitian operators and majorization

Throughout this section we consider a fixed Hilbert space $\mathcal{H}$ with $\dim \mathcal{H} = d$. For a Hermitian operator $A \in \mathrm{Herm}(\mathcal{H})$, we denote by $\lambda_i(A)$ the eigenvalues of $A$ in non-increasing order (counted with multiplicities), i.e., $\lambda_1(A) \geq \cdots \geq \lambda_d(A)$, and we form the vector of eigenvalues $\lambda(A) = (\lambda_i(A))_{i=1}^d$.

We first recall an important result by Weyl concerning the eigenvalues of sums of Hermitian operators. For a proof, see for example Section III in [Bha97].

**Theorem 2.3.6** (Weyl's Monotonicity Theorem)**.**
*For $A, B \in \mathrm{Herm}(\mathcal{H})$ and $i = 1, \ldots, d$,*

$$\lambda_i(A) + \lambda_d(B) \leq \lambda_i(A + B) \leq \lambda_i(A) + \lambda_1(B).$$

*In particular, for $A \in \mathrm{Herm}(\mathcal{H})$ and $H \in \mathcal{P}(\mathcal{H})$ we have $\lambda_i(A + H) \geq \lambda_i(A)$ for all $i$, and the inequality is strict if $H$ is invertible.*

The second fundamental result is by Ky Fan:

**Theorem 2.3.7** (Ky Fan's Maximum Principle [Fan49])**.**
*For $A \in \mathrm{Herm}(\mathcal{H})$ and $k = 1, \ldots, d$,*

$$\sum_{j=1}^{k} \lambda_j(A) = \max\{\mathrm{tr}(PA) \colon P \text{ is a projection on } \mathcal{H} \text{ with } \mathrm{tr}\, P = k\}.$$

Finally, we introduce the concept of *majorization*: For a vector $x = (x_i)_{i=1}^d$ in $\mathbb{R}^d$, denote by $x^{\downarrow} = (x_i^{\downarrow})_{i=1}^d$ the vector obtained from rearranging the components of $x$ in non-increasing order, i.e., $x_1^{\downarrow} \geq \cdots \geq x_d^{\downarrow}$. Given vectors $x, y \in \mathbb{R}^d$, we say that $x$ is *majorized* by $y$, in symbols $x \prec y$, if

$$\sum_{i=1}^{k} x_i^{\downarrow} \leq \sum_{i=1}^{k} y_i^{\downarrow} \quad \text{for } k = 1, \ldots, d, \text{ and} \quad \sum_{i=1}^{d} x_i^{\downarrow} = \sum_{i=1}^{d} y_i^{\downarrow}.$$

We then have the following result by Nielsen and Kempe concerning separable states:

**Theorem 2.3.8** ([NK01])**.** *If $\rho_{AB} \in \mathcal{D}(\mathcal{H}_{AB})$ is separable, then*

$$\lambda(\rho_{AB}) \prec \lambda(\rho_A) \quad \text{and} \quad \lambda(\rho_{AB}) \prec \lambda(\rho_B).$$

# Part I.

# Mathematical properties of entropies

# 3. Sandwiched Rényi divergence

A key quantity in classical information theory is the *Kullback-Leibler divergence* $D(\cdot\|\cdot)$, defined for classical probability distributions $P$ and $Q$ on a finite alphabet $\mathcal{X}$ as

$$D(P\|Q) := \sum_{x\in\mathcal{X}} P(x)\log\frac{P(x)}{Q(x)} \tag{3.1}$$

if $\operatorname{supp} P \subseteq \operatorname{supp} Q$ (i.e., $P(x) = 0$ whenever $Q(x) = 0$), and $+\infty$ otherwise. We do not concern ourselves with operational interpretations of the Kullback-Leibler divergence in this thesis, and merely note that it is as important in classical information theory as the quantum relative entropy is in quantum information theory. This is further substantiated by the fact that for commuting quantum states, the quantum relative entropy reduces to the Kullback-Leibler divergence of the probability distributions given by the eigenvalues of the two states.

In the following, we investigate $D(\cdot\|\cdot)$ from an axiomatic point of view as exhibited in Rényi's seminal paper [Rén61]. Based on work by Faddeev [Fad57] and Feinstein [Fei58], Rényi axiomatized the notion of a relative entropy or generalized divergence through five desiderata (listed below as Axioms I–V), and proved that there are exactly two quantities satisfying these axioms: the Kullback-Leibler divergence $D(\cdot\|\cdot)$ that we defined in (3.1), and the so-called *$\alpha$-Rényi divergence*, whose quantum generalizations are the subject of this chapter. In the following paragraphs, we discuss Rényi's result in more detail.

For a finite set $\mathcal{X}$, a *generalized probability distribution $P$* is defined as a function $P\colon \mathcal{X} \to \mathbb{R}_+$ such that $w(P) := \sum_{x\in\mathcal{X}} P(x) \leq 1$. The support of $P$ is defined as the subset of $\mathcal{X}$ with non-zero probability, $\operatorname{supp} P := \{x \in \mathcal{X} : P(x) \neq 0\}$. For two generalized probability distributions $P$ and $Q$ with $\operatorname{supp} P \subseteq \operatorname{supp} Q$, Rényi considered the following axioms for a relative entropy $\mathcal{D}(\cdot\|\cdot)$:

I. Symmetry: $\mathcal{D}(P^\pi\|Q^\pi) = \mathcal{D}(P\|Q)$, where $P^\pi$ and $Q^\pi$ are obtained from $P$ and $Q$ respectively by permuting the probability vectors $(P(x))_{x\in\mathcal{X}}$ and $(Q(x))_{x\in\mathcal{X}}$ with respect to a permutation $\pi \in \mathcal{S}_{|\mathcal{X}|}$.

II. **Order:** If $P(x) \leq Q(x)$ for all $x \in \mathcal{X}$, then $\mathcal{D}(P\|Q) \geq 0$. Likewise, if $P(x) \geq Q(x)$ for all $x \in \mathcal{X}$, then $\mathcal{D}(P\|Q) \leq 0$.

III. **Normalization:** $\mathcal{D}(\{1\}\|\{1/2\}) = 1$ for the singleton distributions $\{1\}$ and $\{1/2\}$.

IV. **Additivity:** $\mathcal{D}(P \times P'\|Q \times Q') = \mathcal{D}(P\|Q) + \mathcal{D}(P'\|Q')$ where $P'$ and $Q'$ are generalized probability distributions with supp $P' \subseteq$ supp $Q'$.

V. **Generalized mean property:** Let $P'$ and $Q'$ be generalized probability distributions satisfying supp $P' \subseteq$ supp $Q'$, $w(P) + w(P') \leq 1$, and $w(Q) + w(Q') \leq 1$. Then there exists a continuous and strictly increasing function $g \colon \mathbb{R} \to \mathbb{R}$, such that

$$\mathcal{D}(P \oplus P'\|Q \oplus Q') = g^{-1}\left( \frac{w(P)}{w(P) + w(P')} \, g(\mathcal{D}(P\|Q)) + \frac{w(P')}{w(P) + w(P')} \, g(\mathcal{D}(P'\|Q')) \right).$$

Here, the distribution $P \oplus P'$ on the alphabet $\mathcal{X} \oplus \mathcal{X}$ is defined via $(P \oplus P')(x \oplus x') \coloneqq P(x) + P'(x')$, and similarly for $Q \oplus Q'$.

In [Rén61, Theorem 3], Rényi showed that for a relative entropy $\mathcal{D}(\cdot\|\cdot)$ satisfying Axioms I–V, the function $g$ in Axiom V is necessarily a linear or an exponential function. In the linear case, the unique relative entropy satisfying Axioms I–V is the Kullback-Leibler divergence $D(\cdot\|\cdot)$ defined in (3.1). In the exponential case, the unique relative entropy satisfying Axioms I–V is the $\alpha$-*Rényi divergence* $D_\alpha(\cdot\|\cdot)$, defined for generalized probability distributions $P$ and $Q$ with supp $P \subseteq$ supp $Q$ and $\alpha \neq 1$ as

$$D_\alpha(P\|Q) \coloneqq \frac{1}{\alpha - 1} \log \frac{\sum_{x \in \mathcal{X}} P(x)^\alpha Q(x)^{1-\alpha}}{w(P)}. \tag{3.2}$$

Moreover, Rényi noted that $D(\cdot\|\cdot)$ corresponds to the limit of $D_\alpha(\cdot\|\cdot)$ for $\alpha \to 1$,

$$\lim_{\alpha \to 1} D_\alpha(P\|Q) = D(P\|Q). \tag{3.3}$$

The $\alpha$-Rényi divergence can therefore be regarded as a 'deformation' of the Kullback-Leibler divergence.[1]

---

[1] In principle, the definition (3.2) also makes sense for negative values of $\alpha$. However, $D_\alpha(\cdot\|\cdot)$ does not have an apparent information-theoretic meaning for $\alpha < 0$ [Rén61]. To overcome this problem, one can add a sixth axiom to the above list that ensures continuity of $\mathcal{D}(\cdot\|\cdot)$, excluding negative values of $\alpha$. We refrain from doing this here, since requiring the data processing inequality (3.5) to hold for quantum generalizations of $D_\alpha(\cdot\|\cdot)$ has the same effect; that is, it enforces the valid range of $\alpha$ to be included in $\alpha \geq 0$.

In this chapter, we are interested in quantum generalizations of the classical Rényi divergence (3.2). More precisely, we are looking for real-valued functionals $\mathcal{D}_\alpha(\cdot\|\cdot)$ on pairs of positive operators, parametrized by $\alpha \in \mathbb{R}$, and satisfying

$$\mathcal{D}_\alpha(\rho_X\|\sigma_X) = D_\alpha(P\|Q) \tag{3.4}$$

for classical states $\rho_X = \sum_{x \in \mathcal{X}} P(x)|x\rangle\langle x|_X$ and $\sigma_X = \sum_{x \in \mathcal{X}} Q(x)|x\rangle\langle x|_X$ with supp $P \subseteq$ supp $Q$.[2] Furthermore, we require the data processing inequality to hold: For a quantum operation $\Lambda$,

$$\mathcal{D}_\alpha(\rho\|\sigma) \geq \mathcal{D}_\alpha(\Lambda(\rho)\|\Lambda(\sigma)). \tag{3.5}$$

A straightforward quantum generalization of (3.2) is given by

$$D_\alpha(\rho\|\sigma) \coloneqq \frac{1}{\alpha - 1}\log\frac{\operatorname{tr}(\rho^\alpha\sigma^{1-\alpha})}{\operatorname{tr}\rho} \tag{3.6}$$

for $\rho, \sigma \in \mathcal{P}(\mathcal{H})$ with supp $\rho \subseteq$ supp $\sigma$ if $\alpha > 1$ or $\rho \not\perp \sigma$ if $\alpha \in [0, 1)$. The quantity in (3.6) is called the $\alpha$-*relative Rényi entropy* ($\alpha$-RRE), and was for example considered by Petz [Pet86a], who showed that it can be obtained as a so-called $f$-divergence. The $\alpha$-RRE has direct operational interpretations as generalized cut-off rates in quantum hypothesis testing [MH11] and as error exponents in composite hypothesis testing [HT16]. Furthermore, it satisfies the data processing inequality for the range $\alpha \in [0, 2]$ [Lie73; Uhl77; Pet86a]:

$$D_\alpha(\rho\|\sigma) \geq D_\alpha(\Lambda(\rho)\|\Lambda(\sigma)). \tag{3.7}$$

In the classical setting, the form of the $\alpha$-Rényi divergence is determined by Axioms I–V, as shown by Rényi [Rén61]. In contrast, in the quantum setting we have a great deal of freedom in furnishing quantum relative entropies $\mathcal{D}_\alpha(\cdot\|\cdot)$ satisfying (3.4), given that they also satisfy the data processing inequality (3.5). In the following sections, we investigate a different quantum generalization of the $\alpha$-Rényi divergence called the $\alpha$-*sandwiched Rényi divergence* $\widetilde{D}_\alpha(\cdot\|\cdot)$, that was introduced concurrently by Müller-Lennert et al. [MDS13] and Wilde et al. [WWY14]. One of the merits of this particular quantum generalization of (3.2) is its application in proving strong converse theorems in quantum information theory. We defer this discussion to Chapter 6, focusing for now on some of the mathematical properties of the $\alpha$-sandwiched

---

[2] It is also possible to formulate quantum versions of Axioms I–V and look for quantities that satisfy them, cf. [MDS13; Tom16].

Rényi divergence.

In Section 3.1, we define this quantity and review its most important properties. In Section 3.2 we investigate the connection between the $\alpha$-sandwiched Rényi divergence and the $\alpha$-RRE defined in (3.6), and examine special values and limits of the parameter $\alpha$. Section 3.3 is concerned with entropic quantities derived from $\widetilde{D}_\alpha(\cdot\|\cdot)$ and their properties. Finally, in Section 3.4 we take a closer look at the data processing inequality (3.5) for $\widetilde{D}_\alpha(\cdot\|\cdot)$, as stated in Proposition 3.1.2(vi)). We derive a necessary and sufficient condition for equality in (3.5), and give applications of this result to certain entropic bounds.

## 3.1. Definition and properties

Following [MDS+13; WWY14], we define one of the central quantities of this thesis:

**Definition 3.1.1** ($\alpha$-sandwiched Rényi divergence)**.**
For $\alpha \in (0, \infty) \setminus \{1\}$ and $\rho, \sigma \in \mathcal{P}(\mathcal{H})$, we define the trace functional

$$\widetilde{Q}_\alpha(\rho\|\sigma) := \operatorname{tr}\left(\sigma^{(1-\alpha)/2\alpha} \rho \sigma^{(1-\alpha)/2\alpha}\right)^\alpha. \tag{3.8}$$

The *$\alpha$-sandwiched Rényi divergence* ($\alpha$-SRD) is defined as

$$\widetilde{D}_\alpha(\rho\|\sigma) := \begin{cases} \frac{1}{\alpha-1} \log\left[(\operatorname{tr}\rho)^{-1}\widetilde{Q}_\alpha(\rho\|\sigma)\right] & \text{if } \operatorname{supp}\rho \subseteq \operatorname{supp}\sigma \text{ or } (\alpha \in (0,1) \wedge \rho \not\perp \sigma) \\ +\infty & \text{otherwise.} \end{cases}$$

Alternatively, the $\alpha$-SRD can be expressed in terms of Schatten (quasi)norms as

$$\widetilde{D}_\alpha(\rho\|\sigma) = \frac{\alpha}{\alpha-1} \log\left\|\sigma^{(1-\alpha)/2\alpha} \rho \sigma^{(1-\alpha)/2\alpha}\right\|_\alpha - \frac{1}{\alpha-1} \log \operatorname{tr}\rho$$
$$= \frac{2\alpha}{\alpha-1} \log\left\|\rho^{1/2} \sigma^{(1-\alpha)/2\alpha}\right\|_{2\alpha} - \frac{1}{\alpha-1} \log \operatorname{tr}\rho,$$

where the first line follows from a simple rewriting of the definition of $\widetilde{D}_\alpha(\rho\|\sigma)$, and the second line uses the identity (2.7). This expression is useful, since the Schatten norms satisfy Hölder's inequality, Theorem 2.3.2(i); we make use of this fact in Section 3.3. The $\alpha$-SRD has operational interpretations as the strong converse exponent in various settings in quantum hypothesis testing [MO15; CMW14; HT16] and classical-quantum channel coding [MO14]. Its application in proving strong converse theorems is discussed in detail in Chapter 6. In the following Proposition 3.1.2 we collect a couple of properties of the $\alpha$-SRD that are needed throughout this thesis. Note that Proposition 3.1.2 is by no means exhaustive; for example, we do not address the differentiability of the $\alpha$-SRD for $\alpha$ in a neighborhood of 1, or its continuity properties. For a discussion of these issues as well as a plethora of other results, we refer the reader to the textbook [Tom16] or Section III in [MO14].

**Proposition 3.1.2** (Properties of the $\alpha$-SRD).
*Unless specified otherwise, let $\alpha \in (0, \infty) \setminus \{1\}$, and consider $\rho \in \mathcal{D}(\mathcal{H})$ and $\sigma \in \mathcal{P}(\mathcal{H})$ with* supp $\rho \subseteq$ supp $\sigma$ *if $\alpha > 1$ or $\rho \not\perp \sigma$ if $\alpha < 1$.*

(i) *Additivity: Let $\rho_i \in \mathcal{D}(\mathcal{H})$ and $\sigma_i \in \mathcal{P}(\mathcal{H})$ for $i = 1, 2$. Then*

$$\widetilde{D}_\alpha(\rho_1 \otimes \rho_2 \| \sigma_1 \otimes \sigma_2) = \widetilde{D}_\alpha(\rho_1 \| \sigma_1) + \widetilde{D}_\alpha(\rho_2 \| \sigma_2).$$

(ii) *Premetric: Let $\sigma \in \mathcal{P}(\mathcal{H})$ be such that* tr $\sigma = 1$. *Then $\widetilde{D}_\alpha(\rho \| \sigma) \geq 0$, and $\widetilde{D}_\alpha(\rho \| \sigma) = 0$ if and only if $\rho = \sigma$.*

(iii) *Monotonicity in $\alpha$: If $\beta \geq \alpha$, then*

$$\widetilde{D}_\alpha(\rho \| \sigma) \leq \widetilde{D}_\beta(\rho \| \sigma).$$

(iv) *Tensor invariance: Let $\tau \in \mathcal{D}(\mathcal{H})$. Then*

$$\widetilde{Q}_\alpha(\rho \otimes \tau \| \sigma \otimes \tau) = \widetilde{Q}_\alpha(\rho \| \sigma),$$

*and hence also $\widetilde{D}_\alpha(\rho \otimes \tau \| \sigma \otimes \tau) = \widetilde{D}_\alpha(\rho \| \sigma)$.*

(v) *Isometric invariance: Let $V \colon \mathcal{H} \to \mathcal{K}$ be an isometry. Then*

$$\widetilde{Q}_\alpha(\rho \| \sigma) = \widetilde{Q}_\alpha\left(V \rho V^\dagger \,\middle\|\, V \sigma V^\dagger\right),$$

*and hence also $\widetilde{D}_\alpha(\rho \| \sigma) = \widetilde{D}_\alpha\left(V \rho V^\dagger \,\middle\|\, V \sigma V^\dagger\right)$.*

(vi) *Data processing inequality (DPI): For $\alpha \geq 1/2$ and a quantum operation $\Lambda \colon \mathcal{B}(\mathcal{H}) \to \mathcal{B}(\mathcal{K})$,*

$$\widetilde{D}_\alpha(\rho \| \sigma) \geq \widetilde{D}_\alpha(\Lambda(\rho) \| \Lambda(\sigma)).$$

(vii) *Monotonicity in second slot: If $\alpha \geq 1/2$ and $\tau \in \mathcal{P}(\mathcal{H})$ is such that $\sigma \leq \tau$, then*

$$\widetilde{D}_\alpha(\rho\|\sigma) \geq \widetilde{D}_\alpha(\rho\|\tau).$$

*Proof.* (i), (iv) and (v) follow easily from Definition 3.1.1. (ii) was proved in [MDS+13], and (iii) in [MDS+13; Bei13].

The DPI (vi) was first proved for the range $\alpha \in (1, 2]$ by Müller-Lennert et al. [MDS+13] and Wilde et al. [WWY14] using an argument from [Wol12] based on the operator Jensen inequality. Beigi [Bei13] proved (vi) for the range $\alpha > 1$ using the theory of non-commutative $L_p$-spaces (see also [MR15] for an extension to positive maps). In an earlier paper preceding [MDS+13; WWY14], Hiai [Hia13] proved joint concavity of the trace functional $\widetilde{Q}_\alpha(\cdot\|\cdot)$ for the range $\alpha \in [1/2, 1)$ (and thus the DPI for the $\alpha$-SRD and this range of $\alpha$, as explained in Section 3.4). The first proof of the DPI for the full range $\alpha \geq 1/2$ was given by Frank and Lieb [FL13]. We review this proof in Section 3.4.1 below, as part of the proof of a condition for equality in the DPI (vi), which we state in Theorem 3.4.1.

(vii) was proved in [MDS+13]. We give a slightly different proof in the following. Setting $\gamma = (1 - \alpha)/2\alpha$, the trace functional $\widetilde{Q}_\alpha(\rho\|\sigma)$ defined in (3.8) can be expressed as

$$\widetilde{Q}_\alpha(\rho\|\sigma) = \operatorname{tr}\left(\rho^{1/2}\sigma^{2\gamma}\rho^{1/2}\right)^\alpha,$$

which follows from the fact that for any two operators $A$ and $B$ their products $AB$ and $BA$ have the same eigenvalues (see for example [Bha97, Exercise I.3.7]). Let us first consider the case $\alpha \in [1/2, 1)$, such that $2\gamma = (1 - \alpha)/\alpha \in (0, 1]$. By Theorem 2.3.3(ii) the function $x \mapsto x^{2\gamma}$ is operator monotone, and hence $\sigma \leq \tau$ implies $\sigma^{2\gamma} \leq \tau^{2\gamma}$, from which $\rho^{1/2}\sigma^{2\gamma}\rho^{1/2} \leq \rho^{1/2}\tau^{2\gamma}\rho^{1/2}$ follows. Applying Theorem 2.3.4(i) with $g(x) = x^\alpha$ then yields

$$\widetilde{Q}_\alpha(\rho\|\sigma) \leq \widetilde{Q}_\alpha(\rho\|\tau),$$

from which we obtain the claim as $\alpha - 1 < 0$.

For $\alpha > 1$ we have $2\gamma = (1 - \alpha)/\alpha \in [-1, 0)$, and hence the function $x \mapsto x^{2\gamma}$ is *order reversing* by Theorem 2.3.3(i), that is, $\sigma \leq \tau$ implies $\sigma^{2\gamma} \geq \tau^{2\gamma}$. The proof then continues in the same way as in the previous paragraph. □

# 3.2. Limits and special values of the Rényi parameter

## 3.2.1. Relative Rényi entropy and quantum relative entropy

Let $\rho, \sigma \in \mathcal{P}(\mathcal{H})$ with supp $\rho \subseteq$ supp $\sigma$ if $\alpha > 1$ or $\rho \not\perp \sigma$ if $\alpha < 1$, and assume furthermore that tr $\rho = 1$. Comparing the $\alpha$-SRD and the $\alpha$-RRE,

$$\widetilde{D}_\alpha(\rho\|\sigma) = \frac{1}{\alpha - 1} \log \left[ \text{tr} \, (\sigma^\gamma \rho \sigma^\gamma)^\alpha \right]$$

$$D_\alpha(\rho\|\sigma) = \frac{1}{\alpha - 1} \log \text{tr} \left( \rho^\alpha \sigma^{1-\alpha} \right),$$

it is easy to see that

$$[\rho, \sigma] = 0 \quad \text{implies} \quad D_\alpha(\rho\|\sigma) = \widetilde{D}_\alpha(\rho\|\sigma). \tag{3.9}$$

In fact, if $[\rho, \sigma] = 0$, then both the $\alpha$-RRE and the $\alpha$-SRD are equal to the *classical* $\alpha$-Rényi divergence of the probability distributions given by the eigenvalues of $\rho$ and $\sigma$.

Furthermore, the $\alpha$-SRD $\widetilde{D}_\alpha(\cdot\|\cdot)$ and the $\alpha$-RRE $D_\alpha(\cdot\|\cdot)$ share an important property for any pair of quantum states that are not necessarily commuting: both are equal to the quantum relative entropy $D(\cdot\|\cdot)$ in the limit $\alpha \to 1$, and can hence be seen as a 'deformation' of the quantum relative entropy, in analogy to the corresponding statement (3.3) for the classical $\alpha$-Rényi divergence and the Kullback-Leibler divergence. More precisely, we have:

**Proposition 3.2.1** ([Pet86a; MDS+13; WWY14]). *For $\rho, \sigma \in \mathcal{P}(\mathcal{H})$ with supp $\rho \subseteq$ supp $\sigma$,*

$$\lim_{\alpha \to 1} \widetilde{D}_\alpha(\rho\|\sigma) = D(\rho\|\sigma) = \lim_{\alpha \to 1} D_\alpha(\rho\|\sigma).$$

Optimal rates in information theory are typically expressed as entropic quantities such as the von Neumann entropy $S(A)_\rho$ or the quantum conditional entropy $S(A|B)_\rho$, both of which can be derived from the quantum relative entropy from Definition 2.2.6 (see for example Schumacher's quantum coding theorem (5.9) or the optimal rates (6.9) in state redistribution proved by Luo and Devetak [LD09] and Yard and Devetak [YD09]). The importance of Proposition 3.2.1 stems from the fact that it connects Rényi entropic quantities derived from $\widetilde{D}_\alpha(\cdot\|\cdot)$ (which we define in Section 3.3) to the entropic quantities that characterize optimal rates, giving rise to the 'Rényi entropy method' of proving strong converse theorems in information theory. This method is the subject of Chapter 6, for which Proposition 3.2.1 is a crucial ingredient.

To conclude this section, we also mention that the $\alpha$-RRE and the $\alpha$-SRD can be regarded as special cases of a two-parameter family of relative entropies called the $\alpha$-$z$-relative entropies [AD15], defined as

$$D_{\alpha,z}(\rho\|\sigma) := \frac{1}{\alpha-1}\log\left[\operatorname{tr}\left(\rho^{\alpha/z}\sigma^{(1-\alpha)/z}\right)^z\right].$$

It holds that $D_{\alpha,1}(\cdot\|\cdot) = D_\alpha(\cdot\|\cdot)$ and $D_{\alpha,\alpha}(\cdot\|\cdot) = \widetilde{D}_\alpha(\cdot\|\cdot)$.

## 3.2.2. Min- and max-entropies

An important tool in one-shot information theory is the smooth entropy framework (see [Ren05; Tom12; Dat09; TCR10] and references therein), whose central quantities are smoothed versions of the min-entropy $H_{\min}(A|B)_\rho$ and the max-entropy $H_{\max}(A|B)_\rho$. These quantities can be derived from two corresponding relative entropies, the max-relative entropy $D_{\max}(\cdot\|\cdot)$ and the min-relative entropy $D_{\min}(\cdot\|\cdot)$. While the max-relative entropy was introduced in [Dat09], the min-relative entropy was only implicitly used in [KRS09] in an expression for the max-entropy $H_{\max}(A|B)_\rho$ based on a quantity called the decoupling accuracy, whose definition is closely related to the min-relative entropy.

**Definition 3.2.2** (Min- and max-relative entropies).
Let $\rho \in \mathcal{P}(\mathcal{H})$ with $\operatorname{tr}\rho \leq 1$ and $\sigma \in \mathcal{P}(\mathcal{H})$.

(i) If $\rho \not\perp \sigma$, the *min-relative entropy* $D_{\min}(\rho\|\sigma)$ is defined as

$$D_{\min}(\rho\|\sigma) := -2\log F(\rho,\sigma) + 2\log\operatorname{tr}\rho,$$

where $F(\cdot,\cdot)$ is the fidelity from Definition 2.2.3.

(ii) If $\operatorname{supp}\rho \subseteq \operatorname{supp}\sigma$, the *max-relative entropy* $D_{\max}(\rho\|\sigma)$ is defined as

$$D_{\max}(\rho\|\sigma) := \inf\left\{\lambda\colon \rho \leq 2^\lambda\sigma\right\}.$$

Comparing Definition 3.2.2(i) of the min-relative entropy to Definition 3.1.1 of the $\alpha$-SRD, it is easy to see that $D_{\min}(\cdot\|\cdot) = \widetilde{D}_{1/2}(\cdot\|\cdot)$. Hence, $D_{\min}(\cdot\|\cdot)$ inherits all properties from Proposition 3.1.2, as it corresponds to taking $\alpha = 1/2$ therein. In particular, this shows that the

fidelity is non-decreasing under partial trace,[3]

$$F(\rho_{AB}, \sigma_{AB}) \leq F(\rho_A, \sigma_A). \tag{3.10}$$

The max-relative entropy can be obtained from the $\alpha$-SRD as well:

**Proposition 3.2.3** ([MDS+13])**.** *If $\rho, \sigma \in \mathcal{P}(\mathcal{H})$ with supp $\rho \subseteq$ supp $\sigma$, then*

$$\lim_{\alpha \to \infty} \widetilde{D}_\alpha(\rho\|\sigma) = D_{\max}(\rho\|\sigma).$$

### 3.2.3. 0-relative Rényi entropy

In the limit $\alpha \to 0$, the $\alpha$-RRE $D_\alpha(\cdot\|\cdot)$ gives rise to another useful divergence, the 0-RRE:

$$D_0(\rho\|\sigma) := -\log \operatorname{tr}(\Pi_\rho \sigma),$$

where $\Pi_\rho$ denotes the projection onto supp $\rho$. Noting that $X^0 := \lim_{p \to 0} X^p = \Pi_X$ holds for arbitrary normal operators $X$ by the spectral theorem, it follows that $\lim_{\alpha \to 0} D_\alpha(\rho\|\sigma) = D_0(\rho\|\sigma)$. The 0-RRE has an operational interpretation in quantum hypothesis testing, where it is equal to the minimum probability of the type-II error under the condition that the probability of the type-I error is 0 (see the beginning of Chapter 4 for a short introduction to quantum hypothesis testing). Furthermore, in perfect entanglement dilution it determines the entanglement cost of a bipartite state in the one-shot setting [BD11].

Since the $\alpha$-SRD and the $\alpha$-RRE coincide in the limit $\alpha \to 1$ (both recovering the quantum relative entropy, as stated in Proposition 3.2.1), it is natural to ask whether the same is true in the limit $\alpha \to 0$. That is, we are interested in the question whether the 0-RRE is recovered from the $\alpha$-SRD $\widetilde{D}_\alpha(\cdot\|\cdot)$ in the limit $\alpha \to 0$. We answer this question in the following section, which is based on [DL14].

We first prove that the $\alpha$-SRD is bounded from above by the $\alpha$-RRE for all $\alpha \geq 0$, which follows from the Araki-Lieb-Thirring inequalities (Theorem 2.3.5):

**Proposition 3.2.4** ([WWY14; DL14])**.** *Let $\alpha \geq 0$ and $\rho \in \mathcal{D}(\mathcal{H})$, $\sigma \in \mathcal{P}(\mathcal{H})$, with supp $\rho \subseteq$ supp $\sigma$ if $\alpha \geq 1$ or $\rho \not\perp \sigma$ if $\alpha < 1$. Then,*

$$\widetilde{D}_\alpha(\rho\|\sigma) \leq D_\alpha(\rho\|\sigma).$$

---

[3]Note that this can also be proved directly, e.g., via Uhlmann's Theorem 2.2.4.

*Proof.* Let us first consider $\alpha \in [0,1)$, and set $\gamma = (1-\alpha)/2\alpha$. Choosing $q = 1$, $r = \alpha$, $A = \rho$, and $B = \sigma^{2\gamma}$ in Theorem 2.3.5(ii), we obtain

$$\widetilde{Q}_\alpha(\rho\|\sigma) = \operatorname{tr}(\sigma^\gamma \rho \sigma^\gamma)^\alpha \geq \operatorname{tr}\left(\sigma^{(1-\alpha)/2}\rho^\alpha\sigma^{(1-\alpha)/2}\right) = \operatorname{tr}\left(\rho^\alpha\sigma^{1-\alpha}\right),$$

where we used cyclicity of the trace in the last step. This yields the claim, since the logarithm is monotonically increasing and $\alpha - 1 < 0$. In the case $\alpha \geq 1$ we make the same choices as above and use Theorem 2.3.5(i) instead. $\qquad\square$

Taking the limit $\alpha \to 0$ in Proposition 3.2.4 immediately yields the following:

**Corollary 3.2.5.** *If $\rho \in \mathcal{D}(\mathcal{H})$ and $\sigma \in \mathcal{P}(\mathcal{H})$ with $\rho \not\perp \sigma$, then*

$$\widetilde{D}_0(\rho\|\sigma) := \lim_{\alpha\to 0+}\widetilde{D}_\alpha(\rho\|\sigma) \leq D_0(\rho\|\sigma).$$

Hence, in the light of Corollary 3.2.5 we need to investigate if $\widetilde{D}_0(\rho\|\sigma)$ is also bounded from below by $D_0(\rho\|\sigma)$. It turns out that this is true if $\operatorname{supp}\rho = \operatorname{supp}\sigma$:

**Proposition 3.2.6.** *If $\rho \in \mathcal{D}(\mathcal{H})$ and $\sigma \in \mathcal{P}(\mathcal{H})$ with $\operatorname{supp}\rho = \operatorname{supp}\sigma$, then*

$$\widetilde{D}_0(\rho\|\sigma) \geq D_0(\rho\|\sigma).$$

*Proof.* We set $d = \dim\mathcal{H}$ and consider eigenvalue decompositions

$$\rho = \sum_{i=1}^d r_i|\phi_i\rangle\langle\phi_i| \qquad\qquad \sigma = \sum_{j=1}^d s_j|\psi_j\rangle\langle\psi_j|$$

as in Proposition 2.2.8(iv), that is, $\{r_i\}_{i=1}^d$ and $\{s_j\}_{j=1}^d$ are the eigenvalues of $\rho$ and $\sigma$, respectively, in decreasing order and repeated according to their multiplicities. For $j = 1, \ldots, d$ we also define $\mu_j := \sum_{i=1}^d r_i|\langle\phi_i|\psi_j\rangle|^2 = \langle\psi_j|\rho|\psi_j\rangle$, satisfying $\mu_j > 0$ for all $j = 1, \ldots, d$ due to the assumption $\operatorname{supp}\rho = \operatorname{supp}\sigma$.[4] Since we are interested in the limit $\alpha \to 0$, we only consider $\alpha \in (0,1)$ in the following, and set $\gamma = (1-\alpha)/2\alpha$ as before. Using the pinching technique in the form of

---

[4]The $\mu_j$ are related to the Nussbaum-Szkoła distributions of $\rho$ and $\sigma$, that we define in Section 4.2.1. More precisely, we have $\mu_j = \sum_{i=1}^d P_{\rho,\sigma}(i,j)$, where $P_{\rho,\sigma}(i,j) := r_i|\langle\phi_i|\psi_j\rangle|^2$ is one of the Nussbaum-Szkoła distributions on $[d]\times[d]$.

Proposition 2.2.8(iv), we then have

$$\sigma^\gamma \rho \sigma^\gamma \le d \sum_{j=1}^{d} |\psi_j\rangle\langle\psi_j| \sigma^\gamma \rho \sigma^\gamma |\psi_j\rangle\langle\psi_j|$$

$$= d \sum_{j=1}^{d} s_j^{2\gamma} |\psi_j\rangle\langle\psi_j| \rho |\psi_j\rangle\langle\psi_j|$$

$$= d \sum_{j=1}^{d} s_j^{2\gamma} \mu_j |\psi_j\rangle\langle\psi_j|.$$

Since $x \mapsto x^\alpha$ is operator monotone for $\alpha \in (0,1)$ by Theorem 2.3.3(ii), it follows that

$$(\sigma^\gamma \rho \sigma^\gamma)^\alpha \le d^\alpha \left( \sum_{j=1}^{d} s_j^{2\gamma} \mu_j |\psi_j\rangle\langle\psi_j| \right)^\alpha$$

$$= d^\alpha \sum_{j=1}^{d} s_j^{1-\alpha} \mu_j^\alpha |\psi_j\rangle\langle\psi_j|,$$

where in the equality we used the fact that the operator $\sum_{j=1}^{d} s_j^{2\gamma} \mu_j |\psi_j\rangle\langle\psi_j|$ is given in its spectral decomposition. Hence,

$$\widetilde{Q}_\alpha(\rho\|\sigma) = \text{tr}(\sigma^\gamma \rho \sigma^\gamma)^\alpha \le d^\alpha \sum_{j=1}^{d} s_j^{1-\alpha} \mu_j^\alpha,$$

and taking the limit $\alpha \to 0$ (recall that $\mu_j > 0$ for all $j = 1, \ldots, d$) and the negative of the logarithm yields

$$\widetilde{D}_0(\rho\|\sigma) \ge -\log\left( \sum_{j=1}^{d} s_j \right) = -\log(\text{tr}\,\sigma).$$

On the other hand, we have

$$D_0(\rho\|\sigma) = -\log\left[ \text{tr}\left( \Pi_\rho \sigma \right) \right] = -\log(\text{tr}\,\sigma)$$

by the assumption supp $\rho =$ supp $\sigma$, and this yields the claim. $\qquad\square$

Thus, combining Corollary 3.2.5 and Proposition 3.2.6, we have the following result:

**Theorem 3.2.7.** *Let $\rho \in \mathcal{D}(\mathcal{H})$ and $\sigma \in \mathcal{P}(\mathcal{H})$ with $\operatorname{supp} \rho = \operatorname{supp} \sigma$. Then,*

$$\widetilde{D}_0(\rho\|\sigma) = D_0(\rho\|\sigma) = -\log(\operatorname{tr}\sigma).$$

We also mention the following example, which demonstrates that the support condition of Theorem 3.2.7 is indispensable: Consider the states

$$\rho = \begin{pmatrix} 1 & 0 \\ 0 & 0 \end{pmatrix} \qquad\qquad \sigma = \begin{pmatrix} 1 & c \\ c & 1 \end{pmatrix}$$

for $c \in (0, 1)$, which satisfy $\operatorname{supp} \rho \subsetneq \operatorname{supp} \sigma$ and $[\rho, \sigma] \neq 0$ (such that $\widetilde{D}_\alpha(\rho\|\sigma) \neq D_\alpha(\rho\|\sigma)$, cf. (3.9)). Furthermore, $\Pi_\rho = \rho$, and hence,

$$D_0(\rho\|\sigma) = -\log(\operatorname{tr}(\rho\sigma)) = 0.$$

On the other hand, we observe that the eigenvalues of $\sigma^\gamma \rho \sigma^\gamma$ are given by

$$\lambda_1 = \frac{1}{2}\left((1+c)^{2\gamma} + (1-c)^{2\gamma}\right) \qquad\qquad \lambda_2 = 0,$$

and hence, $\widetilde{D}_\alpha(\rho\|\sigma) = \frac{1}{\alpha-1}\log(\lambda_1^\alpha)$. A simple application of l'Hôpital's rule then yields

$$\widetilde{D}_0(\rho\|\sigma) = -\log(1+c) < 0.$$

Note that this example is also valid if we choose a *normalized* state such as $\sigma = |+\rangle\langle+|$ with $|+\rangle = \frac{1}{\sqrt{2}}(|0\rangle + |1\rangle)$.

## 3.3. Derived Rényi entropic quantities

### 3.3.1. Definition and properties

We saw in Definition 2.2.6 in Section 2.2.3 that the von Neumann entropy $S(A)$, the quantum conditional entropy $S(A|B)$, and the quantum mutual information $I(A; B)$ can all be derived from a single parent quantity, the quantum relative entropy $D(\cdot\|\cdot)$. Generalizations of these entropic quantities can hence be defined by replacing the quantum relative entropy in Definition 2.2.6 with a generalized divergence such as the $\alpha$-SRD:

**Definition 3.3.1** (Rényi entropic quantities).
Let $\rho_{AB}$ be a bipartite state with marginal $\rho_A$ and let $\alpha \geq 0$.

(i) The *$\alpha$-Rényi entropy* $S_\alpha(\rho_A)$ is defined as

$$S_\alpha(\rho_A) := -\widetilde{D}_\alpha(\rho_A \| \mathbb{1}_A) = \frac{1}{1-\alpha} \log \left( \operatorname{tr} \rho_A^\alpha \right).$$

We use the notation $S_\alpha(A)_\rho \equiv S_\alpha(\rho_A)$.

(ii) The *$\alpha$-Rényi conditional entropy* $\widetilde{S}_\alpha(A|B)_\rho$ is defined as

$$\widetilde{S}_\alpha(A|B)_\rho := -\min_{\sigma_B} \widetilde{D}_\alpha(\rho_{AB} \| \mathbb{1}_A \otimes \sigma_B),$$

where the minimization is over states $\sigma_B$.

(iii) The *$\alpha$-Rényi mutual information* $\tilde{I}_\alpha(A; B)_\rho$ is defined as

$$\tilde{I}_\alpha(A; B)_\rho := \min_{\sigma_B} \widetilde{D}_\alpha(\rho_{AB} \| \rho_A \otimes \sigma_B),$$

where the minimization is over states $\sigma_B$.

Since $[\rho_A, \mathbb{1}_A] = 0$ for every $\rho_A$, the Rényi entropy in (i) can also be obtained from the $\alpha$-RRE, i.e., $S_\alpha(A)_\rho = -D_\alpha(\rho_A \| \mathbb{1}_A)$. This is the reason why we do not use a tilde notation for the Rényi entropy, as opposed to the Rényi conditional entropy and the Rényi mutual information, which were first defined and investigated in [MDS+13] and [GW15], respectively.

We note that the minimization over states $\sigma_B$ in Definition 3.3.1(ii) of the Rényi conditional entropy is needed so that this quantity satisfies the duality property (iv) of Proposition 3.3.2 below, which we use in the subsequent discussion. For the Rényi mutual information the minimization is usually included for similar reasons. While in the case of the usual quantum conditional entropy $S(A|B)_\rho$ (defined in Definition 2.2.6(ii)) we have

$$S(A|B)_\rho = -\min_{\sigma_B} D(\rho_{AB} \| \mathbb{1}_A \otimes \sigma_B) = -D(\rho_{AB} \| \mathbb{1}_A \otimes \rho_B),$$

the second equality is in general not true anymore when the QRE $D(\cdot \| \cdot)$ is replaced by the $\alpha$-SRD $\widetilde{D}_\alpha(\cdot \| \cdot)$ or the $\alpha$-RRE $D_\alpha(\cdot \| \cdot)$. See [TBH14] for a discussion of different versions of Rényi conditional entropies and how they are related to each other.

In contrast to conditional entropy and mutual information, a sensible definition of a Rényi generalization of the quantum conditional mutual information $I(A; B|C)_\rho$ is not as straight-forward. Berta et al. [BSW15] used an expression of $I(A; B|C)_\rho$ in terms of a single quantum relative entropy term derived in [LR73], together with a generalized Lie-Trotter product formula to define different versions of a Rényi conditional mutual information. In Theorem 3.3.6, we focus on the following definition from [BSW15] for a tripartite state $\rho_{ABC}$ and $\alpha > 0$:

$$\tilde{I}_\alpha(A; B|C)_\rho := \frac{2\alpha}{\alpha - 1} \log \left\| \rho_{ABC}^{1/2} \rho_{AC}^{(1-\alpha)/2\alpha} \rho_C^{(\alpha-1)/2\alpha} \rho_{BC}^{(1-\alpha)/2\alpha} \right\|_{2\alpha}. \tag{3.11}$$

The Rényi entropic quantities from Definition 3.3.1 satisfy the following properties:

**Proposition 3.3.2** (Properties of Rényi entropic quantities).
*Unless specified otherwise, let $\alpha \geq 0$ and $\rho_{AB}$ be a state with marginal $\rho_A$.*

(i) *Positivity and dimension bound: We have $0 \leq S_\alpha(A)_\rho \leq \log |A|$ for all $\alpha \geq 0$. The extremal values are achieved for pure states and completely mixed states, respectively.*

(ii) *Additivity: Let $\sigma_{A'B'}$ be another bipartite state with marginal $\sigma_{A'}$. Then we have*

$$S_\alpha(AA')_{\rho \otimes \sigma} = S_\alpha(A)_\rho + S_\alpha(A')_\sigma.$$

*Furthermore, additivity also holds for the Rényi conditional entropy and the Rényi mutual information for $\alpha \geq 1/2$:*

$$\widetilde{S}_\alpha(AA'|BB')_{\rho \otimes \sigma} = \widetilde{S}_\alpha(A|B)_\rho + \widetilde{S}_\alpha(A'|B')_\sigma$$
$$\tilde{I}_\alpha(AA'; BB')_{\rho \otimes \sigma} = \tilde{I}_\alpha(A; B)_\rho + \tilde{I}_\alpha(A'; B')_\sigma.$$

(iii) *Duality for the Rényi entropy: Let $|\rho\rangle_{AB}$ be a pure state with marginals $\rho_A$ and $\rho_B$, then $S_\alpha(A)_\rho = S_\alpha(B)_\rho$ for all $\alpha \geq 0$.*

(iv) *Duality for the Rényi conditional entropy: Let $|\rho\rangle_{ABC}$ be a pure state with marginals $\rho_{AB}$ and $\rho_{AC}$, and for $\alpha \geq 1/2$ define $\beta$ through $1/\alpha + 1/\beta = 2$. Then*

$$\widetilde{S}_\alpha(A|B)_\rho = -\widetilde{S}_\beta(A|C)_\rho.$$

(v) *Let $\Lambda \colon B \to C$ be a quantum operation, and define $\sigma_{AC} := (\mathrm{id}_A \otimes \Lambda)(\rho_{AB})$. For $\alpha \geq 1/2$, we then have*

$$\widetilde{S}_\alpha(A|B)_\rho \leq \widetilde{S}_\alpha(A|C)_\sigma \qquad\qquad \tilde{I}_\alpha(A; B)_\rho \geq \tilde{I}_\alpha(A; C)_\sigma.$$

(vi) *We have* $\lim_{\alpha \to 1} S_\alpha(A)_\rho = S(A)_\rho$, *and*

$$\lim_{\alpha \to 1} \widetilde{S}_\alpha(A|B)_\rho = S(A|B)_\rho \qquad\qquad \lim_{\alpha \to 1} \widetilde{I}_\alpha(A;B)_\rho = I(A;B)_\rho.$$

*Proof.* (i) can be proved using the method of Lagrange multipliers. The assertions for the Rényi conditional entropy and the Rényi mutual information in (ii) were proved in [Bei13; HT16]. Since the Rényi entropy $S_\alpha(\rho)$ only depends on the eigenvalues of $\rho$, (iii) follows from Schmidt decomposition (Theorem 2.2.1). (iv) was proved independently by Müller-Lennert et al. [MDS+13] and Beigi [Bei13], and (v) is an immediate consequence of the DPI for the $\alpha$-SRD, Proposition 3.1.2(vi). In (vi), the limit for the Rényi entropy follows immediately from its definition and Proposition 3.2.1. Lemma 8 in [CMW14] proves the assertion for the Rényi mutual information, and a straightforward adaption of their argument also proves the one for the Rényi conditional entropy. □

### 3.3.2. Dimension bounds and a useful fidelity bound

In this section we prove additional properties for the Rényi entropic quantities introduced in Definition 3.3.1, further advancing the 'Rényi entropic calculus'. The material of this section is taken from [LWD16, Section 2].

We start with a subadditivity property for the Rényi entropies first proved in [vDH02], for which we give a simplified proof based on the data processing inequality.

**Lemma 3.3.3** (Subadditivity of Rényi entropies [vDH02])**.**
*If $\alpha \geq 0$ and $\rho_{AB} \in \mathcal{D}(\mathcal{H}_{AB})$, then*

$$S_\alpha(A)_\rho - \log|B| \leq S_\alpha(AB)_\rho \leq S_\alpha(A)_\rho + \log|B|.$$

*Proof.* To prove the upper bound on $S_\alpha(AB)_\rho$, observe first that

$$S_\alpha(AB)_\rho = -\widetilde{D}_\alpha(\rho_{AB}\|\mathbb{1}_A \otimes \pi_B) + \log|B| \qquad (3.12)$$

$$= -D_\alpha(\rho_{AB}\|\mathbb{1}_A \otimes \pi_B) + \log|B|. \qquad (3.13)$$

Assuming that $\alpha \geq 1/2$ and using (3.12), we then have

$$S_\alpha(AB)_\rho = -\widetilde{D}_\alpha(\rho_{AB}\|\mathbb{1}_A \otimes \pi_B) + \log|B|$$

$$\leq -\widetilde{D}_\alpha(\rho_A\|\mathbb{1}_A) + \log|B|$$

$$= S_\alpha(A)_\rho + \log|B|,$$

where the inequality follows from the data processing inequality with respect to partial trace over $B$, Proposition 3.1.2(vi). If $\alpha \in [0, 1/2)$, we use relation (3.13) together with the data processing inequality (3.7) for the $\alpha$-Rényi relative entropy instead. The lower bound on $S_\alpha(AB)_\rho$ follows from the upper bound using duality for the Rényi entropies, Proposition 3.3.2(iii), as discussed in [vDH02]. □

Note that both inequalities in Lemma 3.3.3 can be tightened by replacing the log terms with the 0-Rényi entropy $S_0(B)_\rho = \log \mathrm{rk}\, \rho_B$. However, if $\mathcal{H}_B$ is restricted to the support of $\rho_B$, we have $S_0(B)_\rho = \log|B|$.

We proceed with a result concerning dimension bounds on the Rényi conditional entropy and mutual information, as well as invariance properties with respect to tensor product states, which we state in Proposition 3.3.5 below. To prove it, we use the following result from [MDS+13] concerning the Rényi conditional entropy:

**Lemma 3.3.4** ([MDS+13])**.** *For $\alpha > 0$ and $\rho_{ABC} \in \mathcal{D}(\mathcal{H}_{ABC})$, we have*

$$\widetilde{S}_\alpha(A|BC)_\rho \geq \widetilde{S}_\alpha(AC|B)_\rho - \log|C|.$$

Note that the case $\alpha = 1$ is not explicitly proved in [MDS+13], but follows in the exact same way. We can now prove:

**Proposition 3.3.5.** *Let $\alpha \in [1/2, \infty)$.*

(i) *For an arbitrary tripartite state $\rho_{ABC}$ we have*

$$\widetilde{S}_\alpha(A|BC)_\rho + 2\log|C| \geq \widetilde{S}_\alpha(A|B)_\rho \tag{3.14}$$

$$\tilde{I}_\alpha(A;B)_\rho + 2\log|C| \geq \tilde{I}_\alpha(A;BC)_\rho. \tag{3.15}$$

(ii) *For states $\rho_{AB}$ and $\sigma_C$, we have*

$$\widetilde{S}_\alpha(A|BC)_{\rho\otimes\sigma} = \widetilde{S}_\alpha(A|B)_\rho \tag{3.16}$$

$$\tilde{I}_\alpha(A;BC)_{\rho\otimes\sigma} = \tilde{I}_\alpha(A;B)_\rho. \tag{3.17}$$

*Proof.* We first prove (3.14). By Lemma 3.3.4, we have

$$\widetilde{S}_\alpha(A|BC)_\rho \geq \widetilde{S}_\alpha(AC|B)_\rho - \log|C|. \tag{3.18}$$

From duality for the Rényi conditional entropy, Proposition 3.3.2(iv), we obtain that

$$\widetilde{S}_\alpha(AC|B)_\rho = -\widetilde{S}_\beta(AC|D)_\rho,$$

where the state $|\rho\rangle_{ABCD}$ purifies $\rho_{ABC}$, and $\beta$ is such that $1/\alpha + 1/\beta = 2$. By the same reasoning, we find that

$$\widetilde{S}_\beta(A|CD)_\rho \geq \widetilde{S}_\beta(AC|D)_\rho - \log|C|.$$

But from duality, this is the same as

$$\widetilde{S}_\alpha(A|B)_\rho - \log|C| \leq \widetilde{S}_\alpha(AC|B)_\rho.$$

Substituting this in (3.18) then yields the claim.

To prove (3.15), consider the following steps:

$$
\begin{aligned}
\tilde{I}_\alpha(A;CB)_\rho &= \min_{\tau_{CB}} \frac{\alpha}{\alpha-1} \log \left\| \left(\rho_A^{(1-\alpha)/2\alpha} \otimes \tau_{CB}^{(1-\alpha)/2\alpha}\right) \rho_{ABC} \left(\rho_A^{(1-\alpha)/2\alpha} \otimes \tau_{CB}^{(1-\alpha)/2\alpha}\right) \right\|_\alpha \\
&= \min_{\tau_{CB}} \frac{\alpha}{\alpha-1} \log \left\| \tau_{CB}^{(1-\alpha)/2\alpha} \tilde{\rho}_{ABC} \tau_{CB}^{(1-\alpha)/2\alpha} \right\|_\alpha + \frac{\alpha}{\alpha-1} \log \operatorname{tr}\!\left(\rho_A^{(1-\alpha)/\alpha} \rho_{AB}\right) \\
&= -\widetilde{S}_\alpha(A|CB)_{\tilde{\rho}} + \frac{\alpha}{\alpha-1} \log \operatorname{tr}\!\left(\rho_A^{(1-\alpha)/\alpha} \rho_{AB}\right),
\end{aligned}
\tag{3.19}
$$

where we defined the density operator

$$\tilde{\rho}_{ABC} := \frac{1}{\operatorname{tr}\!\left(\rho_A^{(1-\alpha)/\alpha} \rho_{ABC}\right)} \rho_A^{(1-\alpha)/2\alpha} \rho_{ABC} \rho_A^{(1-\alpha)/2\alpha} \tag{3.20}$$

and observed that

$$\operatorname{tr}\!\left(\rho_A^{(1-\alpha)/\alpha} \rho_{ABC}\right) = \operatorname{tr}\!\left(\rho_A^{(1-\alpha)/\alpha} \rho_{AB}\right).$$

Furthermore,

$$\tilde{\rho}_{AB} = \operatorname{tr}_C(\tilde{\rho}_{ABC}) = \frac{1}{\operatorname{tr}\!\left(\rho_A^{(1-\alpha)/\alpha} \rho_{AB}\right)} \rho_A^{(1-\alpha)/2\alpha} \rho_{AB} \rho_A^{(1-\alpha)/2\alpha}.$$

By (3.14), we have

$$-\widetilde{S}_\alpha(A|CB)_{\tilde{\rho}} \leq -\widetilde{S}_\alpha(A|B)_{\tilde{\rho}} + 2\log|C|,$$

and using this in (3.19) yields

$$-\widetilde{S}_\alpha(A|CB)_{\tilde{\rho}} + \frac{\alpha}{\alpha-1} \log \operatorname{tr}\left(\rho_A^{(1-\alpha)/\alpha} \rho_{AB}\right) \leq -\widetilde{S}_\alpha(A|B)_{\tilde{\rho}} + \frac{\alpha}{\alpha-1} \log \operatorname{tr}\left(\rho_A^{(1-\alpha)/\alpha} \rho_{AB}\right) + 2 \log |C|$$
$$= \tilde{I}_\alpha(A;B)_\rho + 2 \log |C|.$$

Thus, we arrive at (3.15).

We continue with the proof of (3.16). From the data processing inequality, Proposition 3.3.2(v), we know that

$$\widetilde{S}_\alpha(A|BC)_{\rho \otimes \sigma} \leq \widetilde{S}_\alpha(A|B)_\rho.$$

On the other hand, we have for all $\theta_B \in \mathcal{D}(\mathcal{H}_B)$ that

$$-\widetilde{S}_\alpha(A|BC)_{\rho \otimes \sigma} = \min_{\tau_{BC}} \widetilde{D}_\alpha(\rho_{AB} \otimes \sigma_C \| \mathbb{1}_A \otimes \tau_{BC})$$
$$\leq \widetilde{D}_\alpha(\rho_{AB} \otimes \sigma_C \| \mathbb{1}_A \otimes \theta_B \otimes \sigma_C)$$
$$= \widetilde{D}_\alpha(\rho_{AB} \| \mathbb{1}_A \otimes \theta_B).$$

Since the inequality holds for all $\theta_B \in \mathcal{D}(\mathcal{H}_B)$, we get that

$$-\widetilde{S}_\alpha(A|BC)_{\rho \otimes \sigma} \leq -\widetilde{S}_\alpha(A|B)_\rho,$$

which yields (3.16).

Finally, we prove (3.17). We know from the data processing inequality, Proposition 3.3.2(v), that

$$\tilde{I}_\alpha(A;BC)_{\rho \otimes \sigma} \geq \tilde{I}_\alpha(A;B)_\rho.$$

On the other hand, we have

$$\tilde{I}_\alpha(A;BC)_{\rho \otimes \sigma} = \min_{\tau_{BC}} \widetilde{D}_\alpha(\rho_{AB} \otimes \sigma_C \| \rho_A \otimes \tau_{BC})$$
$$\leq \widetilde{D}_\alpha(\rho_{AB} \otimes \sigma_C \| \rho_A \otimes \theta_B \otimes \sigma_C)$$
$$= \widetilde{D}_\alpha(\rho_{AB} \| \rho_A \otimes \theta_B).$$

Since the inequality holds for all $\theta_B$, we get that

$$\tilde{I}_\alpha(A;BC)_{\rho \otimes \sigma} \leq \tilde{I}_\alpha(A;B)_\rho,$$

which concludes the proof.                                                             □

The following result, Theorem 3.3.6, is crucial for proving the strong converse theorems in Chapter 6. It bounds the difference of Rényi entropic quantities of two quantum states in terms of their fidelity. Note that the inequality in (i) for the Rényi entropies was originally proved in [vDH02] employing the 'classical' Hölder inequality for the $p$-(vector) norm. We give a different proof of the inequality (i) based on the Hölder inequality for the Schatten $p$-norm (Theorem 2.3.2). The advantage of this method is that it can be used to obtain similar fidelity bounds for the Rényi conditional entropy, the Rényi mutual information, and the Rényi conditional mutual information, stated in the assertions (ii)–(iv) of Theorem 3.3.6 below.

**Theorem 3.3.6** (Fidelity bounds for Rényi entropic quantities)**.**
*Let $\alpha \in (1/2, 1)$ and define $\beta \equiv \beta(\alpha) \coloneqq \alpha/(2\alpha - 1)$. Then the following inequalities hold:*

(i) *For $\rho_A, \sigma_A \in \mathcal{D}(\mathcal{H}_A)$,*

$$S_\alpha(A)_\rho - S_\beta(A)_\sigma \geq \frac{2\alpha}{1 - \alpha} \log F(\rho_A, \sigma_A).$$

(ii) *For $\rho_{AB}, \sigma_{AB} \in \mathcal{D}(\mathcal{H}_{AB})$,*

$$\widetilde{S}_\alpha(A|B)_\rho - \widetilde{S}_\beta(A|B)_\sigma \geq \frac{2\alpha}{1 - \alpha} \log F(\rho_{AB}, \sigma_{AB}).$$

(iii) *For $\rho_{AB}, \sigma_{AB} \in \mathcal{D}(\mathcal{H}_{AB})$ with $\rho_A = \sigma_A$,*

$$\tilde{I}_\beta(A; B)_\rho - \tilde{I}_\alpha(A; B)_\sigma \geq \frac{2\alpha}{1 - \alpha} \log F(\rho_{AB}, \sigma_{AB}).$$

(iv) *Let $\rho_{ABC}, \sigma_{ABC} \in \mathcal{D}(\mathcal{H}_{ABC})$ with $\rho_{AC} = \sigma_{AC}$, $\rho_{BC} = \sigma_{BC}$, and $\rho_C = \sigma_C$, and assume that $\rho_{BC}$ has full support. Then,*

$$\tilde{I}_\beta(A; B|C)_\rho - \tilde{I}_\alpha(A; B|C)_\sigma \geq \frac{2\alpha}{1 - \alpha} \log F(\rho_{ABC}, \sigma_{ABC}).$$

*Proof.* We first observe that (i) follows from (ii) by setting $\mathcal{H}_B = \mathbb{C}$. To prove (ii), let $\tau_B$ be an arbitrary density operator, and let $\varepsilon \in (0, 1)$. Furthermore, let

$$\tau(\varepsilon)_B \coloneqq (1 - \varepsilon)\tau_B + \varepsilon\pi_B, \tag{3.21}$$

and recall that

$$\widetilde{D}_\alpha(\omega\|\theta) \leq \widetilde{D}_\alpha(\omega\|\theta') \tag{3.22}$$

holds for all $\alpha \in [1/2, \infty]$ and $\theta \geq \theta' \geq 0$ (Proposition 3.1.2(vii)). For $c > 0$, we have

$$\widetilde{D}_\alpha(\omega\|c\theta) = \widetilde{D}_\alpha(\omega\|\theta) - \log c. \tag{3.23}$$

Consider then the following chain of inequalities:

$$\begin{aligned}
-\widetilde{D}_\alpha(\rho_{AB}\|\mathbb{1}_A \otimes \tau(\varepsilon)_B) &+ \widetilde{D}_\beta(\sigma_{AB}\|\mathbb{1}_A \otimes \tau_B) - \log(1-\varepsilon) \\
&= -\widetilde{D}_\alpha(\rho_{AB}\|\mathbb{1}_A \otimes \tau(\varepsilon)_B) + \widetilde{D}_\beta(\sigma_{AB}\|\mathbb{1}_A \otimes (1-\varepsilon)\,\tau_B) \\
&\geq -\widetilde{D}_\alpha(\rho_{AB}\|\mathbb{1}_A \otimes \tau(\varepsilon)_B) + \widetilde{D}_\beta(\sigma_{AB}\|\mathbb{1}_A \otimes \tau(\varepsilon)_B) \\
&= \frac{2\alpha}{1-\alpha}\log\left\|\rho_{AB}^{1/2}\tau(\varepsilon)_B^{(1-\alpha)/2\alpha}\right\|_{2\alpha} + \frac{2\beta}{\beta-1}\log\left\|\tau(\varepsilon)_B^{(1-\beta)/2\beta}\sigma_{AB}^{1/2}\right\|_{2\beta} \\
&= \frac{2\alpha}{1-\alpha}\log\left(\left\|\rho_{AB}^{1/2}\tau(\varepsilon)_B^{(1-\alpha)/2\alpha}\right\|_{2\alpha}\left\|\tau(\varepsilon)_B^{(1-\beta)/2\beta}\sigma_{AB}^{1/2}\right\|_{2\beta}\right) \\
&\geq \frac{2\alpha}{1-\alpha}\log\left\|\rho_{AB}^{1/2}\tau(\varepsilon)_B^{(1-\alpha)/2\alpha}\tau(\varepsilon)_B^{(1-\beta)/2\beta}\sigma_{AB}^{1/2}\right\|_1 \\
&= \frac{2\alpha}{1-\alpha}\log\left\|\rho_{AB}^{1/2}\sigma_{AB}^{1/2}\right\|_1 \\
&= \frac{2\alpha}{1-\alpha}\log F(\rho_{AB}, \sigma_{AB}). \tag{3.24}
\end{aligned}$$

The first equality is an application of (3.23). The first inequality is a consequence of (3.22) and the fact that

$$(1-\varepsilon)\,\tau_B \leq (1-\varepsilon)\,\tau_B + \varepsilon\pi_B = \tau(\varepsilon)_B.$$

The second and third equalities follow from the definition of the $\alpha$-SRD, Definition 3.1.1, and the relation

$$\frac{\beta}{\beta-1} = \frac{\alpha}{1-\alpha}. \tag{3.25}$$

The second inequality is an application of Hölder's inequality, Theorem 2.3.2(i), together with the fact that $1/2\alpha + 1/2\beta = 1$. The penultimate equality follows because $\tau(\varepsilon)_B$ is a full rank operator for $\varepsilon \in (0, 1)$, such that $\tau(\varepsilon)_B^{(1-\alpha)/2\alpha}\tau(\varepsilon)_B^{(1-\beta)/2\beta} = \mathbb{1}_B$.

Since (3.24) holds for an arbitrary density operator $\tau_B$, we may choose $\tau_B$ to be the optimizing state for $\widetilde{S}_\beta(A|B)_\sigma$. We can then continue from (3.24) as

$$
\begin{aligned}
\frac{2\alpha}{1-\alpha} \log F\left(\rho_{AB}, \sigma_{AB}\right) &\leq -\widetilde{D}_\alpha(\rho_{AB} \| \mathbb{1}_A \otimes \tau(\varepsilon)_B) + \widetilde{D}_\beta(\sigma_{AB} \| \mathbb{1}_A \otimes \tau_B) - \log\left(1-\varepsilon\right) \\
&\leq \max_{\omega_B \in \mathcal{D}(\mathcal{H}_B)} \left\{ -\widetilde{D}_\alpha(\rho_{AB} \| \mathbb{1}_A \otimes \omega_B) \right\} + \widetilde{D}_\beta(\sigma_{AB} \| \mathbb{1}_A \otimes \tau_B) - \log\left(1-\varepsilon\right) \\
&= \widetilde{S}_\alpha(A|B)_\rho - \widetilde{S}_\beta(A|B)_\sigma - \log\left(1-\varepsilon\right).
\end{aligned}
$$

We have shown that this relation holds for all $\varepsilon \in (0,1)$, and hence taking the limit $\varepsilon \searrow 0$ yields the claim.

We continue with the proof of (iii). For an arbitrary density operator $\tau_B$ and $\varepsilon \in (0,1)$ define the state $\tau(\varepsilon)_B$ as in (3.21) in the proof of (ii). We then have the following chain of inequalities:

$$
\begin{aligned}
\widetilde{D}_\beta&(\rho_{AB} \| \rho_A \otimes \tau_B) - \widetilde{D}_\alpha(\sigma_{AB} \| \sigma_A \otimes \tau(\varepsilon)_B) - \log(1-\varepsilon) \\
&= \widetilde{D}_\beta(\rho_{AB} \| \rho_A \otimes (1-\varepsilon)\tau_B) - \widetilde{D}_\alpha(\sigma_{AB} \| \sigma_A \otimes \tau(\varepsilon)_B) \\
&\geq \widetilde{D}_\beta(\rho_{AB} \| \rho_A \otimes \tau(\varepsilon)_B) - \widetilde{D}_\alpha(\sigma_{AB} \| \sigma_A \otimes \tau(\varepsilon)_B) \\
&= \frac{2\beta}{\beta-1} \log \left\| \rho_{AB}^{1/2} \left( \rho_A^{(1-\beta)/2\beta} \otimes \tau(\varepsilon)_B^{(1-\beta)/2\beta} \right) \right\|_{2\beta} \\
&\qquad - \frac{2\alpha}{\alpha-1} \log \left\| \left( \sigma_A^{(1-\alpha)/2\alpha} \otimes \tau(\varepsilon)_B^{(1-\alpha)/2\alpha} \right) \sigma_{AB}^{1/2} \right\|_{2\alpha} \\
&= \frac{2\alpha}{1-\alpha} \log \left( \left\| \rho_{AB}^{1/2} \left( \rho_A^{(1-\beta)/2\beta} \otimes \tau(\varepsilon)_B^{(1-\beta)/2\beta} \right) \right\|_{2\beta} \left\| \left( \sigma_A^{(1-\alpha)/2\alpha} \otimes \tau(\varepsilon)_B^{(1-\alpha)/2\alpha} \right) \sigma_{AB}^{1/2} \right\|_{2\alpha} \right) \\
&\geq \frac{2\alpha}{1-\alpha} \log \left\| \rho_{AB}^{1/2} \left( \rho_A^{(1-\beta)/2\beta} \otimes \tau(\varepsilon)_B^{(1-\beta)/2\beta} \right) \left( \sigma_A^{(1-\alpha)/2\alpha} \otimes \tau(\varepsilon)_B^{(1-\alpha)/2\alpha} \right) \sigma_{AB}^{1/2} \right\|_1 \\
&= \frac{2\alpha}{1-\alpha} \log \left\| \rho_{AB}^{1/2} \left( \rho_A^{(1-\beta)/2\beta} \sigma_A^{(1-\alpha)/2\alpha} \otimes \tau(\varepsilon)_B^{(1-\beta)/2\beta} \tau(\varepsilon)_B^{(1-\alpha)/2\alpha} \right) \sigma_{AB}^{1/2} \right\|_1 \\
&= \frac{2\alpha}{1-\alpha} \log \left\| \rho_{AB}^{1/2} \sigma_{AB}^{1/2} \right\|_1 \\
&= \frac{2\alpha}{1-\alpha} \log F(\rho_{AB}, \sigma_{AB}).
\end{aligned}
$$

In the first equality and inequality we used (3.23) and (3.22), respectively. The following equalities follow from the definition of the $\alpha$-SRD and (3.25). In the second inequality we applied Hölder's inequality, Theorem 2.3.2(i). For the second-to-last equality we used the fact that $\rho_A = \sigma_A$ by assumption, and that $\tau(\varepsilon)_B$ has full support for $\varepsilon \in (0,1)$, such that $\tau(\varepsilon)_B^{-1} \tau(\varepsilon)_B = \mathbb{1}_B$.

Choosing $\tau_B$ to be the optimizing state for $\tilde{I}_\beta(A;B)_\rho$, we then have

$$\frac{2\alpha}{1-\alpha} \log F(\rho_{AB}, \sigma_{AB}) \leq \widetilde{D}_\beta(\rho_{AB} \| \rho_A \otimes \tau_B) - \widetilde{D}_\alpha(\sigma_{AB} \| \sigma_A \otimes \tau(\varepsilon)_B) - \log(1-\varepsilon)$$

$$\leq \widetilde{D}_\beta(\rho_{AB} \| \rho_A \otimes \tau_B) - \min_{\omega_B \in \mathcal{D}(\mathcal{H}_B)} \widetilde{D}_\alpha(\sigma_{AB} \| \sigma_A \otimes \omega_B) - \log(1-\varepsilon)$$

$$= \tilde{I}_\beta(A;B)_\rho - \tilde{I}_\alpha(A;B)_\sigma - \log(1-\varepsilon).$$

Since this relation holds for all $\varepsilon \in (0, 1)$, we obtain the claim by taking the limit $\varepsilon \searrow 0$.

To prove (iv), we can rewrite the definition (3.11) of $\tilde{I}_\beta(A;B|C)_\rho$ in Section 3.3.1 as

$$\tilde{I}_\beta(A;B|C)_\rho = -\frac{2\alpha}{\alpha-1} \log \left\| \rho_{BC}^{(1-\beta)/2\beta} \rho_C^{(\beta-1)/2\beta} \rho_{AC}^{(1-\beta)/2\beta} \rho_{ABC}^{1/2} \right\|_{2\beta}.$$

Consider then

$$\frac{1-\alpha}{2\alpha} \left( \tilde{I}_\beta(A;B|C)_\rho - \tilde{I}_\alpha(A;B|C)_\sigma \right)$$

$$= \log \left( \left\| \sigma_{ABC}^{1/2} \rho_{AC}^{(1-\alpha)/2\alpha} \rho_C^{(\alpha-1)/2\alpha} \rho_{BC}^{(1-\alpha)/2\alpha} \right\|_{2\alpha} \left\| \rho_{BC}^{(1-\beta)/2\beta} \rho_C^{(\beta-1)/2\beta} \rho_{AC}^{(1-\beta)/2\beta} \rho_{ABC}^{1/2} \right\|_{2\beta} \right)$$

$$\geq \log \left\| \sigma_{ABC}^{1/2} \rho_{AC}^{(1-\alpha)/2\alpha} \rho_C^{(\alpha-1)/2\alpha} \rho_{BC}^{(1-\alpha)/2\alpha} \rho_{BC}^{(1-\beta)/2\beta} \rho_C^{(\beta-1)/2\beta} \rho_{AC}^{(1-\beta)/2\beta} \rho_{ABC}^{1/2} \right\|_1$$

$$= \log \left\| \sigma_{ABC}^{1/2} \rho_{ABC}^{1/2} \right\|_1$$

$$= \log F(\rho_{ABC}, \sigma_{ABC}),$$

where we used the assumptions $\rho_{AC} = \sigma_{AC}$, $\rho_{BC} = \sigma_{BC}$, $\rho_C = \sigma_C$, and the fact that $\rho_{BC}$ has full support in the penultimate equality. This concludes the proof. $\qquad\square$

The last result of this section concerns the Rényi conditional entropy of c-q states. Note that a special case of Proposition 3.3.7(i) for Rényi entropies (i.e., where the system $B$ is trivial) appeared in [Sha14].

**Proposition 3.3.7.** *Let $\rho_{ABX} = \sum_{x \in \mathcal{X}} p_x \rho_{AB}^x \otimes |x\rangle\langle x|_X$ be a c-q state. Then the following properties hold for all $\alpha > 0$:*

(i) *Monotonicity under discarding classical information:*

$$\widetilde{S}_\alpha(AX|B)_\rho \geq \widetilde{S}_\alpha(A|B)_\rho$$

(ii) *Dimension bounds:*

$$\widetilde{S}_\alpha(A|BX)_\rho + \log|X| \geq \widetilde{S}_\alpha(A|B)_\rho \tag{3.26}$$

$$\tilde{I}_\alpha(A;BX)_\rho \leq \log|X| + \tilde{I}_\alpha(A;B)_\rho \tag{3.27}$$

*Proof.* To prove (i), let $\tau_B$ be the optimizing state for $\widetilde{S}_\alpha(A|B)_\rho$, and assume $\alpha \neq 1$. We then have

$$
\begin{aligned}
\widetilde{S}_\alpha(AX|B)_\rho &\geq \frac{1}{1-\alpha} \log \operatorname{tr} \left( \tau_B^{(1-\alpha)/2\alpha} \rho_{ABX} \tau_B^{(1-\alpha)/2\alpha} \right)^\alpha \\
&= \frac{1}{1-\alpha} \log \left( \sum_{x \in \mathcal{X}} \operatorname{tr} \left( \tau_B^{(1-\alpha)/2\alpha} p_x \rho_{AB}^x \tau_B^{(1-\alpha)/2\alpha} \right)^\alpha \right) \\
&= \frac{1}{1-\alpha} \log \left( \sum_{x \in \mathcal{X}} \left\| \tau_B^{(1-\alpha)/2\alpha} p_x \rho_{AB}^x \tau_B^{(1-\alpha)/2\alpha} \right\|_\alpha^\alpha \right) \\
&\geq \frac{1}{1-\alpha} \log \left\| \tau_B^{(1-\alpha)/2\alpha} \sum_{x \in \mathcal{X}} p_x \rho_{AB}^x \tau_B^{(1-\alpha)/2\alpha} \right\|_\alpha^\alpha \\
&= \frac{1}{1-\alpha} \log \left\| \tau_B^{(1-\alpha)/2\alpha} \rho_{AB} \tau_B^{(1-\alpha)/2\alpha} \right\|_\alpha^\alpha \\
&= \widetilde{S}_\alpha(A|B)_\rho,
\end{aligned}
$$

where the first inequality follows from the definition of the Rényi conditional entropy, and the second inequality follows from McCarthy's inequalities (Theorem 2.3.2(ii)), using (2.8) for $\alpha < 1$ and (2.9) for $\alpha > 1$. For $\alpha = 1$, the claim follows easily from Definition 2.2.5.

To prove (3.26), observe that we have

$$\widetilde{S}_\alpha(A|BX)_\rho + \log|X| \geq \widetilde{S}_\alpha(AX|B)_\rho \geq \widetilde{S}_\alpha(A|B)_\rho,$$

where the first inequality follows from Lemma 3.3.4, and the second inequality is (i).

Finally, to prove (3.27) observe that

$$
\begin{aligned}
\tilde{I}_\alpha(A;BX)_\rho &= -\widetilde{S}_\alpha(A|BX)_{\tilde\rho} + \frac{\alpha}{\alpha-1} \log \operatorname{tr} \left( \rho_A^{(1-\alpha)/\alpha} \rho_{AB} \right) \\
&\leq -\widetilde{S}_\alpha(A|B)_{\tilde\rho} + \frac{\alpha}{\alpha-1} \log \operatorname{tr} \left( \rho_A^{(1-\alpha)/\alpha} \rho_{AB} \right) + \log|X| \\
&= \tilde{I}_\alpha(A;B)_\rho + \log|X|,
\end{aligned}
$$

where the first equality is just (3.19) with $\tilde\rho_{ABX}$ as defined in (3.20), and the inequality uses (3.26), which concludes the proof. □

# 3.4. Data processing for the SRD: a condition for equality

As stated in Proposition 3.1.2(vi) in Section 3.1, the $\alpha$-SRD satisfies the DPI, that is, for $\alpha \in [1/2, \infty)$, $\rho \in \mathcal{D}(\mathcal{H})$, $\sigma \in \mathcal{P}(\mathcal{H})$, and a quantum operation $\Lambda \colon \mathcal{B}(\mathcal{H}) \to \mathcal{H}(\mathcal{K})$,

$$\widetilde{D}_\alpha(\rho \| \sigma) \geq \widetilde{D}_\alpha(\Lambda(\rho) \| \Lambda(\sigma)). \tag{3.28}$$

In this section, we derive a necessary and sufficient condition for equality in (3.28), which is stated in Theorem 3.4.1 below and appeared in [LRD16]. Throughout this section, we set

$$\gamma := \frac{1-\alpha}{2\alpha}.$$

**Theorem 3.4.1** (Equality condition in the DPI for the $\alpha$-SRD)**.**
*Let $\alpha \in [1/2, 1) \cup (1, \infty)$ and set $\gamma = (1-\alpha)/2\alpha$. Furthermore, let $\rho \in \mathcal{D}(\mathcal{H})$ and $\sigma \in \mathcal{P}(\mathcal{H})$ with* $\operatorname{supp} \rho \subseteq \operatorname{supp} \sigma$ *if $\alpha > 1$ or $\rho \not\perp \sigma$ if $\alpha \in [1/2, 1)$, and let $\Lambda \colon \mathcal{B}(\mathcal{H}) \to \mathcal{B}(\mathcal{K})$ be a quantum operation. We have equality in the data processing inequality* (3.28)*,*

$$\widetilde{D}_\alpha(\rho \| \sigma) = \widetilde{D}_\alpha(\Lambda(\rho) \| \Lambda(\sigma)),$$

*if and only if*

$$\sigma^\gamma \left(\sigma^\gamma \rho \sigma^\gamma\right)^{\alpha-1} \sigma^\gamma = \Lambda^\dagger \left(\Lambda(\sigma)^\gamma \left(\Lambda(\sigma)^\gamma \Lambda(\rho) \Lambda(\sigma)^\gamma\right)^{\alpha-1} \Lambda(\sigma)^\gamma\right).$$

For $\alpha > 1$ and *positive* trace-preserving maps, Theorem 3.4.1 was also proved using the framework of non-commutative $L_p$-spaces [DJW15]. Hiai and Mosonyi [HM16] and Jenčová [Jen16] discussed the case of equality in the DPI for the $\alpha$-SRD in the light of *sufficiency*. We discuss the notion of sufficiency as well as the connection between Theorem 3.4.1 and the results of [HM16; Jen16] in Section 3.4.3.

The rest of this section is organized as follows: In Section 3.4.1, we recall the proof of the DPI (3.28) for the $\alpha$-SRD as given in [FL13]. We then analyze this proof in greater detail in Section 3.4.2, extracting a necessary and sufficient condition for equality in (3.28) and thus proving Theorem 3.4.1. In Section 3.4.3 we compare our result to the sufficiency results in [HM16; Jen16] mentioned above. Finally, we present applications of Theorem 3.4.1 to bounds on entanglement and distance measures in Section 3.4.4.

### 3.4.1. Review of the proof of the DPI by Frank and Lieb

For the remainder of this section we will assume that $\rho \in \mathcal{D}(\mathcal{H})$ and $\sigma \in \mathcal{P}(\mathcal{H})$ with supp $\rho \subseteq$ supp $\sigma$ if $\alpha > 1$, or $\rho \not\perp \sigma$ if $\alpha \in [1/2, 1)$. First, we note that the DPI (3.28) for the $\alpha$-SRD is equivalent to monotonicity under partial trace, which can be seen as follows. Given a quantum operation $\Lambda \colon \mathcal{B}(\mathcal{H}) \to \mathcal{B}(\mathcal{K})$, the Stinespring Representation Theorem (Theorem 2.2.2) states that there is a Hilbert space $\mathcal{H}'$, a pure state $|\tau\rangle \in \mathcal{H}' \otimes \mathcal{K}$, and a unitary $U$ acting on $\mathcal{H} \otimes \mathcal{H}' \otimes \mathcal{K}$, such that $\Lambda$ acts on $\rho \in \mathcal{B}(\mathcal{H})$ as

$$\Lambda(\rho) = \mathrm{tr}_{12}\big(U(\rho \otimes \tau)U^\dagger\big). \tag{3.29}$$

Here, $\mathrm{tr}_{12}$ denotes the partial trace over $\mathcal{H}$ and $\mathcal{H}'$, that is, the first two tensor factors of $\mathcal{H} \otimes \mathcal{H}' \otimes \mathcal{K}$. As both $\widetilde{Q}_\alpha(\cdot\|\cdot)$ and $\widetilde{D}_\alpha(\cdot\|\cdot)$ are invariant under tensoring with a fixed state (Proposition 3.1.2(iv)) and isometries (Proposition 3.1.2(v)), the representation (3.29) implies that the DPI (3.28) is equivalent to monotonicity of the $\alpha$-SRD under partial trace:

$$\widetilde{D}_\alpha(\rho_{AB}\|\sigma_{AB}) \geq \widetilde{D}_\alpha(\rho_A\|\sigma_A), \tag{3.30}$$

where the subscripts $AB$ and $A$ indicate (generic) Hilbert spaces $\mathcal{H}_{AB} = \mathcal{H}_A \otimes \mathcal{H}_B$ and $\mathcal{H}_A$ on which the density matrices act, and the partial trace is taken over the $B$ system. Since the logarithm is monotonically increasing, the monotonicity of $\widetilde{D}_\alpha(\cdot\|\cdot)$ under partial trace (3.30) is in turn equivalent to the following monotonicity properties of $\widetilde{Q}_\alpha(\cdot\|\cdot)$:

$$\begin{aligned}
\widetilde{Q}_\alpha(\rho_{AB}\|\sigma_{AB}) &\leq \widetilde{Q}_\alpha(\rho_A\|\sigma_A) \quad \text{for } \alpha \in [1/2, 1), \\
\widetilde{Q}_\alpha(\rho_{AB}\|\sigma_{AB}) &\geq \widetilde{Q}_\alpha(\rho_A\|\sigma_A) \quad \text{for } \alpha \in (1, \infty).
\end{aligned} \tag{3.31}$$

We set $d = \dim \mathcal{H}_B$ and let $\{V_i\}_{i=1}^{d^2}$ be a representation of the discrete Heisenberg-Weyl group on $\mathcal{H}_B$, satisfying the following relation for all $\rho_{AB} \in \mathcal{D}(\mathcal{H}_{AB})$ (see for example [Wil16] or [Wol12]):

$$\frac{1}{d^2} \sum_{i=1}^{d^2} (\mathbb{1}_A \otimes V_i)\rho_{AB}\left(\mathbb{1}_A \otimes V_i^\dagger\right) = \rho_A \otimes \pi_B. \tag{3.32}$$

The crucial ingredient in proving (3.31) is the joint concavity/convexity of the trace functional $\widetilde{Q}_\alpha(\cdot\|\cdot)$ for $\alpha \in [1/2, 1)$ and $\alpha > 1$, respectively:

**Proposition 3.4.2** ([FL13]). *The functional $(\rho, \sigma) \mapsto \widetilde{Q}_\alpha(\rho \| \sigma)$ is jointly concave for $\alpha \in [1/2, 1)$ and jointly convex for $\alpha \in (1, \infty)$.*

In order to prove Proposition 3.4.2, we need to introduce matrix differentials. We only cover so much material as to be able to show two identities (Lemma 3.4.3 below) that are needed in the proof of Proposition 3.4.2.

Let $X = (x_{ij})$ be an $n \times n$ matrix with entries in $\mathbb{C}$, and let $f \colon \mathbb{C}^{n^2} \to \mathbb{C}$ be a differentiable function. Regarding $X$ as a vector in $\mathbb{C}^{n^2}$, the differential $df$ is defined as

$$df = \sum_{i,j=1}^{n} \frac{\partial f(X)}{\partial x_{ij}} dx_{ij}, \tag{3.33}$$

and we define $\partial f(X)/\partial X$ to be the $n \times n$ matrix whose $(i, j)$-th element is $\partial f(X)/\partial x_{ij}$.

**Lemma 3.4.3.** *Let $A$ and $S$ be constant matrices, and let $g$ be an analytic function. Then,*

$$\frac{\partial \operatorname{tr}(AX)}{\partial X} = A^T \tag{3.34}$$

$$\frac{\partial \operatorname{tr}(g(SXS))}{\partial X} = \left[ Sg'(SXS)S \right]^T. \tag{3.35}$$

*Proof.* Let us write $dX$ for the matrix with elements $dx_{ij}$. Then it is easy to check (by element-wise computation) that $d$ acting on matrices satisfies the following rules:

  (i) $dA = 0$ for constant $A$.

 (ii) Linearity: $d(aX) = a\,dX$ for constant $a \in \mathbb{C}$, and $d(X + Y) = dX + dY$.

(iii) Product rule: $d(XY) = (dX)Y + X\,dY$.

Using these properties together with linearity of the trace, we can prove (3.34) as follows:

$$d(\operatorname{tr}(AX)) = \operatorname{tr}(d(AX)) = \operatorname{tr}(A\,dX) = \sum_{i,j=1}^{n} A_{ij} dx_{ji},$$

and comparing this with (3.33) yields (3.34).

To prove (3.35), we write the analytic function $g$ as a power series $g(x) = \sum_{i=0}^{\infty} a_i x^i$, such that $g'(x) = \sum_{i=1}^{\infty} i a_i x^{i-1}$. We then have

$$d(\operatorname{tr}(g(SXS))) = d\left( \operatorname{tr}\left( \sum_i a_i (SXS)^i \right) \right)$$

$$= \sum_i a_i \operatorname{tr}\left(d(SXS)^i\right)$$
$$= \sum_i a_i \operatorname{tr}\left(S(dX)S(SXS)^{i-1} + \dots\right)$$
$$= \sum_i i a_i \operatorname{tr}\left(S(SXS)^{i-1}S\,dX\right)$$
$$= \operatorname{tr}\left(S\left(\sum_i i a_i (SXS)^{i-1}\right)S\,dX\right)$$
$$= \operatorname{tr}\left(Sg'(SXS)S\,dX\right),$$

and hence (3.35) follows. □

We can now continue with the proof of Proposition 3.4.2.

*Proof of Proposition 3.4.2.* Assume first that $\alpha > 1$. Then the trace functional $\widetilde{Q}_\alpha(\rho\|\sigma)$ can be rewritten as

$$\widetilde{Q}_\alpha(\rho\|\sigma) = \sup_{H \geq 0}\left\{\alpha \operatorname{tr}(\rho H) - (\alpha - 1)\operatorname{tr}\left(\sigma^{-\gamma}H\sigma^{-\gamma}\right)^{\alpha/(\alpha-1)}\right\}. \tag{3.36}$$

To see this, we define the function

$$f_\alpha(H, \rho, \sigma) := \alpha \operatorname{tr}(\rho H) - (\alpha - 1)\operatorname{tr}\left(\sigma^{-\gamma}H\sigma^{-\gamma}\right)^{\alpha/(\alpha-1)}. \tag{3.37}$$

For fixed $\rho$ and $\sigma$, Lemma 3.4.3 then implies that

$$\frac{\partial f_\alpha(H, \rho, \sigma)}{\partial H} = \alpha\rho^T - \alpha\left[\sigma^{-\gamma}\left(\sigma^{-\gamma}H\sigma^{-\gamma}\right)^{1/(\alpha-1)}\sigma^{-\gamma}\right]^T.$$

A critical point of $f_\alpha(H, \rho, \sigma)$ satisfies $\partial f_\alpha(H, \rho, \sigma)/\partial H = 0$, which is equivalent to

$$\rho = \sigma^{-\gamma}\left(\sigma^{-\gamma}H\sigma^{-\gamma}\right)^{1/(\alpha-1)}\sigma^{-\gamma}.$$

This yields the optimal $H$ in (3.36), which we denote by $\hat{H} \equiv \hat{H}(\rho, \sigma)$:

$$\hat{H} = \sigma^\gamma(\sigma^\gamma\rho\sigma^\gamma)^{\alpha-1}\sigma^\gamma. \tag{3.38}$$

As $H \mapsto f_\alpha(H, \rho, \sigma)$ is concave (this is proved in Proposition 3.4.5 below), it attains its supremum at the optimal $\hat{H}$ in (3.38), and substituting this into (3.37) yields $f_\alpha(\hat{H}, \rho, \sigma) = \widetilde{Q}_\alpha(\rho\|\sigma)$. It remains to show that $(\rho, \sigma) \mapsto f_\alpha(H, \rho, \sigma)$ is jointly convex for fixed $H$, since the joint convexity of $\widetilde{Q}_\alpha(\cdot\|\cdot)$ then follows from the fact that it is the supremum over a family of jointly convex functions. To prove joint convexity of $(\rho, \sigma) \mapsto f_\alpha(H, \rho, \sigma)$, note first that

$\rho \mapsto f_\alpha(H, \rho, \sigma)$ is linear. Furthermore, we have

$$\operatorname{tr}(\sigma^{-\gamma} H \sigma^{-\gamma})^{\alpha/(\alpha-1)} = \operatorname{tr}(H^{1/2} \sigma^{-2\gamma} H^{1/2})^{\alpha/(\alpha-1)},$$

and by Lemma 5 in [FL13] the functional

$$\sigma \mapsto (1 - \alpha) \operatorname{tr}\left(H^{1/2} \sigma^{-2\gamma} H^{1/2}\right)^{\alpha/(\alpha-1)}$$

is convex for $\alpha > 1$, which proves joint convexity of $(\rho, \sigma) \mapsto f_\alpha(H, \rho, \sigma)$.

For $\alpha \in [1/2, 1)$, the trace functional $\widetilde{Q}_\alpha(\rho \| \sigma)$ is instead rewritten as

$$\widetilde{Q}_\alpha(\rho \| \sigma) = \inf_{H \geq 0} \left\{ \alpha \operatorname{tr} \rho H - (\alpha - 1) \operatorname{tr} (\sigma^{-\gamma} H \sigma^{-\gamma})^{\alpha/(\alpha-1)} \right\},$$

which is proved along the same lines as above. The joint concavity of $\widetilde{Q}_\alpha(\cdot \| \cdot)$ now follows from the joint concavity of $(\rho, \sigma) \mapsto f_\alpha(H, \rho, \sigma)$, which in turn follows from the fact that

$$\sigma \mapsto (1 - \alpha) \operatorname{tr}\left(H^{1/2} \sigma^{-2\gamma} H^{1/2}\right)^{\alpha/(\alpha-1)}$$

is concave for $\alpha \in [1/2, 1)$ by Lemma 5 in [FL13]. $\qquad\square$

**Remark 3.4.4.** The joint convexity/concavity of the trace functional $\widetilde{Q}_\alpha(\cdot \| \cdot)$ is a special case of the joint convexity/concavity of a more general trace functional underlying the $\alpha$-$z$-Rényi relative entropies mentioned in Section 3.2.1. This was proved in [Hia13] using the theory of Pick functions. A more accessible proof can be found in the arXiv version of [AD15].

Joint convexity/concavity of the trace functional $\widetilde{Q}_\alpha(\cdot \| \cdot)$ as stated in Proposition 3.4.2 can be used to prove the monotonicity of $\widetilde{Q}_\alpha(\cdot \| \cdot)$ under partial trace (3.31) as follows. Abbreviating $V_i \equiv \mathbb{1}_A \otimes V_i$, we have for $\alpha > 1$ that

$$
\begin{aligned}
\widetilde{Q}_\alpha(\rho_{AB} \| \sigma_{AB}) &= \widetilde{Q}_\alpha\left(V_i \rho_{AB} V_i^\dagger \,\middle\|\, V_i \sigma_{AB} V_i^\dagger\right) \\
&= \frac{1}{d^2} \sum_i \widetilde{Q}_\alpha\left(V_i \rho_{AB} V_i^\dagger \,\middle\|\, V_i \sigma_{AB} V_i^\dagger\right) \\
&\geq \widetilde{Q}_\alpha\left(d^{-2} \sum_i V_i \rho_{AB} V_i^\dagger \,\middle\|\, d^{-2} \sum_i V_i \sigma_{AB} V_i^\dagger\right) \qquad (3.39) \\
&= \widetilde{Q}_\alpha(\rho_A \otimes \pi_B \| \sigma_A \otimes \pi_B) \\
&= \widetilde{Q}_\alpha(\rho_A \| \sigma_A). \qquad (3.40)
\end{aligned}
$$

In the first equality we used the invariance of $\widetilde{Q}_\alpha(\cdot\|\cdot)$ under joint unitaries, Proposition 3.1.2(v). The inequality follows from the joint convexity of $\widetilde{Q}_\alpha(\cdot\|\cdot)$ as proved in Proposition 3.4.2. In the third equality we used property (3.32) of the Heisenberg-Weyl operators, and in the last equality we used the invariance of $\widetilde{Q}_\alpha(\cdot\|\cdot)$ under tensoring with a fixed state, Proposition 3.1.2(iv).

For $\alpha \in [1/2, 1)$, we go through the same steps as above to show that

$$\widetilde{Q}_\alpha(\rho_{AB}\|\sigma_{AB}) \le \widetilde{Q}_\alpha(\rho_A\|\sigma_A),$$

only this time employing the joint concavity of $\widetilde{Q}_\alpha(\cdot\|\cdot)$ from Proposition 3.4.2 in (3.39) instead.

### 3.4.2. Deriving an equality condition

The preceding discussion showed that joint convexity/concavity of the trace functional $\widetilde{Q}_\alpha(\cdot\|\cdot)$ used in (3.39) is the only step in the proof of (3.30) introducing an inequality. To obtain an equality condition for (3.30), we therefore need to analyze when equality holds in the joint convexity/concavity of $\widetilde{Q}_\alpha(\cdot\|\cdot)$.

The proof of the DPI for the $\alpha$-SRD by Frank and Lieb in Section 3.4.1 involves the function

$$f_\alpha(H, \rho, \sigma) := \alpha \operatorname{tr}(\rho H) - (\alpha - 1) \operatorname{tr}(\sigma^{-\gamma} H \sigma^{-\gamma})^{\alpha/(\alpha-1)}.$$

We claim that for fixed $\rho, \sigma$ the function $H \mapsto f_\alpha(H, \rho, \sigma)$ is strictly concave if $\alpha > 1$, and strictly convex if $\alpha \in [1/2, 1)$. To see this, note that $H \mapsto \operatorname{tr}(\sigma^{-\gamma} H \sigma^{-\gamma})^{\alpha/(\alpha-1)}$ is the composition of the linear function $X \mapsto \sigma^{-\gamma} X \sigma^{-\gamma}$ and the functional $A \mapsto \operatorname{tr} A^{\alpha/(\alpha-1)}$, the latter being strictly convex by Theorem 2.3.4(ii) upon choosing $g\colon \mathbb{R}_+ \to \mathbb{R}$, $g(x) = x^{\alpha/(\alpha-1)}$. As $\operatorname{tr}(\rho H)$ is linear in $H$, and the sum of a linear function and a strictly convex (concave) function is strictly convex (concave), we arrive at:

**Proposition 3.4.5.** *For $\alpha > 1$ the function*

$$f_\alpha(H, \rho, \sigma) := \alpha \operatorname{tr}(\rho H) - (\alpha - 1) \operatorname{tr}(\sigma^{-\gamma} H \sigma^{-\gamma})^{\alpha/(\alpha-1)}$$

*is strictly concave in $H$, and therefore attains its unique maximum at*

$$\hat{H} = \sigma^\gamma (\sigma^\gamma \rho \sigma^\gamma)^{\alpha-1} \sigma^\gamma. \tag{3.41}$$

*For $\alpha \in [1/2, 1)$ the function $f_\alpha(H, \rho, \sigma)$ is strictly convex in $H$, and therefore attains its unique minimum at $\hat{H}$ defined in (3.41).*

In the light of Proposition 3.4.5, we revisit the joint convexity of $\widetilde{Q}_\alpha(\cdot\|\cdot)$, assuming first that $\alpha > 1$. Let $\rho_i$ and $\sigma_i$ be arbitrary operators and $\lambda_i \geq 0$ be such that $\sum_i \lambda_i = 1$. Setting $\rho = \sum_i \lambda_i \rho_i$ and $\sigma = \sum_i \lambda_i \sigma_i$, we consider the operators

$$\widetilde{H} := \arg\max_H f_\alpha(H, \rho, \sigma) \qquad\qquad H_i := \arg\max_H f_\alpha(H, \rho_i, \sigma_i),$$

which are well-defined by Proposition 3.4.5. We then have

$$
\begin{aligned}
\widetilde{Q}_\alpha(\rho\|\sigma) &= f_\alpha\big(\widetilde{H}, \rho, \sigma\big) \\
&\leq \sum_i \lambda_i f_\alpha\big(\widetilde{H}, \rho_i, \sigma_i\big) && (3.42) \\
&\leq \sum_i \lambda_i f_\alpha(H_i, \rho_i, \sigma_i) && (3.43) \\
&= \sum_i \lambda_i \widetilde{Q}_\alpha(\rho_i\|\sigma_i),
\end{aligned}
$$

where we used the joint convexity of $(\rho, \sigma) \mapsto f_\alpha(H, \rho, \sigma)$ for fixed $H$ in the first inequality, and the definition of $H_i$ in the second inequality.

Assume now that we have equality in the joint convexity, that is, $\widetilde{Q}_\alpha(\rho\|\sigma) = \sum_i \lambda_i \widetilde{Q}_\alpha(\rho_i\|\sigma_i)$. Then the chain of inequalities above collapses, and in particular we obtain

$$f_\alpha\big(\widetilde{H}, \rho_i, \sigma_i\big) = f_\alpha(H_i, \rho_i, \sigma_i) \quad \text{for every } i.$$

In other words, the operator $\widetilde{H}$ maximizes $f_\alpha(H, \rho_i, \sigma_i)$ for every $i$, and since the maximizing element of $f_\alpha(H, \rho_i, \sigma_i)$ is unique by Proposition 3.4.5, we obtain $\widetilde{H} = H_i$ for every $i$. In the case $\alpha \in [1/2, 1)$, we define $\widetilde{H} := \arg\min_H f_\alpha(H, \rho, \sigma)$ and $H_i := \arg\min_H f_\alpha(H, \rho_i, \sigma_i)$. The inequalities in (3.42) and (3.43) are now reversed due to the joint concavity of $\widetilde{Q}_\alpha(\cdot\|\cdot)$, and since $f_\alpha(H, \rho_i, \sigma_i)$ attains a *minimum* at $H_i$. Again, we obtain $\widetilde{H} = H_i$ for every $i$ due to Proposition 3.4.5.

Let us now particularize the above observation to monotonicity of $\widetilde{Q}_\alpha(\cdot\|\cdot)$ under partial trace as proved in (3.40), where we employed the representation $\{V_i\}_{i=1}^{d^2}$ of the discrete Heisenberg-Weyl group. Recall the following choices for $\rho_i, \sigma_i$, and $\lambda_i$ for every $i = 1, \dots, d^2$:

$$\rho_i = V_i \rho_{AB} V_i^\dagger \qquad\qquad \sigma_i = V_i \sigma_{AB} V_i^\dagger \qquad\qquad \lambda_i = \frac{1}{d^2}. \qquad (3.44)$$

Hence, $\rho = \sum_i \lambda_i \rho_i = \rho_A \otimes \pi_B$ by (3.32), and similarly, $\sigma = \sigma_A \otimes \pi_B$. Substituting the expressions from (3.44) in the explicit form (3.41) of the optimizing $\hat{H}$, we obtain the following identity from $\widetilde{H} = H_i$ for every $i$:

$$\sigma_A^\gamma \left(\sigma_A^\gamma \rho_A \sigma_A^\gamma\right)^{\alpha-1} \sigma_A^\gamma \otimes \pi_B^{2\gamma+(\alpha-1)(2\gamma+1)} = (\mathbb{1}_A \otimes V_i)\sigma_{AB}^\gamma \left(\sigma_{AB}^\gamma \rho_{AB} \sigma_{AB}^\gamma\right)^{\alpha-1} \sigma_{AB}^\gamma \left(\mathbb{1}_A \otimes V_i^\dagger\right). \quad (3.45)$$

Since $2\gamma + (\alpha-1)(2\gamma+1) = 0$, the dimension factor of $\pi_B$ on the left-hand side of (3.45) cancels, and eliminating the unitary $V_i$ yields

$$\sigma_A^\gamma \left(\sigma_A^\gamma \rho_A \sigma_A^\gamma\right)^{\alpha-1} \sigma_A^\gamma \otimes \mathbb{1}_B = \sigma_{AB}^\gamma \left(\sigma_{AB}^\gamma \rho_{AB} \sigma_{AB}^\gamma\right)^{\alpha-1} \sigma_{AB}^\gamma. \quad (3.46)$$

This is a necessary condition for equality in the monotonicity of the trace functional $\widetilde{Q}_\alpha(\cdot\|\cdot)$. Furthermore, it is easy to see that (3.46) is also sufficient, as $\widetilde{Q}_\alpha(\rho_{AB}\|\sigma_{AB}) = \widetilde{Q}_\alpha(\rho_A\|\sigma_A)$ follows from multiplying both sides of (3.46) by $\rho_{AB}$, taking the trace, and using cyclicity of the trace. In summary, we have proved the following:

**Proposition 3.4.6.** *For $\alpha \in [1/2, 1) \cup (1, \infty)$ we have $\widetilde{D}_\alpha(\rho_{AB}\|\sigma_{AB}) = \widetilde{D}_\alpha(\rho_A\|\sigma_A)$ if and only if*

$$\sigma_A^\gamma \left(\sigma_A^\gamma \rho_A \sigma_A^\gamma\right)^{\alpha-1} \sigma_A^\gamma \otimes \mathbb{1}_B = \sigma_{AB}^\gamma \left(\sigma_{AB}^\gamma \rho_{AB} \sigma_{AB}^\gamma\right)^{\alpha-1} \sigma_{AB}^\gamma.$$

We are now in a position to prove the main result of this section, Theorem 3.4.1:

*Proof of Theorem 3.4.1.* For a quantum operation $\Lambda\colon \mathcal{B}(\mathcal{H}) \to \mathcal{B}(\mathcal{K})$, the Stinespring Representation Theorem 2.2.2 asserts that there is a Hilbert space $\mathcal{H}'$, a pure state $|\tau\rangle \in \mathcal{H}' \otimes \mathcal{K}$, and a unitary $U$ acting on $\mathcal{H} \otimes \mathcal{H}' \otimes \mathcal{K}$ such that for every $\rho \in \mathcal{B}(\mathcal{H})$ we have

$$\Lambda(\rho) = \mathrm{tr}_{12}\left(U(\rho \otimes \tau)U^\dagger\right),$$

where $\mathrm{tr}_{12}$ denotes the partial trace over $\mathcal{H}$ and $\mathcal{H}'$, that is, the first two factors of $\mathcal{H} \otimes \mathcal{H}' \otimes \mathcal{K}$. We then have

$$\begin{aligned}
\widetilde{D}_\alpha(\rho\|\sigma) &= \widetilde{D}_\alpha\left(U(\rho \otimes \tau)U^\dagger \,\middle\|\, U(\sigma \otimes \tau)U^\dagger\right) \\
&\geq \widetilde{D}_\alpha\left(\mathrm{tr}_{12}\left(U(\rho \otimes \tau)U^\dagger\right) \,\middle\|\, \mathrm{tr}_{12}\left(U(\sigma \otimes \tau)U^\dagger\right)\right) \\
&= \widetilde{D}_\alpha(\Lambda(\rho)\|\Lambda(\sigma)),
\end{aligned}$$

where the first line follows from Proposition 3.1.2(iv) and (v), and the inequality follows from Proposition 3.1.2(vi). By Proposition 3.4.6 we have equality in the second line if and only if

$$\left(U(\sigma \otimes \tau)U^\dagger\right)^\gamma \left[\left(U(\sigma \otimes \tau)U^\dagger\right)^\gamma U(\rho \otimes \tau)U^\dagger \left(U(\sigma \otimes \tau)U^\dagger\right)^\gamma\right]^{\alpha-1} \left(U(\sigma \otimes \tau)U^\dagger\right)^\gamma$$

$$= \mathbb{1}_{\mathcal{H} \otimes \mathcal{H}'} \otimes \Lambda(\sigma)^{\gamma} \left[ \Lambda(\sigma)^{\gamma} \Lambda(\rho) \Lambda(\sigma)^{\gamma} \right]^{\alpha-1} \Lambda(\sigma)^{\gamma}.$$

Since $f(UXU^{\dagger}) = U f(X) U^{\dagger}$ for every function $f$ and unitary $U$, this is equivalent to

$$U \left( \sigma^{\gamma} \left( \sigma^{\gamma} \rho \sigma^{\gamma} \right)^{\alpha-1} \sigma^{\gamma} \otimes \tau \right) U^{\dagger} = \mathbb{1}_{\mathcal{H} \otimes \mathcal{H}'} \otimes \Lambda(\sigma)^{\gamma} \left[ \Lambda(\sigma)^{\gamma} \Lambda(\rho) \Lambda(\sigma)^{\gamma} \right]^{\alpha-1} \Lambda(\sigma)^{\gamma}.$$

The theorem now follows from the fact that the adjoint of $\Lambda$ is given by

$$\Lambda^{\dagger}(\omega) = V^{\dagger}(\mathbb{1}_{\mathcal{H} \otimes \mathcal{H}'} \otimes \omega)V,$$

where $V = U(\mathbb{1}_{\mathcal{H}} \otimes |\tau\rangle)$ is the Stinespring isometry of $\Lambda$ satisfying $V^{\dagger}V = \mathbb{1}_{\mathcal{H}}$. $\qquad \square$

### 3.4.3. Relation to sufficiency

We have shown in Theorem 3.4.1 that equality in the DPI for the $\alpha$-SRD $\widetilde{D}_{\alpha}(\cdot\|\cdot)$ holds for a quantum operation $\Lambda$, a quantum state $\rho$, and a positive operator $\sigma$ (with suitable support conditions) if and only if the following algebraic condition is satisfied (setting $\gamma = (1-\alpha)/2\alpha$):

$$\sigma^{\gamma} \left( \sigma^{\gamma} \rho \sigma^{\gamma} \right)^{\alpha-1} \sigma^{\gamma} = \Lambda^{\dagger}\left( \Lambda(\sigma)^{\gamma} \left[ \Lambda(\sigma)^{\gamma} \Lambda(\rho) \Lambda(\sigma)^{\gamma} \right]^{\alpha-1} \Lambda(\sigma)^{\gamma} \right). \tag{3.47}$$

In the case of the $\alpha$-RRE $D_{\alpha}(\cdot\|\cdot)$ for $\alpha \in [0,2]$ (which includes the quantum relative entropy corresponding to $\alpha \to 1$), a necessary and sufficient algebraic condition for equality in the DPI (3.7) is given by [Pet86b; Pet88; HMP+11]

$$\Lambda^{\dagger}(\Lambda(\sigma)^{-z}\Lambda(\rho)^{z}) = \sigma^{-z}\rho^{z} \quad \text{for all } z \in \mathbb{C}.$$

This can be rephrased in terms of the existence of a recovery map by an argument detailed in [HMP+11, Theorem 5.1], stating that we have $D_{\alpha}(\rho\|\sigma) = D_{\alpha}(\Lambda(\rho)\|\Lambda(\sigma))$ if and only if there is a quantum operation $\mathcal{R}$, the *recovery map*, satisfying

$$\mathcal{R}(\Lambda(\sigma)) = \sigma \qquad\qquad \mathcal{R}(\Lambda(\rho)) = \rho. \tag{3.48}$$

In general, we call a quantum operation $\Lambda$ *sufficient* for a set $\mathcal{S} \subseteq \mathcal{D}(\mathcal{H})$ of quantum states if there exists a quantum operation $\mathcal{R}$ satisfying $\mathcal{R}(\Lambda(\tau)) = \tau$ for all $\tau \in \mathcal{S}$ (cf. [Pet86b; Pet88; MP04; Mos05; JP06; Jen12]). Hence, (3.48) says that $\Lambda$ is sufficient for $\{\rho, \sigma\}$. Furthermore, there is a recovery map $\mathcal{R}_{\sigma,\Lambda}$ satisfying (3.48) that admits an explicit formula on the support of $\Lambda(\sigma)$:

$$\mathcal{R}_{\sigma,\Lambda}(\cdot) = \sigma^{1/2}\Lambda^{\dagger}\left(\Lambda(\sigma)^{-1/2} \cdot \Lambda(\sigma)^{-1/2}\right)\sigma^{1/2}. \tag{3.49}$$

Since $\mathcal{R}_{\sigma,\Lambda}(\Lambda(\sigma)) = \sigma$ holds by definition of $\mathcal{R}_{\sigma,\Lambda}$ in (3.49), the non-trivial part of (3.48) is the assertion $\mathcal{R}_{\sigma,\Lambda}(\Lambda(\rho)) = \rho$. Note that by a theorem of Petz [Pet88] a quantum channel $\Lambda$ is sufficient for $\{\rho, \sigma\}$ if and only if $\mathcal{R}_{\sigma,\Lambda}(\Lambda(\rho)) = \rho$ holds for the map defined in (3.49). We also observe that the recovery map $\mathcal{R}_{\sigma,\Lambda}$ is *independent of $\alpha$*, and the existence of a map $\mathcal{R}$ satisfying (3.48) forces equality in the DPI for any $\alpha \in [0, 2]$,

$$D_{\alpha}(\rho\|\sigma) \geq D_{\alpha}(\Lambda(\rho)\|\Lambda(\sigma)) \geq D_{\alpha}(\mathcal{R}(\Lambda(\rho))\|\mathcal{R}(\Lambda(\sigma))) = D_{\alpha}(\rho\|\sigma). \tag{3.50}$$

Here, the first inequality follows from applying the DPI with respect to $\Lambda$, and the second follows from applying the DPI with respect to $\mathcal{R}$. Thus, we have equality in the DPI for the $\alpha$-RRE for *all* $\alpha \in [0, 2]$ once it holds for *some* $\alpha \in [0, 2]$.

Taking a closer look at the condition (3.47) for equality in the DPI for the $\alpha$-SRD, it is easy to see that choosing $\alpha = 2$ in (3.47) yields precisely the statement $\mathcal{R}_{\sigma,\Lambda}(\Lambda(\rho)) = \rho$. Hence, in the case $\alpha = 2$ we have equality in the DPI for the 2-SRD if and only if the recovery map $\mathcal{R}_{\sigma,\Lambda}$ defined in (3.49) satisfies (3.48). This was already observed in [DJW15] for positive trace-preserving maps.

The connection between sufficiency and equality in the DPI for the $\alpha$-SRD was investigated by Jenčová [Jen16] and Hiai and Mosonyi [HM16]. The main result of [Jen16] is that a 2-positive trace-preserving map $\Lambda$ is sufficient with respect to $\{\rho, \sigma\}$ if and only if $\widetilde{D}_{\alpha}(\Lambda(\rho)\|\Lambda(\sigma)) = \widetilde{D}_{\alpha}(\rho\|\sigma)$ holds for some $\alpha > 1$. By the theorem of Petz [Pet88] mentioned above, we therefore have equality in the DPI for the $\alpha$-SRD for any $\alpha > 1$ if and only if the map $\mathcal{R}_{\sigma,\Lambda}$ defined in (3.49) satisfies (3.48). Furthermore, a similar argument as in (3.50) for $\widetilde{D}_{\alpha}(\cdot\|\cdot)$ shows that equality holds in the DPI for the $\alpha$-SRD for all $\alpha > 1$ if it holds for some $\alpha > 1$. This result settles the sufficiency question for the $\alpha$-SRD for the range $\alpha > 1$ and 2-positive trace-preserving maps (which include all quantum operations).

In [HM16] sufficiency is analyzed for 2-positive *bistochastic* maps $\Phi$, that is, both $\Phi$ and $\Phi^{\dagger}$ are 2-positive and trace-preserving. The main theorem of [HM16] regarding the $\alpha$-SRD states conditions for sufficiency of $\Lambda$ for certain ranges of $\alpha$ (including the range $\alpha \in [1/2, 1)$) under the additional assumption that one of the two states $\rho$ and $\sigma$ is a *fixed point* of $\Lambda$. In fact, this result is obtained as a corollary of a more general theorem analyzing sufficiency for the $\alpha$-$z$-Rényi relative entropies under similar assumptions.

In view of the results proved in [HM16; Jen16], the main contribution of Theorem 3.4.1 is a *general* necessary and sufficient condition for equality in the DPI for the $\alpha$-SRD in the range $\alpha \in [1/2, 1)$. It is still open whether equality in the DPI for the $\alpha$-SRD in this range is equivalent with sufficiency of $\Lambda$ for $\{\rho, \sigma\}$.

For $\alpha = 1/2$ and $\alpha = \infty$, it is known that such a general sufficiency result cannot hold [MO15]. For $\alpha = 1/2$ this can be seen as follows: Recall from Section 3.2.2 that $\widetilde{D}_{1/2}(\rho \| \sigma) = -2 \log F(\rho, \sigma)$. It is well-known (see for example [Wil16]) that for given $\rho$ and $\sigma$ there exists a measurement $\mathcal{M} = \{M_x\}_{x \in \mathcal{X}}$ for some finite set $\mathcal{X}$ such that the fidelity $F(\rho, \sigma)$ is equal to the *classical* fidelity

$$F(P, Q) = \sum_{x \in \mathcal{X}} \sqrt{P(x)Q(x)}.$$

Here, $P$ and $Q$ are defined through $P(x) := \operatorname{tr}(M_x \rho)$ and $Q(x) := \operatorname{tr}(M_x \sigma)$ for all $x \in \mathcal{X}$, respectively, corresponding to the probability distributions obtained from measuring $\rho$ and $\sigma$ with respect to $\mathcal{M}$. Hence, for any two states $\rho$ and $\sigma$ we have equality in the DPI for the 1/2-SRD with respect to this particular measurement, and in general it is not possible to recover the states $\rho$ and $\sigma$ from the measurement outcomes $\{P(x)\}_{x \in \mathcal{X}}$ and $\{Q(x)\}_{x \in \mathcal{X}}$ alone. This proves that a general sufficiency result as stated above cannot hold for $\alpha = 1/2$, and a similar argument shows the same for $\alpha = \infty$ [MO15].

### 3.4.4. Applications of the equality condition

In this section we discuss applications of Theorem 3.4.1. To begin with, our goal is to generalize a set of results by Carlen and Lieb [CL12] about the Araki-Lieb inequality and the entanglement of formation to the corresponding Rényi quantities. We first state a Rényi version of the Araki-Lieb inequality (Lemma 3.4.8), and analyze the case of equality (Theorem 3.4.9). We then prove a general lower bound on the Rényi entanglement of formation (analogous to the corresponding bound on the entanglement of formation in [CL12]), and show that this lower bound is achieved by states saturating the Rényi version of the Araki-Lieb inequality. These results are presented in Theorem 3.4.13. Finally, we discuss the case of equality in a well-known upper bound on the entanglement fidelity in terms of the usual fidelity, which we state in Proposition 3.4.15.

**Rényi version of Araki-Lieb inequality and the case of equality**

The Araki-Lieb inequality [AL70] states that for every bipartite state $\rho_{AB}$,

$$S(AB)_\rho \geq |S(A)_\rho - S(B)_\rho|. \tag{3.51}$$

There are a few different characterizations for the case of equality in the Araki-Lieb inequality [NC00; ZW11; CL12]. Here, we concentrate on a result by Carlen and Lieb:

**Theorem 3.4.7** (Equality in the Araki-Lieb inequality; [CL12])**.**
*For a bipartite state $\rho_{AB}$ denote by $r_{AB}$, $r_A$, and $r_B$ the ranks of $\rho_{AB}$, $\rho_A$, and $\rho_B$, respectively. The state $\rho_{AB}$ saturates the Araki-Lieb inequality* (3.51),

$$S(AB)_\rho = S(B)_\rho - S(A)_\rho,$$

*if and only if the following conditions are satisfied:*

(i) *$r_B = r_A r_{AB}$*

(ii) *The state $\rho_{AB}$ has a spectral decomposition of the form*

$$\rho_{AB} = \sum_{i=1}^{r_{AB}} \lambda_i |i\rangle\langle i|_{AB},$$

*where the vectors $\{|i\rangle_{AB}\}_{i=1}^{r_{AB}}$ are such that $\mathrm{tr}_B\, |i\rangle\langle j|_{AB} = \delta_{ij}\rho_A$ for $i, j = 1, \dots, r_{AB}$.*

We can regard the Araki-Lieb inequality (3.51) as lower bounds on the conditional entropies:

$$S(A|B)_\rho \geq -S(A)_\rho \qquad\qquad S(B|A)_\rho \geq -S(B)_\rho. \tag{3.52}$$

In the following, we only focus on the bound $S(A|B)_\rho \geq -S(A)_\rho$, noting that all the results we obtain hold for $S(B|A)_\rho$ in an analogous manner. The formulation (3.52) of the Araki-Lieb inequality admits a simple proof as follows. Observe first that for a purification $|\rho\rangle_{ABC}$ of $\rho_{AB}$ we have the duality relation

$$S(A|B)_\rho = -S(A|C)_\rho, \tag{3.53}$$

which can be regarded as a special case of the duality for the Rényi conditional entropy, Proposition 3.3.2(iv), upon choosing $\alpha = 1$ (which implies $\beta = 1$). Note however, that (3.53) can also easily be proved directly. Using the duality relation (3.53), we have

$$S(A|B)_\rho = -S(A|C)_\rho = D(\rho_{AC}\|\mathbb{1}_A \otimes \rho_C) \geq D(\rho_A\|\mathbb{1}_A) = -S(A)_\rho, \quad (3.54)$$

where the inequality follows from the DPI for the quantum relative entropy with $\Lambda = \mathrm{tr}_C$.

Generalizing the Araki-Lieb inequality in the form of (3.51) to Rényi quantities is problematic: as proved in [LMW13], quantities based on linear combinations of Rényi entropies can take arbitrary values, and hence their use is limited. In contrast, the advantage of phrasing the Araki-Lieb inequality in the form of (3.52) is that we can simply replace the conditional entropy $S(A|B)_\rho$ in (3.54) by the Rényi conditional entropy $\widetilde{S}_\alpha(A|B)_\rho$. With $\alpha, \beta \in [1/2, \infty)$ chosen such that $1/\alpha + 1/\beta = 2$, we then have the following:

$$\widetilde{S}_\alpha(A|B)_\rho = -\widetilde{S}_\beta(A|C)_\rho = \widetilde{D}_\beta(\rho_{AC}\|\mathbb{1}_A \otimes \tilde{\sigma}_C) \geq \widetilde{D}_\beta(\rho_A\|\mathbb{1}_A) = -S_\beta(A)_\rho. \quad (3.55)$$

Here, we used Proposition 3.3.2(iv) in the first equality, chose an optimizing state $\tilde{\sigma}_C$ for $\widetilde{S}_\beta(A|C)_\rho$ in the second equality, and the inequality is the DPI for the $\beta$-SRD with respect to $\Lambda = \mathrm{tr}_C$. We also obtain the upper bound $\widetilde{S}_\alpha(A|B)_\rho \leq S_\alpha(A)_\rho$ by a simple application of the DPI with respect to $\Lambda = \mathrm{tr}_B$. We summarize these observations in the following

**Lemma 3.4.8** (Rényi version of the Araki-Lieb inequality).
*Let $\rho_{AB}$ be a bipartite state, and let $\alpha, \beta \in [1/2, \infty)$ be such that $\frac{1}{\alpha} + \frac{1}{\beta} = 2$. Then,*

$$-S_\beta(A)_\rho \leq \widetilde{S}_\alpha(A|B)_\rho \leq S_\alpha(A)_\rho. \quad (3.56)$$

Since the inequality in the lower bound of Lemma 3.4.8 stems from the DPI for $\widetilde{D}_\beta(\cdot\|\cdot)$ (as demonstrated in (3.55)), we can apply Theorem 3.4.1 (in the form of Proposition 3.4.6) to investigate the case of equality. By Proposition 3.4.6 we have $\widetilde{D}_\beta(\rho_{AC}\|\mathbb{1}_A \otimes \tilde{\sigma}_C) = \widetilde{D}_\beta(\rho_A\|\mathbb{1}_A)$ if and only if

$$\rho_A^{\beta-1} \otimes \mathbb{1}_C = \left(\mathbb{1}_A \otimes \tilde{\sigma}_C^\delta\right)\left(\left(\mathbb{1}_A \otimes \tilde{\sigma}_C^\delta\right)\rho_{AC}\left(\mathbb{1}_A \otimes \tilde{\sigma}_C^\delta\right)\right)^{\beta-1}\left(\mathbb{1}_A \otimes \tilde{\sigma}_C^\delta\right), \quad (3.57)$$

where $\delta = (1-\beta)/2\beta$. It is easy to see that (3.57) is equivalent to $\rho_{AC} = \rho_A \otimes \tilde{\sigma}_C$, that is, if $\rho_{ABC}$ is a purification of $\rho_{AB}$, then the marginal $\rho_{AC}$ is of product form. We can then go through the same steps as in the proof of Theorem 1.4 in [CL12] (which we stated as Theorem 3.4.7 above) to arrive at a Rényi generalization of this result, Theorem 3.4.9 below. For the convenience of the reader, we give a proof of this theorem in our notation, essentially reproducing the proof of Theorem 3.4.7 given in [CL12] with only minor adaptations.

**Theorem 3.4.9** (Equality in the Rényi version of the Araki-Lieb inequality).
*Let $\rho_{AB}$ be a bipartite state with purification $\rho_{ABC}$, and let $\alpha, \beta \in [1/2, \infty)$ be such that $1/\alpha + 1/\beta = 2$. Denote by $r_{AB}$, $r_A$, and $r_B$ the ranks of $\rho_{AB}$, $\rho_A$, and $\rho_B$, respectively. We have equality in the Rényi version of the Araki-Lieb inequality,*

$$\widetilde{S}_\alpha(A|B)_\rho = -S_\beta(A)_\rho,$$

*if and only if the following conditions are satisfied:*

(i) $r_B = r_A r_{AB}$.

(ii) *The state $\rho_{AB}$ has a spectral decomposition of the form*

$$\rho_{AB} = \sum_{i=1}^{r_{AB}} \lambda_i |i\rangle\langle i|_{AB},$$

*where the vectors $\{|i\rangle_{AB}\}_{i=1}^{r_{AB}}$ are such that $\mathrm{tr}_B\, |i\rangle\langle j|_{AB} = \delta_{ij}\rho_A$ for $i, j = 1, \ldots, r_{AB}$.*

*Proof.* We first prove that (i) and (ii) imply $\widetilde{S}_\alpha(A|B)_\rho = -S_\beta(A)_\rho$. Using the spectral decomposition of $\rho_{AB}$, we consider a purification $|\rho\rangle_{ABC}$ of $\rho_{AB}$ given by

$$|\rho\rangle_{ABC} := \sum_{i=1}^{r_{AB}} \sqrt{\lambda_i}|i\rangle_{AB} \otimes |i\rangle_C.$$

Since $\mathrm{tr}_B\, |i\rangle\langle j|_{AB} = \delta_{ij}\rho_A$ for $i, j = 1, \ldots, r_{AB}$, we obtain $\rho_{AC} = \rho_A \otimes \rho_C$ with $\rho_C = \sum_{i=1}^{r_{AB}} \lambda_i |i\rangle\langle i|_C$. By duality for the Rényi conditional entropy (Proposition 3.3.2(iv)), additivity of the $\beta$-SRD (Proposition 3.1.2(i)), and the premetric property (Proposition 3.1.2(ii)), it follows that

$$\begin{aligned}
\widetilde{S}_\alpha(A|B)_\rho &= -\widetilde{S}_\beta(A|C)_\rho \\
&= \min_{\sigma_C} \widetilde{D}_\beta(\rho_A \otimes \rho_C \| \mathbb{1}_A \otimes \sigma_C) \\
&= \widetilde{D}_\beta(\rho_A \| \mathbb{1}_A) + \min_{\sigma_C} \widetilde{D}_\beta(\rho_C \| \sigma_C) \\
&= -S_\beta(A)_\rho.
\end{aligned}$$

Conversely, assume that we have equality in the Rényi version of the Araki-Lieb inequality, $\widetilde{S}_\alpha(A|B)_\rho = -S_\beta(A)_\rho$. By the paragraph preceding the theorem, this implies that for a purification

$|\rho\rangle_{ABC}$ of $\rho_{AB}$ we have $\rho_{AC} = \rho_A \otimes \tilde{\sigma}_C$, with the state $\tilde{\sigma}_C$ optimizing the Rényi conditional entropy $\widetilde{S}_\beta(A|C)_\rho$. Setting $s_C = \mathrm{rk}\, \tilde{\sigma}_C$, we have $\mathrm{rk}(\rho_{AC}) = r_A s_C$, and $r_B = \mathrm{rk}(\rho_{AC})$ by Schmidt decomposition (Theorem 2.2.1). It then follows that

$$r_B = r_A s_C. \tag{3.58}$$

We now make the following ansatz for $|\rho\rangle_{ABC}$:

$$|\rho\rangle_{ABC} = \sum_{i,j,k} \theta_{ijk} |i\rangle_A |j\rangle_B |k\rangle_C,$$

where $\{|i\rangle_A\}_{i=1}^{r_A}$ and $\{|k\rangle_C\}_{k=1}^{s_C}$ are eigenbases for $\rho_A$ and $\tilde{\sigma}_C$, respectively, and $\{|j\rangle_B\}_{j=1}^{r_B}$ is an arbitrary orthonormal basis for $\mathcal{H}_B$. We then have

$$\rho_{AC} = \sum_{i,j,k} \sum_{i',k'} \theta_{ijk} \theta_{i'jk'}^* |i\rangle\langle i'|_A \otimes |k\rangle\langle k'|_C, \tag{3.59}$$

and consequently,

$$\rho_A = \sum_{i,j,k,i'} \theta_{ijk} \theta_{i'jk}^* |i\rangle\langle i'|_A = \sum_{i,j,k} |\theta_{ijk}|^2 |i\rangle\langle i|_A = \sum_i \mu_i |i\rangle\langle i|_A, \tag{3.60}$$

where we used the fact that $\rho_A$ is diagonal with respect to its eigenbasis $\{|i\rangle_A\}$, and we defined $\mu_i := \sum_{j,k} |\theta_{ijk}|^2$ in the last step.

By a similar reasoning, we obtain

$$\tilde{\sigma}_C = \sum_k \lambda_k |k\rangle\langle k|_C \tag{3.61}$$

with $\lambda_k := \sum_{i,j} |\theta_{ijk}|^2$. Moreover, since $\rho_{AC} = \rho_A \otimes \tilde{\sigma}_C$, it follows from (3.59), (3.60), and (3.61) that

$$\sum_{i,j,k} \sum_{i',k'} \theta_{ijk} \theta_{i'jk'}^* |i\rangle\langle i'|_A \otimes |k\rangle\langle k'|_C = \sum_{i,k} \mu_i \lambda_k |i\rangle\langle i|_A \otimes |k\rangle\langle k|_C$$
$$= \sum_{i,i',k,k'} \mu_i \delta_{i,i'} \lambda_k \delta_{k,k'} |i\rangle\langle i'|_A \otimes |k\rangle\langle k'|_C, \tag{3.62}$$

and hence, $\sum_j \theta_{ijk} \theta_{i'jk'}^* = \mu_i \delta_{i,i'} \lambda_k \delta_{k,k'}$.

We now define vectors $|e_i f_k\rangle := \sum_j \theta_{ijk} |j\rangle_B$ for $i = 1, \ldots, r_A$ and $k = 1, \ldots, s_C$, and observe that they are orthogonal:

$$\langle e_i f_k | e_{i'} f_{k'} \rangle = \sum_{j,j'} \theta_{ijk} \theta^*_{i'j'k'} \langle j | j' \rangle = \sum_j \theta_{ijk} \theta^*_{i'jk'} = \mu_i \delta_{i,i'} \lambda_k \delta_{k,k'}.$$

Furthermore, $r_B = r_A s_C$, and hence $\{|e_i f_k\rangle_B\}_{i,k}$ spans $\mathcal{H}_B$. Thus, $\{|e_i f_k\rangle_B\}_{i,k}$ constitutes an orthogonal basis for $\mathcal{H}_B$. A similar argument shows that the vectors $|g_k\rangle_{AB} := \sum_{i,j} \theta_{ijk} |i\rangle_A |j\rangle_B$ for $k = 1, \ldots, s_C = r_{AB}$ constitute an orthogonal basis for $\mathcal{H}_{AB}$, with $\langle g_k | g_{k'} \rangle = \delta_{k,k'} \lambda_k$. By construction, we have $\sum_k |g_k\rangle\langle g_k|_{AB} = \rho_{AB}$, and furthermore,

$$\mathrm{tr}_B |g_k\rangle\langle g_{k'}|_{AB} = \sum_{i,i',j} \theta_{ijk} \theta^*_{i'jk'} |i\rangle\langle i'|_A.$$

Hence, it follows from (3.59) and (3.62) that

$$\rho_{AC} = \sum_{k,k'} \mathrm{tr}_B |g_k\rangle\langle g_{k'}|_{AB} \otimes |k\rangle\langle k'|_C = \sum_{i,k,k'} \mu_i \lambda_k \delta_{k,k'} |i\rangle\langle i|_A \otimes |k\rangle\langle k'|_C,$$

which implies that

$$\mathrm{tr}_B |g_k\rangle\langle g_{k'}|_{AB} = \delta_{k,k'} \lambda_k \sum_i \mu_i |i\rangle\langle i|_A = \delta_{k,k'} \lambda_k \rho_A.$$

Setting $|\eta_i\rangle_{AB} := \frac{1}{\sqrt{\lambda_i}} |g_i\rangle_{AB}$ so that $\rho_{AB} = \sum_i \lambda_i |\eta_i\rangle\langle\eta_i|_{AB}$ now proves (ii). Furthermore, considering the canonical purification

$$|\rho\rangle_{ABC} = \sum_i \sqrt{\lambda_i} |\eta_i\rangle_{AB} |i\rangle_C$$

of $\rho_{AB}$, the expression (3.61) for $\tilde{\sigma}_C$ shows that $\tilde{\sigma}_C = \rho_C$, and hence $s_C = r_{AB}$. Substituting this in (3.58) finally yields (i). $\qquad \square$

### Remark 3.4.10.

(i) Both assertions in Theorem 3.4.9 are identical to the ones in Theorem 3.4.7, and in particular independent of $\alpha$. Thus, if the Rényi version of the Araki-Lieb inequality (3.56) is saturated for *some* $\alpha \in [1/2, \infty)$ and $\beta = \alpha/(2\alpha - 1)$ (such that $1/\alpha + 1/\beta = 2$), it is saturated for *all* such $\alpha$ and $\beta$.

(ii) In the following, we give a simple example of a *mixed* state $\rho_{AB}$ satisfying conditions (i)

and (ii) of Theorem 3.4.7 and Theorem 3.4.9, following a 'recipe' for constructing such states provided by Carlen and Lieb [CL12].

Let $\mathcal{H}_A$ and $\mathcal{H}_B$ be Hilbert spaces with $|A| = 2$, $|B| = 4$, and let $\lambda_0, \lambda_1 > 0$ with $\lambda_0 + \lambda_1 = 1$. Consider the state

$$\rho_A = \lambda_0 |0\rangle\langle 0|_A + \lambda_1 |1\rangle\langle 1|_A,$$

satisfying $r_A := \operatorname{rk} \rho_A = 2$. We construct two orthogonal purifications of $\rho_A$,

$$|\eta_0\rangle_{AB} := \sqrt{\lambda_0} |0\rangle_A |0\rangle_B + \sqrt{\lambda_1} |1\rangle_A |1\rangle_B$$
$$|\eta_1\rangle_{AB} := \sqrt{\lambda_0} |0\rangle_A |2\rangle_B + \sqrt{\lambda_1} |1\rangle_A |3\rangle_B,$$

and for $\nu_0, \nu_1 > 0$ with $\nu_0 + \nu_1 = 1$, we set

$$\rho_{AB} := \nu_0 |\eta_0\rangle\langle\eta_0|_{AB} + \nu_1 |\eta_1\rangle\langle\eta_1|_{AB}.$$

Since $\langle\eta_0|\eta_1\rangle = 0$, we have $r_{AB} := \operatorname{rk} \rho_{AB} = 2$, and hence, $\rho_{AB}$ is a mixed state. Furthermore, it is easy to verify that the states $|\eta_i\rangle$ satisfy

$$\operatorname{tr}_B |\eta_i\rangle\langle\eta_j|_{AB} = \delta_{ij}\rho_A \quad \text{for } i, j = 0, 1. \tag{3.63}$$

An easy computation also shows that $r_B := \operatorname{rk} \rho_B = 4$.

Hence, the state $\rho_{AB}$ satisfies conditions (i) and (ii) of Theorem 3.4.7 and Theorem 3.4.9. To verify that the Araki-Lieb inequalities in the von Neumann and Rényi case are indeed satisfied, we consider the following purification of $\rho_{AB}$,

$$|\rho\rangle_{ABC} := \sqrt{\nu_0} |\eta_0\rangle_{AB} |0\rangle_C + \sqrt{\nu_1} |\eta_1\rangle_{AB} |1\rangle_C,$$

such that $\rho_{AC} = \rho_A \otimes \rho_C$ with $\rho_C = \nu_0 |0\rangle\langle 0|_C + \nu_1 |1\rangle\langle 1|_C$ by (3.63). We then compute:

$$S(A|B)_\rho = -S(A|C)_\rho = S(C)_\rho - S(AC)_\rho = S(C)_\rho - S(A)_\rho - S(C)_\rho = -S(A)_\rho,$$

where we used duality for the conditional entropy and the fact that $\rho_{AC} = \rho_A \otimes \rho_C$. In the Rényi case, we have for arbitrary $\alpha \geq 1/2$ and $\beta$ chosen such that $1/\alpha + 1/\beta = 2$,

$$\widetilde{S}_\alpha(A|B)_\rho = -\widetilde{S}_\beta(A|C)_\rho$$

$$= \min_{\sigma_C} \widetilde{D}_\beta(\rho_A \otimes \rho_C \| \mathbb{1}_A \otimes \sigma_C)$$

$$= \widetilde{D}_\beta(\rho_A \| \mathbb{1}_A) + \min_{\sigma_C} \widetilde{D}_\beta(\rho_C \| \sigma_C)$$

$$= -S_\beta(A)_\rho.$$

Here, we used duality for the Rényi conditional entropy (Proposition 3.3.2(iv)), additivity of the $\beta$-SRD (Proposition 3.1.2(i)), and the premetric property of the $\beta$-SRD (Proposition 3.1.2(ii)). This computation also shows that the optimizing state $\tilde{\sigma}_C$ in $\widetilde{S}_\beta(A|C)_\rho$ is equal to $\rho_C$, as shown in the proof of Theorem 3.4.9.

(iii) For the upper bound $\widetilde{S}_\alpha(A|B)_\rho \leq S_\alpha(A)_\rho$ in Lemma 3.4.8, we have equality if and only if $\widetilde{D}_\alpha(\rho_{AB} \| \mathbb{1}_A \otimes \tilde{\sigma}_B) = \widetilde{D}_\alpha(\rho_A \| \mathbb{1}_A)$, where $\tilde{\sigma}_B$ is a state optimizing $\widetilde{S}_\alpha(A|B)_\rho$. Similar to above, we obtain from Proposition 3.4.6 that this is the case if and only if $\rho_{AB} = \rho_A \otimes \tilde{\sigma}_B$.

**Rényi entanglement of formation**

For a bipartite state $\rho_{AB}$ the *entanglement of formation* (EoF) $E_F(\rho_{AB})$ [BDS+96; BBP+96] is defined as the expected entropy of entanglement of an ensemble $\{p_i, \psi_i\}$ of pure states realizing $\rho_{AB}$, minimized over all such ensembles:

$$E_F(\rho_{AB}) := \min_{\{p_i, \psi_i\}} \sum_i p_i S(\mathrm{tr}_B \psi_i). \tag{3.64}$$

This entanglement measure satisfies $E_F(\rho_{AB}) \geq 0$ for all $\rho_{AB}$, and is furthermore *faithful*, that is, $E_F(\rho_{AB}) = 0$ if and only if $\rho_{AB}$ is separable. The EoF is an upper bound on the (two-way) distillable entanglement [BDS+96; BBP+96]. Moreover, its regularized version

$$E_F^\infty(\rho_{AB}) := \lim_{n \to \infty} \frac{E_F(\rho_{AB}^{\otimes n})}{n}$$

is equal to the asymptotic entanglement cost of preparing the state $\rho_{AB}$ [HHT01]. Carlen and Lieb [CL12] proved the following result, which provides a lower bound on $E_F(\rho_{AB})$ that is achieved by states saturating the Araki-Lieb inequality (Theorem 3.4.7):

**Theorem 3.4.11** ([CL12]). *Let $\rho_{AB}$ be a bipartite state. Then*

$$E_F(\rho_{AB}) \geq \max\left\{-S(A|B)_\rho, -S(B|A)_\rho, 0\right\}, \tag{3.65}$$

and this bound is saturated by states satisfying the conditions of Theorem *3.4.7*. That is, for states $\rho_{AB}$ with $S(A|B)_\rho = -S(A)_\rho$ we have $E_F(\rho_{AB}) = -S(A|B)_\rho$.

**Remark 3.4.12.** If $S(A|B)_\rho = -S(A)_\rho$, then $E_F(\rho_{AB}) = -S(A|B)_\rho \geq -S(B|A)_\rho$ by (3.65).

Using the results of the preceding section on a Rényi version of the Araki-Lieb inequality, our goal in this section is to obtain a Rényi generalization of Theorem *3.4.11*. To this end, we consider the *Rényi entanglement of formation* (REoF) $E_{F,\alpha}(\rho_{AB})$ [Vid00; WMV+16], which is obtained from the definition of $E_F(\rho_{AB})$ in (3.64) by replacing the von Neumann entropy with the Rényi entropy of order $\alpha \geq 0$ from Definition 3.3.1(i):

$$E_{F,\alpha}(\rho_{AB}) := \min_{\{p_i, \psi_i\}} \sum_i p_i S_\alpha(\mathrm{tr}_B \, \psi_i).$$

Note that in [SBW14] the authors consider a different Rényi generalization of the EoF based on the $\alpha$-Rényi conditional entropy. As in the von Neumann case, the REoF satisfies $E_{F,\alpha}(\rho_{AB}) \geq 0$ for all $\rho_{AB}$, and it is faithful as well. We prove the following generalization of Theorem *3.4.11* for $\alpha > 1$:

**Theorem 3.4.13.** *Let $\rho_{AB}$ be a bipartite state, and let $\alpha > 1$ and $\beta = \alpha/(2\alpha - 1) \in (1/2, 1)$ such that $1/\alpha + 1/\beta = 2$. Then we have the following bound on the REoF:*

$$E_{F,\alpha}(\rho_{AB}) \geq \max\left\{ -\widetilde{S}_\beta(A|B)_\rho, -\widetilde{S}_\beta(B|A)_\rho, 0 \right\}. \tag{3.66}$$

*If $\rho_{AB}$ saturates the Rényi version (3.56) of the Araki-Lieb inequality with Rényi parameter $\beta$, that is, $\widetilde{S}_\beta(A|B)_\rho = -S_\alpha(A)_\rho$, then*

$$E_{F,\alpha}(\rho_{AB}) = -\widetilde{S}_\beta(A|B)_\rho. \tag{3.67}$$

*Proof.* Let $\{q_i, \phi_i\}_i$ be an ensemble of pure states minimizing the REoF, that is, $E_{F,\alpha}(\rho_{AB}) = \sum_i q_i S_\alpha(\mathrm{tr}_B(\phi_i))$. We define a purification $\rho_{ABC}$ of $\rho_{AB}$ by $|\rho\rangle_{ABC} = \sum_i \sqrt{q_i} |\phi_i\rangle_{AB} |i\rangle_C$, where $\{|i\rangle_C\}_i$ is an orthonormal basis for $\mathcal{H}_C$. Denoting by $\tilde{\sigma}_C$ the state optimizing the Rényi conditional entropy $\widetilde{S}_\alpha(A|C)_\rho$, we have

$$\begin{aligned}
\widetilde{S}_\beta(A|B)_\rho &= -\widetilde{S}_\alpha(A|C)_\rho \\
&= \widetilde{D}_\alpha(\rho_{AC} \| \mathbb{1}_A \otimes \tilde{\sigma}_C) \\
&= \widetilde{D}_\alpha\left( \sum_{i,j} \sqrt{q_i q_j} \, \mathrm{tr}_B \, |\phi_i\rangle\langle\phi_j|_{AB} \otimes |i\rangle\langle j|_C \, \Big\| \, \mathbb{1}_A \otimes \tilde{\sigma}_C \right),
\end{aligned}$$

where the first line follows from Proposition 3.3.2(iv).

We now apply the pinching map $\rho \mapsto \sum_i |i\rangle\langle i|_C \rho |i\rangle\langle i|_C$ (see Definition 2.2.7) to both states and use the DPI for $\widetilde{D}_\alpha(\cdot\|\cdot)$. Setting $\lambda_i = \langle i|\tilde{\sigma}_C|i\rangle$, we obtain

$$
\begin{aligned}
\widetilde{S}_\beta(A|B)_\rho &\geq \widetilde{D}_\alpha\Big(\sum_i q_i \operatorname{tr}_B \phi_i \otimes |i\rangle\langle i|_C \,\Big\|\, \mathbb{1}_A \otimes \sum_i \lambda_i |i\rangle\langle i|_C\Big) \\
&= \frac{1}{\alpha-1} \log\Big(\operatorname{tr}\Big(\sum_i q_i^\alpha \lambda_i^{1-\alpha} \,(\operatorname{tr}_B \phi_i)^\alpha \otimes |i\rangle\langle i|_C\Big)\Big) \\
&= \frac{1}{\alpha-1} \log\Big(\sum_i q_i \,(q_i/\lambda_i)^{\alpha-1} \operatorname{tr}(\operatorname{tr}_B \phi_i)^\alpha\Big) \\
&\geq \frac{1}{\alpha-1} \sum_i q_i \log\Big((q_i/\lambda_i)^{\alpha-1} \operatorname{tr}(\operatorname{tr}_B \phi_i)^\alpha\Big) \\
&= \sum_i q_i \frac{1}{\alpha-1} \log \operatorname{tr}(\operatorname{tr}_B \phi_i)^\alpha + \sum_i q_i \log(q_i/\lambda_i) \\
&= -E_{F,\alpha}(\rho_{AB}) + D(\{q_i\}\|\{\lambda_i\}) \\
&\geq -E_{F,\alpha}(\rho_{AB}).
\end{aligned}
$$

In the first equality we used the fact that the states $\sum_i q_i \operatorname{tr}_B \phi_i \otimes |i\rangle\langle i|_C$ and $\mathbb{1}_A \otimes \sum_i \lambda_i |i\rangle\langle i|_C$ commute (cf. Proposition 2.2.8(ii)), and hence $\widetilde{D}_\alpha(\cdot\|\cdot)$ reduces to the ordinary $\alpha$-RRE (cf. (3.9)). In the second inequality we used concavity of the logarithm together with $\alpha > 1$, and in the last inequality we used non-negativity of the classical Kullback-Leibler divergence for probability distributions $P$ and $Q$ with $P(x) = 0$ whenever $Q(x) = 0$. Note that the latter is satisfied as $\operatorname{supp} \rho_C \subseteq \operatorname{supp} \tilde{\sigma}_C$ holds for the optimizing state $\tilde{\sigma}_C$ of $\widetilde{S}_\alpha(A|C)_\rho$ [MDS+13]. The bound $E_{F,\alpha}(\rho_{AB}) \geq -\widetilde{S}_\beta(B|A)_\rho$ follows in an analogous way, yielding (3.66).

To prove (3.67), we first note that by Theorem 3.4.9 the state $\rho_{AB}$ satisfies $\widetilde{S}_\beta(A|B)_\rho = -S_\alpha(A)_\rho$ if and only if the rank condition of Theorem 3.4.9(i) holds and $\rho_{AB}$ has a spectral decomposition of the form

$$
\rho_{AB} = \sum_i \lambda_i |i\rangle\langle i|_{AB},
$$

where the vectors $\{|i\rangle_{AB}\}$ satisfy $\operatorname{tr}_B |i\rangle\langle j|_{AB} = \delta_{ij}\rho_A$. We can now employ the same argument used in [CL12] in the proof of the second assertion of Theorem 3.4.11 to prove the corresponding assertion of Theorem 3.4.13:

$$
\begin{aligned}
E_{F,\alpha}(\rho_{AB}) &= \min_{\{p_i, \psi_i\}} \sum_i p_i S_\alpha(\operatorname{tr}_B \psi_i) \\
&\leq \sum_i \lambda_i S_\alpha(\operatorname{tr}_B |i\rangle\langle i|_{AB})
\end{aligned}
$$

$$= \sum_i \lambda_i S_\alpha(A)_\rho$$
$$= S_\alpha(A)_\rho$$
$$= -\widetilde{S}_\beta(A|B)_\rho,$$

where in the inequality we chose the particular ensemble $\{\lambda_i, |i\rangle_{AB}\}_i$ realizing $\rho_{AB}$. This upper bound on $E_{F,\alpha}(\rho_{AB})$, together with the general lower bound in (3.66), yields the claim. □

**Remark 3.4.14.**

(i) The proof method of the lower bound (3.66) for $E_{F,\alpha}(\rho_{AB})$ in Theorem 3.4.13 can be specialized to the quantum relative entropy $D(\cdot\|\cdot)$, yielding a new proof of (3.65) in Theorem 3.4.11:

$$S(A|B)_\rho = -S(A|C)_\rho$$
$$= D(\rho_{AC}\|\mathbb{1}_A \otimes \rho_C)$$
$$= D\left(\sum_{i,j} \sqrt{q_i q_j}\, \mathrm{tr}_B\, |\phi_i\rangle\langle\phi_j|_{AB} \otimes |i\rangle\langle j|_C \,\Big\|\, \mathbb{1}_A \otimes \rho_C\right)$$
$$\geq D\left(\sum_i q_i\, \mathrm{tr}_B\, \phi_i \otimes |i\rangle\langle i|_C \,\Big\|\, \sum_i \mathbb{1}_A \otimes \lambda_i |i\rangle\langle i|_C\right)$$
$$= D(\{q_i\}\|\{\lambda_i\}) + \sum_i q_i D(\mathrm{tr}_B\, \phi_i \| \mathbb{1}_A)$$
$$= D(\{q_i\}\|\{\lambda_i\}) - \sum_i q_i S(\mathrm{tr}_B\, \phi_i)$$
$$\geq -E_F(\rho_{AB}).$$

The bound $S(B|A)_\rho \geq -E_F(\rho_{AB})$ can be proved in an analogous way.

(ii) If a state $\rho_{AB}$ satisfies $\widetilde{S}_\beta(A|B)_\rho = -S_\alpha(A)_\rho$ for $a > 1$ and $\beta = \alpha/(2\alpha - 1)$, then $E_{F,\alpha}(\rho_{AB}) = -\widetilde{S}_\beta(A|B)_\rho \geq -\widetilde{S}_\beta(B|A)_\rho$ by (3.66) in Theorem 3.4.13.

**Entanglement fidelity**

For a state $\rho \in \mathcal{D}(\mathcal{H})$ and a quantum channel $\mathcal{N}: \mathcal{B}(\mathcal{H}) \to \mathcal{B}(\mathcal{H})$, the *entanglement fidelity* $F_e(\rho, \mathcal{N})$ [Sch96] is defined as

$$F_e(\rho, \mathcal{N}) := \sqrt{\langle\psi^\rho|(\mathcal{N} \otimes \mathrm{id}_{\mathcal{H}'})(\psi^\rho)|\psi^\rho\rangle}, \tag{3.68}$$

where $|\psi^\rho\rangle \in \mathcal{H} \otimes \mathcal{H}'$ is a purification of $\rho$. Since any two purifications of $\rho$ are related by an isometry acting on the purifying system, the definition (3.68) of the entanglement fidelity is independent of the chosen purification. For a mixed state $\rho$ with spectral decomposition $\rho = \sum_{i=1}^{\mathrm{rk}\,\rho} \lambda_i |i\rangle\langle i|_\mathcal{H}$, a canonical purification is given by

$$|\psi^\rho\rangle = \sum_{i=1}^{\mathrm{rk}\,\rho} \sqrt{\lambda_i} |i\rangle_\mathcal{H} \otimes |i\rangle_{\mathcal{H}'}$$

for suitable orthonormal vectors $\{|i\rangle_{\mathcal{H}'}\}_{i=1}^{\mathrm{rk}\,\rho}$ in $\mathcal{H}'$. Hence, in the following discussion we can assume without loss of generality that $\dim \mathcal{H}' = \mathrm{rk}\,\rho$.

The entanglement fidelity $F_e(\rho, \mathcal{N})$ can be expressed in terms of the usual fidelity $F(\omega, \tau) \coloneqq \|\sqrt{\omega}\sqrt{\tau}\|_1$ as

$$F_e(\rho, \mathcal{N}) = F(\psi^\rho, (\mathcal{N} \otimes \mathrm{id}_{\mathcal{H}'})(\psi^\rho)).$$

By the discussion in Section 3.2.2, the fidelity is related to the 1/2-SRD via

$$F(\omega, \tau) = \widetilde{Q}_{1/2}(\omega\|\tau). \tag{3.69}$$

Hence, it follows from the DPI for $\widetilde{Q}_{1/2}(\cdot\|\cdot)$ that the fidelity is non-decreasing under partial trace, as mentioned in (3.10). This implies the following upper bound on the entanglement fidelity, where we write $\mathcal{N}(\psi^\rho) \equiv (\mathcal{N} \otimes \mathrm{id}_{\mathcal{H}'})(\psi^\rho)$:

$$F_e(\rho, \mathcal{N}) = F(\psi^\rho, \mathcal{N}(\psi^\rho)) \leq F(\rho, \mathcal{N}(\rho)). \tag{3.70}$$

Due to (3.70), the entanglement fidelity provides a more stringent notion of distance between a quantum state and its image under a quantum channel than the fidelity. However, it is clear that we have equality in (3.70) if the state $\rho$ is pure. The condition for equality in the DPI for the 1/2-SRD from Theorem 3.4.1 (in the form of Proposition 3.4.6) shows that this is in fact the only case of equality:

**Proposition 3.4.15.** *For $\rho \in \mathcal{D}(\mathcal{H})$ and a quantum operation $\mathcal{N}\colon \mathcal{B}(\mathcal{H}) \to \mathcal{B}(\mathcal{H})$, we have $F_e(\rho, \mathcal{N}) = F(\rho, \mathcal{N}(\rho))$ if and only if $\rho$ is pure.*

*Proof.* We have already noted above that purity of $\rho$ is sufficient for equality in (3.70). If $F_e(\rho, \mathcal{N}) = F(\rho, \mathcal{N}(\rho))$, then (3.69) implies that we have equality in the DPI for the 1/2-SRD with respect to $\Lambda = \mathrm{tr}_{\mathcal{H}'}$. Hence, Proposition 3.4.6 implies that

$$\mathcal{N}(\rho)^{1/2} \left( \mathcal{N}(\rho)^{1/2} \rho \mathcal{N}(\rho)^{1/2} \right)^{-1/2} \mathcal{N}(\rho)^{1/2} \otimes \mathbb{1}_{\mathcal{H}'}$$
$$= \mathcal{N}(\psi^\rho)^{1/2} \left( \mathcal{N}(\psi^\rho)^{1/2} \psi^\rho \mathcal{N}(\psi^\rho)^{1/2} \right)^{-1/2} \mathcal{N}(\psi^\rho)^{1/2},$$

and evaluating the powers of the unnormalized pure state on the right-hand side yields

$$\rho \otimes \mathbb{1}_{\mathcal{H}'} = c(\rho, \mathcal{N}) \mathcal{N}(\rho)^{-1} \mathcal{N}(\psi^\rho) \psi^\rho \mathcal{N}(\psi^\rho) \mathcal{N}(\rho)^{-1} \tag{3.71}$$

for some constant $c(\rho, \mathcal{N})$. The (unnormalized) state $\mathcal{N}(\rho)^{-1} \mathcal{N}(\psi^\rho) \psi^\rho \mathcal{N}(\psi^\rho) \mathcal{N}(\rho)^{-1}$ on the right-hand side of (3.71) has rank 1, and hence the same applies to $\rho \otimes \mathbb{1}_{\mathcal{H}'}$. But this is only possible if $\rho$ is pure and $\dim \mathcal{H}' = 1$. $\qquad \square$

# 4. Information spectrum relative entropies

This chapter focuses on another functional on pairs of positive operators called the *information spectrum relative entropy* $D_s^\varepsilon(\cdot\|\cdot)$, which was introduced in the quantum setting by Tomamichel and Hayashi [TH13]. Despite its name, the information spectrum relative entropy is not a true relative entropy in the sense of Section 1.3, as it does not satisfy the data processing inequality, i.e., monotonicity under quantum operations, and can be negative when evaluated on pairs of quantum states. However, there is still a good reason to use this quantity, as it possesses a second order asymptotic expansion. We explain this concept in the following with the example of quantum hypothesis testing.

In this task, Alice gives Bob one of two quantum states $\rho$ (the null hypothesis) and $\sigma$ (the alternative hypothesis), and Bob has to decide which one he received using a quantum measurement given by a two-element POVM $\{T, \mathbb{1} - T\}$. The operator $T$ with $0 \le T \le \mathbb{1}$, also called a *test*, corresponds to accepting the null hypothesis, i.e., Bob infers that he received $\rho$. On the other hand, the test $\mathbb{1} - T$ corresponds to accepting the alternative hypothesis, i.e., Bob infers that he received $\sigma$. There are two fundamental errors that Bob can make: accepting the alternative hypothesis $\sigma$ while in fact he received $\rho$ (type-I error), or accepting the null hypothesis while in fact he received $\sigma$ (type-II error). In terms of the POVM $\{T, \mathbb{1} - T\}$, the probability $\alpha(T)$ of a type-I error and the probability $\beta(T)$ of a type-II error are defined as

$$\alpha(T) \coloneqq \operatorname{tr}\left((\mathbb{1} - T)\rho\right) \qquad\qquad \beta(T) \coloneqq \operatorname{tr}\left(T\sigma\right).$$

There is a fundamental trade-off between these two errors, and hence there are different optimal solutions to quantum hypothesis testing depending on the quantity to be optimized. In asymmetric hypothesis testing, one seeks to minimize the probability of a type-II error $\beta(T)$ under the constraint that the probability of a type-I error $\alpha(T)$ does not exceed a certain threshold $\varepsilon \in [0, 1]$.[1] We denote the minimal probability of a type-II error under this constraint by $\beta_\varepsilon(\rho\|\sigma)$,

$$\beta_\varepsilon(\rho\|\sigma) := \min_{0 \leq T \leq \mathbb{1}} \{\beta(T) \colon \alpha(T) \leq \varepsilon\}. \tag{4.1}$$

Hypothesis testing is usually investigated in the i.i.d. scenario, where Alice and Bob are assumed to have access to $n$ i.i.d. copies of the two states $\rho$ and $\sigma$. More precisely, Bob receives either $\rho^{\otimes n}$ or $\sigma^{\otimes n}$ from Alice and applies a POVM $\{T_n, \mathbb{1}_n - T_n\}$ with $0 \leq T_n \leq \mathbb{1}_n$ to distinguish between the two. We are interested in the behavior of $\beta_\varepsilon(\rho^{\otimes n}\|\sigma^{\otimes n})$ in the asymptotic limit $n \to \infty$. Quantum Stein's Lemma [HP91; ON00] tells us that this behavior is determined by the quantum relative entropy: for all $\varepsilon \in [0, 1]$ and states $\rho$ and $\sigma$, it holds that[2]

$$\lim_{n \to \infty} -\frac{\log \beta_\varepsilon(\rho^{\otimes n}\|\sigma^{\otimes n})}{n} = D(\rho\|\sigma). \tag{4.2}$$

Asymmetric hypothesis testing and its asymptotically optimal solution (4.2) can also be recast in terms of a relative entropy called the *hypothesis testing relative entropy* [WR12]. For $\varepsilon \in [0, 1]$ and states $\rho$ and $\sigma$, it is defined as

$$D_H^\varepsilon(\rho\|\sigma) := -\log \left( \min_{0 \leq T \leq \mathbb{1}} \{\operatorname{tr}(T\sigma) \colon \operatorname{tr}(T\rho) \geq 1 - \varepsilon\} \right). \tag{4.3}$$

Comparing (4.3) to (4.1), it is evident that $D_H^\varepsilon(\rho\|\sigma) = -\log \beta_\varepsilon(\rho\|\sigma)$, and Quantum Stein's Lemma (4.2) can be expressed as

$$D_H^\varepsilon(\rho^{\otimes n}\|\sigma^{\otimes n}) = nD(\rho\|\sigma) + o(n) \qquad \text{for all } \varepsilon \in [0, 1]. \tag{4.4}$$

We call the coefficient of $n$ in (4.4) the *first order coefficient* of $D_H^\varepsilon(\rho^{\otimes n}\|\sigma^{\otimes n})$, given by the quantum relative entropy $D(\rho\|\sigma)$. In a *second order (asymptotic)* expansion, one seeks to further analyze the term $o(n)$, which turns out to be of the order $\sqrt{n}$. Indeed, Li [Li14] and Tomamichel

---

[1]A (classical) real-life example of asymmetric hypothesis testing is an HIV test, the null hypothesis being 'HIV positive' and the alternative hypothesis being 'HIV negative'. For obvious reasons, an undetected HIV infection puts the life of the test subject and others at great risk. Hence, it is imperative to keep the probability of a false rejection of the null hypothesis, or *false negative* (viz. type-I error), as small as possible. One can then try to improve the test statistics by minimizing the probability of a *false positive* (viz. type-II error) under this constraint.

[2]Note that (4.2) is independent of $\varepsilon$, and hence asymmetric quantum hypothesis testing satisfies the *strong converse property*. See Chapter 6 for a detailed discussion of this concept.

and Hayashi [TH13] proved independently from each other that (4.4) can be refined to

$$D_H^\varepsilon(\rho^{\otimes n}\|\sigma^{\otimes n}) = nD(\rho\|\sigma) + \sqrt{nV(\rho\|\sigma)}\Phi^{-1}(\varepsilon) + o(\sqrt{n}), \qquad (4.5)$$

where $V(\rho\|\sigma)$ denotes the quantum information variance from Definition 2.2.5, and $\Phi^{-1}$ denotes the inverse of the cumulative distribution function (CDF) of a standard normal random variable. Hence, the *second order coefficient* of $D_H^\varepsilon(\rho^{\otimes n}\|\sigma^{\otimes n})$ is given by $\sqrt{V(\rho\|\sigma)}\Phi^{-1}(\varepsilon)$. In general, an expansion into terms in $n$, $\sqrt{n}$ and $o(\sqrt{n})$ such as (4.5) is called a *second order asymptotic expansion*. Because of the appearance of $\Phi^{-1}$, a second order asymptotic expansion is often also referred to as Gaussian approximation.

Second order asymptotic expansions were first derived by Strassen [Str62] in classical information theory for the tasks of hypothesis testing and source coding. Li [Li14] and Tomamichel and Hayashi [TH13] introduced the concept in quantum information theory by deriving the second order asymptotics of asymmetric quantum hypothesis testing as stated in (4.5). Note that in both [Li14] and [TH13], the term $o(\sqrt{n})$ in (4.5) was further determined to be of order $O(\log n)$, with the former paper establishing that it lies between a constant and $2\log n$. This leads to third order asymptotic expansions, with the third order being logarithmic in $n$. The dependence on $n$ of the first, second, and third orders in typical asymptotic expansions is summarized in Table 4.1. Tomamichel and Hayashi [TH13] also obtained the second order expansion of a 'smooth' version of the max-relative entropy $D_{\max}(\cdot\|\cdot)$ defined in Definition 3.2.2.

| Order | 1st | 2nd | 3rd |
|---|---|---|---|
| $f(n)$ | $n$ | $\sqrt{n}$ | $\log n$ |

Table 4.1: Orders in typical asymptotic expansions in information theory.

Second order asymptotic expansions are useful for a number of reasons. On a mathematical level, they determine the rate of convergence of an averaged quantity to the first order coefficient.[3] In the case of quantum hypothesis testing, (4.5) yields

$$-\frac{\log\beta_n(\varepsilon)}{n} = \frac{D_H^\varepsilon(\rho^{\otimes n}\|\sigma^{\otimes n})}{n} = D(\rho\|\sigma) + \frac{1}{\sqrt{n}}\sqrt{V(\rho\|\sigma)}\Phi^{-1}(\varepsilon) + g(n) \qquad (4.6)$$

---

[3]This relation is analogous to the one between the Central Limit Theorem and the Berry-Esseen Theorem 4.2.2, as the latter determines the rate of convergence in the former. Indeed, the Berry-Esseen Theorem is typically used in deriving second order asymptotic expansions, as demonstrated in Section 5.2.

for sufficiently large $n$, where $g(n) \in O((\log n)/n)$. Hence, (4.6) determines the rate of convergence in Quantum Stein's Lemma (4.2) (and implies it).

On an information-theoretic level, second order asymptotic expansions of relative entropies can be used to determine a corresponding second order expansion of operational quantities such as the minimal compression length in source coding (cf. Chapter 5). These second order expansions of operational quantities provide a useful approximation for finite blocklength $n$, refining optimal rates that typically correspond to the first order coefficient in asymptotic expansions. An example of this refinement in the case of quantum source coding is depicted in Figure 5.3 in Chapter 5. Furthermore, second order expansions of operational quantities can be used to derive strong converse theorems, which we further elucidate in Chapter 6.

In this chapter, we define the information spectrum relative entropy in Section 4.1, and derive its second order asymptotic expansion in Section 4.2 following [TH13]. This result provides the main tool for our analysis of the second order asymptotics of quantum source coding in Chapter 5. In addition to the present chapter, Appendix A discusses variants of the information spectrum relative entropy that we introduced in [DL15]. These quantities have the particular advantage of satisfying the DPI, in contrast to the information spectrum relative entropy defined in [TH13], which does not satisfy this property (as discussed in more detail in Section 4.1 below).

## 4.1. Definition and properties

We first define the main quantity of this chapter:

**Definition 4.1.1** (Information spectrum relative entropy; [TH13])**.**
Let $\rho \in \mathcal{D}(\mathcal{H})$ and $\sigma \in \mathcal{P}(\mathcal{H})$. For $\varepsilon \in [0, 1]$, the *information spectrum relative entropy* is defined as

$$D_s^\varepsilon(\rho \| \sigma) := \sup \left\{ \gamma \in \mathbb{R} \colon \operatorname{tr}\left( \rho \left\{ \rho \leq 2^\gamma \sigma \right\} \right) \leq \varepsilon \right\}.$$

This quantity is a quantum generalization of the *classical information spectrum relative entropy*, which can be seen as follows. Assume that $\rho$ and $\sigma$ commute, that is, there is a common eigenbasis $\{|i\rangle\}_i$ for $\mathcal{H}$ such that $\rho = \sum_i r_i |i\rangle\langle i|$ and $\sigma = \sum_i s_i |i\rangle\langle i|$, and define the random variables $P$ and $Q$ corresponding to the eigenvalue distributions $\{r_i\}_i$ and $\{s_i\}_i$, respectively (i.e., $P$ takes the value $r_i$ with probability $r_i$, and similarly for $Q$). Furthermore, we define a random variable $\log(P/Q) = \log P - \log Q$ that takes the value $\log(r_i/s_i)$ with probability $r_i$. We then have $\text{tr}\,(\rho\,\{\rho \leq 2^\gamma \sigma\}) = \Pr\{\log P - \log Q \leq \gamma\}$, and $D_s^\varepsilon(\rho\|\sigma)$ reduces to the classical information spectrum relative entropy

$$D_s^\varepsilon(P\|Q) := \sup\{\gamma \in \mathbb{R} \colon \Pr\{\log P - \log Q \leq \gamma\} \leq \varepsilon\}.$$

As mentioned before, the information spectrum relative entropy $D_s^\varepsilon(\rho\|\sigma)$ does not satisfy the data processing inequality. By Lemma 2.2.9, $D_s^\varepsilon(\cdot\|\cdot)$ is invariant under joint unitaries and tensoring with a fixed state. Hence, by the Stinespring Representation Theorem 2.2.2, the data processing inequality is equivalent to joint convexity of $D_s^\varepsilon(\cdot\|\cdot)$, and it is easy to find numerical examples that violate this property. Moreover, one can find examples of pairs of states $\rho$ and $\sigma$ and $\varepsilon \in (0, 1)$ such that $D_s^\varepsilon(\rho\|\sigma) < 0$.

For these reasons, the information spectrum relative entropy should only be considered an auxiliary quantity. Its main use lies in the fact that it has a second order asymptotic expansion (Corollary 4.2.8), which is the subject of the next section. Furthermore, one can bound $D_s^\varepsilon(\cdot\|\cdot)$ in terms of the hypothesis testing relative entropy $D_H^\varepsilon(\cdot\|\cdot)$ and the *collision relative entropy* $\widetilde{D}_2(\cdot\|\cdot)$ [Ren05], both of which do satisfy the DPI. Therefore, the second order expansion of the information spectrum relative entropy can be used to derive second order asymptotics for information-processing tasks [TH13; BG14; BDL16]. We give a more detailed account of these and other second order asymptotic results in Chapter 5.

We also note that there are closely related variants of the information spectrum relative entropy that do satisfy the data processing inequality. They were introduced in [DL15] to derive second order asymptotics of quantum source coding, noisy dense-coding, entanglement conversion, and classical-quantum channel coding. We define and analyze these variants in Appendix A.

## 4.2. Second order asymptotic expansion

In this section we derive the second order asymptotic expansion of the information spectrum relative entropy $D_s^\varepsilon(\rho^{\otimes n}\|\sigma^{\otimes n})$, closely following the proof in [TH13]. We first define the classical Nussbaum-Szkoła distributions $P_{\rho,\sigma}$ and $Q_{\rho,\sigma}$ [NS09], which encode useful information about the quantum states $\rho$ and $\sigma$. These distributions are then used to derive a second order asymptotic expansion of the classical quantity $D_s^\varepsilon(P_{\rho,\sigma}^n\|Q_{\rho,\sigma}^n)$, where for a probability distribution $P$ on an alphabet $\mathcal{X}$ we denote by $P^n$ the $n$-fold i.i.d. distribution on $\mathcal{X}^n$, i.e., $P^n(x^n) = \prod_{i=1}^{n} P(x_i)$ for a string $x^n = (x_1, \dots, x_n) \in \mathcal{X}^n$. Finally, the desired asymptotic expansion of the quantum quantity $D_s^{\varepsilon}(\rho^{\otimes n} \| \sigma^{\otimes n})$ is obtained from bounds on $D_s^{\varepsilon}(\rho \| \sigma)$ in terms of $D_s^{\varepsilon}(P_{\rho,\sigma} \| Q_{\rho,\sigma})$.

## 4.2.1. Nussbaum-Szkoła distributions

Given $\rho \in \mathcal{D}(\mathcal{H})$ and $\sigma \in \mathcal{P}(\mathcal{H})$ with $\dim \mathcal{H} = d$ and eigenvalue decompositions

$$\rho = \sum_{i=1}^{d} r_i |e_i\rangle\langle e_i| \qquad\qquad \sigma = \sum_{i=1}^{d} s_i |f_i\rangle\langle f_i|,$$

we define probability distributions $P_{\rho,\sigma}$ and $Q_{\rho,\sigma}$ on $[d] \times [d]$ by[4]

$$P_{\rho,\sigma}(i,j) := r_i |\langle e_i|f_j\rangle|^2 \qquad\qquad Q_{\rho,\sigma}(i,j) := s_j |\langle e_i|f_j\rangle|^2. \tag{4.7}$$

These distributions are called *Nussbaum-Szkoła distributions*. We also recall the definition of the *information variance* $V(P\|Q)$:

$$V(P\|Q) := \mathbb{E}\left((Z - D(P\|Q))^2\right) = \mathbb{E}\left(Z^2\right) - D(P\|Q)^2, \tag{4.8}$$

where for random variables $P$ and $Q$ we define a random variable $Z = \log \frac{P}{Q}$ called the *log-likelihood ratio* that has the same law as $P$. For classical states $\rho = \sum_x P(x)|x\rangle\langle x|$ and $\sigma = \sum_x Q(x)|x\rangle\langle x|$ satisfying $[\rho, \sigma] = 0$, we have $V(\rho\|\sigma) = V(P\|Q)$, that is, the quantum information variance from Definition 2.2.5(ii) reduces to the classical quantity in (4.8).

We can now derive the following properties of the Nussbaum-Szkoła distributions in a straightforward way:

**Proposition 4.2.1** (Properties of Nussbaum-Szkoła distributions)**.**
*Let $\rho \in \mathcal{D}(\mathcal{H})$, $\sigma \in \mathcal{P}(\mathcal{H})$, and let $P_{\rho,\sigma}$ and $Q_{\rho,\sigma}$ be the corresponding Nussbaum-Szkoła distributions as defined in* (4.7). *Then the following properties are satisfied:*

(i) $\operatorname{supp} P_{\rho,\sigma} \subseteq \operatorname{supp} Q_{\rho,\sigma}$ *if and only if* $\operatorname{supp} \rho \subseteq \operatorname{supp} \sigma$.

---

[4]Strictly speaking, $Q_{\rho,\sigma}$ is only normalized, $\sum_{i,j=1}^{d} Q_{\rho,\sigma}(i,j) = 1$, if $\operatorname{tr}\sigma = 1$. However, we still refer to $P_{\rho,\sigma}$ and $Q_{\rho,\sigma}$ collectively as probability distributions even if $\operatorname{tr}\sigma \neq 1$.

(ii) *If* supp $\rho \subseteq$ supp $\sigma$, *then*

$$D\left(P_{\rho,\sigma} \,\middle\|\, Q_{\rho,\sigma}\right) = D(\rho\|\sigma) \qquad\qquad V\left(P_{\rho,\sigma} \,\middle\|\, Q_{\rho,\sigma}\right) = V(\rho\|\sigma).$$

(iii) *For i.i.d. operators $\rho^{\otimes n}$ and $\sigma^{\otimes n}$, the corresponding Nussbaum-Szkoła distributions are also of i.i.d. form:*

$$P_{\rho^{\otimes n},\sigma^{\otimes n}} = P_{\rho,\sigma}^{n} \qquad\qquad Q_{\rho^{\otimes n},\sigma^{\otimes n}} = Q_{\rho,\sigma}^{n}.$$

Proposition 4.2.1(ii) shows that the first two moments of the operators $\rho$ and $\sigma$ coincide with those of the Nussbaum-Szkoła distributions $P_{\rho,\sigma}$ and $Q_{\rho,\sigma}$. This allows us to employ the Nussbaum-Szkoła distributions in the derivation of the second order asymptotic expansion of $D_s^\varepsilon(\rho^{\otimes n}\|\sigma^{\otimes n})$, as exhibited in the next section.

### 4.2.2. Deriving the main result

In view of Proposition 4.2.1 of the preceding section, it is useful to first derive a second order asymptotic expansion of the classical quantity $D_s^\varepsilon(P_{\rho,\sigma}^n\|Q_{\rho,\sigma}^n)$. A key tool in the proof is the well-known Berry-Esseen Theorem. Before we state it, we first recall the Central Limit Theorem.

Let $\{X_i\}_{i\in\mathbb{N}}$ be a sequence of i.i.d. random variables with $\mu := \mathbb{E}(X_i)$ and $\sigma^2 := \mathbb{E}\left((X_i - \mu)^2\right) < \infty$ for all $i$, and define the sequence $\{S_n\}_{n\in\mathbb{N}}$ of random variables $S_n := \frac{1}{n}\sum_{i=1}^{n} X_i$. The Central Limit Theorem states that the random variable $\sqrt{n}(S_n - \mu)/\sigma$ converges in distribution to a standard normal random variable $N(0,1)$, that is,

$$\lim_{n\to\infty} \Pr\left[\frac{\sqrt{n}}{\sigma}(S_n - \mu) \le z\right] = \Phi(z) \quad \text{for all } z \in \mathbb{R},$$

where $\Phi$ denotes the CDF of a standard normal random variable. The Berry-Esseen Theorem determines the rate of convergence in the Central Limit Theorem:

**Theorem 4.2.2** (Berry-Esseen Theorem; [Ber41; Ess42]).
*Let $\{X_i\}_{i\in\mathbb{N}}$ be a sequence of i.i.d. random variables with $\mu := \mathbb{E}(X_i)$, $\sigma^2 := \mathbb{E}\left((X_i - \mu)^2\right)$, and $t^3 := \mathbb{E}\left(|X_i - \mu|^3\right) < \infty$ for all $i \in \mathbb{N}$. Define the random variable*

$$Y_n := \frac{\sum_{i=1}^{n} X_i - n\mu}{\sqrt{n}\sigma},$$

*and let $F_{Y_n}(x) := \Pr(Y_n \leq x)$ be the CDF of $Y_n$. Then there is a constant $C \in (0, 1/2)$ such that for all $z \in \mathbb{R}$ and $n \in \mathbb{N}$ we have*

$$|F_{Y_n}(z) - \Phi(z)| \leq \frac{Ct^3}{\sigma^3\sqrt{n}}.$$

Note that in this thesis all probability distributions corresponding to quantum states have finite support, since we only consider finite-dimensional Hilbert spaces. Therefore, the condition $\mathbb{E}\left(|X_i - \mu|^3\right) < \infty$ in Theorem 4.2.2 is always satisfied in our applications. For arbitrary probability distributions $P$ and $Q$ with $D(P\|Q), V(P\|Q) < \infty$, we now have the following result:

**Proposition 4.2.3** ([TH13]). *Let $P$ and $Q$ be probability distributions with supports on a finite set $\mathcal{X}$ (where $Q$ need not be normalized), and such that $P(x) = 0$ whenever $Q(x) = 0$ for $x \in \mathcal{X}$. Then for $\varepsilon \in (0, 1)$,*

$$D_s^\varepsilon(P^n\|Q^n) = nD(P\|Q) + \sqrt{nV(P\|Q)}\Phi^{-1}(\varepsilon) + O(1).$$

*Proof.* For $i \in \mathbb{N}$, we define the i.i.d. random variables $X_i := \log P - \log Q$ with the same law as $P$, and note that

$$\begin{aligned}
\mu &= \mathbb{E}(X_i) = D(P\|Q) \\
\sigma^2 &= \mathbb{E}\left((X_i - \mu)^2\right) = V(P\|Q).
\end{aligned} \tag{4.9}$$

Forming the random variables $S_n = \frac{1}{n}\sum_{i=1}^n X_i$ and $Y_n = \frac{\sqrt{n}}{\sigma}(S_n - \mu)$ as above, we observe that

$$\begin{aligned}
D_s^\varepsilon(P^n\|Q^n) &= \sup\{\gamma \in \mathbb{R}\colon \Pr\{\log P^n - \log Q^n \leq \gamma\} \leq \varepsilon\} \\
&= n\sup\{\gamma \in \mathbb{R}\colon \Pr\{S_n \leq \gamma\} \leq \varepsilon\} \\
&= nF_{S_n}^{-1}(\varepsilon) \\
&= n\mu + \sqrt{n}\sigma F_{Y_n}^{-1}(\varepsilon),
\end{aligned} \tag{4.10}$$

where for a random variable $Z$ we denote by $F_Z(x) := \Pr(Z \leq x)$ the CDF of $Z$, and its inverse $F_Z^{-1}$ is defined by $F_Z^{-1}(\varepsilon) := \sup\{x \in \mathbb{R}\colon F_Z(x) \leq \varepsilon\}$. The last step of (4.10) follows since $S_n = \frac{\sigma}{\sqrt{n}}Y_n + \mu$. Together with (4.9), we arrive at

$$D_s^\varepsilon(P^n\|Q^n) = nD(P\|Q) + \sqrt{nV(P\|Q)}F_{Y_n}^{-1}(\varepsilon).$$

We now want to relate the term $F_{Y_n}^{-1}(\varepsilon)$ to $\Phi^{-1}(\varepsilon)$. The Berry-Esseen Theorem 4.2.2 gives

$$|F_{Y_n}(z) - \Phi(z)| \leq \frac{C_1}{\sqrt{n}}, \tag{4.11}$$

where $C_1 := Ct^3/\sigma^3$ with $t^3 := \mathbb{E}\left(|X_i - \mu|^3\right)$, which is always finite if $P$ and $Q$ are supported on a finite set. Let $z \in \mathbb{R}$ be such that $F_{Y_n}(z) = \varepsilon$. Then (4.11) yields

$$\Phi(z) \leq F_{Y_n}(z) + \frac{C_1}{\sqrt{n}} = \varepsilon + \frac{C_1}{\sqrt{n}},$$

and hence,

$$\Phi^{-1}\left(\varepsilon + \frac{C_1}{\sqrt{n}}\right) = \sup\left\{z' \in \mathbb{R} \colon \Phi(z') \leq \varepsilon + \frac{C_1}{\sqrt{n}}\right\} \geq z = F_{Y_n}^{-1}(\varepsilon). \tag{4.12}$$

In a similar way, one can show that

$$\Phi^{-1}\left(\varepsilon - \frac{C_1}{\sqrt{n}}\right) \leq F_{Y_n}^{-1}(\varepsilon). \tag{4.13}$$

Using (4.12) and (4.13) together with (4.9) in (4.10), we obtain

$$\begin{aligned}
D_s^\varepsilon(P^n \| Q^n) &\geq nD(P\|Q) + \sqrt{nV(P\|Q)}\Phi^{-1}\left(\varepsilon - \frac{C_1}{\sqrt{n}}\right) \\
D_s^\varepsilon(P^n \| Q^n) &\leq nD(P\|Q) + \sqrt{nV(P\|Q)}\Phi^{-1}\left(\varepsilon + \frac{C_1}{\sqrt{n}}\right).
\end{aligned} \tag{4.14}$$

Since $\Phi^{-1}$ is continuously differentiable, a Taylor expansion around $\varepsilon$ shows that

$$\sqrt{n}\Phi^{-1}\left(\varepsilon \pm \frac{C_1}{\sqrt{n}}\right) = \sqrt{n}\Phi^{-1}(\varepsilon) + O(1),$$

and inserting this in (4.14) yields the claim. $\qquad\square$

An immediate corollary of Proposition 4.2.1 and Proposition 4.2.3 is the following:

**Corollary 4.2.4.** *For $\varepsilon \in (0,1)$, $\rho \in \mathcal{D}(\mathcal{H})$, and $\sigma \in \mathcal{P}(\mathcal{H})$ with $\operatorname{supp}\rho \subseteq \operatorname{supp}\sigma$,*

$$D_s^\varepsilon\left(P_{\rho^{\otimes n}, \sigma^{\otimes n}} \,\middle\|\, Q_{\rho^{\otimes n}, \sigma^{\otimes n}}\right) = nD(\rho\|\sigma) + \sqrt{nV(\rho\|\sigma)}\Phi^{-1}(\varepsilon) + O(1).$$

It remains to bound $D_s^\varepsilon(\rho^{\otimes n}\|\sigma^{\otimes n})$ in terms of $D_s^\varepsilon(P_{\rho^{\otimes n}, \sigma^{\otimes n}}\|Q_{\rho^{\otimes n}, \sigma^{\otimes n}})$, since by Corollary 4.2.4 the second order asymptotic expansion of the latter then implies the one for the former. These bounds were derived in [TH13], and we state them in Proposition 4.2.5 below. As the proof of this proposition is rather involved, we omit it here for the sake of brevity.

**Proposition 4.2.5** ([TH13])**.** *Let $\rho \in \mathcal{D}(\mathcal{H})$ and $\sigma \in \mathcal{P}(\mathcal{H})$ with corresponding Nussbaum-Szkoła distributions $P_{\rho,\sigma}$ and $Q_{\rho,\sigma}$. For $\varepsilon \in (0, 1)$, let $\eta \in (0, \varepsilon)$ and $\delta > 0$ be such that $\delta < \min\{\varepsilon, 1 - \varepsilon\}$ and $\delta + \eta < \varepsilon$. Furthermore, let $\nu := \nu(\sigma)$ denote the number of distinct eigenvalues of $\sigma$. We then have*

$$D_s^{\varepsilon-\eta-\delta}\left(P_{\rho,\sigma} \,\middle\|\, Q_{\rho,\sigma}\right) + \log\frac{\delta\eta}{\nu} \leq D_s^{\varepsilon}(\rho\|\sigma) \leq D_s^{\varepsilon+\delta}\left(P_{\rho,\sigma} \,\middle\|\, Q_{\rho,\sigma}\right) + \log\frac{2^8(\varepsilon+\delta)\nu}{\delta^4(1-\varepsilon-\delta)}.$$

We are now ready to prove the second order asymptotic expansion of $D_s^{\varepsilon}(\rho^{\otimes n}\|\sigma^{\otimes n})$, which is the main goal of this chapter.

**Theorem 4.2.6** (Second order expansion of the information spectrum relative entropy; [TH13])**.** *Let $\rho \in \mathcal{D}(\mathcal{H})$ and $\sigma \in \mathcal{P}(\mathcal{H})$ with supp $\rho \subseteq$ supp $\sigma$. For $\varepsilon \in (0, 1)$, the second order asymptotic expansion of the information spectrum relative entropy is given by*

$$D_s^{\varepsilon}(\rho^{\otimes n}\|\sigma^{\otimes n}) = nD(\rho\|\sigma) + \sqrt{nV(\rho\|\sigma)}\Phi^{-1}(\varepsilon) + O(\log n).$$

*Proof.* First, observe that $\nu(\sigma^{\otimes n})$, equal to the number of distinct eigenvalues of $\sigma^{\otimes n}$, grows at most polynomially in $n$, which follows from a type-counting argument [CK11]. Hence, $\log(\nu(\sigma^{\otimes n})) \in O(\log n)$, and using the upper bound of Proposition 4.2.5 with $\delta = 1/\sqrt{n}$ and setting $\varepsilon_n = \varepsilon + \delta$, we obtain

$$\begin{aligned}
D_s^{\varepsilon}(\rho^{\otimes n}\|\sigma^{\otimes n}) &\leq D_s^{\varepsilon_n}\left(P_{\rho^{\otimes n},\sigma^{\otimes n}} \,\middle\|\, Q_{\rho^{\otimes n},\sigma^{\otimes n}}\right) + O(\log n) \\
&= nD(\rho\|\sigma) + \sqrt{nV(\rho\|\sigma)}\Phi^{-1}(\varepsilon_n) + O(\log n) \\
&= nD(\rho\|\sigma) + \sqrt{nV(\rho\|\sigma)}\Phi^{-1}(\varepsilon) + O(\log n),
\end{aligned} \qquad (4.15)$$

where we used Corollary 4.2.4 in the first equality, and the fact that

$$\sqrt{n}\Phi^{-1}(\varepsilon_n) = \sqrt{n}\Phi^{-1}\left(\varepsilon + \frac{1}{\sqrt{n}}\right) = \sqrt{n}\Phi^{-1}(\varepsilon) + O(1)$$

in the second equality. In a similar way, using the lower bound of Proposition 4.2.5 with

$\delta = \eta = 1/\sqrt{n}$, we obtain

$$D_s^\varepsilon(\rho^{\otimes n}\|\sigma^{\otimes n}) \geq nD(\rho\|\sigma) + \sqrt{nV(\rho\|\sigma)}\Phi^{-1}(\varepsilon) + O(\log n),$$

which together with (4.15) yields the claim. □

**Remark 4.2.7.** Since the quantum relative entropy $D(\cdot\|\cdot)$ is non-negative on pairs of quantum states and satisfies the DPI, it follows from Theorem 4.2.6 that the information spectrum relative entropy $D_s^\varepsilon(\cdot\|\cdot)$ inherits these properties when evaluated on pairs of sufficiently many copies of i.i.d. states: for quantum states $\rho$ and $\sigma$, a quantum operation $\Lambda$, and $\varepsilon \in (0, 1)$, we have the following for sufficiently large $n$:

$$D_s^\varepsilon(\rho^{\otimes n}\|\sigma^{\otimes n}) \geq 0 \qquad\qquad D_s^\varepsilon(\rho^{\otimes n}\|\sigma^{\otimes n}) \geq D_s^\varepsilon(\Lambda(\rho)^{\otimes n}\|\Lambda(\sigma)^{\otimes n}).$$

In Chapter 5, we are mainly interested in the case where $\sigma = \mathbb{1}$ in Theorem 4.2.6. In this case, the quantity $D_s^\varepsilon(\rho\|\mathbb{1})$ reduces to a function of the eigenvalues of $\rho$, i.e., $D_s^\varepsilon(\rho\|\mathbb{1})$ is a classical quantity. Indeed, the following Corollary 4.2.8 was derived by Strassen [Str62] as part of his work on second order expansions for classical hypothesis testing and source coding. It admits a straightforward proof based on the Berry-Esseen Theorem, which we recapitulate in the following.

**Corollary 4.2.8** ([Str62])**.** *Let* $\rho \in \mathcal{D}(\mathcal{H})$, *and set* $S := S(\rho)$ *and* $\sigma := \sigma(\rho)$ *as defined in* (2.3). *Then, for any* $a > 0$ *and* $C \in \mathbb{R}$, *we have*

$$\lim_{n\to\infty} \mathrm{tr}\left(\rho^{\otimes n}\left\{\rho^{\otimes n} \leq 2^{-na+\sqrt{n}C}\mathbb{1}_n\right\}\right) = \begin{cases} 0 & \text{if } S < a, \\ \Phi\left(C/\sigma\right) & \text{if } S = a, \\ 1 & \text{if } S > a. \end{cases}$$

*Proof.* By definition (4.7) of the Nussbaum-Szkoła distributions, the distribution $P_\rho \equiv P_{\rho,\mathbb{1}}$ consists of the eigenvalues of $\rho$, and it is easy to see that $P_{\rho^{\otimes n}} = P_\rho^n$ (see also Proposition 4.2.1(iii)). Furthermore, for the random variable $\log P_\rho$ with distribution $P_\rho$, Proposition 4.2.1(ii) and Definition 2.2.5 imply that

$$\mu = \mathbb{E}(\log P_\rho) = -S(\rho) \qquad\qquad \sigma^2 = \mathbb{E}\left((\log P_\rho - \mu)^2\right) = \sigma(\rho)^2.$$

Hence, for arbitrary $\gamma \in \mathbb{R}$,

$$
\begin{aligned}
\operatorname{tr}\left(\rho^{\otimes n}\left\{\rho^{\otimes n} \leq 2^{\gamma}\mathbb{1}_n\right\}\right) &= \operatorname{Pr}\left\{P_{\rho}^n \leq 2^{\gamma}\right\} \\
&= \operatorname{Pr}\left\{\log P_{\rho} \leq \frac{\gamma}{n}\right\} \\
&= \operatorname{Pr}\left\{Y_n \leq \frac{1}{\sqrt{n}\sigma}(\gamma - n\mu)\right\} \\
&= F_{Y_n}\left(\frac{1}{\sqrt{n}\sigma}(\gamma - n\mu)\right),
\end{aligned}
$$

where we defined $Y_n = \frac{\sqrt{n}}{\sigma}(\log P_{\rho} - \mu)$ just as in the proof of Proposition 4.2.3. Choosing $\gamma = n\mu + \sqrt{n}C = -nS + \sqrt{n}C$ and using the Berry-Esseen Theorem 4.2.2 yields

$$
\left|\operatorname{tr}\left(\rho^{\otimes n}\left\{\rho^{\otimes n} \leq 2^{-nS + \sqrt{n}C}\mathbb{1}_n\right\}\right) - \Phi\left(\frac{C}{\sigma}\right)\right| \leq \frac{K}{\sqrt{n}}, \tag{4.16}
$$

where $K$ is a suitable constant independent of $n$. This proves the claim in the case $S = a$.

In the case $S < a$, define $f_n := \sqrt{n}(a - S)$ and note that $f_n \to \infty$ for $n \to \infty$ by assumption. We then obtain the following from (4.16) for any $C \in \mathbb{R}$:

$$
\begin{aligned}
\operatorname{tr}\left(\rho^{\otimes n}\left\{\rho^{\otimes n} \leq 2^{-na + \sqrt{n}C}\mathbb{1}_n\right\}\right) &= \operatorname{tr}\left(\rho^{\otimes n}\left\{\rho^{\otimes n} \leq 2^{-nS + \sqrt{n}(C - f_n)}\mathbb{1}_n\right\}\right) \\
&\leq \Phi\left(\frac{C - f_n}{\sigma}\right) + \frac{K}{\sqrt{n}} \xrightarrow{n \to \infty} 0,
\end{aligned}
$$

where we used $\lim_{x \to -\infty} \Phi(x) = 0$. The case $S > a$ is proved along similar lines. $\qquad\square$

# Part II.

# Characterizing information-theoretic tasks

# 5. Second order asymptotics of quantum source coding

In the previous chapter, we discussed second order asymptotic expansions of relative entropies such as the hypothesis testing relative entropy and the information spectrum relative entropy. The aim of the present chapter is to use the second order expansion of the latter quantity to determine the second order asymptotics of quantum source coding. In this task, the goal is to compress signals emitted from a quantum source such that they can later be retrieved with an error less than a given threshold value (see Section 5.1 for a detailed discussion of the operational setting). The main operational quantity in quantum source coding is the *minimal compression length* $\log M_n$, where $M_n$ denotes the dimension of the Hilbert space that accommodates the compressed signals emitted on $n$ uses of the source.

Determining the second order asymptotics of quantum source coding amounts to deriving an expansion of $\log M_n$ of the form

$$\log M_n = an + b\sqrt{n} + o(\sqrt{n}), \tag{5.1}$$

where the first order asymptotic rate $a$ is equal to the optimal rate of quantum source coding, and the second order asymptotic rate $b$ is to be determined. In the case of a memoryless source, the first order rate $a$ is given by the von Neumann entropy, as discussed at the end of Section 5.1.2.[1]

There are at least two ways to derive expansions such as (5.1). One option is to first prove so-called 'one-shot' bounds on the quantity $\log M$ (where $M \equiv M_1$) in terms of relative entropies whose second order asymptotic expansion is known. Combining the one-shot bounds for $\log M_n$ with these expansions then yields the result. This method is for instance employed in [DL15], where (5.1) is derived for a memoryless source (amongst other results).

---

[1] Here, we have in fact used Winter's strong converse theorem for quantum source coding [Win99b] to infer that $a$ is independent of the error $\varepsilon$ incurred in the protocol. We will see in Theorem 5.2.8 and Theorem 5.2.10 that this is not the case for quantum source coding using a mixed source.

Another option is to determine the optimal achievable second order asymptotic rate $b$ in (5.1) 'directly' under a given error constraint, using the second order asymptotic expansion of the information spectrum entropy in the form of Corollary 4.2.8. This method was employed in [LD16], on which the present chapter is based. We derive the second order asymptotics of visible quantum source coding using a mixed source (which we define in Section 5.1), obtaining the results for a memoryless source from [DL15] as a corollary.

The first results about second order asymptotic expansions in quantum information theory were obtained by Li [Li14] and Tomamichel and Hayashi [TH13], who independently from each other derived the second order asymptotics of quantum hypothesis testing. In both papers, this result was obtained by deriving a second order asymptotic expansion of the hypothesis testing relative entropy $D_H^\varepsilon(\rho^{\otimes n} \| \sigma^{\otimes n})$, which we discussed previously in Chapter 4. In [TH13], the authors also used the asymptotic expansion of $D_H^\varepsilon(\rho^{\otimes n} \| \sigma^{\otimes n})$ to derive the second order asymptotics of classical data compression with quantum side information and randomness extraction. Furthermore, [TT15] used a refined asymptotic expansion of the hypothesis testing relative entropy to derive the second order asymptotics of classical-quantum channel coding. Second order asymptotic achievability bounds for quantum hypothesis testing, classical-quantum channel coding, and classical data compression with quantum side information were rederived by Beigi and Gohari [BG14]. Their proof method is based on the pretty good measurement decoding and the second order asymptotic expansion of the information spectrum relative entropy (Theorem 4.2.6).

Kumagai and Hayashi [KH13] and the authors of [DL15] derived second order asymptotic characterizations of entanglement concentration and entanglement dilution. The goal of these tasks is to convert $n$ copies of a given pure bipartite state $\psi_{AB}$ into an MES of Schmidt rank $M_n$ (entanglement concentration), or vice versa (entanglement dilution), using local operations and classical communication. For both tasks, the optimal rate is given by the entanglement entropy of $\psi_{AB}$, which is equal to the von Neumann entropy of one of the marginals:

$$\lim_{n \to \infty} \frac{\log M_n}{n} = S(\rho_A),$$

where $\rho_A = \text{tr}_B \psi_{AB}$.

Since the optimal rates of the two tasks coincide, it was long assumed that entanglement concentration and dilution are asymptotically reversible. However, both [KH13] and [DL15] used second order asymptotic considerations to show that the two tasks are in fact irreversible. Kumagai and Hayashi [KH13] considered the total error of a concatenated concentration-dilution protocol and showed that this error always converges to 1 if $\psi_{AB}$ is not maximally entangled. In [DL15], we proved the following second order expansions of the operational quantities $\log M_n^c$ and $\log M_n^d$ denoting the Schmidt ranks of the MES in entanglement concentration and entanglement dilution, respectively:

$$\log M_n^c = nS(\rho_A) + \sqrt{n}\sigma(\rho_A)\Phi^{-1}(\varepsilon) + O(\log n)$$
$$\log M_n^d = nS(\rho_A) - \sqrt{n}\sigma(\rho_A)\Phi^{-1}(\varepsilon) + O(\log n),$$

where $\varepsilon \in (0, 1)$ is the error incurred in the protocol. If $\varepsilon \in (0, 1/2)$, we have $\Phi^{-1}(\varepsilon) < 0$ and hence $\log M_n^c < \log M_n^d$ for sufficiently large $n$. This gap can be used to prove the asymptotic irreversibility of entanglement conversion.

In [DL15], the authors also derived the second order asymptotics of quantum source coding using a memoryless source in both the visible and the blind encoding setting, which we define in Section 5.1.2. We will obtain this result as a corollary of the second order asymptotics of source coding using a mixed source, which is the main result of this chapter. Other instances of second order asymptotic characterizations of information-processing tasks include noisy dense-coding [DL15], and achievability bounds on the coding rate for entanglement-assisted communication [DTW16], the quantum communication cost in state redistribution [DHO16] (see Section 6.1.1 for a detailed description of this protocol), and the quantum capacity [BDL16; TBR16].

Common to all second order asymptotic results mentioned so far is the assumption that the underlying resource is memoryless. For example, in the case of quantum source coding using a memoryless source, the resource is the *source state* of the quantum source (to be defined in Section 5.1.1), which is assumed to be of i.i.d. form, that is, given by $\rho^{\otimes n}$. Accordingly, the second order asymptotic expansions that we discussed in Chapter 4 were derived for relative entropies evaluated on i.i.d. states.

Obtaining second order asymptotic expansions for any information-processing task employing resources with memory is a more challenging undertaking. The first foray in this direction was made in classical information theory by Polyanskiy, Poor and Verdú [PPV11], who obtained second order expansions for the capacity of a classical mixed channel (see also [TT14]). In [NH13], Nomura and Han evaluated second order optimal rates for fixed-length source coding using a classical mixed source (see also [Hay08]). Yagi et al. [YHN16] derived the second order coding rate, or channel dispersion, of a mixed classical channel under the assumption that the channel is *well-ordered* (see Definition 3 in [YHN16]).

The information-processing tasks mentioned above involving a mixed channel or a mixed source are simple yet instructive examples of tasks employing resources with memory. As a quantum generalization of the results in [NH13], the present chapter treats quantum source coding using a mixed quantum source based on [LD16]. In Section 5.1, we introduce memoryless and mixed quantum sources, and define the information-theoretic task of quantum source coding. In Section 5.2, we then derive the main result, which marked the first second order asymptotic analysis of a quantum information-theoretic task with memory. In this context, we also mention the recent work [DPR16], in which the authors derive second order asymptotic expansions for quantum hypothesis testing involving non-i.i.d. hypotheses. This includes for example the Gibbs states of quantum spin systems at high temperatures.

## 5.1. Operational setting

### 5.1.1. Quantum sources

Imagine a physical system that emits signals described by pure states in a Hilbert space $\mathcal{H}$, e.g., a highly attenuated laser emitting single photons. We assume that these signals are taken from a set $\{|\psi_i\rangle\}_{i=1}^m \subseteq \mathcal{H}$ of pure states of cardinality $m \in \mathbb{N}$, and each signal $|\psi_i\rangle$ is emitted from the source with probability $p_i$, i.e., $\{p_i\}_{i=1}^m$ forms a probability distribution. Such a physical system is called a *quantum source*. Note that in general the pure states $|\psi_i\rangle$ are not orthogonal to each other, that is, $\langle\psi_i|\psi_j\rangle \neq 0$ for at least one pair of indices $i, j = 1, \ldots, m$ with $i \neq j$.[2]

More abstractly, to every quantum source as described above we associate the *source ensemble* $\mathfrak{E}$ of pure states,

$$\mathfrak{E} = \{p_i; |\psi_i\rangle\}_{i=1}^m, \tag{5.2}$$

where for $i = 1, \ldots, m$ the pure state $|\psi_i\rangle$ is emitted from the source with probability $p_i$. We

---

[2]In the case of orthogonal states $\{|\psi_i\rangle\}_{i=1}^m$ satisfying $\langle\psi_i|\psi_j\rangle = \delta_{ij}$ for all $i, j = 1, \ldots, m$, quantum source coding as defined in Section 5.1.2 reduces to classical source coding. Of course, in this situation we must have $\dim \mathcal{H} \geq m$.

usually drop the cardinality, $m$, of the ensemble in the sequel. The ensemble average state

$$\rho = \sum_i p_i |\psi_i\rangle\langle\psi_i|$$

is called the *source state* of a quantum source. We will see at various points in this chapter that quantum source coding can be characterized solely in terms of (the eigenvalues of) the source state (cf. Schumacher's noiseless coding theorem (5.9), Theorem 5.2.8, and Corollary 5.2.11).

Assume now that a quantum source emits a *sequence* of signals $|\psi_i\rangle$ of length $n$. In this thesis, we restrict ourselves to the case where these sequences are given as tensor products of the pure states $\{|\psi_i\rangle\}_i$ from the source ensemble (5.2), that is, we only consider sequences of signals of the form

$$|\psi_{i^n}\rangle := |\psi_{i_1}\rangle \otimes \ldots \otimes |\psi_{i_n}\rangle, \tag{5.3}$$

where $i^n = (i_1, \ldots, i_n) \in [m]^n$ denotes a sequence of indices of length $n$. As for the underlying probability distribution of the signals $|\psi_{i^n}\rangle$, we consider two important scenarios: memoryless sources and mixed sources consisting of memoryless sources.

**Memoryless source**

A source is called *memoryless* if there are no correlations between successive signals emitted by the source. More precisely, for $j = 1, \ldots, n$, the $j$-th signal $|\psi_{i_j}\rangle$ in the sequence (5.3) is independent and identically distributed (i.i.d.), and drawn from the set $\{|\psi_i\rangle\}_i$ with probability $p_{i_j}$. Consequently, we can characterize $n$ uses of a memoryless source $\mathfrak{E}$ by the source ensemble

$$\mathfrak{E}^n = \{p_{i^n}; |\psi_{i^n}\rangle\}_{i^n \in [m]^n}, \tag{5.4}$$

where for a sequence $i^n = (i_1, \ldots, i_n) \in [m]^n$ we define $p_{i^n} := \prod_{j=1}^n p_{i_j}$. The corresponding source state is given by the i.i.d. state

$$\rho^{\otimes n} = \sum_{i^n} p_{i^n} |\psi_{i^n}\rangle\langle\psi_{i^n}|.$$

We refer to a memoryless source simply by its source ensemble $\mathfrak{E} = \{p_i; |\psi_i\rangle\}_i$.

**Mixed sources**

We now construct a mixed source consisting of memoryless sources. To this end, let $\Lambda$ be an arbitrary parameter space with a normalized measure $\mu$, i.e., $\int_\Lambda d\mu(\lambda) = 1$. For a fixed set $\{|\varphi_i\rangle\}_i$ of pure states and $\lambda \in \Lambda$, let $\mathfrak{E}_\lambda = \left\{ q_i^{(\lambda)}; |\varphi_i\rangle \right\}_i$ be the source ensemble of a memoryless source, with corresponding source state $\rho_\lambda = \sum_i q_i^{(\lambda)} |\varphi_i\rangle\langle\varphi_i|$. We define the mixed source to be the one that initially selects the memoryless source $\mathfrak{E}_\lambda$ according to the probability measure $\mu$, and emits all successive signals from this memoryless source according to the probability distribution $\left\{ q_i^{(\lambda)} \right\}_i$.[3] Hence, $n$ uses of this mixed source are described by the ensemble

$$\mathfrak{E}_{\text{mix}}^{(n)} := \left\{ d\mu(\lambda) q_{i^n}^{(\lambda)}; |\varphi_{i^n}\rangle \right\}_{\lambda \in \Lambda, \, i^n \in [m]^n}, \tag{5.5}$$

where $i^n$ again denotes a sequence of indices of length $n$. The corresponding source state $\rho^{(n)}$ of the mixed source is given by

$$\rho^{(n)} = \int_\Lambda d\mu(\lambda) \rho_\lambda^{\otimes n} = \int_\Lambda d\mu(\lambda) \sum_{i^n} q_{i^n}^{(\lambda)} |\varphi_{i^n}\rangle\langle\varphi_{i^n}|. \tag{5.6}$$

We denote the mixed source obtained from this construction by $(\mathfrak{E}_\lambda, \mu)_{\lambda \in \Lambda}$. Note that the source state (5.6) of a mixed source is a convex combination of the i.i.d. source states $\rho_\lambda^{\otimes n}$ of the individual memoryless sources that constitute the mixed source. This fact is crucial in the derivation of the second order asymptotics of mixed source coding in Section 5.2.3. Furthermore, we observe that for a singleton space $\Lambda = \{\lambda\}$ with $\mu(\lambda) = 1$, the corresponding mixed source reduces to a memoryless source as described above.

Let us consider the special case where the measure $\mu$ has finite support on points $\lambda_1, \ldots, \lambda_k \in \Lambda$, corresponding to a discrete probability distribution $\{t_j\}_{j=1}^k$. Hence, we have $k$ memoryless quantum information sources with source ensembles $\mathfrak{E}_j = \left\{ q_i^{(j)}, |\varphi_i\rangle \right\}_i$ and source states $\rho_j = \sum_i q_i^{(j)} \varphi_i$ for $j = 1, \ldots, k$. The source ensemble for $n$ uses of this mixed source is

$$\mathfrak{E}_{\text{mix}}^{(n)} := \left\{ t_j q_{i^n}^{(j)}; |\varphi_{i^n}\rangle \right\}_{j \in [k], \, i^n \in [m]^n},$$

---

[3]This is indeed the most general situation for a mixed source, which can be seen as follows: Consider a family of memoryless sources parametrized by $\lambda \in \Lambda$, where each source has an underlying ensemble $\{q_i^{(\lambda)}; |\varphi_i^{(\lambda)}\rangle\}_i$. We take the union over all $\lambda \in \Lambda$ of the sets $\{|\varphi_i^{(\lambda)}\rangle\}_i$ of pure states, and write the resulting set as $\{|\varphi_i\rangle\}_i$ (where the index $i$ now runs over a possibly larger set). Padding the probability distributions $\{q_i^{(\lambda)}\}_i$ for each $\lambda \in \Lambda$ with zeros if necessary, they can be regarded as probability distributions over the set $\{|\varphi_i\rangle\}_i$. Each memoryless source parametrized by $\lambda \in \Lambda$ now corresponds to a probability distribution $\{q_i^{(\lambda)}\}_i$ over this set, with source state $\rho_\lambda = \sum_i q_i^{(\lambda)} |\varphi_i\rangle\langle\varphi_i|$.

and the source state (5.6) reduces to

$$\rho^{(n)} = \sum_{j=1}^{k} t_j \rho_j^{\otimes n}.$$

We denote such a discrete mixed source consisting of $k$ memoryless sources $\mathfrak{E}_1, \ldots, \mathfrak{E}_k$ by $(\mathfrak{E}_j, t_j)_{j=1}^{k}$. In the special case of two memoryless sources, $k = 2$, we set $t \equiv t_1$ (such that $t_2 = 1 - t$) and write $(\mathfrak{E}_1, \mathfrak{E}_2, t)$ for the resulting mixed source. The source state for $n$ uses of the mixed source $(\mathfrak{E}_1, \mathfrak{E}_2, t)$ is given by $\rho^{(n)} = t\rho_1^{\otimes n} + (1 - t)\rho_2^{\otimes n}$. The parameter $t$ is also referred to as the *mixing parameter*. Figure 5.1 shows a schematic description of such a source.

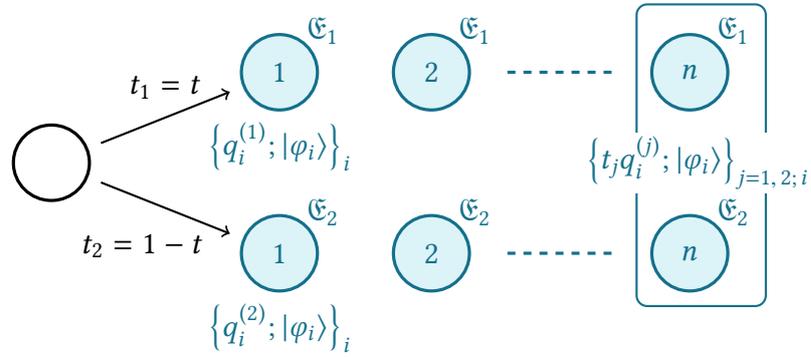

Figure 5.1: Schematic description of $n$ uses of a mixed source $(\mathfrak{E}_1, \mathfrak{E}_2, t)$ with mixing parameter $t \in (0, 1)$ (that is, $t_1 = t$ and $t_2 = 1 - t$), consisting of two memoryless sources $\mathfrak{E}_1 = \{q_i^{(1)}; |\varphi_i\rangle\}_i$ and $\mathfrak{E}_2 = \{q_i^{(2)}; |\varphi_i\rangle\}_i$.

## 5.1.2. Quantum source coding

The aim of fixed-length quantum source coding is to compress the information emitted by a quantum source by storing it in a compressed Hilbert space $\mathcal{H}_c \subset \mathcal{H}$ with $\dim \mathcal{H}_c < \dim \mathcal{H}$. That is, the compression part of the protocol is the reduction of the dimensionality of the space needed to describe the signals emitted by the source. However, the compression should allow us to decompress these compressed signals at a later stage such that they are sufficiently close to the original signals with respect to a chosen distance measure.

For the compression (or encoding) part of the quantum source coding protocol, there are two different settings: *blind* and *visible* encoding [Win99b; BCF+01; Hay02]. In the blind encoding setting, the encoder, Alice, cannot infer which (sequence of) signals she received from the source. That is, Alice receives an unknown quantum state (the source state $\rho$ described in Section 5.1.1), and to encode this in a compressed state she applies a quantum operation $\mathcal{E} \colon \mathcal{D}(\mathcal{H}) \to \mathcal{D}(\mathcal{H}_c)$ to $\rho$ (see Figure 6.6 in Chapter 6 for a schematic description of blind quantum source coding obtained as a reduction from the state redistribution protocol, which is discussed in Section 6.1).

In the visible encoding setting, Alice is assumed to be able to identify each signal $|\psi_i\rangle$ emitted by the source. Hence, upon receiving a signal $|\psi_i\rangle$ from the source, she essentially possesses classical information (the index $i$) about the signals. She then uses an *arbitrary* map $\mathcal{V} \colon [m] \to \mathcal{D}(\mathcal{H}_c)$ to *prepare* a compressed signal $\mathcal{V}(i)$. We refer to such maps as *visible encoders*. In this thesis, we restrict ourselves to the visible encoding setting, as it allows us to obtain a complete characterization of the second order asymptotics of quantum source coding. In Section 5.2.4, we compare our results obtained in the visible setting to the (partial) second order asymptotic results for quantum source coding in the blind encoding setting obtained in [DL15].

In the decompression (or decoding) part of quantum source coding, Bob uses a decoding given by a quantum operation $\mathcal{D} \colon \mathcal{D}(\mathcal{H}_c) \to \mathcal{D}(\mathcal{H})$ to obtain signals that should be close to the original signals according to a suitably chosen distance measure.

Based on the discussion above, for $n \in \mathbb{N}$ uses of a quantum source, we define a *quantum source code* $\{C_n\}_{n\in\mathbb{N}}$ as a sequence of triples $C_n = (\mathcal{V}_n, \mathcal{D}_n, M_n)$, where $\mathcal{V}_n \colon [m]^n \to \mathcal{D}(\mathcal{H}_c)$ is a visible encoding map, $\mathcal{D}_n \colon \mathcal{D}(\mathcal{H}_c) \to \mathcal{D}(\mathcal{H}^{\otimes n})$ is a decoding quantum operation, and $M_n := \dim \mathcal{H}_c$ is the dimension of the compressed Hilbert space $\mathcal{H}_c \subset \mathcal{H}^{\otimes n}$. We choose the ensemble average fidelity $\bar{F}\left(\mathfrak{E}^{(n)}, C_n\right)$ as the distance measure to compare the decoded state to the original state, where $\mathfrak{E}^{(n)}$ denotes the source ensemble for $n$ uses of the source. In the case of a memoryless source, for which $n$ uses are described by the source ensemble $\mathfrak{E}^n$ as defined in (5.4), we set

$$\bar{F}\left(\mathfrak{E}^n, C_n\right) := \sum_{i^n} p_{i^n} \operatorname{tr}\left((\mathcal{D}_n \circ \mathcal{V}_n)(i^n)\psi_{i^n}\right). \tag{5.7}$$

In the case of a mixed source, for which $\mathfrak{E}^{(n)} = \mathfrak{E}^{(n)}_{\mathrm{mix}}$ as defined in (5.5), we set

$$\bar{F}\left(\mathfrak{E}^{(n)}_{\mathrm{mix}}, C_n\right) := \int_\Lambda d\mu(\lambda) \sum_{i^n} q^{(\lambda)}_{i^n} \operatorname{tr}\left((\mathcal{D}_n \circ \mathcal{V}_n)(i^n)\varphi_{i^n}\right). \tag{5.8}$$

It is easy to see that upon choosing $\Lambda$ to be a singleton space, the ensemble average fidelity for a mixed source given by (5.8) reduces to the one of a memoryless source given by (5.7).

We now define the central operational quantities of this chapter:

**Definition 5.1.1** (Achievable rates in source coding).
Let $\mathfrak{C}^{(n)}$ be the source ensemble for $n$ uses of a quantum source $\mathfrak{S}$. Furthermore, let $\varepsilon \in (0, 1)$ be an error parameter.

(i) A positive real number $R$ is called an *$\varepsilon$-achievable first order rate*, if there exists a sequence $\{C_n\}_{n \in \mathbb{N}}$ of quantum source codes $C_n = (\mathcal{V}_n, \mathcal{D}_n, M_n)$ such that

$$\liminf_{n \to \infty} \bar{F}\left(\mathfrak{C}^{(n)}, C_n\right) \geq 1 - \varepsilon \qquad \limsup_{n \to \infty} \frac{\log M_n}{n} \leq R.$$

The *first order asymptotic rate $a(\varepsilon|\mathfrak{S})$* of the source $\mathfrak{S}$ is defined as the infimum over all $\varepsilon$-achievable first order rates.

(ii) Given $R > 0$, a real number $r$ is called an *$(R, \varepsilon)$-achievable second order rate*, if there exists a sequence $\{C_n\}_{n \in \mathbb{N}}$ of quantum source codes $C_n = (\mathcal{V}_n, \mathcal{D}_n, M_n)$ such that

$$\liminf_{n \to \infty} \bar{F}\left(\mathfrak{C}^{(n)}, C_n\right) \geq 1 - \varepsilon \qquad \limsup_{n \to \infty} \frac{\log M_n - nR}{\sqrt{n}} \leq r.$$

The *second order asymptotic rate $b(R, \varepsilon|\mathfrak{S})$* of the source $\mathfrak{S}$ is defined as the infimum over all $(R, \varepsilon)$-achievable second order rates.

In the blind encoding setting, Schumacher [Sch95] proved the *noiseless quantum coding theorem*, establishing for a memoryless quantum source $\mathfrak{C} = \{p_i; |\psi_i\rangle\}_i$ with source state $\rho = \sum_i p_i \psi_i$ and $\varepsilon \in (0, 1)$ that

$$a(\varepsilon|\mathfrak{C}) = S(\rho). \tag{5.9}$$

Winter [Win99b] extended Schumacher's quantum coding theorem, proving that (5.9) also holds in the visible encoding setting. Furthermore, he refined (5.9) to

$$\log M_n = nS(\rho) + O(\sqrt{n}) \quad \text{for all } \varepsilon \in (0, 1), \tag{5.10}$$

which constitutes a strong converse for quantum source coding.

It follows from Definition 5.1.1 and an expansion of $\log M_n$ as in (5.10) that the second order asymptotic rate $b(R, \varepsilon|\mathfrak{S})$ of a quantum source $\mathfrak{S}$ is only finite if the parameter $R$ is equal to the first order asymptotic rate, $R = a(\varepsilon|\mathfrak{S})$. To see this, set $a = a(\varepsilon|\mathfrak{S})$ and consider that, using

(5.10), we have

$$\limsup_{n \to \infty} \frac{\log M_n - nR}{\sqrt{n}} = \limsup_{n \to \infty} \sqrt{n}(a - R) + O(1) = \begin{cases} +\infty & \text{if } a > R \\ -\infty & \text{if } a < R. \end{cases} \quad (5.11)$$

Hence, we have $b(a, \varepsilon | \mathfrak{S}) = +\infty$ if $a > R$, and $b(a, \varepsilon | \mathfrak{S}) = -\infty$ if $a < R$.

In the following section, we derive the second order asymptotic rate $b(a, \varepsilon | \mathfrak{M})$ of a mixed source $\mathfrak{M} = (\mathfrak{E}_\lambda, \mu)_{\lambda \in \Lambda}$, giving rise to a refinement of (5.10) in the form of

$$\log M_n = na + \sqrt{n} b + o(\sqrt{n}), \quad (5.12)$$

where $a = a(\varepsilon | \mathfrak{M})$, $b = b(a, \varepsilon | \mathfrak{M})$, and $\varepsilon \in (0, 1)$. As a corollary of our results, the first order rate of mixed source coding depends on $\varepsilon$, in contrast to memoryless source coding. This implies that the optimal rate of mixed source coding does not satisfy the strong converse property (see Chapter 6 for a more detailed discussion). We will not concern ourselves with the dependence on $n$ of the third order in (5.12) here, and merely note that it can be shown to be of the order $O(\log n)$ (see [DL15] for a proof of this fact in the case of quantum source coding using a single memoryless source).

## 5.2. Deriving the second order asymptotics

In this section we derive our main results of this chapter: expressions for the second order asymptotic rates of quantum source coding using mixed sources (Theorem 5.2.8 in Section 5.2.3) and memoryless sources (Corollary 5.2.11 in Section 5.2.4). Before proving these results, we first construct universal quantum source codes achieving second order rates in Section 5.2.1. In Section 5.2.2 we derive an upper bound on the ensemble average fidelity needed in the converse part of our main result. Throughout this section, we abbreviate $\rho^n \equiv \rho^{\otimes n}$.

### 5.2.1. Universal codes achieving second order rate

We now construct a universal source code that, given parameters $a \in \mathbb{R}$ (which is to be chosen later as the first order rate) and $\varepsilon \in (0, 1)$, achieves a second order asymptotic rate $b(a, \varepsilon | \mathfrak{E})$ for $n$ uses of a memoryless source $\mathfrak{E}$ with source state $\rho \in \mathcal{D}(\mathcal{H})$. Our construction is based on ideas by Jozsa et al. [JHH+98] and Hayashi [Hay08]: we combine the construction of a universal quantum source code achieving first order rates from [JHH+98] with universal classical source codes achieving second order rates from [Hay08], yielding the desired universal quantum source codes achieving second order rates. This result appeared in [LD16].

Let $\mathcal{X}$ be a finite alphabet with $|\mathcal{X}| = d$. The type $t_{x^n}$ of a sequence $x^n = (x_1, \ldots, x_n) \in \mathcal{X}^n$ is the empirical distribution of the letters of $\mathcal{X}$ in $x^n$, that is, $t_{x^n}(x) = \frac{1}{n} \sum_{i=1}^{n} \delta_{x_i, x}$ for all $x \in \mathcal{X}$. We denote by $\mathcal{T}_n$ the set of all types, and for a type $t \in \mathcal{T}_n$ we denote by $T_t^n \subset \mathcal{X}^n$ the set of sequences of type $t$. Following [Hay08], for $a, b \in \mathbb{R}$ we define

$$T_n(a, b) := \bigcup \left\{ T_t^n : t \in \mathcal{T}_n \text{ with } |T_t^n| \leq 2^{an + b\sqrt{n}} \right\} \subset \mathcal{X}^n.$$

A simple type-counting argument [CK11] shows that

$$|T_n(a, b)| \leq (n+1)^d 2^{an + b\sqrt{n}}.$$

Let now $\mathcal{H}$ be a $d$-dimensional Hilbert space, set $\mathcal{X} = [d]$, and let $B = \{|e_1\rangle, \ldots, |e_d\rangle\}$ be a basis of $\mathcal{H}$. In analogy to [JHH+98], we define the subspace

$$\Xi_{a,b}^n(B) := \mathrm{span} \left\{ |e_{x^n}\rangle \in B^{\otimes n} : x^n \in T_n(a, b) \right\},$$

that is, $\Xi_{a,b}^n(B)$ is the span of those basis vectors of the product basis $B^{\otimes n}$ of $\mathcal{H}^{\otimes n}$ labeled by sequences in $T_n(a, b)$. The code space $\Upsilon_{a,b}^n$ of the universal source code is now obtained by varying $B$ over all bases of $\mathcal{H}$. More precisely, we define $\Upsilon_{a,b}^n$ as the smallest subspace of $\mathcal{H}^{\otimes n}$ containing $\Xi_{a,b}^n(B)$ for every basis $B$ of $\mathcal{H}$. To estimate the size of $\Upsilon_{a,b}^n$, we use the following

**Lemma 5.2.1** ([JHH+98]). *Let $|\phi\rangle \in \mathcal{H}^{\otimes n}$ with $\dim \mathcal{H} = d$, and let*

$$\mathcal{H}_\phi := \mathrm{span} \left\{ A^{\otimes n} |\phi\rangle : A \in \mathcal{B}(\mathcal{H}) \right\}.$$

*We then have $\dim \mathcal{H}_\phi \leq (n+1)^{d^2}$.*

**Lemma 5.2.2.** *With the above definitions, the dimension of the code space $\Upsilon_{a,b}^n \subseteq \mathcal{H}^{\otimes n}$ can be bounded as*

$$\dim \Upsilon_{a,b}^n \leq (n+1)^{d^2 + d} 2^{an + b\sqrt{n}}.$$

*Proof.* We closely follow an argument in [JHH+98]. First, let $B_0$ be a fixed basis of $\mathcal{H}$. Then any other basis $B$ can be obtained from $B_0$ by applying some unitary $U$ to the elements of $B_0$.

As $\Xi_{a,b}^n(B)$ is the span of tensor products of elements in $B$, we have

$$\Xi_{a,b}^n(B) = \left\{ U^{\otimes n}|\phi\rangle \colon |\phi\rangle \in \Xi_{a,b}^n(B_0) \right\}.$$

Hence, the following holds for the code space $\Upsilon_{a,b}^n$:

$$\Upsilon_{a,b}^n = \mathrm{span}\left\{ U^{\otimes n}|\phi\rangle \colon U \in \mathcal{U}_d, |\phi\rangle \in \Xi_{a,b}^n(B_0) \right\}$$
$$\subset \mathrm{span}\left\{ A^{\otimes n}|\phi\rangle \colon A \in \mathcal{B}(\mathcal{H}), |\phi\rangle \in \Xi_{a,b}^n(B_0) \right\}.$$

As $\dim \Xi_{a,b}^n(B_0) = |T_n(a,b)| \leq (n+1)^{d^2} 2^{an+b\sqrt{n}}$, the claim now follows from Lemma 5.2.1. $\quad\square$

**Proposition 5.2.3** (Universal code achieving second order rate)**.**
*Let $\mathfrak{E} = \{p_i; |\psi_i\rangle\}_i$ be the pure-state ensemble of an arbitrary memoryless quantum source with associated source state $\rho \in \mathcal{D}(\mathcal{H})$, and abbreviate $S \equiv S(\rho)$ and $\sigma \equiv \sigma(\rho)$. For $b \in \mathbb{R}$ and $n$ uses of the source $\mathfrak{E}$, let $\Pi_n$ be the projection onto the code space $\Upsilon_{S,b}^n$ defined as above, and consider the visible encoding map*

$$\mathcal{V}_n \colon i^n \longmapsto \frac{\Pi_n \psi_{i^n} \Pi_n}{\mathrm{tr}(\Pi_n \psi_{i^n})}. \tag{5.13}$$

*We set $M_n := \dim \Upsilon_{S,b}^n$, and define the decoding operation $\mathcal{D}_n \colon \Upsilon_{S,b}^n \to \mathcal{H}^{\otimes n}$ to be the trivial embedding. The code $\{C_n\}_{n\in\mathbb{N}}$ with $C_n = (\mathcal{V}_n, \mathcal{D}_n, M_n)$ then achieves the second order rate $b$ with $\varepsilon = 1 - \Phi(b/\sigma)$.*

*Proof.* Lemma 5.2.2 immediately yields

$$\limsup_{n\to\infty} \frac{\log M_n - Sn}{\sqrt{n}} \leq \limsup_{n\to\infty} \frac{(d^2 + d)\log(n+1)}{\sqrt{n}} + b = b.$$

With the visible encoding given by (5.13) and the trivial embedding as the decoding, we can express the ensemble average fidelity $\bar{F}(\mathfrak{E}^n, C_n)$ in the following form:

$$\bar{F}(\mathfrak{E}^n, C_n) = \sum_{i^n} p_{i^n} \, \mathrm{tr}\left((\mathcal{D}_n \circ \mathcal{V}_n)(i^n)\psi_{i^n}\right)$$
$$= \sum_{i^n} p_{i^n} \frac{1}{\mathrm{tr}(\Pi_n \psi_{i^n})} \, \mathrm{tr}(\Pi_n \psi_{i^n} \Pi_n \psi_{i^n})$$
$$= \sum_{i^n} p_{i^n} \frac{\langle \psi_{i^n} | \Pi_n | \psi_{i^n} \rangle^2}{\langle \psi_{i^n} | \Pi_n | \psi_{i^n} \rangle}$$
$$= \sum_{i^n} p_{i^n} \langle \psi_{i^n} | \Pi_n | \psi_{i^n} \rangle$$

$$= \operatorname{tr}(\rho^n \Pi_n). \tag{5.14}$$

We now employ a relation proved by Hayashi [Hay08] in the context of classical fixed-length source coding, which holds for arbitrary $a, b \in \mathbb{R}$ and probability distributions $P^n$ on $\mathcal{X}^n$:

$$S_n(a, b) := \left\{ x^n \in \mathcal{X}^n : -\log P^n(x^n) < na + \sqrt{n}b \right\} \subseteq T_n(a, b). \tag{5.15}$$

To prove (5.15), fix a probability distribution $P^n$ and note that $S_n(a, b)$ can be written as

$$S_n(a, b) = \bigcup \left\{ T_t^n : P^n(\omega) > 2^{-na - \sqrt{n}b} \text{ for } \omega \in T_t^n \right\}.$$

Let $t \in \mathcal{T}_n$ be a type satisfying $P^n(\omega) > 2^{-na - \sqrt{n}b}$ for all $\omega \in T_t^n$. We then have

$$1 \geq P^n(T_t^n) = \sum_{\omega \in T_t^n} P^n(\omega) > |T_t^n| 2^{-na - \sqrt{n}b},$$

and hence $|T_t^n| < 2^{na + \sqrt{n}b}$, which implies that $T_t^n \subseteq T_n(a, b)$. Thus, we obtain (5.15).

Consider now an arbitrary state $\omega \in \mathcal{D}(\mathcal{H})$ with spectral decomposition $\omega = \sum_x r_x |\varphi_x\rangle\langle\varphi_x|$, and set $P_\omega = \{r_x\}_x$ and $B_\omega = \{|\varphi_x\rangle\}_x$. Observe that the projection $\left\{ \omega^n > 2^{-na - \sqrt{n}b} \mathbb{1}_n \right\}$ projects onto eigenvectors of $\omega^n$ labeled by elements of $S_n(a, b)$, upon choosing $P = P_\omega$. Since the code space $\Upsilon_{a,b}^n$ includes the subspace $\Xi_{a,b}^n(B_\omega)$, we have the operator inequality

$$\Pi_n \geq \left\{ \omega^n > 2^{-na - \sqrt{n}b} \mathbb{1}_n \right\}. \tag{5.16}$$

We now set $\omega = \rho$ and $a = S$ in (5.16), and substitute it in (5.14). Taking the limit inferior on both sides of (5.14), we obtain

$$\begin{aligned}
\liminf_{n \to \infty} \bar{F}(\mathfrak{C}^n, \mathcal{C}_n) &= \liminf_{n \to \infty} \operatorname{tr}(\rho^n \Pi_n) \\
&\geq \liminf_{n \to \infty} \operatorname{tr}\left(\rho^n \left\{ \rho^n > 2^{-nS - \sqrt{n}b} \mathbb{1}_n \right\}\right) \\
&= 1 - \limsup_{n \to \infty} \operatorname{tr}\left(\rho^n \left\{ \rho^n \leq 2^{-nS - \sqrt{n}b} \mathbb{1}_n \right\}\right) \\
&= 1 - \Phi\left(-\frac{b}{\sigma}\right) \\
&= \Phi\left(\frac{b}{\sigma}\right),
\end{aligned}$$

where we used Corollary 4.2.8 in the third equality. Setting $\varepsilon := 1 - \Phi(b/\sigma)$ now yields the

claim. □

## 5.2.2. Bounding the ensemble average fidelity

In this section, we derive a bound on the ensemble average fidelity that we need for proving the converse bound of Theorem 5.2.8 in Section 5.2.3. It is based on a result by Hayashi [Hay02], whose proof we reproduce in our notation in the following for the convenience of the reader. First, we record the following observation:

**Lemma 5.2.4** ([NK01; Hay02]). *Let $\rho_{AB} \in \mathcal{D}(\mathcal{H}_{AB})$ be a separable state as in (2.1). Then for any $k \in \mathbb{N}$, we have*

$$\max \left\{ \mathrm{tr}(P \rho_A) \colon P \text{ is a projection on } \mathcal{H}_A \text{ with } \mathrm{tr}\, P = k \right\}$$

$$\geq \max \left\{ \mathrm{tr}(P \rho_{AB}) \colon P \text{ is a projection on } \mathcal{H}_{AB} \text{ with } \mathrm{tr}\, P = k \right\}.$$

*Proof.* By Ky Fan's Maximum Principle, Theorem 2.3.7, we have

$$\sum_{i=1}^{k} \lambda_i(\rho_{AB}) = \max \left\{ \mathrm{tr}(P \rho_{AB}) \colon P \text{ is a projection on } \mathcal{H}_{AB} \text{ with } \mathrm{tr}\, P = k \right\}.$$

Since $\rho_{AB}$ is separable, Theorem 2.3.8 yields $\lambda(\rho_{AB}) \prec \lambda(\rho_A)$, and in particular

$$\sum_{i=1}^{k} \lambda_i(\rho_{AB}) \leq \sum_{i=1}^{k} \lambda_i(\rho_A)$$

$$= \max \left\{ \mathrm{tr}(P \rho_A) \colon P \text{ is a projection on } \mathcal{H}_A \text{ with } \mathrm{tr}\, P = k \right\},$$

where the equality follows once again from Ky Fan's Maximum Principle. □

The following crucial proposition is due to Hayashi [Hay02]. We give a slightly modified, but essentially identical proof.

**Proposition 5.2.5** ([Hay02]). *Consider a quantum information source with source ensemble $\mathfrak{E} = \{p_i; \psi_i\}_i$ and source state $\rho = \sum_i p_i \psi_i \in \mathcal{D}(\mathcal{H})$. Let $C = (\mathcal{V}, \mathcal{D}, M)$ be a quantum source code, where $\mathcal{V} \colon [m] \to \mathcal{D}(\mathcal{H}_c)$ is a visible encoding map into a compressed Hilbert space $\mathcal{H}_c$, $\mathcal{D} \colon \mathcal{D}(\mathcal{H}_c) \to \mathcal{D}(\mathcal{H})$ is a decoding operation, and $M = \dim \mathcal{H}_c$ is the dimension of the compressed*

*Hilbert space $\mathcal{H}_c$. Then,*

$$\bar{F}(\mathfrak{E}, C) \leq \max \{\operatorname{tr} P\rho : P \text{ is a projection on } \mathcal{H} \text{ with } \operatorname{tr} P = M\}.$$

*Proof.* Let $\mathcal{V}(i) = \sum_j \mu_{i,j} |\varphi_{i,j}\rangle\langle\varphi_{i,j}|$ be the eigenvalue decomposition of the encoded signal $\mathcal{V}(i)$. Using linearity of $\mathcal{D}$ and the trace, we have

$$\bar{F}(\mathfrak{E}, C) = \sum_{i,j} p_i \mu_{i,j} \operatorname{tr}\left(\mathcal{D}\left(\varphi_{i,j}\right)\psi_i\right). \tag{5.17}$$

Furthermore, let $U\colon \mathcal{H}_c \to \mathcal{H} \otimes \mathcal{H}'$ be a Stinespring isometry of the decoding quantum operation $\mathcal{D}\colon \mathcal{D}(\mathcal{H}_c) \to \mathcal{D}(\mathcal{H})$, and note that $UU^\dagger$ is the projection onto the image of $U$ in $\mathcal{H} \otimes \mathcal{H}'$. Consider now for fixed $i$ and $j$ the pure state

$$\psi'_{i,j} := \frac{(\psi_i \otimes \mathbb{1}_{\mathcal{H}'}) U \varphi_{i,j} U^\dagger (\psi_i \otimes \mathbb{1}_{\mathcal{H}'})}{\operatorname{tr}\left(U\varphi_{i,j}U^\dagger(\psi_i \otimes \mathbb{1}_{\mathcal{H}'})\right)} \in \mathcal{D}(\mathcal{H} \otimes \mathcal{H}'), \tag{5.18}$$

which satisfies

$$
\begin{aligned}
\operatorname{tr}\left(U\varphi_{i,j}U^\dagger \psi'_{i,j}\right) &= \frac{\operatorname{tr}(U\varphi_{i,j}U^\dagger(\psi_i \otimes \mathbb{1}_{\mathcal{H}'})U\varphi_{i,j}U^\dagger(\psi_i \otimes \mathbb{1}_{\mathcal{H}'}))}{\operatorname{tr}\left(U\varphi_{i,j}U^\dagger(\psi_i \otimes \mathbb{1}_{\mathcal{H}'})\right)} \\
&= \frac{\langle\varphi_{i,j}|U^\dagger(\psi_i \otimes \mathbb{1}_{\mathcal{H}'})U|\varphi_{i,j}\rangle^2}{\langle\varphi_{i,j}|U^\dagger(\psi_i \otimes \mathbb{1}_{\mathcal{H}'})U|\varphi_{i,j}\rangle} \\
&= \langle\varphi_{i,j}|U^\dagger(\psi_i \otimes \mathbb{1}_{\mathcal{H}'})U|\varphi_{i,j}\rangle \\
&= \operatorname{tr}\left(U\varphi_{i,j}U^\dagger(\psi_i \otimes \mathbb{1}_{\mathcal{H}'})\right) \\
&= \operatorname{tr}\left(\mathcal{D}(\varphi_{i,j})\psi_i\right). \tag{5.19}
\end{aligned}
$$

We claim that for each $j$, the state $\psi'_{i,j}$ is a purification of $\psi_i$, that is,

$$\operatorname{tr}_{\mathcal{H}'} \psi'_{i,j} = \psi_i. \tag{5.20}$$

This can be seen as follows. Firstly, for the denominator of (5.18) we have

$$\operatorname{tr}\left(U\varphi_{i,j}U^\dagger(\psi_i \otimes \mathbb{1}_{\mathcal{H}'})\right) = \operatorname{tr}\left(\mathcal{D}(\varphi_{i,j})\psi_i\right) = \langle\psi_i|\mathcal{D}(\varphi_{i,j})|\psi_i\rangle. \tag{5.21}$$

Secondly, using an orthonormal basis $\{|k\rangle\}_k$ of $\mathcal{H}'$, the numerator of (5.18) can be rewritten as

$$(\psi_i \otimes \mathbb{1}_{\mathcal{H}'}) U \varphi_{i,j} U^\dagger (\psi_i \otimes \mathbb{1}_{\mathcal{H}'}) = \left(|\psi_i\rangle\langle\psi_i| \otimes \sum_k |k\rangle\langle k|\right) U \varphi_{i,j} U^\dagger \left(|\psi_i\rangle\langle\psi_i| \otimes \sum_m |m\rangle\langle m|\right)$$

$$= \sum_{k,m} \left[ \left( \langle \psi_i | \otimes \langle k | \right) U \varphi_{i,j} U^\dagger \left( |\psi_i\rangle \otimes |m\rangle \right) \right] |\psi_i\rangle\langle\psi_i| \otimes |k\rangle\langle m|.$$

Hence,

$$\mathrm{tr}_{\mathcal{H}'} \left( (\psi_i \otimes \mathbb{1}_{\mathcal{H}'}) U \varphi_{i,j} U^\dagger (\psi_i \otimes \mathbb{1}_{\mathcal{H}'}) \right)$$

$$= \mathrm{tr}_{\mathcal{H}'} \left( \sum_{k,m} \left[ \left( \langle\psi_i| \otimes \langle k| \right) U \varphi_{i,j} U^\dagger \left( |\psi_i\rangle \otimes |m\rangle \right) \right] |\psi_i\rangle\langle\psi_i| \otimes |k\rangle\langle m| \right)$$

$$= \sum_k \left[ \left( \langle\psi_i| \otimes \langle k| \right) U \varphi_{i,j} U^\dagger \left( |\psi_i\rangle \otimes |k\rangle \right) \right] |\psi_i\rangle\langle\psi_i|$$

$$= \langle\psi_i| \mathcal{D}(\varphi_{i,j}) |\psi_i\rangle \psi_i, \tag{5.22}$$

where the last equality follows from the definition of the partial trace. Using (5.21) and (5.22) in (5.18) yields (5.20).

Substituting (5.19) in (5.17), we obtain

$$\bar{F}(\mathfrak{E}, C) = \sum_{i,j} p_i \mu_{i,j} \, \mathrm{tr} \left( \mathcal{D}\left( \varphi_{i,j} \right) \psi_i \right)$$

$$= \sum_{i,j} p_i \mu_{i,j} \, \mathrm{tr} \left( U \varphi_{i,j} U^\dagger \psi'_{i,j} \right)$$

$$\leq \sum_{i,j} p_i \mu_{i,j} \, \mathrm{tr} \left( U U^\dagger \psi'_{i,j} \right), \tag{5.23}$$

where in the inequality we used $\varphi_{i,j} \leq \mathbb{1}_{\mathcal{H}_c}$ for all $i$ and $j$, and hence $U \varphi_{i,j} U^\dagger \leq U U^\dagger$. For fixed $i$, let $\chi_i$ be one of the $\psi'_{i,j}$ that maximizes $\mathrm{tr} \left( U U^\dagger \psi'_{i,j} \right)$, and note that $\chi_i$ is a purification of $\psi_i$ by (5.20). Hence, there exists a pure state $|\phi_i\rangle \in \mathcal{H}'$ such that $\chi_i = \psi_i \otimes \phi_i$. It follows that

$$\rho' := \sum_i p_i \chi_i = \sum_i p_i \psi_i \otimes \phi_i \in \mathcal{D}(\mathcal{H} \otimes \mathcal{H}')$$

is separable and satisfies $\mathrm{tr}_{\mathcal{H}'} \rho' = \rho$.

Thus, we can apply Lemma 5.2.4 to infer that we have

$$\max\{\mathrm{tr}(P\rho') \colon P \text{ proj. on } \mathcal{H} \otimes \mathcal{H}' \text{ with } \mathrm{tr}\, P = M\}$$

$$\leq \max\{\mathrm{tr}(P\rho) \colon P \text{ proj. on } \mathcal{H} \text{ with } \mathrm{tr}\, P = M\}, \tag{5.24}$$

and we observe that the projection $U U^\dagger \in \mathcal{B}(\mathcal{H} \otimes \mathcal{H}')$ with

$$\mathrm{tr}\, U U^\dagger = \mathrm{tr}\, U^\dagger U = \mathrm{tr}\, \mathbb{1}_{\mathcal{H}_c} = M$$

is feasible for the left-hand side of (5.24). Using the above considerations, we continue from (5.23) as

$$\sum_{i,j} p_i \mu_{i,j} \operatorname{tr}\left(UU^\dagger \psi'_{i,j}\right) \le \sum_i p_i \left(\sum_j \mu_{i,j}\right) \operatorname{tr}\left(UU^\dagger \chi_i\right)$$
$$= \operatorname{tr}\left(UU^\dagger \rho'\right)$$
$$\le \max\{\operatorname{tr}(P\rho') : P \text{ is a projection on } \mathcal{H} \otimes \mathcal{H}' \text{ with } \operatorname{tr} P = M\}$$
$$\le \max\{\operatorname{tr}(P\rho) : P \text{ is a projection on } \mathcal{H} \text{ with } \operatorname{tr} P = M\},$$

which concludes the proof. $\qquad\square$

We can now prove the promised bound on the ensemble average fidelity:

**Lemma 5.2.6.** *Consider $n$ uses of a mixed source $(\mathfrak{E}_\lambda, \mu)_{\lambda \in \Lambda}$ with source ensemble $\mathfrak{E}_{\mathrm{mix}}^{(n)}$ as defined in (5.5) and source state $\rho^{(n)}$ as defined in (5.6). Furthermore, let $\{C_n\}_{n \in \mathbb{N}}$ be a source code with $C_n = (\mathcal{V}_n, \mathcal{D}_n, M_n)$. For every $n \in \mathbb{N}$ and $\gamma \in \mathbb{R}$, the ensemble average fidelity satisfies*

$$\bar{F}\left(\mathfrak{E}_{\mathrm{mix}}^{(n)}, C_n\right) \le 1 - \int_\Lambda d\mu(\lambda) \operatorname{tr}\left(\rho_\lambda^n \left\{\rho_\lambda^n \le 2^{-\gamma} \mathbb{1}_n\right\}\right) + 2^{-\gamma + \log M_n}.$$

*Proof.* We first note that Proposition 5.2.5 also holds for continuous ensembles such as $\mathfrak{E}_{\mathrm{mix}}^{(n)}$. It then guarantees the existence of a projection $Q$ with $\operatorname{tr} Q = M_n$ such that $\bar{F}\left(\mathfrak{E}_{\mathrm{mix}}^{(n)}, C_n\right) \le \operatorname{tr}(Q\rho^n)$. For arbitrary $\gamma \in \mathbb{R}$, we then compute:

$$\bar{F}\left(\mathfrak{E}_{\mathrm{mix}}^{(n)}, C_n\right) \le \operatorname{tr}\left(Q\rho^n\right)$$
$$= \int_\Lambda d\mu(\lambda) \operatorname{tr}\left(Q\rho_\lambda^n\right)$$
$$= \int_\Lambda d\mu(\lambda) \operatorname{tr}\left(Q\left(\rho_\lambda^n - 2^{-\gamma}\mathbb{1}_n\right)\right) + 2^{-\gamma} \operatorname{tr} Q$$
$$\le \int_\Lambda d\mu(\lambda) \operatorname{tr}\left(\left\{\rho_\lambda^n > 2^{-\gamma}\mathbb{1}_n\right\}\left(\rho_\lambda^n - 2^{-\gamma}\mathbb{1}_n\right)\right) + 2^{-\gamma + \log M_n}$$
$$= 1 - 2^{-\gamma} \operatorname{tr} \mathbb{1}_n - \int_\Lambda d\mu(\lambda) \operatorname{tr}\left(\left\{\rho_\lambda^n \le 2^{-\gamma}\mathbb{1}_n\right\}\left(\rho_\lambda^n - 2^{-\gamma}\mathbb{1}_n\right)\right) + 2^{-\gamma + \log M_n}$$
$$= 1 - 2^{-\gamma} \operatorname{tr} \mathbb{1}_n - \int_\Lambda d\mu(\lambda) \operatorname{tr}\left(\rho_\lambda^n \left\{\rho_\lambda^n \le 2^{-\gamma}\mathbb{1}_n\right\}\right)$$
$$\qquad + 2^{-\gamma} \int_\Lambda d\mu(\lambda) \operatorname{tr}\left\{\rho_\lambda^n \le 2^{-\gamma}\mathbb{1}_n\right\} + 2^{-\gamma + \log M_n}$$

$$\leq 1 - \int_\Lambda d\mu(\lambda) \operatorname{tr} \left( \rho_\lambda^n \left\{ \rho_\lambda^n \leq 2^{-\gamma} \mathbb{1}_n \right\} \right) + 2^{-\gamma + \log M_n}$$

where we used Lemma 2.2.10 in the second inequality, the identity $\left\{ \rho_\lambda^n > 2^{-\gamma} \mathbb{1}_n \right\} = \mathbb{1}_n - \left\{ \rho_\lambda^n \leq 2^{-\gamma} \mathbb{1}_n \right\}$ in the third equality, and $\left\{ \rho_\lambda^n \leq 2^{-\gamma} \mathbb{1}_n \right\} \leq \mathbb{1}_n$ in the last inequality. □

## 5.2.3. Mixed visible quantum source coding

The main result of this chapter is the derivation of the second order asymptotic rate of a mixed source $(\mathfrak{E}_\lambda, \mu)_{\lambda \in \Lambda}$ with source state $\rho^{(n)} = \int_\Lambda \rho_\lambda^{\otimes n} d\mu(\lambda)$ as defined in (5.6) in Section 5.1.1. In order to state this result in Theorem 5.2.8 below, we define the following sets for a fixed $a > 0$, which partition the set $\Lambda$:

$$\Lambda_=(a) := \{ \lambda \in \Lambda \colon S(\rho_\lambda) = a \}$$
$$\Lambda_<(a) := \{ \lambda \in \Lambda \colon S(\rho_\lambda) < a \}$$
$$\Lambda_>(a) := \{ \lambda \in \Lambda \colon S(\rho_\lambda) > a \}.$$

Furthermore, recall from (2.3) that we set $\sigma(\rho) = \sqrt{V(\rho \| \mathbb{1})}$ for $\rho \in \mathcal{D}(\mathcal{H})$. In the proof of Theorem 5.2.8 we also invoke the Dominated Convergence Theorem, which we state here for the reader's convenience:

**Theorem 5.2.7** (Dominated Convergence Theorem).
*Let $(\Lambda, \Sigma, \mu)$ be a measure space, that is, $\Sigma$ is a $\sigma$-algebra over $\Lambda$ and $\mu$ is a measure on $\Sigma$. Let $\{f_n\}_{n \in \mathbb{N}}$ be a sequence of real-valued measurable functions $f_n \colon \Lambda \to \mathbb{R}$ converging pointwise to a function $f$. If there is an integrable function $g \colon \Lambda \to \mathbb{R}$ (that is, $\int_\Lambda d\mu(\lambda) |g(\lambda)| < \infty$) such that $|f_n(\lambda)| \leq g(\lambda)$ for all $\lambda \in \Lambda$ and $n \in \mathbb{N}$, then*

$$\lim_{n \to \infty} \int_\Lambda d\mu(\lambda) f_n(\lambda) = \int_\Lambda d\mu(\lambda) \lim_{n \to \infty} f_n(\lambda) = \int_\Lambda d\mu(\lambda) f(\lambda).$$

The main result of this chapter is now as follows:

**Theorem 5.2.8** (First and second order asymptotic rates of mixed quantum source coding).
*Let $\Lambda$ be an arbitrary parameter space with a normalized measure $\mu$, that is, $\int_\Lambda d\mu(\lambda) = 1$, and let $\mathfrak{M} = (\mathfrak{E}_\lambda, \mu)_{\lambda \in \Lambda}$ be a mixed source as defined in Section 5.1.1. Furthermore, let $a > 0$, $\varepsilon \in (0,1)$, and set $\sigma_\lambda = \sigma(\rho_\lambda)$ for $\lambda \in \Lambda$. Then the second order asymptotic rate $b(a, \varepsilon | \mathfrak{M})$ of the mixed source*

$\mathfrak{M}$ *is the solution of the equation*

$$\int_{\Lambda_=(a)} \Phi\left(\frac{b}{\sigma_\lambda}\right) d\mu(\lambda) + \int_{\Lambda_<(a)} d\mu(\lambda) = 1 - \varepsilon. \qquad (5.25)$$

*We have $|b(a, \varepsilon|\rho)| < \infty$ if and only if $a$ equals the first order rate $a(\varepsilon|\mathfrak{M})$ of the mixed source $\mathfrak{M}$, specified by the conditions*

$$\int_{\Lambda_>(a)} d\mu(\lambda) < \varepsilon \qquad (5.26a)$$

$$\int_{\Lambda_>(a)} d\mu(\lambda) + \int_{\Lambda_=(a)} d\mu(\lambda) > \varepsilon. \qquad (5.26b)$$

We note that the conditions (5.26) uniquely determine the first order rate $a(\varepsilon|\mathfrak{M})$ of the mixed source $\mathfrak{M}$. Furthermore, in the limit $\varepsilon \to 0$ the conditions (5.26) imply that the first order rate of mixed source coding is given by the $\mu$-essential supremum over $\{S(\rho_\lambda)\}_{\lambda \in \Lambda}$. In the classical case, this was proved by Han [Han03]. In the quantum case, for a mixed source consisting of two memoryless sources the first order rate was derived by [BD06b] (see also Theorem 5.2.10 below).

*Proof of Theorem 5.2.8.* We first observe that the function $G(b) \coloneqq \int_{\Lambda_=(a)} \Phi\left(\frac{b}{\sigma_\lambda}\right) d\mu(\lambda)$ is continuous and monotonically increasing by the Dominated Convergence Theorem 5.2.7 and the corresponding properties of $\Phi(\cdot)$. Moreover, since $\Phi(z) \in [0, 1]$ for all $z \in \mathbb{R}$, we also have $G(b) \in [0, 1]$ for all $b \in \mathbb{R}$. Hence, if $\int_{\Lambda_<(a)} d\mu(\lambda) < 1 - \varepsilon$, then there is a unique solution of (5.25). If $\int_{\Lambda_<(a)} d\mu(\lambda) \geq 1 - \varepsilon$, we set $b(a, \varepsilon|\mathfrak{M}) = -\infty$ in accordance with (5.11).

Denoting the unique solution of (5.25) by $b^*$, we first prove the converse statement, that is, $b(a, \varepsilon|\mathfrak{M}) \geq b^*$. To this end, assume that $r < b^*$ is an $(a, \varepsilon)$-achievable second order rate, i.e., for $n$ uses of the mixed source $\mathfrak{M} = (\mathfrak{E}_\lambda, \mu)_{\lambda \in \Lambda}$ with source ensemble $\mathfrak{E}_{\text{mix}}^{(n)}$ as in (5.5), there is a source code $\{C_n\}_{n \in \mathbb{N}}$ with $C_n = (\mathcal{V}_n, \mathcal{D}_n, M_n)$ such that

$$\liminf_{n \to \infty} \bar{F}\left(\mathfrak{E}_{\text{mix}}^{(n)}, C_n\right) \geq 1 - \varepsilon, \qquad (5.27a)$$

$$\limsup_{n \to \infty} \frac{\log M_n - na}{\sqrt{n}} \leq r. \qquad (5.27b)$$

Choose $\delta > 0$ such that $r + 2\delta < b^*$. By (5.27b), we have for sufficiently large $n$ that

$$\log M_n < na + \sqrt{n}(r + \delta). \tag{5.28}$$

Lemma 5.2.6 yields the following bound on the fidelity $\bar{F}\left(\mathfrak{E}_{\mathrm{mix}}^{(n)}, C_n\right)$ for arbitrary $\gamma \in \mathbb{R}$:

$$\bar{F}\left(\mathfrak{E}_{\mathrm{mix}}^{(n)}, C_n\right) \leq 1 - \int_\Lambda \mathrm{tr}\left(\rho_\lambda^n \left\{\rho_\lambda^n \leq 2^{-\gamma} \mathbb{1}_n\right\}\right) d\mu(\lambda) + 2^{-\gamma + \log M_n}.$$

For $n \in \mathbb{N}$ we set $\gamma = \log M_n + \sqrt{n}\delta$, such that by (5.28) we have

$$\gamma < na + \sqrt{n}(r + 2\delta).$$

Hence, Lemma 2.2.12 yields

$$\begin{aligned}
\bar{F}\left(\mathfrak{E}_{\mathrm{mix}}^{(n)}, C_n\right) &\leq 1 - \int_\Lambda \mathrm{tr}\left(\rho_\lambda^n \left\{\rho_\lambda^n \leq 2^{-na-\sqrt{n}(r+2\delta)} \mathbb{1}_n\right\}\right) d\mu(\lambda) + 2^{-\sqrt{n}\delta} \\
&= 1 + 2^{-\sqrt{n}\delta} - \int_{\Lambda_=(a)} \mathrm{tr}\left(\rho_\lambda^n \left\{\rho_\lambda^n \leq 2^{-na-\sqrt{n}(r+2\delta)} \mathbb{1}_n\right\}\right) d\mu(\lambda) \\
&\quad - \int_{\Lambda_<(a)} \mathrm{tr}\left(\rho_\lambda^n \left\{\rho_\lambda^n \leq 2^{-na-\sqrt{n}(r+2\delta)} \mathbb{1}_n\right\}\right) d\mu(\lambda) \\
&\quad - \int_{\Lambda_>(a)} \mathrm{tr}\left(\rho_\lambda^n \left\{\rho_\lambda^n \leq 2^{-na-\sqrt{n}(r+2\delta)} \mathbb{1}_n\right\}\right) d\mu(\lambda) \tag{5.29}
\end{aligned}$$

where we defined $\Lambda_>(a) := \{\lambda \in \Lambda \colon S(\rho_\lambda) > a\}$. For the three integrands on the right-hand side of (5.29), Corollary 4.2.8 implies the following in the limit $n \to \infty$:

$$\lim_{n \to \infty} \mathrm{tr}\left(\rho_\lambda^n \left\{\rho_\lambda^n \leq 2^{-na-\sqrt{n}(r+2\delta)} \mathbb{1}_n\right\}\right) = \begin{cases} \Phi\left(\frac{-(r+2\delta)}{\sigma_\lambda}\right) & \text{if } S(\rho_\lambda) = a \\ 1 & \text{if } S(\rho_\lambda) > a \\ 0 & \text{if } S(\rho_\lambda) < a. \end{cases} \tag{5.30}$$

Taking the limit inferior on both sides of (5.29), noting that we can exchange limit and integral by the Dominated Convergence Theorem 5.2.7, and using (5.30), we therefore obtain

$$\begin{aligned}
\liminf_{n \to \infty} \bar{F}\left(\mathfrak{E}_{\mathrm{mix}}^{(n)}, C_n\right) &\leq 1 - \int_{\Lambda_=(a)} \Phi\left(\frac{-(r+2\delta)}{\sigma_\lambda}\right) d\mu(\lambda) - \int_{\Lambda_>(a)} d\mu(\lambda) \\
&= 1 - \int_{\Lambda_=(a)} d\mu(\lambda) + \int_{\Lambda_=(a)} \Phi\left(\frac{r+2\delta}{\sigma_\lambda}\right) d\mu(\lambda) - \int_{\Lambda_>(a)} d\mu(\lambda)
\end{aligned}$$

$$= \int_{\Lambda_<(a)} d\mu(\lambda) + \int_{\Lambda_=(a)} \Phi\left(\frac{r + 2\delta}{\sigma_\lambda}\right) d\mu(\lambda)$$

$$< \int_{\Lambda_<(a)} d\mu(\lambda) + \int_{\Lambda_=(a)} \Phi\left(\frac{b^*}{\sigma_\lambda}\right) d\mu(\lambda)$$

$$= 1 - \varepsilon. \tag{5.31}$$

Here, we used the relation $\Phi(-x) = 1 - \Phi(x)$ in the first equality, the fact that $\mu$ is a normalized measure on $\Lambda = \Lambda_=(a) \cup \Lambda_<(a) \cup \Lambda_>(a)$ in the second equality, the assumption $r + 2\delta < b^*$ in the strict inequality, and the fact that $b^*$ is defined as the solution of the equation (5.25) in the last equality. The bound (5.31) is a contradiction to (5.27a), and hence, $b(a, \varepsilon|\mathfrak{M}) \geq b^*$.

We now use the universal source code $\{C_n\}_{n \in \mathbb{N}}$ with $C_n := \{\mathcal{V}_n, \mathcal{D}_n, M_n\}$ as defined in Proposition 5.2.3 to prove that the second order rate $b^*$ is also achievable. To this end, consider $n$ uses of the mixed source $\mathfrak{M} = (\mathfrak{E}_\lambda, \mu)_{\lambda \in \Lambda}$ with source ensemble $\mathfrak{E}_{\text{mix}}^{(n)}$ as defined in (5.5) and source state $\rho^{(n)}$ as defined in (5.6). Recall that $\Pi_n$ denotes the projection onto the code space $\Upsilon_{a,b}^n$ defined in Section 5.2.1, the decoder $\mathcal{D}_n$ is the trivial embedding, and we set

$$\mathcal{V}_n : i^n \longmapsto \frac{\Pi_n \varphi_{i^n} \Pi_n}{\text{tr}(\Pi_n \varphi_{i^n})}.$$

For arbitrary $a > 0$, we use definition (5.8) of the ensemble average fidelity to compute:

$$\bar{F}\left(\mathfrak{E}_{\text{mix}}^{(n)}, C_n\right) = \int_\Lambda d\mu(\lambda) \sum_{i^n} q_{i^n}^{(\lambda)} \text{tr}\left((\mathcal{D}_n \circ \mathcal{V}_n)(i^n) \varphi_{i^n}\right)$$

$$= \int_\Lambda d\mu(\lambda) \sum_{i^n} q_{i^n}^{(\lambda)} \frac{1}{\text{tr}(\Pi_n \varphi_{i^n})} \text{tr}(\Pi_n \varphi_{i^n} \Pi_n \varphi_{i^n})$$

$$= \int_\Lambda d\mu(\lambda) \sum_{i^n} q_{i^n}^{(\lambda)} \frac{\langle \varphi_{i^n} | \Pi_n | \varphi_{i^n} \rangle^2}{\langle \varphi_{i^n} | \Pi_n | \varphi_{i^n} \rangle}$$

$$= \int_\Lambda d\mu(\lambda) \sum_{i^n} q_{i^n}^{(\lambda)} \langle \varphi_{i^n} | \Pi_n | \varphi_{i^n} \rangle$$

$$= \int_\Lambda d\mu(\lambda) \, \text{tr}\left(\rho_\lambda^n \Pi_n\right)$$

$$\geq \int_\Lambda d\mu(\lambda) \, \text{tr}\left(\rho_\lambda^n \left\{\rho_\lambda^n > 2^{-na - \sqrt{n}b} \mathbb{1}_n\right\}\right)$$

$$= 1 - \int_\Lambda d\mu(\lambda) \, \text{tr}\left(\rho_\lambda^n \left\{\rho_\lambda^n \leq 2^{-na - \sqrt{n}b} \mathbb{1}_n\right\}\right), \tag{5.32}$$

where the inequality follows from (5.16). We set $b = b^*$, where $b^*$ is once again defined as the

solution of the relation

$$\int_{\Lambda_=(a)} \Phi\left(\frac{b}{\sigma_\lambda}\right) d\mu(\lambda) + \int_{\Lambda_<(a)} d\mu(\lambda) = 1 - \varepsilon.$$

As before, Corollary 4.2.8 implies that

$$\lim_{n\to\infty} \operatorname{tr}\left(\rho_\lambda^n \left\{\rho_\lambda^n \leq 2^{-na-\sqrt{n}b^*} \mathbb{1}_n\right\}\right) = \begin{cases} \Phi\left(\frac{-b^*}{\sigma_\lambda}\right) & \text{if } S(\rho_\lambda) = a \\ 1 & \text{if } S(\rho_\lambda) > a \\ 0 & \text{if } S(\rho_\lambda) < a. \end{cases}$$

Following analogous steps as the ones leading up to (5.31), this yields

$$1 - \lim_{n\to\infty} \int_\Lambda \operatorname{tr}\left(\rho_\lambda^n \left\{\rho_\lambda^n \leq 2^{-na-\sqrt{n}b^*} \mathbb{1}_n\right\}\right) d\mu(\lambda) = \int_{\Lambda_=(a)} \Phi\left(\frac{b^*}{\sigma_\lambda}\right) d\mu(\lambda) + \int_{\Lambda_<(a)} d\mu(\lambda)$$
$$= 1 - \varepsilon, \tag{5.33}$$

where the exchange of limit and integral is permitted by the Dominated Convergence Theorem 5.2.7, and we used the definition of $b^*$ in the last equality. Hence, taking the limit inferior in (5.32) and using (5.33), we obtain $\liminf_{n\to\infty} \bar{F}\left(\mathfrak{E}_{\text{mix}}^{(n)}, C_n\right) \geq 1 - \varepsilon$. Moreover, Lemma 5.2.2 implies that the universal source code $\{C_n\}_{n\in\mathbb{N}}$ satisfies

$$\limsup_{n\to\infty} \frac{\log M_n - na}{\sqrt{n}} \leq b^*.$$

Hence, the rate $b^*$ is $(a, \varepsilon)$-achievable, and we obtain $b(a, \varepsilon|\mathfrak{M}) \leq b^*$. Together with $b(a, \varepsilon|\mathfrak{M}) \geq b^*$ as shown above, this proves (5.25).

Assume now that the solution $b^*$ of (5.25) is finite, $|b^*| < \infty$. We then have $\Phi\left(\frac{b^*}{\sigma_\lambda}\right) \in (0, 1)$ for all $\lambda \in \Lambda$. Hence, using $\Phi\left(\frac{b^*}{\sigma_\lambda}\right) > 0$, the relation (5.25) implies that

$$1 - \varepsilon > \int_{\Lambda_<(a)} d\mu(\lambda) = 1 - \int_{\Lambda_=(a)} d\mu(\lambda) - \int_{\Lambda_>(a)} d\mu(\lambda),$$

which is the inequality (5.26b). Using $\Phi\left(\frac{b^*}{\sigma_\lambda}\right) < 1$, we get from (5.25) that

$$1 - \varepsilon < \int_{\Lambda_=(a)} d\mu(\lambda) + \int_{\Lambda_<(a)} d\mu(\lambda) = 1 - \int_{\Lambda_>(a)} d\mu(\lambda),$$

which gives the inequality (5.26a). This proves Theorem 5.2.8. □

If the measure $\mu$ has finite support on points $\lambda_1, \ldots, \lambda_k \in \Lambda$, Theorem 5.2.8 reduces to:

**Corollary 5.2.9.** *Consider a mixed source $\mathfrak{M}_k = (\mathfrak{E}_j, t_j)_{j=1}^{k}$, and set $S_j = S(\rho_j)$ and $\sigma_j = \sigma(\rho_j)$ for $j = 1, \ldots, k$. For $a > 0$ and $\varepsilon \in (0, 1)$, the second order asymptotic rate $b(a, \varepsilon | \mathfrak{M}_k)$ of the mixed source $\mathfrak{M}_k$ is given by the solution of the equation*

$$\sum_{i: \, S_i = a} t_i \Phi\left(\frac{b}{\sigma_i}\right) + \sum_{i: \, S_i < a} t_i = 1 - \varepsilon. \tag{5.34}$$

*We have $|b(a, \varepsilon | \rho)| < \infty$ if and only if $a$ equals the first order rate $a(\varepsilon | \mathfrak{M}_k)$ of the mixed source $\mathfrak{M}_k$, specified by the conditions*

$$\sum_{i: \, S_i > a} t_i < \varepsilon \qquad\qquad \sum_{i: \, S_i > a} t_i + \sum_{i: \, S_i = a} t_i > \varepsilon. \tag{5.35}$$

Finally, we consider the special case of a mixed source $\mathfrak{M}_2 = (\mathfrak{E}_1, \mathfrak{E}_2, t)$ consisting of two memoryless sources $\mathfrak{E}_1$ and $\mathfrak{E}_2$ with source states $\rho_1, \rho_2 \in \mathcal{D}(\mathcal{H})$, respectively, and mixing parameter $t \in (0, 1)$. The source state for $n$ uses of $\mathfrak{M}_2$ is given by

$$\rho^{(n)} = t\rho_1^{\otimes n} + (1-t)\rho_2^{\otimes n}.$$

We adhere to the discussion of classical mixed source coding by Nomura and Han [NH13], considering the following three cases and abbreviating $S_i \equiv S(\rho_i)$ and $\sigma_i \equiv \sigma(\rho_i)$ for $i = 1, 2$:

Case 1: $S_1 = S_2$

Case 2: $S_1 > S_2, t > \varepsilon$

Case 3: $S_1 > S_2, t < \varepsilon$

Note that the assumption $S_1 > S_2$ can be made without loss of generality. We state the first and second order rates in each of the three cases in the following theorem. The first order rates have previously been derived by Bowen and Datta [BD06b].

**Theorem 5.2.10.** *Consider a mixed source $\mathfrak{M}_2 = (\mathfrak{E}_1, \mathfrak{E}_2, t)$ with source states $\rho_1, \rho_2 \in \mathcal{D}(\mathcal{H})$ and $t \in (0, 1)$, and set $S_i := S(\rho_i)$ and $\sigma_i := \sigma(\rho_i)$ for $i = 1, 2$. For $\varepsilon \in (0, 1)$, the first order rate $a(\varepsilon | \mathfrak{M}_2)$ and the second order asymptotic rate $b(a, \varepsilon | \mathfrak{M}_2)$ of the mixed source $\mathfrak{M}_2$ are given by the following expressions:*

(i) *For $S_1 = S_2 \equiv S$, we have $a(\varepsilon | \mathfrak{M}_2) = S$ and $b(S, \varepsilon | \mathfrak{M}_2) = b^*$, where $b^*$ is the solution of the equation*

$$t\Phi\left(\frac{b}{\sigma_1}\right) + (1-t)\Phi\left(\frac{b}{\sigma_2}\right) = 1 - \varepsilon. \tag{5.36}$$

(ii) *For $S_1 > S_2$ and $t > \varepsilon$, we have $a(\varepsilon | \mathfrak{M}_2) = S_1$ and*

$$b(S_1, \varepsilon | \mathfrak{M}_2) = -\sigma_1 \Phi^{-1}\left(\frac{\varepsilon}{t}\right).$$

(iii) *For $S_1 > S_2$ and $t < \varepsilon$, we have $a(\varepsilon | \mathfrak{M}_2) = S_2$ and*

$$b(S_2, \varepsilon | \mathfrak{M}_2) = -\sigma_2 \Phi^{-1}\left(\frac{\varepsilon - t}{1 - t}\right).$$

*Proof.* In (i), where $S_1 = S_2 = S$, the relations (5.35) imply that the first order rate $a(\varepsilon | \mathfrak{M}_2)$ is equal to $S$. Substituting this for $a$ in (5.34) immediately yields

$$t\Phi\left(\frac{b}{\sigma_1}\right) + (1-t)\Phi\left(\frac{b}{\sigma_2}\right) = 1 - \varepsilon.$$

In (ii), we have $S_1 > S_2$ and $t > \varepsilon$. The conditions (5.35) then imply that $a(\varepsilon | \mathfrak{M}_2) = S_1$, and substituting this for $a$ in (5.34) yields

$$t\Phi\left(\frac{b}{\sigma_1}\right) + 1 - t = 1 - \varepsilon,$$

from which we obtain

$$b = \sigma_1 \Phi^{-1}\left(1 - \frac{\varepsilon}{t}\right) = -\sigma_1 \Phi^{-1}\left(\frac{\varepsilon}{t}\right).$$

Finally, we consider (iii), where $S_1 > S_2$ and $t < \varepsilon$. In this case, (5.35) yields $a(\varepsilon | \mathfrak{M}_2) = S_2$, and plugging this into (5.34) yields

$$b = \sigma_2 \Phi^{-1}\left(\frac{1 - \varepsilon}{1 - t}\right) = \sigma_2 \Phi^{-1}\left(1 - \frac{\varepsilon - t}{1 - t}\right) = -\sigma_2 \Phi^{-1}\left(\frac{\varepsilon - t}{1 - t}\right),$$

which proves the claim. □

Theorem 5.2.10(ii) and (iii) show that for a mixed source consisting of two memoryless sources with unequal von Neumann entropies, the first order rate is given by the source with higher entropy for small values of the error parameter $\varepsilon$. That is, in the low-error regime, the source with higher entropy is a 'bottleneck' for the quantum source coding task. The compression rate can be lowered to the entropy of the other source at the expense of incurring a higher error, with the threshold given by the mixing parameter $t$. In both cases, the second order rate is given in terms of the quantum information variance of the source determining the first order rate.

Naturally, if both sources have the same entropy $S$ (Theorem 5.2.10(i)), the first order rate of the quantum source coding task is also equal to $S$. To determine the range of the second order rate $b$ in this case, assume without loss of generality that $\sigma_1 < \sigma_2$. Then, using properties of the CDF $\Phi$ of a normal distribution and definition (5.36) of $b$, it follows that

$$
\begin{aligned}
b &\in \left[-\sigma_1 \Phi^{-1}(\varepsilon), -\sigma_2 \Phi^{-1}(\varepsilon)\right] &&\text{if } \varepsilon \in (0, 1/2), \\
b &\in \left[-\sigma_2 \Phi^{-1}(\varepsilon), -\sigma_1 \Phi^{-1}(\varepsilon)\right] &&\text{if } \varepsilon \in (1/2, 1),
\end{aligned}
\tag{5.37}
$$

and $b = 0$ for $\varepsilon = 1/2$. See Figure 5.2 for a plot showing a typical example of this.

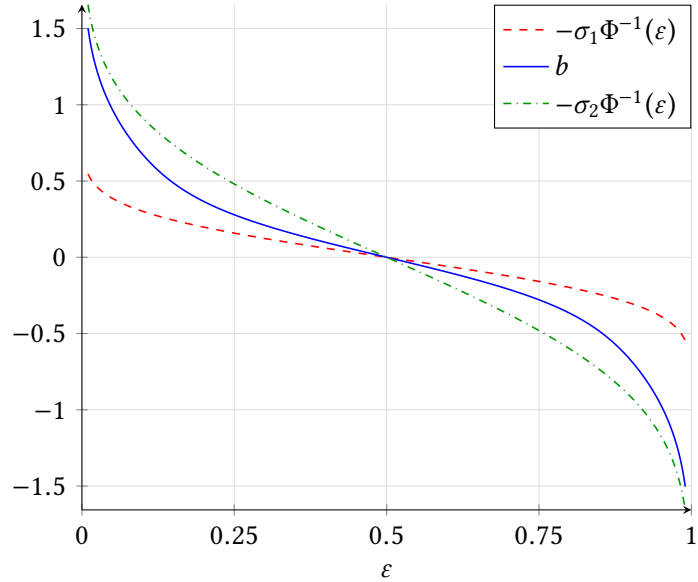

Figure 5.2: Plot of the second order asymptotic rate $b = b(S, \varepsilon | \mathfrak{M}_2)$ (blue-solid) defined in (5.36) and bounds on $b$ in terms of $-\sigma_1 \Phi^{-1}(\varepsilon)$ (red-dashed) and $-\sigma_2 \Phi^{-1}(\varepsilon)$ (green-dash-dotted) given in (5.37) for a mixed quantum source $\mathfrak{M}_2 = (\mathfrak{C}_1, \mathfrak{C}_2, t)$ with equal von Neumann entropies $S(\rho_1) = S(\rho_2) = S$, $\sigma_1 = 0.235$, $\sigma_2 = 0.712$, and $t = 0.425$.

## 5.2.4. Memoryless sources and the blind encoding setting

As mentioned in Section 5.1, a mixed source $(\mathfrak{E}_\lambda, \mu)_{\lambda \in \Lambda}$ reduces to a single memoryless source upon choosing $\Lambda$ to be a singleton space. Correspondingly, the ensemble average fidelity (5.8) for quantum source coding using a mixed source reduces to the one using a single memoryless source, as given in (5.7). Hence, Theorem 5.2.8 yields the following result, which was derived in [DL15] using the information spectrum relative entropies defined in Appendix A:

**Corollary 5.2.11** (Second order rate for quantum source coding using a memoryless source). *For a memoryless quantum source $\mathfrak{E}$ with source state $\rho$, the second order asymptotic rate for $\varepsilon \in (0, 1)$ is given by*

$$b(S(\rho), \varepsilon | \mathfrak{E}) = -\sigma(\rho)\Phi^{-1}(\varepsilon)\,.$$

We can use Corollary 5.2.11 to expand the quantity $\log M_n^*$, denoting the code size of an optimal quantum source code, as follows:

$$\log M_n^* = nS(\rho) - \sqrt{n}\sigma(\rho)\Phi^{-1}(\varepsilon) + o(\sqrt{n}). \tag{5.38}$$

In [DL15], the $o(\sqrt{n})$ term in (5.38) was refined to the third order term $O(\log n)$. There, we first proved one-shot bounds on the quantity $\log M^* \equiv \log M_1^*$ in terms of an information spectrum entropy, and then used the second order expansion of the latter to infer (5.38). Figure 5.3 shows a comparison between the second order asymptotic expansion (5.38) and these one-shot bounds on the minimal compression length $\log M_n$ in (a), as well as the second order expansion (5.38) for different values of $\varepsilon$ in (b).

In the same paper, we also considered the blind encoding setting as described in Section 5.1.2, deriving the following second order asymptotic bounds on $\log M_n^*$:

$$nS(\rho) - \sqrt{n}\sigma(\rho)\Phi^{-1}(\varepsilon) + O(\log n) \leq \log M_n^* \leq nS(\rho) - \sqrt{n}\sigma(\rho)\Phi^{-1}\left(\frac{\varepsilon}{2}\right) + O(\log n).$$

Since the coefficients of the $\sqrt{n}$-terms do not match, we cannot give an exact expression for the second order asymptotic rate $b_{\mathrm{bl}}(S(\rho), \varepsilon | \mathfrak{E})$ in the blind encoding setting, and we merely have

$$-\sigma(\rho)\Phi^{-1}(\varepsilon) \leq b_{\mathrm{bl}}(S(\rho), \varepsilon | \mathfrak{E}) \leq -\sigma(\rho)\Phi^{-1}\left(\frac{\varepsilon}{2}\right).$$

It remains an open problem to derive a closed expression of the second order asymptotic rate in blind quantum source coding using a memoryless source.

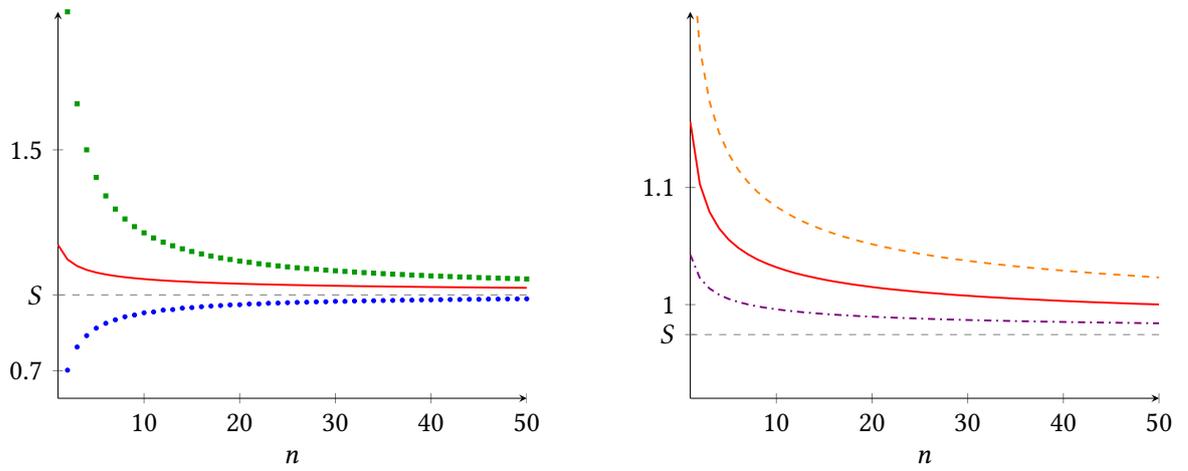

(a) Comparison of the second order asymptotic approximation $S - \sigma \Phi^{-1}(\varepsilon) / \sqrt{n}$ (solid red line) and normalized one-shot bounds (blue dots, green squares) on the minimal compression rate $\log M_n^* / n$ for $\varepsilon = 0.25$.

(b) Second order asymptotic expansions for different error parameters $\varepsilon_1 = 0.1$ (orange/dashed), $\varepsilon_2 = 0.25$ (red/solid), and $\varepsilon_3 = 0.4$ (violet/dash-dotted).

Figure 5.3: Plot of second order asymptotic expansions and one-shot bounds (derived in [DL15]) for a memoryless quantum source with entropy $S \equiv S(\rho) = 0.9744$ and quantum information variance $\sigma \equiv \sigma(\rho) = 0.2693$.

# 6. Strong converse theorems

Coding theorems for information-theoretic tasks establish an entropic quantity as the optimal rate of a task. While the previous chapter focused on finite blocklength approximations of the optimal rate, the current chapter seeks to further refine the converse part of a coding theorem by strengthening it to a *strong converse*. In the following, we explain this concept with the example of an information-theoretic task whose optimal rate corresponds to a *cost* (such as the optimal compression length in quantum source coding).[1]

The achievability part or direct part of a coding theorem states that for any rate above the optimal rate, there is a corresponding code for accomplishing the task successfully. To be precise, if we denote the error incurred in the protocol for $n$ uses of the underlying resource by $\varepsilon_n$, then for any rate above the optimal rate there exists a code, whose error $\varepsilon_n$ vanishes in the asymptotic limit $n \to \infty$. Such rates are called *achievable*, and the optimal rate is defined as the infimum over all achievable rates. In contrast, the converse part of a coding theorem states that for any code with rate below the optimal rate, the error does not vanish asymptotically, that is, it is bounded away from 0 in the asymptotic limit $n \to \infty$.

The converse part of a coding theorem described in the preceding paragraph is called a *weak converse*. In principle, it leaves open the possibility of a trade-off between error and rate of a protocol. For example, even though the error $\varepsilon_n$ cannot vanish asymptotically for rates below the optimal rate, it might still be possible to, say, find a constant upper bound $\varepsilon_n \leq c < 1$ for all $n \in \mathbb{N}$ for such codes. If the *strong converse* holds for a task, then such a trade-off (or the existence of such $c$) is not possible.

To be precise, a strong converse states that for all codes with rate below the optimal rate, the error $\varepsilon_n$ incurred in the protocol converges to 1. In other words, such protocols become worse with increasing blocklength $n$, and eventually fail with certainty in the asymptotic limit

---

[1] To obtain the corresponding statements for tasks with an optimal gain (such as the capacity of a channel), simply exchange 'above' and 'below' in every instance of the following paragraphs, and replace infimum with supremum.

$n \to \infty$. Moreover, one often requires this convergence to unity to be exponential, that is,

$$\varepsilon_n \geq 1 - \exp(-Kn) \tag{6.1}$$

for some positive constant $K$. Figure 6.1 shows a schematic description of the difference between weak and strong converse.

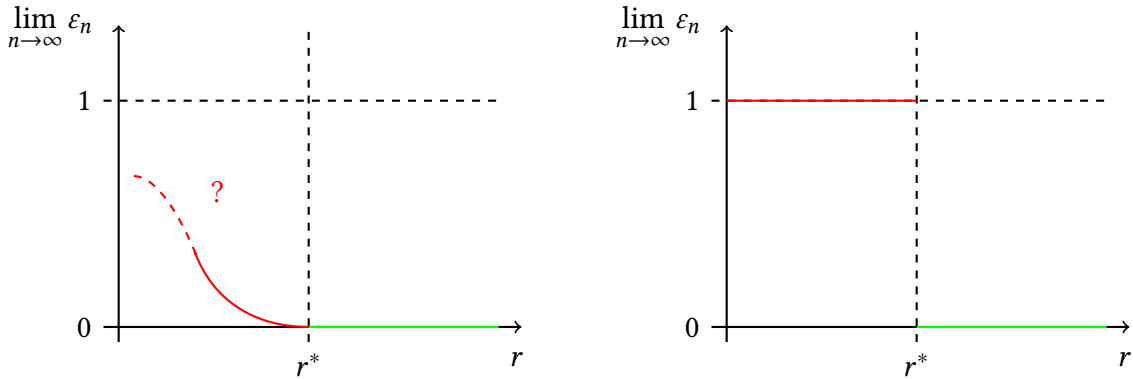

Figure 6.1: Weak (left) vs. strong converse (right), where $r$ denotes the rate of a protocol with optimal rate $r^*$ (corresponding to a cost) and error $\varepsilon_n$.

If the strong converse is known for the optimal rate of an information-theoretic task, we say that this optimal rate satisfies the *strong converse property*. A *strong converse theorem* establishes the strong converse property of the optimal rate of an information-theoretic task, and hence serves to identify the latter as a sharp rate threshold for the task. Alternatively, one may define the *strong converse rate* as the supremum over all rates below which any code fails with certainty. The strong converse property holds if this strong converse rate coincides with the optimal rate.

For information transmission through classical noisy channels, the strong converse theorem was first proved by Wolfowitz [Wol61]. An alternate proof of this theorem was later given by Arimoto [Ari73], who used the properties of a quantity sometimes referred to as the Gallager function [PV10]. Ogawa and Nagaoka [ON99] extended this method to the quantum setting to prove the strong converse property of the capacity of a classical-quantum channel, which was also proved concurrently by Winter [Win99a] using the method of types. Nagaoka [Nag01] further developed Arimoto's idea to give a new proof of this result. To this end, he employed a Rényi divergence and its data processing inequality (i.e., monotonicity under completely posi-tive, trace-preserving maps), establishing what we refer to as the 'Rényi entropy method' from now on. Later, Polyanskiy and Verdú [PV10] realized that it is possible to establish converse bounds by employing any divergence satisfying the data processing inequality. Fong and Tan [FT16] used the Rényi entropy method to obtain strong converse theorems in network information theory. In quantum information theory, the Rényi entropy method has been successfully employed to prove strong converse theorems for classical channel coding with entangled inputs for a large class of quantum channels with additive Holevo capacity [KW09]. Moreover, strong converse theorems were proved for classical information transmission through entanglement-breaking and Hadamard channels [WWY14] and quantum information transmission through generalized dephasing channels [TWW15].

Application of the Rényi entropy method to prove the strong converse property for an information-theoretic task involving local operations and classical communication (LOCC) between two parties was considered by Hayashi et al. [HKM+02] in the context of entanglement concentration (see also [Hay06b]). More recently, Sharma [Sha14] used the Rényi entropy method to establish the strong converse theorem for the task of *state merging*: Alice and Bob initially share a bipartite state and the aim is for Alice to transfer her part of the state to Bob by sending information to him through a noiseless classical channel. Both Alice and Bob are also allowed to make use of prior shared entanglement between them, to assist them in achieving this task. In this case monotonicity of a relevant Rényi divergence under LOCC plays a pivotal role in establishing the strong converse for the optimal entanglement cost. However, note that the strong converse for state merging also follows from an 'operational' argument [Win14]. Furthermore, it was previously proved [Ber08] using the smooth entropy framework (see [Ren05; TCR09; Dat09; Tom12] and references therein) and the quantum asymptotic equipartition property (QAEP) of the smooth min- and max-entropies [TCR10; Tom12].

In view of the second order asymptotic results from Chapter 5, it is worth noting that one possible route to proving strong converse theorems for an information-theoretic task is via second order asymptotic expansions of the optimal rate. Let us demonstrate this with the example of quantum source coding using a memoryless source, as discussed in Section 5.2.4. There, we stated the second order asymptotic expansion of the code size $\log M_n^*$ of an optimal quantum source code in (5.38). The converse part of this expansion is the bound

$$\log M_n^* \geq nS(\rho) - \sqrt{n}\sigma(\rho)\Phi^{-1}(\varepsilon) + f(n), \tag{6.2}$$

where $\rho$ denotes the source state of the quantum source, $\varepsilon \in (0, 1)$ denotes the error incurred in the source coding protocol, and $f(n) \in o(\sqrt{n})$. Rearranging (6.2) and using monotonicity of $\Phi$ yields

$$\varepsilon \geq \Phi\left(\frac{\sqrt{n}}{\sigma(\rho)}\left(S(\rho) - \frac{1}{n}\log M_n^*\right) + g(n)\right), \tag{6.3}$$

where $g(n)$ satisfies $\lim_{n \to \infty} g(n) = 0$.[2] In the converse regime, where we assume

$$\frac{1}{n}\log M_n^* < S(\rho) \quad \text{for all } n \in \mathbb{N},$$

the argument of $\Phi$ in (6.3) diverges to $+\infty$ as $n \to \infty$. Hence, the strong converse for source coding follows from the fact that $\lim_{x \to +\infty} \Phi(x) = 1$. However, unlike (6.1) the convergence of $\varepsilon$ to 1 in (6.3) is not exponential in $n$. In Section 6.1.4 below, we obtain a strong converse theorem for (blind) quantum source coding with exponential convergence of the error to 1 (Corollary 6.1.5).

We also mention that quantum source coding using a mixed source, as discussed in Chapter 5, is an example of an information-processing task for which the strong converse property does *not* hold. For simplicity, let us assume that the mixed source consists of two memoryless sources with mixing parameter $t \in (0, 1)$ and source states $\rho_1$ and $\rho_2$, respectively (see Section 5.1.1 for this terminology). Bowen and Datta [BD06b] proved that the optimal rate of quantum source coding is given by $\max\{S(\rho_1), S(\rho_2)\}$, whereas the strong converse rate is given by $\min\{S(\rho_1), S(\rho_2)\}$. Hence, for two memoryless sources such that $S(\rho_1) \neq S(\rho_2)$, the optimal rate and strong converse rate of quantum source coding using the corresponding mixed source do not coincide, so that the strong converse property is not satisfied. The fact that the optimal and strong converse rate are given by $\max\{S(\rho_1), S(\rho_2)\}$ and $\min\{S(\rho_1), S(\rho_2)\}$, respectively, can also be deduced from our Theorem 5.2.10(iii), together with the discussion in the preceding paragraph.

In this chapter, which is based on [LWD16], we use Nagaoka's Rényi entropy method to establish strong converse theorems for state redistribution with and without feedback (Section 6.1) and measurement compression with quantum side information (QSI) (Section 6.2). The strong converse theorem for state redistribution also yields (previously known) strong converses for blind quantum source coding (Corollary 6.1.5), coherent state merging (Corollary 6.1.6), and quantum state splitting (Corollary 6.1.7), as these protocols can be obtained by considering special cases of state redistribution. Note that our method also yields new proofs of the known strong converse theorems for randomness extraction against QSI and data compression with QSI (both proved by Tomamichel [Tom12] using the smooth entropy framework together with the QAEP). These results can be found in [LWD16].

---

[2]In fact, for quantum source coding $g(n)$ is of the order $(\log n)/\sqrt{n}$.

## 6.1. State redistribution

We first discuss the operational setting of state redistribution (without feedback) in Section 6.1.1. Section 6.1.2 contains the main result of this section, a strong converse theorem for state redistribution. In Section 6.1.3, we introduce the feedback version of the state redistribution protocol devised by Berta et al. [BCT16], and extend the strong converse theorem from the previous section to this setting. Finally, in Section 6.1.4 we show how to obtain various quantum information-processing tasks as special cases of state redistribution. As a corollary, we obtain (previously known) strong converse theorems for these tasks as well.

### 6.1.1. Operational setting

Consider a tripartite state $\rho_{ABC}$ shared between Alice and Bob, with the systems $A$ and $C$ being with Alice and the system $B$ being with Bob. Let $\psi_{ABCR}$ denote a purification of $\rho_{ABC}$, where $R$ is an inaccessible, purifying reference system. Furthermore, Alice and Bob share entanglement in the form of an MES $\Phi^k_{T_A T_B}$ of Schmidt rank $k$, with the systems $T_A$ and $T_B$ being with Alice and Bob, respectively. The goal of the state redistribution protocol is to transfer the system $A$ from Alice to Bob, while preserving its correlations with the other systems. In the process, the shared entanglement is transformed to an MES $\Phi^m_{T'_A T'_B}$ of Schmidt rank $m$, where $T'_A$ and $T'_B$ are with Alice and Bob, respectively. The initial state and the target state are shown in Figure 6.2.

In achieving this goal, Alice and Bob are allowed to use local encoding and decoding operations on the systems in their possession. In addition, Alice is allowed to send qubits to Bob (through a noiseless quantum channel). A general state redistribution protocol $(\rho, \Lambda)$ with $\Lambda \equiv \mathcal{D} \circ \mathcal{E}$ and $\rho = \rho_{ABC}$ therefore consists of the following steps (cf. Figure 6.3):

1. Alice applies an encoding CPTP map $\mathcal{E} \colon ACT_A \to C'T'_A Q$ and sends the system $Q$ to Bob through the noiseless quantum channel.

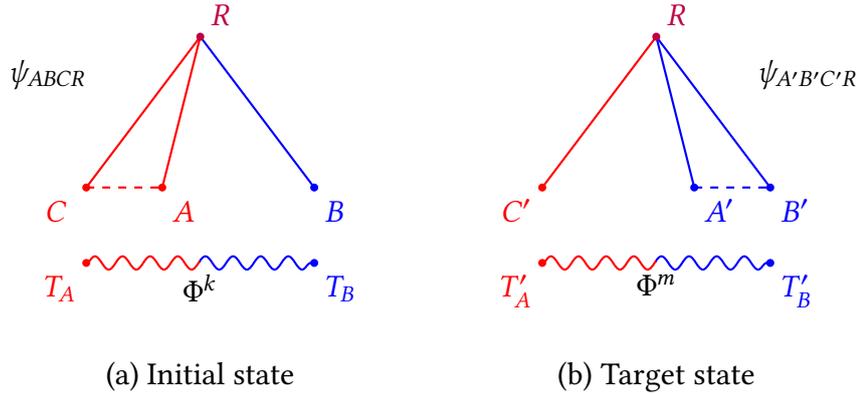

(a) Initial state          (b) Target state

Figure 6.2: State redistribution protocol that transfers Alice's system $A$ to Bob. Starting with the initial state $\Phi^k \otimes \psi$ depicted in (a), the protocol outputs a state that is close in fidelity to the target state $\Phi^m \otimes \psi$ depicted in (b).

2. Upon receiving the system $Q$, Bob applies a decoding CPTP map $\mathcal{D}: QBT_B \to T_B'A'B'$, where $\mathcal{H}_{T_B'} \cong \mathcal{H}_{T_A'}$, $\mathcal{H}_{A'} \cong \mathcal{H}_A$, and $\mathcal{H}_{B'} \cong \mathcal{H}_B$.

The initial state shared between Alice, Bob, and the reference is

$$\Omega \equiv \Omega_{T_A T_B ABCR} \coloneqq \Phi^k_{T_A T_B} \otimes \psi_{ABCR},$$

the state after Alice's encoding operation is

$$\omega \equiv \omega_{T_A' T_B C' QBR} \coloneqq (\mathcal{E} \otimes \mathrm{id}_{BRT_B})(\Omega), \tag{6.4}$$

and the final state of the protocol $(\rho, \Lambda)$ is given by

$$\sigma \equiv \sigma_{T_A' T_B' A'B'C'R} \coloneqq (\Lambda \otimes \mathrm{id}_R)(\Omega) = (\mathcal{D} \circ \mathcal{E} \otimes \mathrm{id}_R)(\Omega). \tag{6.5}$$

The aim is to obtain a state $\sigma$ that is close to the target state

$$\widehat{\Omega} \equiv \widehat{\Omega}_{T_A' T_B' A'B'C'R} \coloneqq \Phi^m_{T_A' T_B'} \otimes \psi_{A'B'C'R},$$

where $\psi_{A'B'C'R} = \psi_{ABCR}$. The figure of merit of the protocol is the fidelity $F(\sigma, \widehat{\Omega})$. The number of qubits that Alice sends to Bob is given by $\log|Q|$, whereas the number of ebits consumed in the protocol is given by $\log k - \log m = \log|T_A| - \log|T_A'|$. If $k < m$, then ebits are *gained* in the protocol.

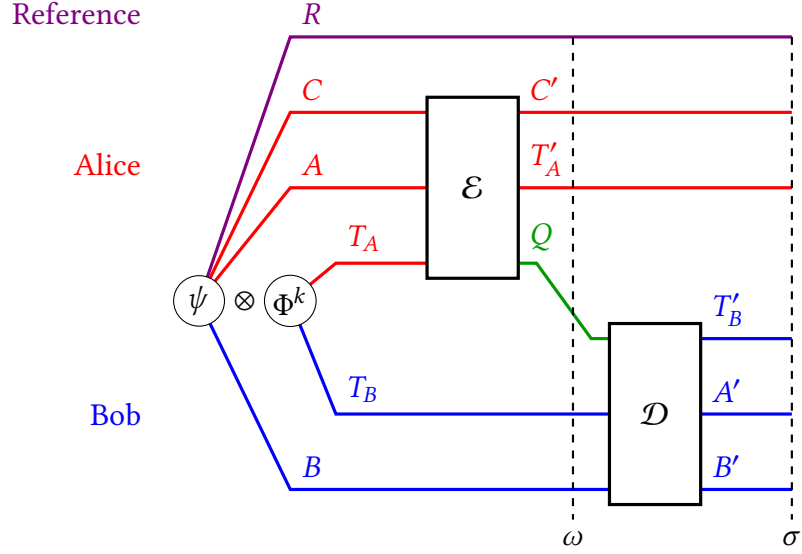

Figure 6.3: Schematic description of a state redistribution protocol between Alice (red), Bob (blue), and a reference system (violet), with quantum communication (green) from Alice to Bob. For a detailed description of the protocol, see Section 6.1.1.

We consider state redistribution in the asymptotic, memoryless setting, where Alice and Bob start with $n$ i.i.d. copies of the initial resource $\rho$, and the strong converse property is established in the limit $n \to \infty$. In this case, Alice and Bob initially share $n$ identical copies of the state $\rho_{ABC}$ with purification $\psi_{ABCR}$, i.e., they share the state $\rho_{ABC}^{\otimes n}$ with purification $\psi_{ABCR}^{\otimes n}$. Moreover, they share an MES $\Phi_{T_A^n T_B^n}^{k_n}$ of Schmidt rank $k_n$. We then consider state redistribution protocols $(\rho^{\otimes n}, \Lambda_n)$ with the figure of merit

$$F_n := F\big(\sigma_n, \Phi^{m_n} \otimes \psi^{\otimes n}\big), \qquad (6.6)$$

where $\Phi_{T_A'^{m_n} T_B'^{m_n}}^{m_n}$ is a maximally entangled state of Schmidt rank $m_n$, and $\sigma_n := (\Lambda_n \otimes \mathrm{id}_{R^n})(\Phi^{k_n} \otimes \psi^{\otimes n})$, where $\Lambda_n : A^n C^n T_A^n \otimes B^n T_B^n \to C'^n T_A'^n \otimes T_B'^n A'^n B'^n$ with $Q^n$ being sent from Alice to Bob. The two operational quantities of interest are:

- the quantum communication cost of the protocol $(\rho^{\otimes n}, \Lambda_n)$, given by

$$q\left(\rho^{\otimes n}, \Lambda_n\right) := \frac{1}{n} \log |Q^n|; \qquad (6.7)$$

- the entanglement cost of the protocol $(\rho^{\otimes n}, \Lambda_n)$, given by

$$e\left(\rho^{\otimes n}, \Lambda_n\right) \coloneqq \frac{1}{n}\left(\log k_n - \log m_n\right). \tag{6.8}$$

A pair $(e, q) \in \mathbb{R}^2$ with $q \geq 0$ is said to be an achievable rate pair for state redistribution of a state $\rho_{ABC}$, if there exists a sequence of protocols $\{(\rho^{\otimes n}, \Lambda_n)\}_{n \in \mathbb{N}}$, satisfying

$$\limsup_{n \to \infty} q\left(\rho^{\otimes n}, \Lambda_n\right) = q \qquad \limsup_{n \to \infty} e\left(\rho^{\otimes n}, \Lambda_n\right) = e \qquad \liminf_{n \to \infty} F_n = 1.$$

Luo and Devetak [LD09] and Yard and Devetak [YD09] (see also [DY08]) proved that a pair $(e, q)$ is an achievable rate pair for state redistribution of a state $\rho_{ABC}$ with purification $|\psi\rangle_{ABCR}$ if and only if it lies in the region defined by

$$q \geq \frac{1}{2}I(A; R|B)_\psi \qquad\qquad q + e \geq S(A|B)_\rho, \tag{6.9}$$

and depicted as the green shaded area in Figure 6.4.

A strong converse theorem for the quantum communication cost was proved by Berta et al. [BCT15] using the smooth entropy framework. This theorem, however, did not prove the strong converse property for the entire boundary of the achievable rate region given by (6.9), that is, it did not prove the strong converse property for $q + e \geq S(A|B)_\rho$. Theorem 6.1.2 fills this gap, and furthermore provides an alternative proof of the strong converse theorem of [BCT15] for the quantum communication cost $q$. Eventually, based on the proof method of Theorem 6.1.2 and discussions with the author of this thesis, Berta et al. [BCT16] extended their strong converse theorem to also include the boundary of the achievable rate region corresponding to $q + e$. Hence, both Theorem 6.1.2 and [BCT16] provide a complete strong converse theorem for state redistribution, as depicted in Figure 6.4.

## 6.1.2. Strong converse theorem

We start with the following 'one-shot' result.

**Lemma 6.1.1.** *Let* $\rho \equiv \rho_{ABC}$ *be a tripartite state with purification* $|\psi\rangle_{ABCR}$*, and let* $(\rho, \Lambda)$ *be a state redistribution protocol where* $\Lambda \equiv \mathcal{D} \circ \mathcal{E}$ *with* $\mathcal{E} \colon ACT_A \to C'T'_A Q$ *and* $\mathcal{D} \colon QBT_B \to T'_B A'B'$

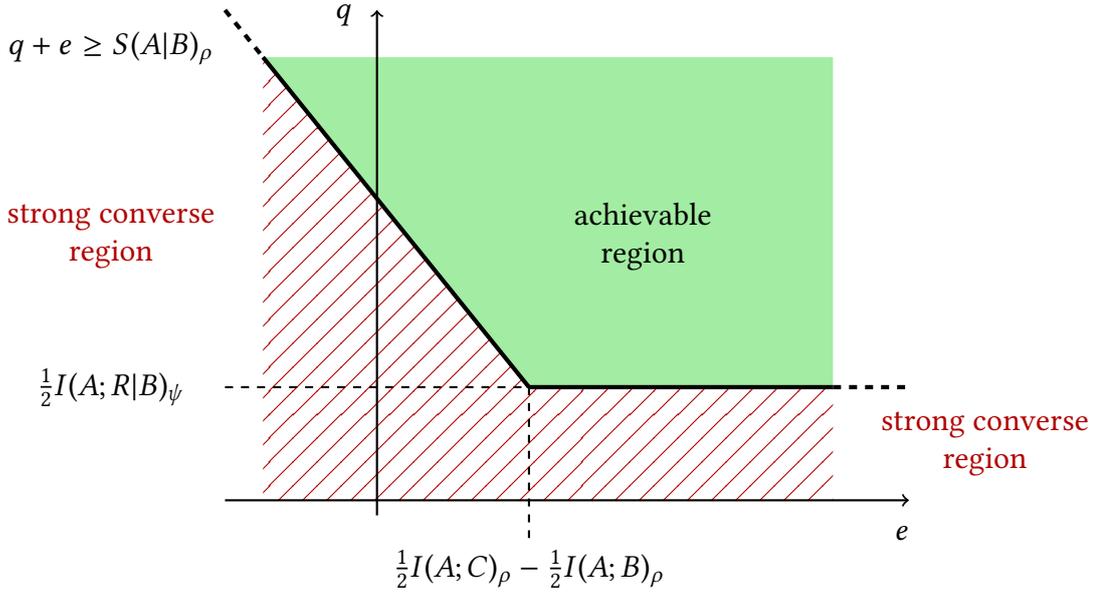

Figure 6.4: Plot of the plane of rate pairs $(e, q)$ for state redistribution, where $e$ is the entanglement cost (6.8) and $q$ is the quantum communication cost (6.7). The shaded area (green) is the region of achievable rate pairs defined by $\{(e, q) : q + e \geq S(A|B)_\rho \text{ and } q \geq \frac{1}{2}I(A;R|B)_\psi\}$. The hatched area (red) is the strong converse region, as proved in Theorem 6.1.2 as well as Berta et al. [BCT16].

*as defined in Section 6.1.1. Furthermore, set*

$$F := F\left(\Phi^m_{T'_A T'_B} \otimes \psi_{A'B'C'R}, (\mathcal{D} \circ \mathcal{E} \otimes \mathrm{id}_R)\left(\Phi^k_{T_A T_B} \otimes \psi_{ABCR}\right)\right).$$

*We have the following bounds on $F$ for $\alpha \in (1/2, 1)$ and $\beta \equiv \beta(\alpha) = \alpha/(2\alpha - 1)$:*

$$\log F \leq \frac{1-\alpha}{2\alpha}\left(\log |Q| + \log |T_A| - \log |T'_A| - S_\beta(AB)_\rho + S_\alpha(B)_\rho\right) \tag{6.10}$$

$$\log F \leq \frac{1-\alpha}{2\alpha}\left(2\log |Q| - \widetilde{S}_\beta(R|B)_\psi + \widetilde{S}_\alpha(R|AB)_\psi\right). \tag{6.11}$$

*We also have the following alternative bound to (6.11):*

$$\log F \leq \frac{1-\alpha}{2\alpha}\left(2\log |Q| - \tilde{I}_\alpha(R;AB)_\psi + \tilde{I}_\beta(R;B)_\psi\right). \tag{6.12}$$

*Proof.* We first prove (6.10). Denote by $U_\mathcal{E} : \mathcal{H}_{ACT_A} \to \mathcal{H}_{C'T'_A QE_1}$ and $U_\mathcal{D} : \mathcal{H}_{QBT_B} \to \mathcal{H}_{T'_B A'B'E_2}$ the Stinespring isometries of the maps $\mathcal{E}$ and $\mathcal{D}$ respectively, and define the pure states

$$
\begin{aligned}
|\omega\rangle_{C'T_A'QT_BBRE_1} &:= U_{\mathcal{E}}\left(|\Phi^k\rangle_{T_AT_B} \otimes |\psi\rangle_{ABCR}\right) \\
|\sigma\rangle_{T_A'T_B'A'B'C'RE_2} &:= U_{\mathcal{D}}|\omega\rangle_{C'T_A'QT_BBRE_1}
\end{aligned}
\tag{6.13}
$$

that purify the mixed states $\omega$ and $\sigma$ defined in (6.4) and (6.5), respectively. We then have

$$
S_\alpha(QT_BB)_\omega \le \log|Q| + \log|T_A| + S_\alpha(B)_\rho,
\tag{6.14}
$$

where we used subadditivity of the Rényi entropies (Lemma 3.3.3) twice, as well as the fact that $T_A$ is the same size as $T_B$. For the fidelity $F := F(\Phi^m_{T_A'T_B'} \otimes \psi_{A'B'C'R}, \sigma_{T_A'T_B'A'B'C'R})$, we know by Uhlmann's Theorem 2.2.4 that there exists a pure state $\phi_{E_1E_2}$ such that

$$
\begin{aligned}
F &= F\left(\Phi^m_{T_A'T_B'} \otimes \psi_{A'B'C'R} \otimes \phi_{E_1E_2}, \sigma_{T_A'T_B'A'B'C'RE_1E_2}\right) \\
&\le F\left(\pi^m_{T_B'} \otimes \rho_{A'B'} \otimes \phi_{E_2}, \sigma_{T_B'A'B'E_2}\right),
\end{aligned}
\tag{6.15}
$$

where $|\sigma\rangle_{T_A'T_B'A'B'C'RE_1E_2}$ is the pure state defined in (6.13). The inequality follows from the monotonicity of the fidelity under partial trace, (3.10).

Hence, setting $\beta = \alpha/(2\alpha - 1)$ we obtain the following bound:

$$
\begin{aligned}
S_\alpha(QT_BB)_\omega = S_\alpha(T_B'A'B'E_2)_\sigma \\
&\ge S_\beta(T_B'A'B'E_2)_{\pi^m\otimes\rho\otimes\phi} + \frac{2\alpha}{1-\alpha}\log F \\
&\ge \log|T_A'| + S_\beta(A'B')_\rho + \frac{2\alpha}{1-\alpha}\log F,
\end{aligned}
\tag{6.16}
$$

where we used the invariance of the Rényi entropies under the isometry $U_{\mathcal{D}}$ (Proposition 3.1.2(v)) in the first equality, Theorem 3.3.6(i) and (6.15) in the first inequality, and positivity and additivity of the Rényi entropies (Proposition 3.3.2(i) and (ii)) in the second inequality. Combining (6.14) and (6.16) then yields

$$
\log|Q| + \log|T_A| + S_\alpha(B)_\rho \ge \log|T_A'| + S_\beta(AB)_\rho + \frac{2\alpha}{1-\alpha}\log F,
$$

which is equivalent to (6.10).

To prove (6.11), we first observe that

$$\frac{2\alpha}{1-\alpha} \log F \leq \frac{2\alpha}{1-\alpha} \log F\left(\pi_{T_B'}^m \otimes \psi_{A'B'R}, \sigma_{T_B'A'B'R}\right)$$
$$\leq \widetilde{S}_\alpha(R|T_B'A'B')_{\pi^m \otimes \psi} - \widetilde{S}_\beta(R|T_B'A'B')_\sigma$$
$$= \widetilde{S}_\alpha(R|A'B')_\psi - \widetilde{S}_\beta(R|T_B'A'B')_\sigma, \tag{6.17}$$

where the first inequality follows from monotonicity of the fidelity under partial trace, (3.10), the second inequality follows from Theorem 3.3.6(ii), and the equality follows from Proposition 3.3.5(ii). We bound the second term on the right-hand side of (6.17) as follows:

$$-\widetilde{S}_\beta(R|T_B'A'B')_\sigma \leq -\widetilde{S}_\beta(R|QBT_B)_\omega$$
$$\leq -\widetilde{S}_\beta(R|BT_B)_\omega + 2\log|Q|$$
$$= -\widetilde{S}_\beta(R|BT_B)_{\pi^k \otimes \psi} + 2\log|Q|$$
$$= -\widetilde{S}_\beta(R|B)_\psi + 2\log|Q|, \tag{6.18}$$

where we used data processing (Proposition 3.3.2(v)) in the first inequality, and (3.14) and (3.16) of Proposition 3.3.5 in the second inequality and the second equality, respectively. Substituting (6.18) in (6.17) now yields (6.11).

The bound (6.12) follows from similar arguments as those used for the proof of (6.11), relying on (3.15) and (3.17) of Proposition 3.3.5 and Theorem 3.3.6(iii) instead. □

Lemma 6.1.1 immediately implies the following strong converse theorem:

**Theorem 6.1.2** (Strong converse for state redistribution).
*Let $\rho \equiv \rho_{ABC}$ be a tripartite state with purification $|\psi\rangle_{ABCR}$, and let $\{(\rho^{\otimes n}, \Lambda_n)\}_{n \in \mathbb{N}}$ be a sequence of state redistribution protocols as described in Section 6.1.1, with figure of merit $F_n$ as defined in (6.6). For all $n \in \mathbb{N}$ we have the following bounds on $F_n$ for $\alpha \in (1/2, 1)$ and $\beta = \alpha/(2\alpha - 1)$:*

$$F_n \leq \exp\left\{-n\kappa(\alpha)\left[S_\beta(AB)_\rho - S_\alpha(B)_\rho - (q+e)\right]\right\} \tag{6.19}$$

$$F_n \leq \exp\left\{-n\kappa(\alpha)\left[\widetilde{S}_\beta(R|B)_\psi - \widetilde{S}_\alpha(R|AB)_\psi - 2q\right]\right\} \tag{6.20}$$

$$F_n \leq \exp\left\{-n\kappa(\alpha)\left[\tilde{I}_\alpha(R;AB)_\psi - \tilde{I}_\beta(R;B)_\psi - 2q\right]\right\}, \tag{6.21}$$

*where $\kappa(\alpha) = (1-\alpha)/(2\alpha)$, and $q \equiv q(\rho^{\otimes n}, \Lambda_n)$ and $e \equiv e(\rho^{\otimes n}, \Lambda_n)$ are the quantum communication cost and entanglement cost defined in (6.7) and (6.8), respectively.*

In the following, we show how a strong converse as stated in (6.1) can be obtained from the bounds in Theorem 6.1.2. Consider the bound (6.19), which involves the rate $q + e$, the sum of quantum communication cost and entanglement cost. By (6.9), the optimal rate for $q + e$ is the conditional entropy $S(A|B)_\rho$. Assume therefore that $q + e < S(A|B)_\rho$, and choose $\delta > 0$ such that $q + e + \delta < S(A|B)_\rho$. Since $\beta = \alpha/(2\alpha - 1) \to 1$ as $\alpha \to 1$, we have from Proposition 3.3.2(vi) that

$$\lim_{\alpha \to 1} \left[ S_\beta(AB)_\rho - S_\alpha(B)_\rho \right] = S(AB)_\rho - S(B)_\rho = S(A|B)_\rho.$$

Hence, there is an $\alpha^* < 1$ such that $S_{\beta^*}(AB)_\rho - S_{\alpha^*}(B)_\rho > S(A|B)_\rho - \delta$, where $\beta^* = \alpha^*/(2\alpha^* - 1)$. It follows that $q + e < S(A|B)_\rho - \delta < S_{\beta^*}(AB)_\rho - S_{\alpha^*}(B)_\rho$, and hence,

$$K := \kappa(\alpha^*)[S_{\beta^*}(AB)_\rho - S_{\alpha^*}(B)_\rho - (q + e)] > 0.$$

Substituting this expression in (6.19) yields (6.1), and analogous arguments show how to obtain (6.1) also from (6.20) or (6.21).

Having the above argument in mind, we refer to theorems of the form of Theorem 6.1.2 as strong converse theorems in the sequel.

### 6.1.3. State redistribution with feedback

We now consider state redistribution with feedback [BCT16], where the state redistribution protocol consists of $M$ rounds of forward and backward quantum communication between Alice and Bob. The initial state of the protocol is again the pure state $\Phi^k_{T_A T_B} \otimes \psi_{ABCR}$, where systems $A$ and $C$ are with Alice, $B$ is with Bob, $R$ is an inaccessible reference system, and $\Phi^k_{T_A T_B}$ is an MES of Schmidt rank $k$ shared between Alice ($T_A$) and Bob ($T_B$). As before, the goal is for Alice to transfer the $A$ system to Bob, while preserving its correlations with the other systems.

The main difference to single-round state redistribution as described in Section 6.1.1 is that now backward quantum communication from Bob to Alice is possible. Furthermore, we allow for $M$ rounds of communication in the following way (cf. Figure 6.5): Alice first applies an encoding operation $\mathcal{E}_1 : ACT_A \to Q_1 A_1$ to the initial state and sends $Q_1$ to Bob, who applies a decoding operation $\mathcal{D}_1 : Q_1 BT_B \to Q_1' B_1$. The system $Q_1'$ is the quantum communication register that he sends back to Alice. She then applies the encoding $\mathcal{E}_2 : Q_1' A_1 \to Q_2 A_2$ and sends $Q_2$ to Bob, who applies the decoding $\mathcal{D}_2 : Q_2 B_1 \to Q_2' B_2$ and sends $Q_2'$ back, and so forth. In the $i$-th round, we denote by $\omega^i$ and $\sigma^i$ the states shared between Alice, Bob, and the reference, after applying the encoding $\mathcal{E}_i$ and decoding $\mathcal{D}_i$, respectively. In the final round, Alice applies the encoding $\mathcal{E}_M \colon Q'_{M-1}A_{M-1} \to Q_M C'T'_A$ and sends $Q_M$ to Bob, who applies the decoding $\mathcal{D}_M \colon Q_M B_{M-1} \to A'B'T'_B$. The protocol succeeds if the final state is close in fidelity to the pure target state $\Phi^m_{T'_A T'_B} \otimes \psi_{A'B'C'R}$, where $\psi_{A'B'C'R} = \psi_{ABCR}$ and $\Phi^m_{T_A T'_B}$ is an MES of Schmidt rank $m$ shared between Alice and Bob.

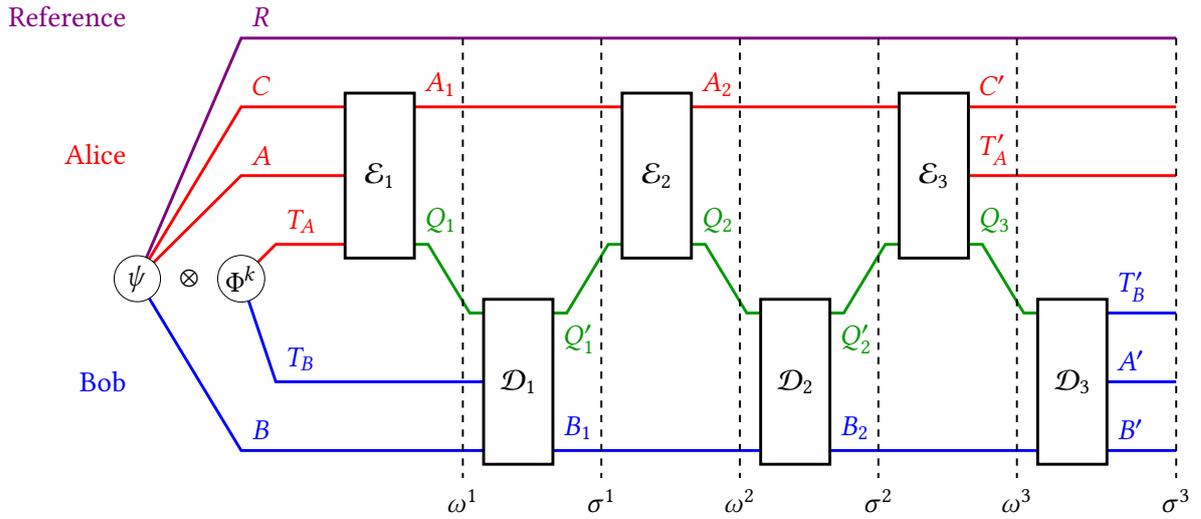

Figure 6.5: Schematic description of a state redistribution protocol with feedback ($M = 3$ rounds) between Alice (red), Bob (blue), and a reference system (violet), with quantum communication (green) between Alice and Bob. For a detailed description of the protocol, see Section 6.1.3.

For a protocol acting on $n$ i.i.d. copies of the initial state $\rho_{ABC}$ with purification $\psi_{ABCR}$ and a shared maximally entangled state $\Phi^{k_n}_{T^n_A T^n_B}$ of Schmidt rank $k_n$, the target state is again $n$ i.i.d. copies of the state $\psi_{A'B'C'R} = \psi_{ABCR}$, together with a maximally entangled state $\Phi^{m_n}_{T'^n_A T'^n_B}$ of Schmidt rank $m_n$. Denoting the total state redistribution protocol with $M$ rounds of communication by $\Lambda^M_n$, the figure of merit is the fidelity

$$F_n := F\left(\Phi^{m_n}_{T'^n_A T'^n_B} \otimes \psi^{\otimes n}_{A'B'C'R}, \left(\Lambda^M_n \otimes \mathrm{id}_{R^n}\right)\left(\Phi^{k_n}_{T^n_A T^n_B} \otimes \psi^{\otimes n}_{ABCR}\right)\right). \tag{6.22}$$

With these definitions, we define the entanglement cost $e\left(\rho^{\otimes n}, \Lambda^M_n\right)$, the forward quantum communication cost $q_\to\left(\rho^{\otimes n}, \Lambda^M_n\right)$, and the total quantum communication cost $q_\leftrightarrow\left(\rho^{\otimes n}, \Lambda^M_n\right)$

(equal to forward plus backward communication) as

$$e\left(\rho^{\otimes n}, \Lambda_n^M\right) := \frac{1}{n}(\log k_n - \log m_n) = \frac{1}{n}\left(\log|T_A^n| - \log|T_A'^n|\right) \qquad (6.23)$$

$$q_\rightarrow\left(\rho^{\otimes n}, \Lambda_n^M\right) := \frac{1}{n}\sum_{i=1}^{M}\log|Q_i^n| \qquad (6.24)$$

$$q_\leftrightarrow\left(\rho^{\otimes n}, \Lambda_n^M\right) := q_\rightarrow + \frac{1}{n}\sum_{i=1}^{M-1}\log|Q_i'^n|. \qquad (6.25)$$

Similar to Section 6.1.1, we call a triple $(e, q_\rightarrow, q_\leftrightarrow) \in \mathbb{R}^3$ achievable for state redistribution of a state $\rho_{ABC}$ with $M$ rounds of communication, if there exists a sequence of protocols $\{(\rho^{\otimes n}, \Lambda_n^M)\}_{n\in\mathbb{N}}$ such that $\liminf_{n\to\infty} F_n = 1$ and

$$\limsup_{n\to\infty} e\left(\rho^{\otimes n}, \Lambda_n^M\right) = e$$

$$\limsup_{n\to\infty} q_\rightarrow\left(\rho^{\otimes n}, \Lambda_n^M\right) = q_\rightarrow$$

$$\limsup_{n\to\infty} q_\leftrightarrow\left(\rho^{\otimes n}, \Lambda_n^M\right) = q_\leftrightarrow.$$

Using the smooth entropy framework, Berta et al. [BCT16] proved that a triple $(e, q_\rightarrow, q_\leftrightarrow)$ is achievable if and only if

$$q_\rightarrow \geq \frac{1}{2}I(A;R|B)_\psi \qquad\qquad q_\leftrightarrow + e \geq S(A|B)_\rho. \qquad (6.26)$$

We first note that both conditions in (6.26) are independent of the number $M$ of communication rounds. Secondly, let us compare them to the conditions for single-round state redistribution in (6.9). The achievable region for forward quantum communication coincides with that of (6.9), as shown in [BCT16]. However, for state redistribution with feedback the condition for the entanglement cost involves the *total* quantum communication between the two parties. This arises from the fact that quantum communication from Bob to Alice can introduce additional entanglement between them. In comparison to a single-round state redistribution protocol without feedback, the optimal rate of the overall entanglement cost $e$ is *lowered* by the amount of backward quantum communication, since we have $q_\leftrightarrow \geq q_\rightarrow$.

Using the Rényi entropy method, we derive a strong converse theorem for state redistribution with feedback, Theorem 6.1.4 below, which originally appeared in [BCT16]. Similar to Section 6.1.2, we first prove the following 'one-shot' result:

**Lemma 6.1.3.** *Let $\rho_{ABC}$ be a tripartite state with purification $|\psi\rangle_{ABCR}$, and let $(\rho, \Lambda^M)$ be a state redistribution protocol with M rounds of communication as described above. Furthermore, set*

$$F := F\left(\Phi^m_{T'_A T'_B} \otimes \psi_{A'B'C'R}, \left(\Lambda^M \otimes \mathrm{id}_R\right)\left(\Phi^k_{T_A T_B} \otimes \psi_{ABCR}\right)\right). \tag{6.27}$$

*For $\alpha \in (1/2, 1)$ and $\beta = \alpha/(2\alpha - 1)$, we have the following bounds on F:*

$$\log F \le \frac{1-\alpha}{2\alpha}\left(\log|T_A| - \log|T'_A| + \sum_{i=1}^{M} \log|Q_i| + \sum_{i=1}^{M-1} \log|Q'_i| - S_\beta(AB)_\rho + S_\alpha(B)_\rho\right) \tag{6.28}$$

$$\log F \le \frac{1-\alpha}{2\alpha}\left(2\sum_{i=1}^{M} \log|Q_i| - \widetilde{S}_\beta(R|B)_\psi + \widetilde{S}_\alpha(R|AB)_\psi\right) \tag{6.29}$$

$$\log F \le \frac{1-\alpha}{2\alpha}\left(2\sum_{i=1}^{M} \log|Q_i| - \tilde{I}_\alpha(R;AB)_\psi + \tilde{I}_\beta(R;B)_\psi\right). \tag{6.30}$$

*Proof.* We first prove (6.29). For $\alpha \in (1/2, 1)$ and $\beta = \alpha/(2\alpha - 1)$, we can bound the fidelity $F$ (defined in (6.27)) from above by

$$\begin{aligned}
\frac{2\alpha}{1-\alpha}\log F &\le \frac{2\alpha}{1-\alpha}\log F\left(\pi^m_{T'_B} \otimes \psi_{A'B'R}, \sigma^M_{T'_B A'B'R}\right)\\
&\le \widetilde{S}_\alpha(R|T'_B A'B')_{\pi^m \otimes \psi} - \widetilde{S}_\beta(R|T'_B A'B')_{\sigma^M}\\
&= \widetilde{S}_\alpha(R|AB)_\psi - \widetilde{S}_\beta(R|T'_B A'B')_{\sigma^M}, \tag{6.31}
\end{aligned}$$

where we used the monotonicity of the fidelity under partial trace, (3.10), in the first inequality, Theorem 3.3.6(ii) in the second inequality, and (3.16) of Proposition 3.3.5 together with the fact that $\rho_{A'B'R} = \rho_{ABR}$ in the equality.

For the second term on the right-hand side of (6.31), we then consider the following chain of inequalities, which should be compared to the schematic description of state redistribution with feedback in Figure 6.5:

$$\begin{aligned}
-\widetilde{S}_\beta(R|T'_B A'B')_{\sigma^M} &\le -\widetilde{S}_\beta(R|Q_M B_{M-1})_{\omega^M}\\
&\le -\widetilde{S}_\beta(R|B_{M-1})_{\omega^M} + 2\log|Q_M|\\
&= -\widetilde{S}_\beta(R|B_{M-1})_{\sigma^{M-1}} + 2\log|Q_M|\\
&\le -\widetilde{S}_\beta(R|Q'_{M-1} B_{M-1})_{\sigma^{M-1}} + 2\log|Q_M|\\
&\le -\widetilde{S}_\beta(R|Q_{M-1} B_{M-2})_{\omega^{M-1}} + 2\log|Q_M|
\end{aligned}$$

$$\vdots$$

$$\leq -\widetilde{S}_\beta(R|T_B B)_{\omega^1} + 2 \sum_{i=1}^M \log |Q_i|$$

$$= -\widetilde{S}_\beta(R|T_B B)_{\pi^k \otimes \psi} + 2 \sum_{i=1}^M \log |Q_i|$$

$$= -\widetilde{S}_\beta(R|B)_\psi + 2 \sum_{i=1}^M \log |Q_i|. \tag{6.32}$$

In the first inequality we used data processing with respect to the decoding map $\mathcal{D}_M$ (Proposition 3.3.2(v)). The second inequality follows from the dimension bound (3.14) for the Rényi conditional entropy in Proposition 3.3.5. In the first equality we used the fact that the system $B_{M-1}$ is not affected by the encoding $\mathcal{E}_M$. The third inequality is data processing for the Rényi conditional entropy with respect to the partial trace over $Q'_{M-1}$. We then iteratively apply these steps until we reach the last inequality. The subsequent equality follows from the fact that the encoding $\mathcal{E}_1$ does not act on the systems $B$ and $T_B$. In the last step we used (3.16) of Proposition 3.3.5. Combining (6.31) and (6.32) now yields (6.29). The proof of the bound in (6.30) follows in a similar manner, and we therefore omit it.

To prove (6.28), we consider Stinespring isometries $U_{\mathcal{E}_i}$ and $U_{\mathcal{D}_i}$ of the encoding and decoding maps $\mathcal{E}_i$ and $\mathcal{D}_i$ with environments $E_i$ and $D_i$, respectively. Moreover, in the following calculations we denote by $\omega^i$ and $\sigma^i$ the *pure states* obtained from applying the isometries $U_{\mathcal{E}_i}$ and $U_{\mathcal{D}_i}$ to the initial state $\Phi^k \otimes \psi$, respectively. The final state of the protocol is then the pure state

$$|\sigma^M\rangle_{T'_A T'_B A'B'C'RE_1 \dots E_M D_1 \dots D_M} = (U_{\mathcal{D}_M} U_{\mathcal{E}_M} \dots U_{\mathcal{D}_1} U_{\mathcal{E}_1} \otimes \mathbb{1}_R)\left(|\Phi^k\rangle_{T_A T_B} \otimes |\psi\rangle_{ABCR}\right).$$

By Uhlmann's Theorem 2.2.4 there exists a pure state $\chi_{E_1 \dots E_M D_1 \dots D_M}$ such that the following holds for $\alpha \in (1/2, 1)$ and $\beta = \alpha/(2\alpha - 1)$:

$$\begin{aligned}
\frac{2\alpha}{1-\alpha} \log F &= \frac{2\alpha}{1-\alpha} \log F\left(\sigma^M_{T'_A T'_B A'B'C'RE_1 \dots E_M D_1 \dots D_M}, \Phi^m_{T'_A T'_B} \otimes \psi_{A'B'C'R} \otimes \chi_{E_1 \dots E_M D_1 \dots D_M}\right) \\
&\leq \frac{2\alpha}{1-\alpha} \log F\left(\sigma^M_{T'_B A'B'D_1 \dots D_M}, \pi^m_{T'_B} \otimes \rho_{A'B'} \otimes \chi_{D_1 \dots D_M}\right) \\
&\leq S_\alpha(T'_B A'B'D_1 \dots D_M)_{\sigma^M} - S_\beta(T'_B A'B'D_1 \dots D_M)_{\pi^m \otimes \rho \otimes \chi} \\
&\leq S_\alpha(T'_B A'B'D_1 \dots D_M)_{\sigma^M} - S_\beta(AB)_\rho - \log |T'_B|, \tag{6.33}
\end{aligned}$$

where the first inequality follows from the monotonicity of the fidelity under partial trace, (3.10), the second inequality follows from Theorem 3.3.6(i), and the third inequality follows from Proposition 3.3.2(i) and (ii).

For the first term of the right-hand side of (6.33), consider the following steps:

$$
\begin{aligned}
S_\alpha(T_B'A'B'D_1\ldots D_M)_{\sigma^M} &= S_\alpha(Q_MB_{M-1}D_1\ldots D_{M-1})_{\omega^M} \\
&\leq S_\alpha(B_{M-1}D_1\ldots D_{M-1})_{\omega^M} + \log|Q_M| \\
&= S_\alpha(RQ_MC'T_A'E_1\ldots E_M)_{\omega^M} + \log|Q_M| \\
&= S_\alpha(RQ_{M-1}'A_{M-1}E_1\ldots E_{M-1})_{\sigma^{M-1}} + \log|Q_M| \\
&\leq S_\alpha(RA_{M-1}E_1\ldots E_{M-1})_{\sigma^{M-1}} + \log|Q_M| + \log|Q_{M-1}'| \\
&= S_\alpha(Q_{M-1}'B_{M-1}D_1\ldots D_{M-1})_{\sigma^{M-1}} + \log|Q_M| + \log|Q_{M-1}'| \\
&= S_\alpha(Q_{M-1}B_{M-2}D_1\ldots D_{M-2})_{\omega^{M-1}} + \log|Q_M| + \log|Q_{M-1}'| \\
&\;\;\vdots \\
&\leq S_\alpha(T_BB)_{\omega^1} + \sum_{i=1}^{M}\log|Q_i| + \sum_{i=1}^{M-1}\log|Q_i'| \\
&= S_\alpha(T_BB)_{\pi^k\otimes\rho} + \sum_{i=1}^{M}\log|Q_i| + \sum_{i=1}^{M-1}\log|Q_i'| \\
&= S_\alpha(B)_\rho + \log|T_B| + \sum_{i=1}^{M}\log|Q_i| + \sum_{i=1}^{M-1}\log|Q_i'|.
\end{aligned}
\tag{6.34}
$$

In the first equality we used invariance of the Rényi entropy under the isometry $U_{\mathcal{D}_M}$ (Proposition 3.1.2(v)). In the first inequality we used subadditivity (Lemma 3.3.3), and in the second equality we used the duality of the Rényi entropy (Proposition 3.3.2(iii)) for the pure state $|\omega^M\rangle$. The third equality follows from the invariance of the Rényi entropy under $U_{\mathcal{E}_M}$. We then follow the same steps iteratively, passing from $\omega^M$ to $\sigma^{M-1}$ and $\omega^{M-1}$ and so on, until we reach $\omega_{T_BB}^1 = \pi_{T_B}^k \otimes \rho_B$. Substituting (6.34) in (6.33) then yields (6.28), and we are done. $\qquad\square$

**Theorem 6.1.4** (Strong converse for state redistribution with feedback).
*Let $\rho_{ABC}$ be a tripartite state with purification $|\psi\rangle_{ABCR}$, and let $\Lambda_n^M$ be a state redistribution protocol with $M$ rounds of communication as described above. For $\alpha \in (1/2, 1)$ and $\beta = \alpha/(2\alpha - 1)$ we have the following bounds on $F_n$ as defined in (6.22):*

$$
F_n \leq \exp\left\{-n\kappa(\alpha)\left[S_\beta(AB)_\rho - S_\alpha(B)_\rho - (q_\leftrightarrow + e)\right]\right\}
$$

$$F_n \leq \exp\left\{-n\kappa(\alpha)\left[\widetilde{S}_\beta(R|B)_\psi - \widetilde{S}_\alpha(R|AB)_\psi - 2q_\rightarrow\right]\right\}$$

$$F_n \leq \exp\left\{-n\kappa(\alpha)\left[\tilde{I}_\alpha(R;AB)_\psi - \tilde{I}_\beta(R;B)_\psi - 2q_\rightarrow\right]\right\},$$

*where* $\kappa(\alpha) = (1-\alpha)/(2\alpha)$, *and* $e = e\left(\rho^{\otimes n}, \Lambda_n^M\right)$, $q_\rightarrow = q_\rightarrow\left(\rho^{\otimes n}, \Lambda_n^M\right)$, *and* $q_\leftrightarrow = q_\leftrightarrow\left(\rho^{\otimes n}, \Lambda_n^M\right)$ *are the entanglement cost* (6.23), *forward quantum communication cost* (6.24), *and total quantum communication cost* (6.25), *respectively.*

## 6.1.4. Reduction to other protocols

In this section, we show how to obtain a range of quantum information-processing protocols by considering special cases of the state redistribution protocol described in Section 6.1.1. For these reductions, which were first discussed in [LD09] and [YD09], we obtain (previously known) strong converse theorems as corollaries of our strong converse theorem for state redistribution (Theorem 6.1.2 in Section 6.1.2).

**Blind quantum source coding**

Assume that the systems $C$ and $B$ in the initial state $\psi_{ABCR}$ of the state redistribution protocol from Section 6.1.1 are trivial systems. In the absence of shared entanglement in both the initial and target state, state redistribution can then be understood as blind quantum source coding, or Schumacher compression [Sch95], where the quantum communication system $Q$ after the encoding $\mathcal{E}: A \to Q$ holds the compressed state $\mathcal{E}(\rho_A)$. Here, we assume that the quantum state $\rho_A$ with purification $\psi_{AR}$ is the source state $\rho_A = \sum_i p_i \phi_i$ of a quantum source with underlying ensemble $\{p_i; \phi_i\}_i$ (cf. Section 5.1.1 in Chapter 5). Since the encoding map $\mathcal{E}$ in the state redistribution protocol from Section 6.1.1 is a linear CPTP map, it corresponds to a blind encoding (cf. Section 5.1.2). The full protocol is depicted in Figure 6.6.

In the reduction of state redistribution to blind quantum source coding outlined above, the corresponding figure of merit for $n$ copies of $\rho_A$ (i.e., the source is assumed to be memoryless) is the fidelity

$$F_n = F\left(\psi_{A'R}^{\otimes n}, (\mathcal{D}_n \circ \mathcal{E}_n \otimes \mathrm{id}_{R^n})\left(\psi_{AR}^{\otimes n}\right)\right), \tag{6.35}$$

where $\mathcal{E}_n: A^n \to Q^n$ and $\mathcal{D}_n: Q^n \to A'^n$ denote the blind encoding and decoding map, respectively. Note that the fidelity $F_n$ in (6.35) is equal to the entanglement fidelity $F_e(\rho^{\otimes n}, \mathcal{D}_n \circ \mathcal{E}_n)$ that we defined in (3.68) in Section 3.4.4. This figure of merit for quantum source coding was first considered by Winter [Win99b], and also used in [DL15]. The rate of the protocol is given in terms of the compressed system dimension $|Q^n|$, which we called $M_n$ in Section 5.1.2:

$$q := \frac{1}{n} \log |Q^n| = \frac{1}{n} \log M_n. \tag{6.36}$$

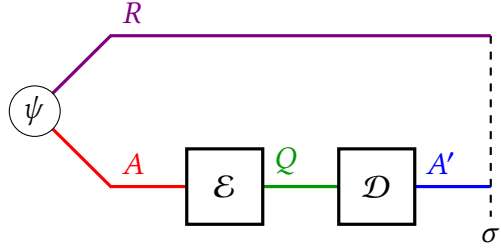

Figure 6.6: Schematic description of blind quantum source coding with blind encoding map $\mathcal{E}$ and decoding map $\mathcal{D}$. The source state is given by the state $\rho_A$ with purification $\psi_{AR}$, and the compressed state on the quantum system $Q$ (green) is $\mathcal{E}(\rho_A)$.

The strong converse theorem for state redistribution, Theorem 6.1.2, now immediately yields a strong converse theorem for blind quantum source coding, first proved by Winter [Win99b] using the method of types. The following formulation of the strong converse theorem based on the Rényi entropy method was first proved by Hayashi [Hay02] for both the visible and the blind encoding setting (see also [Sha14]).

**Corollary 6.1.5** (Strong converse for blind quantum source coding).
*Let $\rho_A = \sum_i p_i \phi_i$ be the source state of a quantum source with source ensemble $\{p_i; |\phi_i\rangle\}_i$, and assume that $\rho_A$ is purified by $|\psi\rangle_{AR}$. Furthermore, for all $n \in \mathbb{N}$ let $F_n$ be defined as in (6.35). We then have the following bound on $F_n$ for all $n \in \mathbb{N}$ and $\beta \in (1, \infty)$:*

$$F_n \leq \exp\left\{ -n\iota(\beta) \left[ S_\beta(A)_\rho - q \right] \right\},$$

*where $\iota(\beta) = (\beta - 1)/\beta$, and $q$ is the compression rate defined in (6.36).*

*Proof.* For $\alpha, \beta \geq 1/2$ such that $1/\alpha + 1\beta = 2$, and a pure state $\psi_{AR}$, we have

$$\widetilde{S}_\alpha(R|A)_\psi = -S_\beta(R)_\psi = -S_\beta(A)_\rho,$$

where the first and second equality use duality for the Rényi conditional entropy (Proposi-tion 3.3.2(iv)) and Rényi entropy (Proposition 3.3.2(iii)), respectively. The desired bound now follows from (6.20) in Theorem 6.1.2. □

**Coherent state merging**

Coherent state merging [ADH+09; Dev06] is the task in which Alice wants to transfer the *A*-part of a bipartite state $\rho_{AB}$ to Bob (who holds the *B* system), while at the same time generating entanglement between them. We assume that $\rho_{AB}$ is purified by an inaccessible reference system *R*. To achieve their goal, Alice and Bob are allowed to perform local operations on the systems in their possession as well as noiseless quantum communication. This protocol (also known as 'Fully Quantum Slepian Wolf' (FQSW) protocol [ADH+09]) is a special case of the state redistribution protocol from Section 6.1.1 where the system *C* is absent (or equivalently taken to be a trivial one-dimensional system), and Alice and Bob do not share any entanglement prior to commencing the protocol (hence, the systems $T_A$ and $T_B$ are trivial and entanglement is always *gained* in the course of the protocol). See Figure 6.7 for the initial and target state of coherent state merging (when read from left to right).

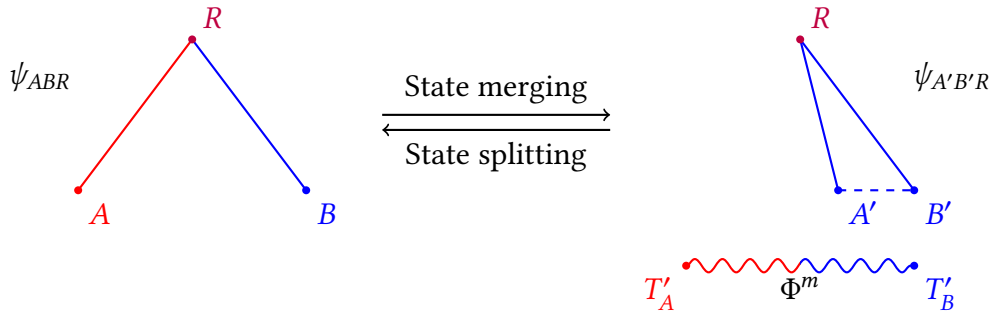

Figure 6.7: Schematic description of coherent state merging (from left to right) and quantum state splitting (from right to left). For the correspondence between the labels used in the description of quantum state splitting and the ones used here, see (6.39).

Let Alice and Bob share *n* identical copies of the state $\rho_{ABR}$ with purification $\psi_{ABR}$, i.e., the state $\rho_{AB}^{\otimes n}$ with purification $\psi_{ABR}^{\otimes n}$. A general coherent state merging protocol $(\rho^{\otimes n}, \Lambda_n)$ is given by a joint quantum operation $\Lambda_n \equiv \mathcal{D}_n \circ \mathcal{E}_n$ where $\mathcal{E}_n \colon A^n \to T_A'^n Q^n$ is Alice's encoding map, the system $Q^n$ is sent to Bob, and $\mathcal{D}_n \colon Q^n B^n \to T_B'^m A'^n B'^m$ is Bob's decoding map. Here, $\mathcal{H}_{A'^n} \cong \mathcal{H}_{A^n}$, $\mathcal{H}_{T_B'^m} \cong \mathcal{H}_{T_A'^m}$, and $\mathcal{H}_{B'^n} \cong \mathcal{H}_{B^n}$. Denoting the final state of the protocol by $\sigma_n = (\Lambda_n \otimes \mathrm{id}_{R^n})(\psi_{ABR}^{\otimes n})$, the figure of merit is chosen to be the fidelity

$$F_n := F\left(\sigma_n, \Phi_{T_A'^n T_B'^n}^{m_n} \otimes \psi_{A'B'R}^{\otimes n}\right), \tag{6.37}$$

where the second argument of the fidelity is the target state of the protocol $(\rho^{\otimes n}, \Lambda_n)$.

The *quantum communication cost* $q_{\mathrm{csm}}(\rho^{\otimes n}, \Lambda_n)$ and the *entanglement gain* $e_{\mathrm{csm}}(\rho^{\otimes n}, \Lambda_n)$ are defined in analogy to Section 6.1.1:

$$q_{\mathrm{csm}}(\rho^{\otimes n}, \Lambda_n) := \frac{1}{n} \log |Q^n| \qquad\qquad e_{\mathrm{csm}}(\rho^{\otimes n}, \Lambda_n) := \frac{1}{n} \log |T_A'^n|. \tag{6.38}$$

Note that $e_{\mathrm{csm}}(\rho^{\otimes n}, \Lambda_n)$ differs by a minus sign from the entanglement cost for state redistribution defined in (6.8) in Section 6.1.1, as $e_{\mathrm{csm}}(\rho^{\otimes n}, \Lambda_n)$ quantifies a gain instead of a cost. A pair $(e, q)$ with $e, q \geq 0$ is said to be an achievable rate pair for coherent state merging of a state $\rho_{AB}$, if there exists a sequence of protocols $\{(\rho^{\otimes n}, \Lambda_n)\}_{n \in \mathbb{N}}$ such that

$$\limsup_{n \to \infty} q_{\mathrm{csm}}(\rho^{\otimes n}, \Lambda_n) = q \qquad \liminf_{n \to \infty} e_{\mathrm{csm}}(\rho^{\otimes n}, \Lambda_n) = e \qquad \liminf_{n \to \infty} F_n = 1.$$

Coherent state merging was introduced by Abeyesinghe et al. [ADH+09] and further investigated in [BCR11; DH11]. In [ADH+09], it was proved that a rate pair $(e, q)$ is achievable if and only if $e$ and $q$ satisfy the conditions

$$q \geq \frac{1}{2} I(A; R)_\psi \qquad\qquad q - e \geq S(A|B)_\rho.$$

As mentioned above, every coherent state merging protocol can be seen as a special case of a state redistribution protocol where the systems $C$ and $T_A$ are trivial. In this case, $I(A; R|B)_\psi = I(A; R)_\psi$, and we obtain the following strong converse theorem from Theorem 6.1.2:

**Corollary 6.1.6** (Strong converse for coherent state merging).
*Let $\rho \equiv \rho_{AB}$ be a bipartite state with purification $|\psi\rangle_{ABR}$, and let $\{(\rho^{\otimes n}, \Lambda_n)\}_{n \in \mathbb{N}}$ be a sequence of coherent state merging protocols as described above, with figure of merit $F_n$ as defined in (6.37). For all $n \in \mathbb{N}$ we have the following bounds on the fidelity $F_n$ for $\alpha \in (1/2, 1)$ and $\beta = \alpha/(2\alpha - 1)$:*

$$F_n \leq \exp\left\{-n\kappa(\alpha)\left[S_\beta(AB)_\rho - S_\alpha(B)_\rho - q_{\mathrm{csm}} + e_{\mathrm{csm}}\right]\right\}$$
$$F_n \leq \exp\left\{-n\kappa(\alpha)\left[S_\beta(R)_\psi - \widetilde{S}_\alpha(R|A)_\psi - 2q_{\mathrm{csm}}\right]\right\},$$

*where $\kappa(\alpha) = (1 - \alpha)/(2\alpha)$, and $q_{\mathrm{csm}} \equiv q_{\mathrm{csm}}(\rho^{\otimes n}, \Lambda_n)$ and $e_{\mathrm{csm}} \equiv e_{\mathrm{csm}}(\rho^{\otimes n}, \Lambda_n)$ denote the*

*quantum communication cost and entanglement gain, respectively, as defined in* (6.38).

**Quantum state splitting**

Quantum state splitting is the task in which Alice holds a bipartite state $\rho_{AC}$ and wants to split it between her and Bob by transferring the $A$ system to him. As before, we assume that $\rho_{AC}$ is purified by an inaccessible reference system $R$, i.e., we consider a purification $\psi_{ACR}$ of $\rho_{AC}$. To accomplish the task, Alice and Bob are allowed to use prior shared entanglement and do local operations on the systems they possess or receive. This protocol (also known as 'Fully Quantum Reverse Shannon' (FQRS) protocol [ADH+09; Dev06]) is dual to the coherent state merging protocol under time reversal [Dev06]. Hence, the quantum state splitting protocol can also be obtained as a special case from the state redistribution protocol, if the systems $B$ and $T'_A$ are taken to be trivial. That is, Bob does not possess a share of the input state of the protocol, and the target state does not consist of an MES shared between Alice and Bob (i.e., the protocol always *consumes* entanglement). See Figure 6.7 for the initial and target state of the quantum state splitting protocol (when read from right to left). Note that Figure 6.7 uses different labels for the quantum systems involved, which stem from depicting quantum state splitting as the time reversal of coherent state merging. The labels used in the left columns of (6.39) correspond to the ones used in Figure 6.7, shown in the right columns of (6.39), as follows:

$$
\begin{aligned}
A &\leftrightarrow A' & C' &\leftrightarrow B \\
C &\leftrightarrow B' & T_A &\leftrightarrow T'_B \\
B' &\leftrightarrow A & T_B &\leftrightarrow T'_A.
\end{aligned}
\tag{6.39}
$$

Let Alice and Bob share $n$ identical copies of the state $\rho_{AC}$ with purification $\psi_{ACR}$, i.e., the state $\rho_{AC}^{\otimes n}$ with purification $\psi_{ACR}^{\otimes n}$. A general quantum state splitting protocol $(\rho^{\otimes n}, \Lambda_n)$ is given by a joint quantum operation $\Lambda_n = \mathcal{D}_n \circ \mathcal{E}_n$ where $\mathcal{E}_n \colon A^n C^n T_A^n \to C'^n Q^n$ is Alice's encoding map, the system $Q^n$ is sent to Bob, and $\mathcal{D}_n \colon Q^n T_B^n \to A'^n$ is Bob's decoding map. Here, $\mathcal{H}_{A'^n} \cong \mathcal{H}_{A^n}$ and $\mathcal{H}_{C'^n} \cong \mathcal{H}_{C^n}$. Denote the final state of the protocol by $\sigma_n = (\Lambda_n \otimes \mathrm{id}_{R^n})(\Omega^n)$ where $\Omega^n \equiv \Phi_{T_A^n T_B^n}^{k_n} \otimes \psi_{ACR}^{\otimes n}$ is the initial state shared between Alice and Bob. With $\psi \equiv \psi_{A'C'R}$, the figure of merit is chosen to be the fidelity

$$
F_n \coloneqq F\left(\sigma_n, \psi^{\otimes n}\right).
\tag{6.40}
$$

The *quantum communication cost* $q_{\mathrm{qss}}(\rho^{\otimes n}, \Lambda_n)$ and the *entanglement cost* $e_{\mathrm{qss}}(\rho^{\otimes n}, \Lambda_n)$ are defined in analogy to Section 6.1.1:

$$q_{\mathrm{qss}}(\rho^{\otimes n}, \Lambda_n) := \frac{1}{n} \log |Q^n| \qquad\qquad e_{\mathrm{qss}}(\rho^{\otimes n}, \Lambda_n) := \frac{1}{n} \log |T_A^n|. \qquad (6.41)$$

A pair $(e, q)$ with $e, q \geq 0$ is said to be an achievable rate pair for quantum state splitting of a state $\rho_{AC}$, if there exists a sequence of protocols $\{(\rho^{\otimes n}, \Lambda_n)\}_{n \in \mathbb{N}}$ such that

$$\liminf_{n \to \infty} F_n = 1 \qquad \limsup_{n \to \infty} e_{\mathrm{qss}}(\rho^{\otimes n}, \Lambda_n) = e \qquad \limsup_{n \to \infty} q_{\mathrm{qss}}(\rho^{\otimes n}, \Lambda_n) = q.$$

The optimal rates of entanglement cost and quantum communication cost for quantum state splitting were investigated in [ADH+09; BDH+14; BCR11]: A rate pair $(e, q)$ is achievable if and only if $e$ and $q$ satisfy

$$q \geq \frac{1}{2} I(A; R)_\psi \qquad\qquad q + e \geq S(A)_\rho.$$

One-shot bounds characterizing the quantum communication cost and entanglement cost for quantum state splitting were derived by Berta et al. [BCR11] as a building block in their proof of the Quantum Reverse Shannon theorem based on smooth entropies.

As mentioned in the beginning of this section, quantum state splitting is a special case of state redistribution with the choices $|B| = |T_A'| = 1$. In this case, $I(A; R|B)_\rho = I(A; R)$, and Theorem 6.1.2 yields the following strong converse theorem for quantum state splitting:

**Corollary 6.1.7** (Strong converse for quantum state splitting).
*Let $\rho \equiv \rho_{AB}$ be a bipartite state with purification $|\psi\rangle_{ABR}$, and let $\{(\rho^{\otimes n}, \Lambda_n)\}_{n \in \mathbb{N}}$ be a sequence of quantum state splitting protocols as described above, with figure of merit $F_n$ as defined in (6.40). For all $n \in \mathbb{N}$ we have the following bounds on the fidelity $F_n$ for $\alpha \in (1/2, 1)$ and $\beta = \alpha/(2\alpha - 1)$:*

$$F_n \leq \exp\left\{ -n\kappa(\alpha) \left[ S_\beta(A)_\rho - (q_{\mathrm{qss}} + e_{\mathrm{qss}}) \right] \right\}$$
$$F_n \leq \exp\left\{ -n\kappa(\alpha) \left[ S_\beta(R)_\psi - \widetilde{S}_\alpha(R|A)_\psi - 2q_{\mathrm{qss}} \right] \right\}$$
$$F_n \leq \exp\left\{ -n\kappa(\alpha) \left[ \widetilde{I}_\alpha(R; A)_\psi - 2q_{\mathrm{qss}} \right] \right\},$$

*where $\kappa(\alpha) = (1 - \alpha)/(2\alpha)$, and $q_{\mathrm{qss}} \equiv q_{\mathrm{qss}}(\rho^{\otimes n}, \Lambda_n)$ and $e_{\mathrm{qss}} \equiv e_{\mathrm{qss}}(\rho^{\otimes n}, \Lambda_n)$ denote the quantum communication cost and entanglement cost defined in (6.41), respectively.*

## 6.2. Measurement compression with QSI

### 6.2.1. Operational setting

Consider a bipartite state $\rho_{AB}$ between two parties (say, Alice and Bob), a POVM $\Lambda = \{\Lambda_x\}_{x \in \mathcal{X}}$ on the $A$ system, and a classical register $X$ corresponding to the outcomes of $\Lambda$. Suppose that Alice wants to communicate $X$ to Bob via classical communication. A simple solution is of course for Alice to apply the POVM $\Lambda$ and send the outcome to Bob, requiring $\log |X|$ bits of communication. In measurement compression with quantum side information (QSI), Alice and Bob want to reduce this communication cost by *simulating* the POVM $\Lambda$ using shared randomness, Bob's QSI $B$, and sending $\log |L|$ bits of classical communication with $|L| \leq |X|$. This information-theoretic task was introduced in [WHB+12] as an extension of Winter's original formulation of measurement compression [Win04]. In the following, we explain this protocol in more detail.

Given $\rho_{AB} \in \mathcal{D}(\mathcal{H}_{AB})$ and a POVM $\Lambda = \{\Lambda_x\}_{x \in \mathcal{X}}$ on the $A$ system with outcome $X$, a general protocol for measurement compression with QSI consists of the following steps (cf. Figure 6.8): Alice applies a quantum operation $\mathcal{E} \colon AM_A \to \bar{X}L$ to her shares of a purification $\psi_{RAB}$ of the initial state $\rho_{AB}$ (with $R$ being an inaccessible reference system) and the shared randomness $\chi_{M_A M_B}$. This produces a classical register $\bar{X}$ that holds her copy of the simulated outcome of the measurement, and a classical register $L$. She sends the latter to Bob, who then applies a quantum operation $\mathcal{D} \colon LBM_B \to \hat{X}B'$ to $L$ and his shares of $\psi_{RAB}$ and $\chi_{M_A M_B}$, producing a quantum output $B'$ and the classical output $\hat{X}$, which represents the simulated outcome of the measurement. We denote the overall state of the protocol after applying $\mathcal{E}$ and $\mathcal{D}$ by $\omega$ and $\sigma$, respectively.

We compare the final state $\sigma$ of the protocol outlined above to the ideal state $\varphi_{RXX'B}$ that would result from Alice applying the POVM $\Lambda$ to her system $A$ yielding the outcome $X$, and sending a copy $X' \cong X$ uncompressed to Bob. That is, for $\zeta_A \in \mathcal{D}(\mathcal{H}_A)$ we define the measurement channel $\mathcal{M}_\Lambda(\zeta_A) := \sum_{x \in \mathcal{X}} \mathrm{tr}(\Lambda_x \zeta_A) |x\rangle\langle x|_X \otimes |x\rangle\langle x|_{X'}$ associated to the POVM $\Lambda$ (cf. Section 2.2.2), and set

$$\varphi_{RXX'B} := (\mathrm{id}_{RB} \otimes \mathcal{M}_\Lambda)(\psi_{RAB}). \tag{6.42}$$

The aim of the measurement compression protocol is to assure that $\sigma$ is close in fidelity to $\varphi$.

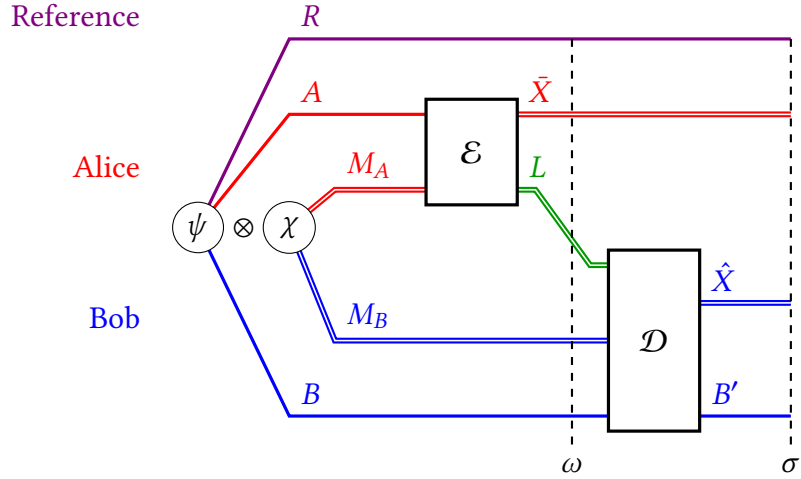

Figure 6.8: Schematic description of a measurement compression protocol with QSI between Alice (red), Bob (blue), and a reference system (violet), with classical communication (green) from Alice to Bob. Double lines indicate classical systems. For a detailed description of the protocol, see Section 6.2.1.

Hence, the figure of merit of the protocol is the fidelity

$$F := F\left(\varphi_{RXX'B}, \sigma_{R\bar{X}\hat{X}B'}\right) = F\left(\varphi_{RXX'B}, (\mathrm{id}_R \otimes \mathcal{D} \circ \mathcal{E})(\psi_{RAB} \otimes \chi_{M_A M_B})\right).$$

Given $n$ identical copies of the input state $\rho_{AB}$ with purification $\psi_{RAB}$, and shared randomness $\chi_{M_A^n M_B^n}$ between Alice and Bob, we consider a measurement compression protocol $(\mathcal{E}_n, \mathcal{D}_n)$ for the POVM $\Lambda^{\otimes n}$, where $\mathcal{E}_n \colon A^n M_A^n \to \bar{X}^n L^n$ and $\mathcal{D}_n \colon L^n B^n M_B^n \to \hat{X}^n B'^n$ are the encoding and decoding maps, respectively, and $L^n$ is a classical register corresponding to the classical communication between Alice and Bob. The figure of merit is then given by the fidelity

$$F_n := F\left(\varphi_n, (\mathrm{id}_{R^n} \otimes \mathcal{D}_n \circ \mathcal{E}_n)\left(\psi_{RAB}^{\otimes n} \otimes \chi_{M_A^n M_B^n}\right)\right), \tag{6.43}$$

where the ideal state $\varphi_n$ is obtained by Alice applying the POVM $\Lambda^{\otimes n}$ to the $A^n$ part of $\rho^{\otimes n}$ yielding the outcome $X^n$, and sending a copy $X'^n \cong X^n$ uncompressed to Bob. Our discussion is centered on the following operational quantities:

- The *classical communication cost*, given by

$$c\left(\rho^{\otimes n}, \Lambda^{\otimes n}\right) := \frac{1}{n}\log|L^n|; \tag{6.44}$$

- the *randomness cost*, given by

$$r\left(\rho^{\otimes n}, \Lambda^{\otimes n}\right) := \frac{1}{n}\log|M_A^n|.$$

A rate pair $(c, r)$ with $c, r \geq 0$ is called achievable if there exists a sequence $\{(\mathcal{E}_n, \mathcal{D}_n)\}_{n \in \mathbb{N}}$ of measurement compression protocols such that

$$\limsup_{n \to \infty} c\left(\rho^{\otimes n}, \Lambda^{\otimes n}\right) = c \qquad \limsup_{n \to \infty} r\left(\rho^{\otimes n}, \Lambda^{\otimes n}\right) = r \qquad \liminf_{n \to \infty} F_n = 1.$$

Wilde et al. [WHB+12] proved that $(c, r)$ is achievable if and only if

$$c \geq I(X; R|B)_\varphi \tag{6.45a}$$

$$c + r \geq S(X|B)_\varphi, \tag{6.45b}$$

where $\varphi_{RXX'B}$ is the ideal state of the protocol defined in (6.42).

## 6.2.2. Strong converse theorem

In this section, we strengthen the weak converse result obtained from (6.45a) for the classical communication cost in measurement compression with QSI to a strong converse theorem. Similar to Section 6.1, we first derive the following 'one-shot' result:

**Lemma 6.2.1.** *Let $\rho_{AB}$ be a bipartite state with purification $\psi_{RAB}$, and let $\Lambda$ be a POVM on $A$. Furthermore, let $(\mathcal{E}, \mathcal{D})$ be a measurement compression protocol as defined in Section 6.2.1 with figure of merit*

$$F := F\left(\varphi_{RXX'B}, (\mathrm{id}_R \otimes \mathcal{D} \circ \mathcal{E})(\psi_{RAB} \otimes \chi_{M_A M_B})\right).$$

*We have the following bound on $F$ for $\alpha \in (1/2, 1)$ and $\beta \equiv \beta(\alpha) = \alpha/(2\alpha - 1)$:*

$$\log F \leq \frac{1-\alpha}{2\alpha}\left(\log|L| - \widetilde{S}_\beta(R|B)_\varphi + \widetilde{S}_\alpha(R|XB)_\varphi\right). \tag{6.46}$$

*Proof.* We define the states

$$\omega_{R\bar{X}LBM_B} := (\mathrm{id}_{RBM_B} \otimes \mathcal{E})(\psi_{RAB} \otimes \chi_{M_A M_B})$$

$$\sigma_{R\bar{X}\hat{X}B'} := (\mathrm{id}_R \otimes \mathcal{D} \circ \mathcal{E})(\psi_{RAB} \otimes \chi_{M_A M_B}).$$

To prove (6.46), consider the following bound for $\alpha \in (1/2, 1)$ and $\beta = \alpha/(2\alpha - 1)$:

$$\frac{2\alpha}{1-\alpha} \log F \leq \frac{2\alpha}{1-\alpha} \log F\left(\sigma_{R\hat{X}B'}, \varphi_{RXB}\right)$$
$$\leq \widetilde{S}_\alpha(R|XB)_\varphi - \widetilde{S}_\beta(R|\hat{X}B')_\sigma, \tag{6.47}$$

where the first inequality follows from the monotonicity of the fidelity under partial trace, and the second inequality follows from Theorem 3.3.6(ii).

We continue to bound the second term on the right-hand side of (6.47):

$$-\widetilde{S}_\beta(R|\hat{X}B')_\sigma \leq -\widetilde{S}_\beta(R|LBM_B)_\omega$$
$$\leq \log |L| - \widetilde{S}_\beta(R|BM_B)_\omega$$
$$= \log |L| - \widetilde{S}_\beta(R|BM_B)_{\psi \otimes \chi}$$
$$= \log |L| - \widetilde{S}_\beta(R|B)_\varphi. \tag{6.48}$$

The first inequality follows from data processing with respect to the quantum operation $\mathcal{D} \colon LM_BB \to \hat{X}B'$ (Proposition 3.3.2(v)), the second inequality follows from (3.26) in Proposition 3.3.7, and the first equality follows from the fact that $\omega_{RBM_B} = \psi_{RB} \otimes \chi_{M_B}$. In the last equality we used (3.16) in Proposition 3.3.5, and the fact that $\psi_{RB} = \varphi_{RB}$. Substituting (6.48) in (6.47) then yields the claim. □

This immediately implies the following strong converse theorem:

**Theorem 6.2.2** (Strong converse theorem for measurement compression with QSI).
*Let $\rho_{AB}$ be a bipartite state, $\Lambda$ a POVM on $A$, and $\{(\mathcal{E}_n, \mathcal{D}_n)\}_{n \in \mathbb{N}}$ be a sequence of measurement compression protocols as described in Section 6.2.1, with figure of merit $F_n$ as defined in (6.43). For all $n \in \mathbb{N}$ we have the following bound on $F_n$ for $\alpha \in (1/2, 1)$ and $\beta = \alpha/(2\alpha - 1)$:*

$$F_n \leq \exp\left\{-n\kappa(\alpha)\left[\widetilde{S}_\beta(R|B)_\varphi - \widetilde{S}_\alpha(R|XB)_\varphi - c\right]\right\},$$

*where $\kappa(\alpha) = (1-\alpha)/(2\alpha)$, and $c \equiv c(\rho^{\otimes n}, \Lambda^{\otimes n})$ is the classical communication cost defined in* (6.44).

The achievable rate region in the $(c, r)$-plane is determined by the two boundaries $c \geq I(X; R|B)_\varphi$ and $c + r \geq S(X|B)_\varphi$, as stated in (6.45) (compare this to the similar situation for state redistribution discussed in Section 6.1). Theorem 6.2.2 only proves the strong converse property for the $c$-boundary of the achievable rate region, and it remains open to prove the strong converse property also for the $(c + r)$-boundary as stated in (6.45b). While the proof of (6.46) in Lemma 6.2.1 closely follows that of (6.11) in Lemma 6.1.1, it seems that the proof method of (6.10) does not immediately carry over to show the desired bound for $c + r$ in measurement compression. In other words, it is unclear whether a bound of the form

$$\log F \overset{?}{\leq} f(\alpha)(\log |L| + \log |M_A| - S_{\beta(\alpha)}(XB)_\varphi + S_\alpha(B)_\varphi)$$

holds for some functions $f(\alpha)$ and $\beta(\alpha)$ satisfying $f(\alpha) > 0$ for all $\alpha$ in some open interval whose boundary contains 1, and $\lim_{\alpha \to 1} \beta(\alpha) = 1$.

# 7. Conclusion

## 7.1. Summary of the main results

This thesis investigated mathematical properties of the sandwiched Rényi divergence (SRD) and the information spectrum relative entropy, which were then used to further refine the optimal rates of quantum information-processing tasks in the following way: firstly, we derived the second order asymptotics for quantum source coding using a general mixed source, obtaining the case of a memoryless source as a corollary. While the second order asymptotic expansion of the optimal compression length provides a useful and easily computable approximation for finite blocklengths, it also determines the rate of convergence to the optimal rate (equal to the von Neumann entropy in the case of a memoryless source). Secondly, we proved strong converse theorems for state redistribution (with and without feedback) and measurement compression with quantum side information. These theorems establish the optimal rates of these tasks as sharp thresholds: any code at a rate beyond the optimal rate fails with certainty in the asymptotic limit.

In Chapter 3, we discussed the $\alpha$-SRD $\widetilde{D}_\alpha(\cdot\|\cdot)$. We first investigated the limit $\alpha \to 0$, and found that the quantity $\lim_{\alpha\to 0} \widetilde{D}_\alpha(\rho\|\sigma)$ is equal to the well-known 0-relative Rényi entropy $D_0(\rho\|\sigma)$ only if the supports of $\rho$ and $\sigma$ are equal (Theorem 3.2.7). We then considered entropic quantities derived from the SRD, and extended the 'Rényi entropic calculus' by proving various properties for the Rényi conditional entropy and the Rényi mutual information (Proposition 3.3.5, Theorem 3.3.6, and Proposition 3.3.7). Most notably, we proved bounds on the fidelity of two quantum states in terms of Rényi entropic quantities, which were crucial in the proofs of the strong converse theorems in Chapter 6. Finally, we discussed Frank and Lieb's proof of the data processing inequality for $\widetilde{D}_\alpha(\cdot\|\cdot)$ for the range $\alpha \geq 1/2$ [FL13], and derived an algebraic necessary and sufficient condition for equality (Theorem 3.4.1). We also gave applications of this result by proving equality conditions for certain entropic bounds (Theorem 3.4.9, Theorem 3.4.13, and Proposition 3.4.15).

Chapter 4 reviewed the derivation of the second order asymptotic expansion of the information spectrum relative entropy $D_s^\varepsilon(\cdot \| \cdot)$ by Tomamichel and Hayashi [TH13], which is based on the Berry-Esseen Theorem. We also gave a direct proof for the second order asymptotic expansion of $D_s^\varepsilon(\cdot \| \cdot)$ in the special case where the second operator is the identity (Corollary 4.2.8).

In the information-theoretic part of this thesis, we used the mathematical results from Chapters 3 and 4 in the analysis of optimal rates of quantum information-theoretic tasks. In Chapter 5, we derived the second order asymptotic rate of quantum source coding using a mixed source (Theorem 5.2.8). For the achievability part of this result, we constructed universal source codes for memoryless quantum sources achieving any given second order rate if the first order rate is equal to the von Neumann entropy of the source (Proposition 5.2.3). We also specialized the main result to a finite mixture of memoryless sources (Corollary 5.2.9) and the case of two memoryless sources, where an exhaustive description is possible (Theorem 5.2.10). Furthermore, in Corollary 5.2.11 we recovered the second order asymptotics of visible quantum source coding using a single memoryless source, which were proved in [DL15].

In Chapter 6, we proved strong converse theorems for both the quantum communication cost and the entanglement cost of state redistribution (Theorem 6.1.2), covering also the feedback case where quantum back-communication from Bob to Alice as well as multiple communication rounds are allowed (Theorem 6.1.4). We showed how the tasks of blind quantum source coding, coherent state merging, and quantum state splitting can be regarded as special cases of the state redistribution protocol, hence rederiving known strong converse theorems for these tasks as well (Corollary 6.1.5, Corollary 6.1.6, and Corollary 6.1.7). We also proved a strong converse theorem for the classical communication cost in measurement compression with quantum side information (Theorem 6.2.2).

## 7.2. Open problems and future research directions

In the following, we collect the open problems that were mentioned in conjunction with the results obtained in this thesis. We also formulate some more general open problems.

### 7.2.1. Sufficiency and the sandwiched Rényi divergence

We derived an algebraic necessary and sufficient condition for equality in the DPI for the $\alpha$-SRD in the full range $\alpha \in [1/2, 1) \cup (1, \infty)$. Jenčová [Jen16] proved that for $\alpha > 1$ equality in the DPI for the $\alpha$-SRD is equivalent to sufficiency of the quantum operation $\Phi$ for the states involved, with the recovery map given by the well-known Petz recovery map. Furthermore, Hiai and Mosonyi [HM16] proved similar results for the range $\alpha \in (1/2, 1)$ under more special circumstances (assuming that the CPTP map $\Phi$ is also unital, and that one of the states is a fixed point of $\Phi$). It would be interesting to have a general sufficiency result for the case of equality in the DPI for the $\alpha$-SRD in the range $\alpha \in (1/2, 1)$. Note that for $\alpha = 1/2$ and $\alpha = \infty$, it is known that such a sufficiency result cannot hold (see Section 3.4.3).

## 7.2.2. Second order asymptotics of tasks with memory

Quantum source coding using a mixed source is a simple toy model for an information-theoretic task employing a resource with memory, and our derivation of its second order asymptotics (Theorem 5.2.8) marks the first such result beyond the memoryless regime. As the assumption of a memoryless resource might not be justified in realistic scenarios, it is certainly desirable to investigate other information-theoretic tasks that employ resources with memory.

A natural candidate for such an investigation is mixed c-q channel coding. In this task, we assume that Alice and Bob are connected by a channel $W^{(n)}$ given as a general mixture of memoryless c-q channels. That is, the channel is of the form $W^{(n)} = \int_\Lambda d\mu(\lambda) W_\lambda^{\otimes n}$, where $\Lambda$ is a parameter space with a normalized measure $\mu$, and for each $\lambda \in \Lambda$ the channel $W_\lambda$ is c-q. In the classical setting, the second order asymptotics of mixed channel coding were investigated by Polyanskiy et al. [PPV11], Tomamichel and Tan [TT15], and Yagi et al. [YHN16]. One would seek to find a quantum generalization of these results in the same way as our Theorem 5.2.8 generalizes the results by Nomura and Han [NH13] for classical mixed source coding.

In analogy to our discussion in Chapter 5, a partial result towards determining the second order asymptotics of mixed c-q channel coding might be the construction of universal codes achieving second order rates for memoryless c-q channels. Universal codes for memoryless c-q channels achieving first order rates were found by Hayashi [Hay09] and Bjelaković and Boche [BB09]. The latter paper derived them to discuss channel coding using compound and averaged quantum channels, tasks that are closely related to mixed channel coding.

At this point, we also mention once more the problem of determining the exact second order asymptotics of quantum source coding using a memoryless source in the blind encoding setting. As discussed in Section 5.2.4, the known second order asymptotic bounds on the compression length derived in [DL15] do not match in the second order coefficient, and closing this gap remains an open problem.

## 7.2.3. Strong converse exponents of state redistribution

Our main result in Chapter 6 is a strong converse for state redistribution, which follows from exponential bounds on the fidelity $F_n$ (defined in (6.6)) between final and target state of the protocol in terms of the operational quantities and Rényi generalizations of the optimal rates. More precisely, in Theorem 6.1.2 we proved the following bounds for a tripartite state $\rho_{ABC}$ with purification $|\psi\rangle_{ABCR}$, $\alpha \in (1/2, 1)$, $\beta = \alpha/(2\alpha - 1)$, and $n \in \mathbb{N}$:

$$F_n \leq \exp\left\{-n\kappa(\alpha)\left[S_\beta(AB)_\rho - S_\alpha(B)_\rho - (q + e)\right]\right\} \tag{7.1}$$

$$F_n \leq \exp\left\{-n\kappa(\alpha)\left[\widetilde{S}_\beta(R|B)_\psi - \widetilde{S}_\alpha(R|AB)_\psi - 2q\right]\right\} \tag{7.2}$$

$$F_n \leq \exp\left\{-n\kappa(\alpha)\left[\tilde{I}_\alpha(R;AB)_\psi - \tilde{I}_\beta(R;B)_\psi - 2q\right]\right\}, \tag{7.3}$$

where $\kappa(\alpha) = (1 - \alpha)/(2\alpha)$, and $q$ and $e$ denote the quantum communication cost and entanglement cost, respectively. The Rényi quantities appearing in these bounds are Rényi generalizations of the optimal rate. For example, the Rényi quantity in (7.1) converges to the optimal rate $S(A|B)_\rho$ for $q + e$, that is, $\lim_{\alpha \to 1} S_\beta(AB)_\rho - S_\alpha(B)_\rho = S(A|B)_\rho$. It would be interesting to determine the exact exponent with which the fidelity decays to 0 in the converse regime, which is known as the *strong converse exponent*. This would amount to proving exponential lower bounds on $F_n$ in terms of the same quantities as in (7.1), (7.2), and (7.3).

Here, we also specifically mention quantum state splitting, for which (7.3) reduces to

$$F_n \leq \exp\left\{-n\kappa(\alpha)\left[\tilde{I}_\alpha(R;A)_\psi - 2q_{\text{qss}}\right]\right\},$$

with $q_{\text{qss}}$ denoting the quantum communication cost. Establishing $\tilde{I}_\alpha(R;A)_\psi$ as a strong converse exponent for the quantum communication cost in quantum state splitting would constitute a new operational interpretation of the Rényi mutual information. Hayashi and Tomamichel [HT16] established the Rényi mutual information as the strong converse exponent in binary hypothesis testing with a composite alternative hypothesis.

### 7.2.4. More strong converses using the Rényi entropy method

The crucial ingredients in the proofs of the strong converse theorems of Chapter 6 are the bounds on the fidelity of two quantum states in terms of Rényi entropic quantities that we derived in Theorem 3.3.6. These bounds could potentially give way to strong converse theorems for other quantum information-processing tasks. A prominent open problem in quantum information theory is to prove the strong converse property for the quantum capacity of degradable channels. In the following paragraphs, we introduce the reader to this problem and show how Theorem 3.3.6 may be applied.

We consider the task of *entanglement generation*: Alice (the sender) and Bob (the receiver) are connected by a quantum channel $\mathcal{N} : A \to B$, and their goal is to generate an MES between them as follows. First, Alice locally prepares a pure state $|\psi\rangle_{RA_1...A_n}$, where $A_i \cong A$ for $i = 1, \ldots, n$. She sends the $A_i$-parts through the channel $\mathcal{N}^{\otimes n}$ (corresponding to $n$ uses of the channel $\mathcal{N}$) to Bob. Upon receiving the channel output, Bob applies a decoding quantum operation $\mathcal{D}_n : B_1 \ldots B_n \to A_1 \ldots A_n$, where $B_i \cong B$ for $i = 1, \ldots, n$, to obtain the final state $\sigma_n := (\mathrm{id}_R \otimes \mathcal{D}_n \circ \mathcal{N}^{\otimes n})(\psi_{RA_1...A_n})$. The goal of the protocol is that $\sigma_n$ be close in fidelity to an MES $\Phi_{RA^n}^{k_n}$ of Schmidt rank $k_n$, where $A^n \equiv A_1 \ldots A_n$. More precisely, defining

$$F_n := F\left(\Phi_{RA^n}^{k_n}, \sigma_n\right),$$ (7.4)

we call a rate $R \geq 0$ *achievable* if there is a sequence of entanglement generation protocols with

$$\liminf_{n \to \infty} \frac{\log k_n}{n} \geq R \quad \text{and} \quad \lim_{n \to \infty} F_n = 1.$$

The *quantum capacity* $Q(\mathcal{N})$ of $\mathcal{N}$ is defined as the supremum over all achievable rates.

According to the Lloyd-Shor-Devetak Theorem [Llo97; Sho02; Dev05], the quantum capacity is given by the regularized formula

$$Q(\mathcal{N}) = \lim_{n \to \infty} \frac{1}{n} Q^{(1)}(\mathcal{N}^{\otimes n}),$$ (7.5)

where $Q^{(1)}(\mathcal{N}) := \max_{\psi_{RA}} I^c(R\rangle B)_{(\mathrm{id}_R \otimes \mathcal{N})(\psi)}$ is the channel coherent information with the maximization over pure states $|\psi\rangle_{RA}$, and $I^c(A\rangle B)_\rho := S(B)_\rho - S(AB)_\rho$ is the coherent information of a bipartite state $\rho_{AB}$. There are examples of channels for which the channel coherent information is super-additive [DSS98], and hence, the regularization in (7.5) is indeed necessary. This renders the quantum capacity effectively uncomputable in most cases. However, for certain quantum channels such as *degradable* quantum channels [DS05], the quantity $Q^{(1)}(\cdot)$ is in fact additive.

To define degradability, we need to introduce the concept of a complementary channel: for a quantum channel $\mathcal{N}\colon A \to B$ let $V\colon \mathcal{H}_A \to \mathcal{H}_{BE}$ be the Stinespring isometry from Theorem 2.2.2, i.e., $\mathcal{N}(\cdot) = \mathrm{tr}_E(V \cdot V^\dagger)$. The complementary channel $\mathcal{N}^c\colon A \to E$ is defined as $\mathcal{N}^c(\cdot) = \mathrm{tr}_B(V \cdot V^\dagger)$. A channel $\mathcal{N}$ is degradable if there is another quantum channel $\mathcal{M}\colon B \to E$ such that $\mathcal{N}^c = \mathcal{M} \circ \mathcal{N}$. For a degradable channel $\mathcal{N}$ we have $Q^{(1)}(\mathcal{N}^{\otimes n}) = nQ^{(1)}(\mathcal{N})$ [DS05], and hence $Q(\mathcal{N}) = Q^{(1)}(\mathcal{N})$ by (7.5).

Inspired by the quest for a strong converse theorem for the (single-letter) quantum capacity of degradable channels, let us apply Theorem 3.3.6(ii) to entanglement generation as described above. To this end, consider a sequence $\{(\psi_{RA^n}, \mathcal{D}_n)\}_{n\in\mathbb{N}}$ of entanglement generation protocols, and let the fidelity $F_n$ be defined as in (7.4). For $\alpha \in (1/2, 1)$ and $\beta = \alpha/(2\alpha - 1)$, we have the following by Theorem 3.3.6(ii):

$$\frac{2\alpha}{1-\alpha} \log F_n \leq \widetilde{S}_\alpha(R|A^n)_{\Phi^{k_n}} - \widetilde{S}_\beta(R|A^n)_{\sigma_n} \tag{7.6}$$

$$\leq -\log k_n - \widetilde{S}_\beta(R|B^n)_{(\mathrm{id}_R \otimes \mathcal{N}^{\otimes n})(\psi_{RA^n})}$$

$$\leq -\log k_n + \max_{\phi_{RA^n}}\left\{-\widetilde{S}_\beta(R|B^n)_{(\mathrm{id}_R \otimes \mathcal{N}^{\otimes n})(\phi_{RA^n})}\right\}$$

$$= -\log k_n + \widetilde{Q}_\beta^{(1)}(\mathcal{N}^{\otimes n}). \tag{7.7}$$

In the second line, we used data processing for the Rényi conditional entropy (Proposition 3.3.2(v)), and the fact that the Rényi conditional entropy is equal to $-\log k_n$ for an MES of Schmidt rank $k_n$. In the third line, we optimized over all pure states $\phi_{RA^n}$, whose $A^n$ part is the input to $\mathcal{N}^{\otimes n}$, and in the fourth line we defined a Rényi generalization of the channel coherent information of a quantum channel $\mathcal{N}$ for $\alpha > 0$ as

$$\widetilde{Q}_\alpha^{(1)}(\mathcal{N}) := \max_{\phi_{RA}}\left\{-\widetilde{S}_\alpha(R|B)_{(\mathrm{id}_R \otimes \mathcal{N})(\phi_{RA})}\right\}, \tag{7.8}$$

where the maximization is over pure states $\phi_{RA}$.

The bound (7.7) was obtained by Wilde and Winter [WW14] using a different proof method.[1] There, it was pointed out that (7.7) would imply a strong converse for entanglement generation,

---

[1]The explicit form of (7.7) in [WW14] is stated in terms of a different Rényi generalization of the channel coherent information, which is obtained by replacing the Rényi conditional entropy derived from the $\alpha$-SRD $\widetilde{D}_\alpha(\cdot\|\cdot)$ in (7.8) with the one derived from the $\alpha$-RRE $D_\alpha(\cdot\|\cdot)$. However, as pointed out by the authors of [WW14], their proof method also applies when using the $\alpha$-SRD, which then leads to (7.7).

if one could show the following subadditivity property for $\beta > 1$:

$$\widetilde{Q}_\beta^{(1)}(\mathcal{N}^{\otimes n}) \overset{?}{\leq} n\widetilde{Q}_\beta^{(1)}(\mathcal{N}) + o(n). \tag{7.9}$$

If (7.9) were indeed true, then substituting it in (7.7) would yield the bound

$$\frac{2\alpha}{1-\alpha}\log F_n \overset{?}{\leq} -n\left(\frac{\log k_n}{n} - \widetilde{Q}_\beta^{(1)}(\mathcal{N}) + f(n)\right), \tag{7.10}$$

where $\lim_{n\to\infty} f(n) = 0$. The strong converse for the quantum capacity of degradable quantum channels would then follow from (7.10) by the same arguments as those used after Theorem 6.1.2 in Section 6.1.2.

A proof of (7.9) for all degradable channels, and more generally a proof of the strong converse for the quantum capacity of degradable quantum channels, has remained elusive so far. The best known result in this direction is the "pretty strong converse" derived by Morgan and Winter [MW14]. Using the smooth entropy framework, they proved that for all degradable quantum channels and entanglement generation codes with rates above capacity, the fidelity (7.4) makes a discontinuous jump from 1 to at most $1/\sqrt{2}$ in the asymptotic limit. It would be interesting to find an alternative proof of this result using the Rényi entropy method, starting from (7.6). To this end, a set of chain rules for the Rényi conditional entropy derived by Dupuis [Dup15] might prove useful.

Finally, we note that for measurement compression with QSI, Theorem 6.2.2 only provides a partial strong converse, since it only proves the strong converse property of the optimal classical communication cost. It remains open to also establish the strong converse property for the optimal shared randomness cost. Interestingly, to the best of our knowledge this cannot be achieved with the same proof method as the one used for Theorem 6.2.2.

# A. Two variants of the information spectrum relative entropy

In this appendix, we introduce two variants of the information spectrum relative entropy $D_s^\varepsilon(\cdot \| \cdot)$ that we defined and discussed in Chapter 4. The main advantage of these variants over $D_s^\varepsilon(\cdot \| \cdot)$ is that they satisfy the data processing inequality (see Proposition A.1.3(iii) below). Moreover, they also possess a second order asymptotic expansion (Proposition A.2.2). The content of this chapter is taken from [DL15], where these variants of the information spectrum relative entropy were used to derive second order asymptotics of various information-processing tasks (see Chapter 5 for an overview of these results).

## A.1. Definition and properties

**Definition A.1.1.** For $\rho \in \mathcal{D}(\mathcal{H})$, $\sigma \in \mathcal{P}(\mathcal{H})$ and $\varepsilon \in (0, 1)$, we define the following variants of the information spectrum relative entropy considered in Chapter 4:

$$\underline{D}_s^\varepsilon(\rho \| \sigma) := \sup \left\{ \gamma : \operatorname{tr} (\rho - 2^\gamma \sigma)_+ \geq 1 - \varepsilon \right\}$$
$$\overline{D}_s^\varepsilon(\rho \| \sigma) := \inf \left\{ \gamma : \operatorname{tr} (\rho - 2^\gamma \sigma)_+ \leq \varepsilon \right\}.$$

We first prove the following lemma stating that the supremum and infimum in Definition A.1.1 are attained, and moreover unique under a mild support condition.

**Lemma A.1.2.**

(i) *For $\gamma \in \mathbb{R}$ let $A(\gamma) \in \operatorname{Herm}(\mathcal{H})$ be a one-parameter family of Hermitian operators such that $\gamma \mapsto \|A(\gamma)\|$ is continuous. Then the function $\gamma \mapsto \operatorname{tr}(A(\gamma))_+$ is continuous.*

(ii) *For $\rho, \sigma \in \mathcal{P}(\mathcal{H})$, the function $\gamma \mapsto \operatorname{tr}(\rho - 2^\gamma \sigma)_+$ is continuous and monotonically decreasing, and strictly so if $\operatorname{supp} \rho \subseteq \operatorname{supp} \sigma$.*

(iii) *The supremum and infimum in Definition A.1.1 are attained, and unique if* supp $\rho \subseteq$ supp $\sigma$.

*Proof.* (i) First, we observe that the trace of the positive part of a Hermitian operator is the sum of its non-negative eigenvalues, the eigenvalues being the roots of the characteristic polynomial. Since the determinant is a continuous function with respect to the operator norm on $\mathcal{B}(\mathcal{H})$, the coefficients of the characteristic polynomial of $A(\gamma)$ continuously depend on $\gamma$, and by factorizing the polynomial into linear factors we see that this carries over to the roots. Composing the sum over the roots with the continuous maximum function $\max(.,0)$, we obtain that the function $\gamma \mapsto \text{tr}(A(\gamma))_+$ is a composition of continuous functions and thus continuous itself.

(ii) Since $\gamma \mapsto \rho - 2^\gamma \sigma$ is continuous with respect to the operator norm, the function $\gamma \mapsto \text{tr}(\rho - 2^\gamma \sigma)_+$ is continuous by (i). To prove that $\gamma \mapsto \text{tr}(\rho - 2^\gamma \sigma)_+$ is decreasing, let $\gamma \leq \gamma'$ and observe that

$$\rho - 2^\gamma \sigma = \rho - 2^{\gamma'} \sigma + \left(2^{\gamma'} - 2^\gamma\right)\sigma,$$

where $\left(2^{\gamma'} - 2^\gamma\right)\sigma \geq 0$. Hence, Weyl's Monotonicity Theorem 2.3.6 implies that

$$\lambda_j\left(\rho - 2^\gamma \sigma\right) \geq \lambda_j\left(\rho - 2^{\gamma'}\sigma\right) \quad \text{for all } j, \tag{A.1}$$

where for a Hermitian operator $A$ we write $\lambda_j(A)$ to denote the $j$-th largest eigenvalue of $A$. It then follows from (A.1) that $\text{tr}\left(\rho - 2^\gamma \sigma\right)_+ \geq \text{tr}\left(\rho - 2^{\gamma'}\sigma\right)_+$.

To prove strict monotonicity, let $\gamma < \gamma'$. Without loss of generality, we restrict $\mathcal{H}$ to the support of $\sigma$ (which is possible due to the assumption supp $\rho \subseteq$ supp $\sigma$), such that $\sigma$ has strictly positive eigenvalues. Consequently, $\left(2^{\gamma'} - 2^\gamma\right)\sigma > 0$, and the inequality in (A.1) is a strict one for all $j$, from which we obtain that $\text{tr}\left(\rho - 2^\gamma \sigma\right)_+ > \text{tr}\left(\rho - 2^{\gamma'}\sigma\right)_+$.

(iii) Since $\lim_{\gamma \to -\infty} \text{tr}(\rho - 2^\gamma \sigma)_+ = \text{tr}(\rho) = 1$ and $\lim_{\gamma \to \infty} \text{tr}(\rho - 2^\gamma \sigma)_+ = 0$, we infer by (ii) and the Intermediate Value Theorem that the supremum in the definition of $\underline{D}_s^\varepsilon(\rho \| \sigma)$ as well as the infimum in the definition of $\overline{D}_s^\varepsilon(\rho \| \sigma)$ are attained. If supp $\rho \subseteq$ supp $\sigma$, then they are moreover unique by the strict monotonicity of $\gamma \mapsto \text{tr}(\rho - 2^\gamma \sigma)_+$. □

The information spectrum relative entropies satisfy the following properties:

**Proposition A.1.3.** *Let $\varepsilon \in (0, 1)$, $\rho \in \mathcal{D}(\mathcal{H})$ and $\sigma \in \mathcal{P}(\mathcal{H})$ with $\operatorname{supp} \rho \subseteq \operatorname{supp} \sigma$. Then the following properties hold:*

(i) $\underline{D}_s^\varepsilon(\rho \| \sigma) \leq D_s^\varepsilon(\rho \| \sigma)$

(ii) $\underline{D}_s^\varepsilon(\rho \| \sigma) = \overline{D}_s^{1-\varepsilon}(\rho \| \sigma)$

(iii) *Data processing inequality: For any quantum operation $\Lambda \colon \mathcal{B}(\mathcal{H}) \to \mathcal{B}(\mathcal{K})$, we have*

$$\underline{D}_s^\varepsilon(\rho \| \sigma) \geq \underline{D}_s^\varepsilon(\Lambda(\rho) \| \Lambda(\sigma)) \qquad \overline{D}_s^\varepsilon(\rho \| \sigma) \geq \overline{D}_s^\varepsilon(\Lambda(\rho) \| \Lambda(\sigma)).$$

(iv) *Monotonicity in $\varepsilon$: For $\varepsilon' \geq \varepsilon$, we have*

$$\underline{D}_s^\varepsilon(\rho \| \sigma) \leq \underline{D}_s^{\varepsilon'}(\rho \| \sigma) \qquad \overline{D}_s^\varepsilon(\rho \| \sigma) \geq \overline{D}_s^{\varepsilon'}(\rho \| \sigma).$$

(v) *For $\sigma' \geq 0$ with $\sigma \leq \sigma'$,*

$$\underline{D}_s^\varepsilon(\rho \| \sigma) \geq \underline{D}_s^\varepsilon(\rho \| \sigma') \qquad \overline{D}_s^\varepsilon(\rho \| \sigma) \geq \overline{D}_s^\varepsilon(\rho \| \sigma').$$

(vi) *For $c > 0$,*

$$\underline{D}_s^\varepsilon(\rho \| c\sigma) = \underline{D}_s^\varepsilon(\rho \| \sigma) - \log c \qquad \overline{D}_s^\varepsilon(\rho \| c\sigma) = \overline{D}_s^\varepsilon(\rho \| \sigma) - \log c.$$

(vii) *Let $\delta > 0$ with $\delta < \min\{\varepsilon, 1 - \varepsilon\}$, and let $\rho' \in \mathcal{D}(\mathcal{H})$ with $T(\rho, \rho') \leq \delta$. Then,*

$$\underline{D}_s^\varepsilon(\rho' \| \sigma) \leq \underline{D}_s^{\varepsilon+\delta}(\rho \| \sigma) \qquad \overline{D}_s^\varepsilon(\rho' \| \sigma) \leq \overline{D}_s^{\varepsilon-\delta}(\rho \| \sigma).$$

*Proof.* (i) Let $\gamma = \underline{D}_s^\varepsilon(\rho \| \sigma)$. Then by the definition of $\underline{D}_s^\varepsilon(\rho \| \sigma)$ we have

$$1 - \varepsilon = \operatorname{tr}\left((\rho - 2^\gamma \sigma)\{\rho > 2^\gamma \sigma\}\right) \leq \operatorname{tr}\left(\rho\{\rho > 2^\gamma \sigma\}\right).$$

Hence,

$$\operatorname{tr}(\rho\{\rho \leq 2^\gamma \sigma\}) = \operatorname{tr}\rho - \operatorname{tr}(\rho\{\rho > 2^\gamma \sigma\}) \leq \operatorname{tr}\rho - (1 - \varepsilon) = \varepsilon,$$

since $\operatorname{tr}\rho = 1$ by assumption. Therefore, $\gamma$ is feasible for $D_s^\varepsilon(\rho \| \sigma)$, and we have

$$D_s^\varepsilon(\rho \| \sigma) \geq \gamma = \underline{D}_s^\varepsilon(\rho \| \sigma),$$

which yields the claim.

(ii) Let $\gamma = \overline{D}_s^{1-\varepsilon}(\rho\|\sigma)$. Then by the definition of $\overline{D}_s^{1-\varepsilon}(\rho\|\sigma)$, we have

$$\operatorname{tr}(\rho - 2^\gamma \sigma)_+ = 1 - \varepsilon.$$

Hence, $\gamma$ is feasible for $\underline{D}_s^\varepsilon(\rho\|\sigma)$, and we obtain

$$\underline{D}_s^\varepsilon(\rho\|\sigma) \geq \gamma = \overline{D}_s^{1-\varepsilon}(\rho\|\sigma).$$

Assume now that

$$\gamma = \overline{D}_s^{1-\varepsilon}(\rho\|\sigma) = \underline{D}_s^\varepsilon(\rho\|\sigma) - \delta$$

holds for some $\delta > 0$, i.e., $\overline{D}_s^{1-\varepsilon}(\rho\|\sigma) < \underline{D}_s^\varepsilon(\rho\|\sigma)$. By the monotonicity of $\gamma \mapsto \operatorname{tr}(\rho - 2^\gamma \sigma)_+$, we have

$$\operatorname{tr}(\rho - 2^\gamma \sigma)_+ > 1 - \varepsilon.$$

On the other hand, $\operatorname{tr}(\rho - 2^\gamma \sigma)_+ = 1 - \varepsilon$ by definition of $\overline{D}_s^{1-\varepsilon}(\rho\|\sigma)$. This leads to a contradiction, yielding $\underline{D}_s^\varepsilon(\rho\|\sigma) = \overline{D}_s^{1-\varepsilon}(\rho\|\sigma)$.

By (ii), we only need to prove (iii)–(vii) for either $\underline{D}_s^\varepsilon(\cdot\|\cdot)$ or $\overline{D}_s^\varepsilon(\cdot\|\cdot)$, since the corresponding assertion for the other quantity then follows from $\underline{D}_s^\varepsilon(\cdot\|\cdot) = \overline{D}_s^{1-\varepsilon}(\cdot\|\cdot)$ (note however the change of direction in the inequality in (iv)). We choose $\underline{D}_s^\varepsilon(\cdot\|\cdot)$ in the following.

(iii) For $\gamma = \underline{D}_s^\varepsilon(\Lambda(\rho)\|\Lambda(\sigma))$, Lemma 2.2.11 implies that

$$\operatorname{tr}(\rho - 2^\gamma \sigma)_+ \geq \operatorname{tr}(\Lambda(\rho) - 2^\gamma \Lambda(\sigma))_+ = 1 - \varepsilon.$$

Hence, $\gamma$ is feasible for $\underline{D}_s^\varepsilon(\rho\|\sigma)$, and we obtain

$$\underline{D}_s^\varepsilon(\rho\|\sigma) \geq \gamma = \underline{D}_s^\varepsilon(\Lambda(\rho)\|\Lambda(\sigma)).$$

(iv) Let $\gamma = \underline{D}_s^\varepsilon(\rho\|\sigma)$. Then,

$$\operatorname{tr}(\rho - 2^\gamma \sigma)_+ = 1 - \varepsilon \geq 1 - \varepsilon'.$$

Hence, $\gamma$ is feasible for $\underline{D}_s^{\varepsilon'}(\rho\|\sigma)$, and consequently,

$$\underline{D}_s^{\varepsilon'}(\rho\|\sigma) \geq \gamma = \underline{D}_s^\varepsilon(\rho\|\sigma).$$

(v) For $\gamma = \underline{D}_s^\varepsilon(\rho\|\sigma')$ and $Q = \{\rho > 2^\gamma \sigma'\}$, we compute:

$$1 - \varepsilon = \operatorname{tr}\left(Q\left(\rho - 2^\gamma \sigma'\right)\right)$$
$$= \operatorname{tr} Q\rho - 2^\gamma \operatorname{tr} Q\sigma'$$
$$\leq \operatorname{tr} Q\rho - 2^\gamma \operatorname{tr} Q\sigma$$
$$\leq \operatorname{tr}(\rho - 2^\gamma \sigma)_+,$$

where the first inequality follows from $\sigma \leq \sigma'$ and the second inequality follows from Lemma 2.2.10. Hence, $\gamma$ is feasible for $\underline{D}_s^\varepsilon(\rho\|\sigma)$, and we obtain $\underline{D}_s^\varepsilon(\rho\|\sigma) \geq \underline{D}_s^\varepsilon(\rho\|\sigma')$.

(vi) For $\gamma = \underline{D}_s^\varepsilon(\rho\|c\sigma)$ with $c > 0$, we have

$$1 - \varepsilon = \operatorname{tr}(\rho - 2^\gamma c\sigma)_+ = \operatorname{tr}(\rho - 2^{\gamma + \log c}\sigma)_+,$$

and hence,

$$\underline{D}_s^\varepsilon(\rho\|\sigma) \geq \gamma + \log c = \underline{D}_s^\varepsilon(\rho\|c\sigma) + \log c.$$

Conversely, let $\gamma = \underline{D}_s^\varepsilon(\rho\|\sigma)$. Then

$$1 - \varepsilon = \operatorname{tr}(\rho - 2^\gamma \sigma)_+ = \operatorname{tr}(\rho - 2^{\gamma - \log c}c\sigma)_+,$$

and hence,

$$\underline{D}_s^\varepsilon(\rho\|c\sigma) \geq \gamma - \log c = \underline{D}_s^\varepsilon(\rho\|\sigma) - \log c,$$

which proves the claim.

(vii) Let $\gamma = \underline{D}_s^\varepsilon(\rho'\|\sigma)$ and $Q = \{\rho' \geq 2^\gamma \sigma\}$. Then

$$1 - \varepsilon = \operatorname{tr}((\rho' - 2^\gamma \sigma)Q)$$
$$= \operatorname{tr}((\rho' - \rho)Q) + \operatorname{tr}((\rho - 2^\gamma \sigma)Q)$$
$$\leq \operatorname{tr}(\rho' - \rho)_+ + \operatorname{tr}(\rho - 2^\gamma \sigma)_+$$
$$\leq \delta + \operatorname{tr}(\rho - 2^\gamma \sigma)_+,$$

where the first inequality follows from Lemma 2.2.10 and the second inequality follows from the fact that $\operatorname{tr}(\rho - \rho')_+ \leq T(\rho, \rho') \leq \delta$ by assumption, which proves the claim. $\qquad \square$

In the light of Proposition A.1.3(ii), one can in principle drop one of the two information spectrum relative entropies in Definition A.1.1. However, given their close relationship to the quantum spectral inf- and sup-divergence that we investigate in Section A.3, we keep both definitions.

## A.2. Second order asymptotic expansion

Similar to the variant $D_s^\varepsilon(\cdot\|\cdot)$ introduced in [TH13], the information spectrum relative entropies $\underline{D}_s^\varepsilon(\cdot\|\cdot)$ and $\overline{D}_s^\varepsilon(\cdot\|\cdot)$ have a second order asymptotic expansion, which makes them useful for deriving second order expansions of information-processing tasks (cf. [DL15]).

We first relate the information spectrum relative entropy $\underline{D}_s^\varepsilon(\cdot\|\cdot)$ to the hypothesis testing relative entropy $D_H^\varepsilon(\cdot\|\cdot)$ defined in (4.3) in Chapter 4. The following proposition and its proof are similar to the corresponding bounds between the information spectrum relative entropy $D_s^\varepsilon(\cdot\|\cdot)$ and $D_H^\varepsilon(\cdot\|\cdot)$ proved in [TH13].

**Proposition A.2.1.** *Let $\varepsilon \in (0, 1)$ and $\delta \in (0, \varepsilon)$, and $\rho, \sigma \in \mathcal{P}(\mathcal{H})$ with* tr $\rho \leq 1$. *Then,*

$$D_H^{\varepsilon-\delta}(\rho\|\sigma) + \log\delta \leq \underline{D}_s^\varepsilon(\rho\|\sigma) \leq D_H^\varepsilon(\rho\|\sigma).$$

*Proof.* To prove the upper bound, let $\gamma = \underline{D}_s^\varepsilon(\rho\|\sigma)$ and $Q := \{\rho \geq 2^\gamma \sigma\}$. Since

$$\mathrm{tr}(\rho Q) \geq \mathrm{tr}\left((\rho - 2^\gamma \sigma)\, Q\right) \geq 1 - \varepsilon$$

holds by the definition of $\underline{D}_s^\varepsilon(\rho\|\sigma)$, we see that $Q$ is feasible for $D_H^\varepsilon(\rho\|\sigma)$. Moreover,

$$\mathrm{tr}(\sigma Q) = \mathrm{tr}\left(\sigma\{\rho \geq 2^\gamma \sigma\}\right) \leq 2^{-\gamma}\,\mathrm{tr}\left(\rho\{\rho \geq 2^\gamma \sigma\}\right) \leq 2^{-\gamma}.$$

Hence, by the definition of $D_H^\varepsilon(\rho\|\sigma)$, we obtain

$$D_H^\varepsilon(\rho\|\sigma) \geq \gamma = \underline{D}_s^\varepsilon(\rho\|\sigma).$$

Conversely, let $0 \leq Q \leq \mathbb{1}$ be optimal for $D_H^\varepsilon(\rho\|\sigma)$, and set $\mu = \log\delta + D_H^\varepsilon(\rho\|\sigma)$. Then

$$\mathrm{tr}(\rho - 2^\mu\sigma)_+ \geq \mathrm{tr}\left(Q(\rho - 2^\mu\sigma)\right) = \mathrm{tr}(Q\rho) - 2^\mu\,\mathrm{tr}(Q\sigma) \geq 1 - (\varepsilon + \delta),$$

where the first inequality follows from Lemma 2.2.10. Hence, $\mu$ is feasible for $\underline{D}_s^{\varepsilon+\delta}(\rho\|\sigma)$ and

we obtain

$$\underline{D}_s^{\varepsilon+\delta}(\rho\|\sigma) \geq D_H^{\varepsilon}(\rho\|\sigma) + \log\delta,$$

which concludes the proof. □

Proposition A.2.1 together with the second order expansion of $D_H^{\varepsilon}(\rho^{\otimes n}\|\sigma^{\otimes n})$ in (4.5) (with $o(\sqrt{n})$ replaced by $O(\log n)$, as proved in both [Li14] and [TH13]) immediately yields second order expansions for the information spectrum relative entropies $\underline{D}_s^{\varepsilon}(\rho^{\otimes n}\|\sigma^{\otimes n})$ and $\overline{D}_s^{\varepsilon}(\rho^{\otimes n}\|\sigma^{\otimes n})$ (for the latter, we use once more Proposition A.1.3(ii))). We state these in Proposition A.2.2 below, which was the main tool in deriving the second order asymptotic results of [DL15].

**Proposition A.2.2.** *Let $\varepsilon \in (0,1)$, $\rho \in \mathcal{D}(\mathcal{H})$ and $\sigma \in \mathcal{P}(\mathcal{H})$ with supp $\rho \subseteq$ supp $\sigma$. Then,*

$$\underline{D}_s^{\varepsilon}\left(\rho^{\otimes n} \,\big\|\, \sigma^{\otimes n}\right) = nD(\rho\|\sigma) + \sqrt{nV(\rho\|\sigma)}\Phi^{-1}(\varepsilon) + O(\log n)$$

$$\overline{D}_s^{\varepsilon}\left(\rho^{\otimes n} \,\big\|\, \sigma^{\otimes n}\right) = nD(\rho\|\sigma) - \sqrt{nV(\rho\|\sigma)}\Phi^{-1}(\varepsilon) + O(\log n).$$

## A.3. Relation to information spectrum approach

Quantum Shannon theory is concerned with deriving optimal rates for information-processing tasks in the asymptotic, memoryless (or i.i.d.) setting. To push the analysis of these optimal rates beyond this rather restrictive (but nevertheless useful) setting, two different methods have received considerable interest.

On the one hand, tasks can be characterized in the 'one-shot' setting, where a resource (such as a channel or a source) is only used once, and therefore one completely avoids any assumptions on the structure of the resource when used multiple times. In the one-shot setting, the main quantities used to derive bounds on operational quantities are (smoothed versions of) the min- and max-entropies that we defined in Section 3.2.2. For this reason, the one-shot setting is sometimes also referred to as smooth entropy framework.

On the other hand, the information spectrum approach deals with arbitrary sequences of states, assuming no special structure of the particular resources used in the information-processing task. The main quantities considered in this approach are the spectral divergence rates, which we define below in Definition A.3.1. The information spectrum approach was introduced in classical information theory by Han and Verdú [HV93], and subsequently gener-alized to the quantum setting by Ogawa and Nagaoka [ON00], Hayashi and Nagaoka [HN03], Hayashi [Hay06a], Nagaoka and Hayashi [NH07], and Bowen and Datta [BD06a].

Datta and Renner [DR09] proved the equivalence of the information spectrum approach to the smooth entropy framework by showing that the spectral divergence rates can be obtained from certain limits of the smooth min- and max-entropies. We show in this section that, similarly, the spectral divergence rates can be recovered as limits of our information spectrum relative entropies defined in Definition A.1.1. Note that a similar result for the information spectrum relative entropy $D_s^\varepsilon(\cdot\|\cdot)$ from Chapter 4 was obtained by Tomamichel and Hayashi [TH13]. First, we define the main quantities of the information spectrum approach:

**Definition A.3.1** (Spectral divergence rates).
Let $\hat\rho = \{\rho_n\}_{n\in\mathbb{N}}$ be an arbitrary sequence of states with $\rho_n \in \mathcal{D}(\mathcal{H}^{\otimes n})$, and let $\hat\omega = \{\omega_n\}_{n\in\mathbb{N}}$ be an arbitrary sequence of positive operators with $\omega_n \in \mathcal{P}(\mathcal{H}^{\otimes n})$.

(i) The *quantum spectral inf-divergence rate* is defined as

$$\underline{D}(\hat\rho\|\hat\omega) := \sup\left\{\gamma\colon \liminf_{n\to\infty} \operatorname{tr}(\rho_n - 2^{n\gamma}\omega_n)_+ = 1\right\}.$$

(ii) The *quantum spectral sup-divergence rate* is defined as

$$\overline{D}(\hat\rho\|\hat\omega) := \inf\left\{\gamma\colon \limsup_{n\to\infty} \operatorname{tr}(\rho_n - 2^{n\gamma}\omega_n)_+ = 0\right\}.$$

The above quantities differ slightly from the spectral divergence rates originally considered in [HN03]. However, as proved in [BD06a], the quantities in Definition A.3.1 are in fact equal to the ones defined in [HN03].

The quantum spectral divergence rates $\underline{D}(\hat\rho\|\hat\omega)$ and $\overline{D}(\hat\rho\|\hat\omega)$ can be recovered from the information spectrum relative entropies $\underline{D}_s^\varepsilon(\rho\|\sigma)$ and $\overline{D}_s^\varepsilon(\rho\|\sigma)$, respectively:

**Proposition A.3.2.** *Let $\hat\rho = \{\rho_n\}_{n\in\mathbb{N}}$ be an arbitrary sequence of states with $\rho_n \in \mathcal{D}(\mathcal{H}^{\otimes n})$, and let $\hat\omega = \{\omega_n\}_{n\in\mathbb{N}}$ be an arbitrary sequence of positive operators with $\omega_n \in \mathcal{P}(\mathcal{H}^{\otimes n})$. Then the following relations hold:*

(i) $\displaystyle\lim_{\varepsilon\to 0} \liminf_{n\to\infty} \frac{1}{n} \underline{D}_s^\varepsilon(\rho_n\|\omega_n) = \underline{D}(\hat\rho\|\hat\omega)$

(ii) $\displaystyle\lim_{\varepsilon\to 0} \limsup_{n\to\infty} \frac{1}{n} \overline{D}_s^\varepsilon(\rho_n\|\omega_n) = \overline{D}(\hat\rho\|\hat\omega)$

*Proof.* (i) We first show

$$\lim_{\varepsilon \to 0} \liminf_{n \to \infty} \frac{1}{n} \underline{D}_s^\varepsilon(\rho_n \| \omega_n) \leq \underline{D}(\hat{\rho} \| \hat{\omega}). \tag{A.2}$$

To this end, let $\gamma_n = \underline{D}_s^\varepsilon(\rho_n \| \omega_n)$ for $n \in \mathbb{N}$, and set $c(\varepsilon) := \liminf_{n \to \infty} \frac{\gamma_n}{n}$. By Proposition A.1.3(iv) the function $c(\varepsilon)$ is monotonically increasing in $\varepsilon$, and hence the limit $c := \lim_{\varepsilon \to 0} c(\varepsilon)$ exists in $\mathbb{R} \cup \{-\infty\}$. Let us assume first that $|c| < \infty$. It then follows from the definition of the limit that for all $\eta > 0$ there exists an $\varepsilon_0$ such that $|c - c(\varepsilon)| \leq \eta$ holds for all $\varepsilon \leq \varepsilon_0$. Moreover, by definition of the limit inferior, for all $\delta > 0$ there exists an $N \in \mathbb{N}$ such that $\frac{\gamma_n}{n} > c(\varepsilon) - \delta$ for all $n > N$, or equivalently,

$$\gamma_n > n(c(\varepsilon) - \delta) \geq n(c - \eta - \delta)$$

for $\varepsilon \leq \varepsilon_0$. Hence, by definition of $\underline{D}_s^\varepsilon(\rho_n \| \omega_n)$, we have

$$\mathrm{tr}\left(\rho_n - 2^{n(c-\eta-\delta)} \omega_n\right)_+ \geq 1 - \varepsilon \tag{A.3}$$

for all $n > N$. Since (A.3) holds for arbitrarily small $\varepsilon \leq \varepsilon_0$,

$$\liminf_{n \to \infty} \mathrm{tr}\left(\rho_n - 2^{n(c-\eta-\delta)} \omega_n\right)_+ = 1,$$

which implies

$$\underline{D}(\hat{\rho} \| \hat{\omega}) \geq c - \eta - \delta$$

by the definition of $\underline{D}(\hat{\rho} \| \hat{\omega})$. As $\eta$ and $\delta$ were arbitrary, we obtain (A.2).

Conversely, let $\gamma = \underline{D}(\hat{\rho} \| \hat{\omega})$. By the definition of the limit inferior, for all $\varepsilon > 0$ there exists an $N \in \mathbb{N}$ such that for all $n > N$ we have

$$\mathrm{tr}\left(\rho_n - 2^{n\gamma} \omega_n\right)_+ \geq 1 - \varepsilon.$$

Hence, $n\gamma \leq \underline{D}_s^\varepsilon(\rho_n \| \omega_n)$ for all $n > N$ by the definition of the information spectrum relative entropy, and consequently,

$$\gamma \leq \lim_{\varepsilon \to 0} \liminf_{n \to \infty} \frac{1}{n} \underline{D}_s^\varepsilon(\rho_n \| \omega_n), \tag{A.4}$$

which yields the lower bound in the proposition.

Finally, in the case $c = -\infty$, the bound (A.4) shows that we also have $\gamma = \underline{D}(\hat{\rho}\|\hat{\omega}) = -\infty$, and hence the assertion of the proposition is trivially true.

(ii) To prove the upper bound, let $\gamma = \overline{D}(\hat{\rho}\|\hat{\omega})$. By definition of $\overline{D}(\hat{\rho}\|\hat{\omega})$ it holds that

$$\limsup_{n\to\infty} \operatorname{tr}(\rho_n - 2^{n\gamma}\omega_n)_+ = 0,$$

that is, for every $\varepsilon > 0$ there exists an $N \in \mathbb{N}$ such that for all $n > N$ we have

$$\operatorname{tr}(\rho_n - 2^{n\gamma}\omega_n)_+ < \varepsilon.$$

By definition of the information spectrum relative entropy, this implies that $\overline{D}_s^\varepsilon(\rho_n\|\sigma_n) \leq n\gamma$ for all $n > N$, and hence,

$$\lim_{\varepsilon\to 0}\limsup_{n\to\infty} \frac{1}{n}\overline{D}_s^\varepsilon(\rho_n\|\omega_n) \leq \gamma = \overline{D}(\hat{\rho}\|\hat{\omega}), \tag{A.5}$$

which proves the upper bound.

Conversely, let $\gamma_n = \overline{D}_s^\varepsilon(\rho_n\|\omega_n)$ and set $c(\varepsilon) \coloneqq \limsup_{n\to\infty} \frac{\gamma_n}{n}$ and $c \coloneqq \lim_{\varepsilon\to 0} c(\varepsilon)$, which exists in $\mathbb{R} \cup \{\infty\}$ due to Proposition A.1.3(iv). If $|c| < \infty$, then by definition of the limit for all $\eta > 0$ there exists an $\varepsilon_0$ such that $|c - c(\varepsilon)| \leq \eta$ holds for all $\varepsilon \leq \varepsilon_0$. Moreover, by the characterization of the limit superior, for all $\delta > 0$ there exists an $N$ such that $\frac{\gamma_n}{n} < c(\varepsilon) + \delta$ for all $n > N$, or equivalently,

$$\gamma_n < n(c(\varepsilon) + \delta) < n(c + \eta + \delta)$$

for all $\varepsilon \leq \varepsilon_0$. This implies

$$\operatorname{tr}(\rho_n - 2^{n(c+\eta+\delta)}\omega_n)_+ \leq \varepsilon$$

for all $n > N$, and hence, $\limsup_{n\to\infty} \operatorname{tr}(\rho_n - 2^{n(c+\eta+\delta)}\omega_n)_+ = 0$ as $\varepsilon$ is arbitrarily small. Therefore, $\overline{D}(\hat{\rho}\|\hat{\omega}) \leq c + \eta + \delta$, which yields the result, since $\delta$ and $\eta$ were arbitrary. If $c = \infty$, then (A.5) shows that also $\gamma = \overline{D}(\hat{\rho}\|\hat{\omega}) = \infty$, in which case the assertion of the proposition is trivially true. $\qquad\square$